\documentclass[12pt, onecolumn, twoside]{report} 
\usepackage{graphicx}
\usepackage[a4paper,top=4cm, bottom=3cm, left=3.2cm, right=3.2cm]{geometry}
\usepackage[compact]{titlesec}
\usepackage[ansinew]{inputenc}
\usepackage{amsmath}
\usepackage{subfig}
\usepackage{amsfonts}
\usepackage{lmodern}
\usepackage{amssymb}
\usepackage{tocvsec2}
\usepackage{bbold}
\usepackage{hhline}
\usepackage{appendix}
\usepackage{fancyhdr}
\usepackage{german}
\usepackage{cite}
\usepackage{comment}
\usepackage[english]{babel}
\usepackage{sectsty}
\usepackage{multirow}
\usepackage{slashed}
\usepackage[titles,subfigure]{tocloft}
\usepackage{ifpdf}
	\ifpdf
	\usepackage[bookmarksopen,bookmarksnumbered,bookmarksopenlevel=2]{hyperref}
	\fi
\usepackage{enumitem}
\usepackage{wrapfig}
\usepackage{float}
\usepackage[indent=10pt]{caption}
\usepackage{sidecap}
\usepackage{booktabs}
\usepackage[usenames,dvipsnames]{xcolor}
\hypersetup{
    unicode=false,          
    pdftoolbar=true,        
    pdfmenubar=true,        
    pdffitwindow=false,     
    pdfstartview={FitH},    
    pdftitle={My title},    
    pdfauthor={Author},     
    pdfsubject={Subject},   
    pdfcreator={Creator},   
    pdfproducer={Producer}, 
    pdfkeywords={keyword1} {key2} {key3}, 
    pdfnewwindow=true,      
    colorlinks=false,       
    linkcolor=Plum,         
    citecolor=BlueViolet,        
    filecolor=magenta,      
    urlcolor=BlueViolet           
}

\titlespacing{\section}{0pt}{14pt}{9pt}
\titlespacing{\subsection}{0pt}{9pt}{4pt}
\titleformat{\section} [hang] {\LARGE\mdseries}{\thesection\quad}{0pt}{}
\titleformat{\subsection}[hang]{\Large\mdseries}{\thesubsection\quad}{0pt}{}
\titleformat{\subsubsection}[hang]{\large\mdseries}{\thesubsubsection\quad}{0pt}{}
\author{Max Kerstan}

\newcommand{\be}{\begin{equation}}
\newcommand{\ee}{\end{equation}}

\newcommand{\ch}{\mathrm{ch}}

\newcommand{\R}{\mathrm{Re}}
\newcommand{\rT}{\mathrm{T}}
\newcommand{\I}{\mathrm{Im}}

\newcommand{\bZ}{\mathbb{Z}}
\newcommand{\bR}{\mathbb{R}}
\newcommand{\bC}{\mathbb{C}}
\newcommand{\bP}{\mathbb{P}}

\newcommand{\cD}{\mathcal{D}}
\newcommand{\cN}{\mathcal{N}}
\newcommand{\cB}{\mathcal{B}}
\newcommand{\cW}{\mathcal{W}}
\newcommand{\cF}{\mathcal{F}}

\newcommand{\cO}{\mathcal{O}}
\newcommand{\cC}{\mathcal{C}}
\newcommand{\cV}{\mathcal{V}}
\newcommand{\cK}{\mathcal{K}}
\newcommand{\cT}{\mathcal{T}}
\newcommand{\cM}{\mathcal{M}}
\newcommand{\cH}{\mathcal{H}}

\newcommand{\tw}{\text{w}}
\newcommand{\tv}{\text{v}}

\newcommand{\ov}[1]{\overline{#1}}

\begin{document}

\vspace{3cm}
\begin{center}        
  {\LARGE   Abelian gauge symmetries and fluxed instantons \\[0.3cm]
in compactifications of type IIB and F-theory }
\end{center}

\vspace{0.75cm}
\begin{center}        
 Max Kerstan
\end{center}

\vspace{0.15cm}
\begin{center}
	
  \emph{Institut f\"ur Theoretische Physik, Ruprecht-Karls-Universit\"at Heidelberg, \\
             Philosophenweg 19, 69120 Heidelberg, Germany}
             \\[0.15cm]

\end{center}

\vspace{2cm}

\begin{center} \textbf{Abstract} \end{center}
We discuss the role of Abelian gauge symmetries in type IIB orientifold compactifications and their F-theory uplift. Particular emphasis is placed on $U(1)$s which become massive through the geometric St\"uckelberg mechanism in type IIB. We present a proposal on how to take such geometrically massive $U(1)$s and the associated fluxes into account in the Kaluza-Klein reduction of F-theory with the help of non-harmonic forms. Evidence for this proposal is obtained by working out the F-theory effective action including such non-harmonic forms and matching the results with the known type IIB expressions. We furthermore discuss how world-volume fluxes on $D3$-brane instantons affect the instanton charge with respect to $U(1)$ gauge symmetries and the chiral zero mode spectrum. The classical partition function of $M5$-instantons in F-theory is discussed and compared with the type IIB results for $D3$-brane instantons. The type IIB match allows us to determine the correct $M5$ partition function. Selection rules for the absence of chiral charged zero modes on $M5$-instantons in backgrounds with $G_4$ flux are discussed and compared with the type IIB results. The dimensional reduction of the democratic formulation of M-theory is presented in the appendix. This article is based on the author's PhD thesis but includes minor modifications.

\tableofcontents
\newpage

\pagestyle{fancy}
\fancypagestyle{plain}{\fancyhf{}\renewcommand{\headrulewidth}{0pt}\fancyfoot[C]{\thepage}}
\renewcommand{\headrulewidth}{0.4pt}
\renewcommand{\chaptermark}[1]{\markboth{\textsc{{\normalsize{Chapter \thechapter:\ #1}}}}{}}
\renewcommand{\sectionmark}[1]{\markright{\textsc{\normalsize\thesection\ #1}}{}}
\fancyhf{}
\fancyhead[LE,RO]{\thepage}
\fancyhead[RE]{\textit{\nouppercase{\leftmark}}}
\fancyhead[LO]{\textit{\nouppercase{\rightmark}}}

\clearpage

\chapter{Introduction and summary}
String theory has been amongst the most-studied subjects in fundamental theoretical physics for over four decades. The original interest in studying a theory based on quantised strings was sparked mainly by two key observations. One is the fact that the extended nature of the string worldsheet allows for the formulation of a consistent scattering theory that avoids the UV divergences ubiquitous in quantum field theory. The second crucial point is the observation that the spectrum of massless states obtained by quantising the string oscillations always includes states that can be understood as belonging to a symmetric 2-tensor, the graviton~\cite{Green:87, Polchinski:1998, Becker:2007zj}. This tantalisingly suggests that string theory has the power to reconcile the subjects of quantum mechanics and general relativity, and to this day string theory remains arguably the most promising candidate to describe such a theory of quantum gravity.

Over the years, a large number of significant discoveries have transformed our understanding of string theory. In fact, starting with the basic idea of a quantised string a number of quite different approaches have been developed to construct ever more phenomenologically viable theories. These approaches have led to several distinct string theories which on the surface appear to be quite different from one another, making string theory as a whole an extremely broad subject. More recently, evidence of striking duality relations between the different string theories has been found. These discoveries not only pull the subject area closer together again, but also intriguingly hint at the presence of an as yet undiscovered fundamental theory unifying the different manifestations of string theory~\cite{Becker:2007zj}.

While the first investigations into string theory focused on a purely bosonic formulation, it is clear that fermionic degrees of freedom must also be included if string theory is to have any chance of describing our visible universe. Various constructions to achieve this were developed in the 1970s and 1980s. These constructions differ in the field content and quantum spectrum of the string worldsheet theory and lead to 5 different branches of string theory. Two of these branches are the so-called heterotic string theories with gauge group $SO(32)$ and $E_8\times E_8$, while the remaining three are known as type I, type IIA and type IIB string theory. All of these theories exhibit supersymmetry relations between the bosonic and fermionic parts of the spectrum\footnote{In fact, the discovery of supersymmetry is closely tied to string theory because string worldsheet actions involving fermions were amongst the first systems in which supersymmetry was identified and consistently implemented~\cite{Schwarz:2000dm}. Of course, it was subsequently realised that supersymmetric quantum field theories can be studied without making recourse to string theory.}, and are thus collectively known as superstring theories. 

The first string theories to attract widespread attention in the physics community were the heterotic and type I string theories. This interest was due to the fact that these theories naturally include gauge symmetries and thus offer a chance to not only describe a viable theory of quantum gravity, but to combine gravity with the gauge theories of the Standard Model of particle physics in one unified theoretical picture~\cite{Green:87, Polchinski:1998, Becker:2007zj}. In the heterotic formulation, gauge symmetries arise directly from symmetries of the worldsheet field content of oriented closed strings~\cite{Gross:1984dd, Gross:1985fr, Gross:1985rr}. In type I and type II theories on the other hand, gauge symmetries appear when open strings are taken into account\footnote{Open strings are in general not possible in the heterotic setting, see however~\cite{Polchinski:2005bg}.}. When quantising open strings it is necessary to specify suitable boundary conditions, which can be of Dirichlet or von Neumann type. Allowing for Dirichlet boundary conditions clearly breaks spacetime Poincar\'e invariance and effectively introduces surfaces known as Dirichlet-branes or D-branes on which the open strings end~\cite{Polchinski:1998, Becker:2007zj}. To avoid breaking Poincar\'e invariance, string theorists initially focused on open strings with von Neumann boundary conditions only, or in other words on situations in which the D-branes fill spacetime completely. In 1984, Green and Schwarz~\cite{Green:1984sg} made the crucial discovery that anomaly cancellation, which is required for a consistent theory, occurs in the context of type I string theory if and only if the open string gauge group is $SO(32) \cong Spin(32)/\bZ_2$. Together with the construction of the first heterotic string theories, which took place around the same time, this discovery sparked the so-called first superstring revolution and initiated a period of intense study of type I and heterotic string theory.

A striking feature of all the superstring theories mentioned above is that their mathematical consistency requires the background spacetime to be 10-dimensional ~\cite{Green:87, Polchinski:1998, Becker:2007zj}. To reconcile this with the fact that our universe appears to be 4-dimensional at experimentally accessible energy scales, it is necessary to consider a process known as compactification. Essentially, this amounts to assuming that the background spacetime $M_{10}$ in which the strings propagate is given by a (possibly warped) product of a non-compact, maximally symmetric 4-dimensional spacetime $M_4$ and a compact 6-dimensional 'internal' space $M_6$
\be
\label{compactification_ansatz_schematic}
M_{10} \sim M_4 \times M_6.
\ee 
In such a scenario, the size of $M_6$ must be sufficiently small in order to explain why it has so far escaped direct experimental detection. 

A crucial aspect of such compactifications is that many features of the effective 4-dimensional theory, which is obtained after integrating out the internal space, depend on the geometry of the compactification manifold $M_6$. A famous early result along these lines is that the effective 4-dimensional theory exhibits unbroken supersymmetry at the compactification scale if and only if $M_6$ is a Calabi-Yau space~\cite{Candelas:1985en}. Strictly speaking this holds only for the simplest backgrounds in which other background fields besides the metric are taken to be trivial. More generally one may consider backgrounds involving topologically non-trivial field configurations known as fluxes or sources for the fields such as D-branes. In this case, requiring 4-dimensional supersymmetry leads to corrections both to the Calabi-Yau geometry of $M_6$ and to the pure product structure of the ansatz~\eqref{compactification_ansatz_schematic}~\cite{Becker:2007zj}. While many of these effects can be treated as small corrections and neglected in a first approximation, let us already mention one result which will play an important role in the following. Namely, in the presence of D-branes the consistency of the theory requires $M_6$ to be a geometric quotient of a Calabi-Yau manifold known as a Calabi-Yau orientifold~\cite{Polchinski:1995mt}. Of course, it is in principle also possible to consider compactifications to a non-supersymmetric 4-dimensional theory. However, the assumption of supersymmetry is very helpful from a technical perspective as it greatly simplifies the process of finding suitable compactification geometries and furthermore guarantees the stability of the vacua. Furthermore, there is well-documented phenomenological motivation e.g. from gauge coupling unification or the so-called hierarchy problem to consider models in which supersymmetry is only broken at a scale some way below the compactification scale\footnote{The compactification scale can vary from model to model, but throughout this thesis we assume that it lies no lower than in the so-called LARGE volume models~\cite{Balasubramanian:2005zx}. In these models the compactification scale can be estimated to lie at about $10^{11}$ GeV, see section~\ref{sec:IIBreduction}.}~\cite{Martin:1997ns}. The continued absence of any explicit experimental hints for low energy supersymmetry at the LHC may lead to a change in perspective regarding this point in the coming years, but throughout this thesis we follow the majority of the historical string theory literature and assume unbroken supersymmetry at the compactification scale. Let us also emphasise that even if the LHC were to rule out low energy supersymmetry, this would only imply that supersymmetry breaking must occur at a higher scale but would not rule out supersymmetry at the compactification scale.

From the discussion above it is clear that studying explicit string compactifications forms an essential part of the field of string phenomenology, which focuses on deducing the effective 4-dimensional consequences of string theory. The conventional approach towards deriving such 4-dimensional predictions from string theory, which we will follow throughout this thesis, can actually be broken into three distinct steps. The first step is to derive an effective action which encodes the low-energy dynamics of string theory in the sense that it reproduces the string theory scattering amplitudes of the lightest states. At the level of the massless string states and the lowest levels of string perturbation theory one obtains the famous 10-dimensional type I, type IIA and type IIB supergravity actions and their heterotic cousins~\cite{Green:87, Polchinski:1998, Becker:2007zj}. In principle, corrections to the effective actions can be computed systematically by considering higher orders in string perturbation theory, but in this thesis we will largely stick to the leading terms in the supergravity action. The second step on the way to a 4-dimensional effective theory is to specify a consistent compactification background. As discussed in more detail in chapter~\ref{chapter:IIBtoFtheory}, this effectively means that one must specify the background flux and D-brane configurations as well as the compactification manifold $M_6$. Finally, the effective 4-dimensional action can then be derived by performing a Kaluza-Klein reduction of the fields in the 10-dimensional supergravity action and integrating out the internal space $M_6$.

As the metric is one of the dynamical degrees of freedom of string theory, the background spacetime must be a solution of the string equations of motion\footnote{More precisely, this is true for the entire vacuum configuration including also other background string fields beyond the metric.}~\cite{Green:87, Polchinski:1998, Becker:2007zj}. This means that conceptionally the compactification ansatz~\eqref{compactification_ansatz_schematic} should actually be considered as arising from a dynamical process, and not as an outside input into the theory. From a theoretical perspective this fact can be simultaneously viewed as one of string theory's biggest strengths and one of its biggest weaknesses. The reason is that, as mentioned above, the couplings and parameters of the 4-dimensional effective theory are determined largely by the compactification geometry. At least in principle, if one was able to understand how precisely the various geometric parameters of the compactification are fixed dynamically\footnote{This problem is known as the problem of moduli stabilisation.} one would thus be able to essentially predict the 4-dimensional parameters. However, it has become clear over the years that the vacuum structure of string theory is so rich that much of the direct predictive power of string theory is lost. More precisely, huge numbers of different 4-dimensional theories can be obtained depending on which string vacuum is chosen for the compactification. In light of this fact, some authors have advocated following a statistical approach and attempting to study the probabilities of finding certain parameter values in the landscape of string theory vacua~\cite{Bousso:2000xa, Douglas:2003qc}. A complementary approach, which will be followed throughout this thesis, is to approach the problem from a model building perspective. By this we mean that rather than attempting to derive general string theory predictions from arbitrary compactifications one explicitly constructs classes of vacua which exhibit certain phenomenologically desirable features. Examples of such features may be low energy supersymmetry, a gauge group which accommodates that of the Standard Model as a subgroup or a potential for moduli stabilisation. Within these restricted classes of vacua, one may then look for generic predictions for the low energy theory.

When aiming to construct a phenomenologically viable 4-dimensional model from a string theory compactification, it is clear that one of the first things to consider is the effective gauge group. As mentioned above, gauge groups arise naturally in the context of type I and heterotic superstring theories, and initially most model building efforts focused on these sectors of string theory. It had also been known for a long time that gauge theories arise also in type II string theories if lower-dimensional D-branes\footnote{By lower-dimensional we mean that the D-branes do not span all 10 spacetime dimensions. A consistent, anomaly free type II string theory including only D-branes spanning all 10 dimensions is identical to type I string theory~\cite{Polchinski:1995mt}.} are included. However, such lower-dimensional D-branes were typically disregarded for model building because they were thought to be unstable and the introduction of lower-dimensional D-branes seemed somewhat ad hoc. The picture fundamentally changed following the discovery by Polchinski that D-branes form essential components of type II string vacua, because they carry conserved charges with respect to a class of fields of type II supergravity known as the Ramond-Ramond (R-R) form fields~\cite{Polchinski:1995mt}. This also showed that D-branes could be identified with a specific class of solutions of the supergravity equations of motion involving non-trivial profiles of the R-R fields. When including D-branes in compactifications to 4-dimensions, consistency requires R-R charge cancellation on the internal space. This can be achieved by considering compactifications on Calabi-Yau orientifolds, because such spaces involve so-called orientifold planes which carry R-R charges of opposite sign to those of the D-branes~\cite{Polchinski:1995mt}. 

The discovery of the importance of D-branes in 1995 had major repercussions for the field of string theory model building, and many string phenomenologists subsequently switched their attentions to type II orientifold compactifications with D-branes. The advantage of such models in comparison with compactifications of heterotic or type I string theory lies in the localisation of the gauge degrees of freedom to the D-brane world-volumes. This localisation makes it relatively easy to tune the 4-dimensional gauge group by specifying how the D-branes lie inside the compactification space $M_6$. In particular, it is reasonably straightforward to construct models with the Standard Model gauge group or a closely related extension, without having to take the detour via $SO(32)$ or $E_8$ as in heterotic or type I string theory~\cite{Cremades:2002te, Kokorelis:2002qi}. Both type IIA and type IIB compactifications have been investigated extensively from a model building perspective in recent years~\cite{Blumenhagen:2006ci}. The type IIB setting has a slight advantage in that the D-brane geometry in type IIB compactifications is mathematically slightly simpler than in the type IIA setting. This is also the reason why many investigations into moduli stabilisation have focused on the IIB sector~\cite{Kachru:2003aw, Balasubramanian:2005zx}. Throughout this thesis, we focus on the corner of the string landscape comprising of type IIB vacua.  

At around the same time as the importance of D-branes became clear, a number of duality relations were discovered which could be used to relate compactifications of different types of string theory\footnote{A famous example is T-duality, which relates type IIB with D-branes to type IIA with slightly different D-branes.}. This means that in many cases the specific 10-dimensional superstring theory and compactification manifold used to obtain a given 4-dimensional model is actually only a matter of choice and computational convenience, and the same 4-dimensional theory could also have been obtained starting with a different theory. This realisation, together with the expansion of the type II picture to include D-branes, led to a major paradigm shift in the field of string model building which has become known as the second superstring revolution. The appearance of the web of string dualities has been interpreted as a tantalising hint that there may exist an as yet unknown fundamental theory of which the various superstring theories are simply different limits~\cite{Becker:2007zj}. A further hint in this direction comes from the fact that some of the dualities involve the (unique) 11-dimensional supergravity theory. Clearly, for simple dimensional reasons this supergravity cannot be understood as a low energy description of a superstring theory, as was the case for the various 10-dimensional supergravity theories. Nevertheless, the analogy strongly suggests that 11-dimensional supergravity should also be connected to some fundamental quantum theory which has been named M-theory~\cite{Witten:1995em}. Various attempts have been made over the years to construct this theory, for example as a theory of quantised membranes or as a matrix theory, but a complete picture remains elusive to this day~\cite{Becker:2007zj}. In this thesis we will use the name M-theory to simply denote the 11-dimensional supergravity theory.

The most important duality for the purposes of this thesis is a duality between type IIB compactifications and compactifications of M-theory to three dimensions on a specific type of 8-dimensional manifold $M_8$. More precisely, the duality appears if $M_8$ is elliptically fibered over a 6-dimensional base. In this case, after taking the limit of vanishing volume of the elliptic fiber one obtains a 4-dimensional theory which can be regarded as a type IIB compactification on the 6-dimensional base. The 4-dimensional theories constructed in this manner via the M-theory duality are known as F-theory vacua~\cite{Vafa:1996xn}. Although they are effectively type IIB vacua, a different name has established itself for F-theory vacua because they include some strong-coupling features which are not directly visible in standard perturbative IIB orientifold compactifications~\cite{Denef:2008wq, Weigand:2010wm, Maharana:2012tu}. One such non-perturbative F-theory effect, which is largely responsible for their recent phenomenological popularity, is that F-theory models combine the localisation of gauge degrees of freedom of type II theories with D-branes with the appearance of exceptional gauge symmetries familiar from heterotic and type I models. This in particular makes it possible to construct realistic grand unified (GUT) models based e.g. on the gauge group $SU(5)$ in F-theory. In perturbative type IIB vacua with $SU(5)$ GUT gauge group on the other hand certain phenomenologically required quark couplings are forced to vanish~\cite{Maharana:2012tu}. This is the most famous example where the non-perturbative effects that are automatically encoded in the F-theory construction have a crucial impact on the effective 4-dimensional theory which cannot be seen directly in perturbative type IIB compactifications. 

The discovery of the apparent suitability of the F-theory construction for GUT model building has led to a considerable amount of work being put into the study of F-theory models in recent years, as reviewed e.g. in~\cite{Weigand:2010wm, Maharana:2012tu}. Because F-theory can be viewed as a sort of non-perturbative extension of type IIB, it is conceptionally clear that many of the effects familiar from type IIB orientifolds should have F-theoretic analogues. However, from the point of view of the actual computational formalism there are considerable differences between the IIB orientifold construction and the F-theory construction via an M-theory compactification on an elliptically fibered space. For this reason, the F-theoretic understanding of some aspects which are relatively easy to understand in the type IIB setting has remained incomplete to this day. The topic of this thesis is a discussion of the F-theory uplift of two such classes of well-known type IIB effects, relating to massive and massless Abelian gauge symmetries and D-brane instantons.

Arguably the most important aspect of type IIB vacua whose F-theory uplift has remained poorly understood until recently is the appearance of Abelian $U(1)$ gauge symmetries in the effective action~\cite{Weigand:2010wm, Maharana:2012tu}. Such Abelian gauge symmetries are extremely common in type IIB orientifold compactifications, because a stack of $N$ D-branes leads to a unitary gauge group\footnote{Strictly speaking orthogonal or symplectic groups are also possible depending on the precise location of the D-branes, but this will not be relevant for the moment.} $U(N)= U(1)\times SU(N)$. The description of non-Abelian gauge groups in F-theory, in this case the $SU(N)$ factor, is related to the degeneration of the elliptic fibration and has largely been known since the early days of F-theory~\cite{Witten:1995ex, Morrison:1996na}. However, the situation regarding the Abelian gauge factors is far less clear. Following earlier work in the context of compactifications to 6 dimensions~\cite{Morrison:1996na, Morrison:1996pp, Morrison:2012ei}, the precise aspects of the geometry of the elliptic fibration that are responsible for the appearance of massless $U(1)$ gauge factors\footnote{Here we mean additional massless $U(1)$s that are not associated with Cartan generators of a non-Abelian part of the gauge group.} in F-theory have only been identified relatively recently~\cite{Mayrhofer:2012zy, Braun:2013yti, Borchmann:2013jwa, Cvetic:2013nia, Braun:2013nqa, Cvetic:2013uta}. Furthermore, it is well-known that in the type IIB setting even $U(1)$ symmetries which become massive via the so-called geometric St\"uckelberg mechanism can have important consequences for the low energy theory~\cite{Blumenhagen:2006ci}. The F-theory uplift of such massive $U(1)$s is particularly complicated because they do not appear in the usual harmonic Kaluza-Klein reduction of M-theory. The role of massive Abelian gauge symmetries and the associated fluxes in F-theory was investigated in reference~\cite{Grimm:2011tb}, which forms the basis of large parts of chapter~\ref{sec:u(1)inIIBandFtheory} of this thesis.

A further important ingredient of type IIB vacua whose F-theory uplift is not yet completely understood is given by D-brane instantons. D-brane instantons are D-branes which lie fully inside the compactification manifold $M_6$ and are thus pointlike from the point of view of the external spacetime $M_4$. D-brane instantons yield corrections to the low energy effective action which are non-perturbative in the string coupling. At weak coupling these corrections are therefore generically negligible in comparison with perturbative corrections~\cite{Blumenhagen:2009qh}. Instantons can nevertheless play a crucial role for certain quantities in the low energy action which are protected from perturbative corrections due to $U(1)$ symmetries or axionic shift symmetries. A famous example of this is that D-brane instantons can induce Yukawa-type couplings between charged matter fields which are forbidden at the perturbative level due to $U(1)$ symmetries~\cite{Blumenhagen:2006xt, Ibanez:2006da, Florea:2006si, Blumenhagen:2007zk}. Furthermore, instantons play a crucial role in many moduli stabilisation scenarios~\cite{Kachru:2003aw, Balasubramanian:2005zx, Blumenhagen:2007sm, Bobkov:2010rf}, due to the fact that they involve a class of moduli fields which cannot appear in the perturbative potential as a consequence of axionic shift symmetries~\cite{Wen:1985jz}. As is familiar from instantons in quantum field theory, fermionic zero modes can cause the instanton contribution to the effective action to vanish. Therefore evaluating the zero mode spectrum is one of the most important tasks when computing an instanton contribution. In reference~\cite{Blumenhagen:2007sm} it was argued that in many phenomenologically interesting IIB compactifications charged zero modes would affect instanton contributions in such a way that the success of moduli stabilisation is endangered. However, we have shown in~\cite{Grimm:2011dj} that gauge flux on the instanton world-volume can alleviate this problem by removing some of the charged zero modes.

The findings of~\cite{Grimm:2011dj} reinforce the fact that it is crucial to take into account the full partition sum over all instanton configurations, and in particular over world-volume fluxes on the instanton, when evaluating the instanton contribution to the low energy effective action. As shown in~\cite{Kerstan:2012cy}, thinking in terms of the instanton partition function rather than a single instanton configuration actually also helps to understand the F-theory uplift of these type IIB effects. In particular, reference~\cite{Kerstan:2012cy} finds a one-to-one correspondence between the classical partition function of $M5$-instantons in F-theory and the classical type IIB instanton partition function\footnote{More precisely, the match of the partition functions is obtained for a certain class of type IIB instantons known as $O(1)$ instantons.}. This correspondence can in fact be used to identify the $M5$ partition function in F-theory, which is non-trivial to accomplish directly in the F-theory picture. For model building purposes, it is often sufficient to demonstrate that a certain type of instanton contribution is present in the model without having to evaluate the entire instanton partition sum. Essentially, this boils down to evaluating certain selection rules which guarantee the absence of fermionic zero modes which could cause the instanton contribution to vanish. The selection rules for the absence of chiral charged zero modes on $M5$-instantons in F-theory are discussed in~\cite{Kerstan:2012cy} and compared with the corresponding type IIB results. Just as in the type IIB setting, an interesting interplay between the instanton zero modes and the massless and massive Abelian gauge symmetries emerges. This makes it clear that further investigations into both topics will be important for future progress in F-theory model building and moduli stabilisation.

In the following we briefly outline the structure and the main results of the remainder of this thesis. We begin in section~\ref{sec:string_quantisation} with a very brief review of the basic definition of string theory using a quantised worldsheet action, before proceeding in section~\ref{sec:IIBsugra_compactification} to an overview of type IIB orientifold compactifications with D-branes. As a motivation for the F-theory construction we discuss the problem of D-brane back-reaction on the IIB compactification geometry in section~\ref{sec:FtheoryAsNonperturbIIB}. A review of the construction of F-theory vacua via the compactification of M-theory on elliptically fibered spaces is given in section~\ref{sec:intro_to_ftheory}.

Chapters~\ref{sec:u(1)inIIBandFtheory} and~\ref{sec:chap_instantons} form the core of this thesis and focus on similarities and differences between the type IIB and F-theory descriptions of Abelian gauge symmetries and fluxed instantons, respectively. The role of $U(1)$ symmetries in type IIB compactifications is discussed in section~\ref{sec:U(1)inIIB}, while section~\ref{sec:U(1)inFtheory} contains a detailed discussion of how the relevant effects can be replicated in the F-theory derivation of the effective action. The central result of section~\ref{sec:U(1)inFtheory}, which is based on~\cite{Grimm:2011tb}, is that geometrically massive $U(1)$s can be explicitly made visible in F-theory compactifications by including certain non-harmonic forms in the dimensional reduction. Strong evidence for this proposal is given by a detailed discussion of the F-theory limit of a compactification including such non-harmonic forms, which exhibits exactly the features that our IIB intuition leads us to expect. A brief discussion of the geometric interpretation of both massless and massive $U(1)$s in F-theory is given in section~\ref{sec:U(1)geometry}. Chapter~\ref{sec:chap_instantons} begins with a general review of the role of D-brane instantons in type IIB string theory. Special emphasis is placed on the role of world-volume flux, which was shown in~\cite{Grimm:2011dj} to influence the charged zero mode structure and to allow some instanton configurations to contribute to moduli stabilisation even though the contribution of the unfluxed instanton vanishes. In section~\ref{sec:fluxedM5sInFtheory} we consider fluxed $M5$-instantons in F-theory and their relation to fluxed instantons in type IIB. The main result, which was first given in~\cite{Kerstan:2012cy}, is that the classical partition function of the $M5$-instantons can be identified unambiguously using the duality with type IIB. We extend the discussion of~\cite{Kerstan:2012cy} slightly by presenting additional evidence for the form of the F-term supersymmetry condition for the $M5$ flux which was conjectured in~\cite{Kerstan:2012cy}. We then proceed to discuss the role of charged zero modes on $M5$-instantons in F-theory and consider how the findings can be understood from the type IIB perspective. The analysis provides convincing evidence that selection rules associated with geometrically massive $U(1)$ symmetries in type IIB continue to be relevant in F-theory, underlining the importance of studying massive $U(1)$s in F-theory as in chapter~\ref{sec:u(1)inIIBandFtheory}~\cite{Kerstan:2012cy}. Finally, the main text is supplemented by two appendices. In appendix~\ref{sec:app_maths}, we summarise our conventions and present a number of mathematical definitions and theorems that are used in the earlier parts of the thesis. Appendix~\ref{sec:app_democ_Mtheory} contains an explicit discussion of the Kaluza-Klein reduction of the so-called democratic formulation of the M-theory supergravity on an elliptically fibered complex fourfold. While appendix~\ref{sec:app_democ_Mtheory} is in part based on~\cite{Kerstan:2012cy}, we significantly extend the analysis presented there, which in particular allows us to explicitly identify the imaginary part of one set of chiral fields associated with the K\"ahler moduli of the fourfold. This imaginary part, which cannot be explicitly computed in the dimensional reduction of the standard formulation of M-theory, is crucial for a precise match with the type IIB theory and in order to determine the exact moduli dependence of the effective action of $M5$-instantons.

\chapter{Low energy theories from compactifications of type IIB string theory and F-theory}
\chaptermark{4D effective actions from IIB and F-theory}
\label{chapter:IIBtoFtheory}

In this chapter we begin with a very brief review of the fundamentals of string theory. Many further details on the points we are only able to touch upon can be found in standard textbooks on string theory such as references~\cite{Green:87, Polchinski:1998, Johnson:2003gi, Becker:2007zj}, which we roughly follow in our presentation. We then specialise to type IIB string theory and its orientifold compactifications, before reviewing how the geometric picture of F-theory arises in this context and finally proceeding to more general features of F-theory that are not visible in the perturbative type IIB setting. While clearly a comprehensive review of these very extensive subjects is beyond the scope of this thesis, we aim to introduce the main tools and concepts that will be needed in the analysis of the later chapters.

\section{From quantised strings to effective actions}
\label{sec:string_quantisation}
The motion of a point particle in $D$-dimensional spacetime with metric $g_{\mu\nu}$ can be described by its world-line. Such a world-line is a 1-dimensional submanifold of spacetime which can be parametrised e.g. by the eigentime $\tau$ of the particle and is embedded into spacetime by a set of functions $X^\mu(\tau), \ \mu = 0,...,D-1$. The action describing the motion of a (free) point particle is then simply the length of this world-line
\be
S= - m \int d\tau \sqrt{- g_{\mu\nu}(X) \frac{\partial X^\mu}{\partial\tau} \frac{\partial X^\nu}{\partial\tau}}.
\label{action_point}
\ee
Here $m$ is the mass of the particle.

Bosonic string theory can be seen as arising from the obvious generalisation of this description to the case of a 1-dimensional object or string. The motion of such a string clearly traces out a 2-dimensional submanifold of spacetime - the worldsheet or world-volume. The generalisation of~\eqref{action_point} is the Nambu-Goto action measuring the area of this worldsheet
\be
S_{NG} = - T \int d^2 \sigma \sqrt{-\det\left(g_{\mu\nu} \frac{\partial X^\mu}{\partial\sigma^\alpha} \frac{\partial X^\nu}{\partial\sigma^\beta}\right)}.
\label{S_NG_bosonic}
\ee
As before, the functions $X^\mu(\sigma^\alpha)$ parametrise the embedding of the world-volume - parametrised by the worldsheet coordinates $(\sigma^\alpha)=\tau,\ \sigma$ - into $D$-dimensional spacetime. As suggested by the notation, $\tau\equiv\sigma^0$ is taken to be a timelike coordinate on the world-sheet while $\sigma\equiv \sigma^1$ is spacelike. The parameter $T$ describes the string tension and is the only dimensionful parameter appearing in the fundamental definition of string theory. Instead of the string tension we will often use the string length scale $\ell_s$ or the Regge slope $\alpha'$, which are related to $T$ by\footnote{Throughout this thesis we always work in conventions where $\hbar = c = 1$.}
\be
T = \frac{1}{2\pi\alpha'} = \frac{2\pi}{\ell_s^2}.
\ee

Although the Nambu-Goto action has the advantage of having a very clear geometric interpretation as the area of the string worldsheet, the presence of the square root leads to problems when attempting to quantise the theory. To overcome this difficulty, one typically introduces an auxiliary world-sheet metric $h_{\alpha\beta}$ and considers the Polyakov action
\be
S_{P} = -\frac{T}{2} \int d^2 \sigma \sqrt{-\det h} \  h^{\alpha\beta} g_{\mu\nu} \frac{\partial X^\mu}{\partial\sigma^\alpha} \frac{\partial X^\nu}{\partial\sigma^\beta},
\label{S_pol_bosonic}
\ee
which is classically equivalent to the Nambu-Goto action.

To specify the classical dynamics completely, the actions must still be supplemented with suitable boundary conditions chosen in such a way that the variation of the action vanishes. There are two qualitatively different ways to achieve this, corresponding to open or closed strings. In the closed string case the variable $\sigma$ is a periodic variable whose fundamental domain can be chosen to be $0\leq \sigma\leq \pi$ without loss of generality, and the boundary conditions are 
\be
X^\mu(\tau,\sigma=0) = X^\mu(\tau,\sigma=\pi).
\ee
In the open string case we may take the same domain for $\sigma$, but there are now two possible boundary conditions that allow for the variation of the action to vanish. One option is to fix the position of the string endpoints in space by imposing
\be
X^\mu(\tau,\sigma = 0) = X^\mu_0(\tau)\ , \qquad X^\mu(\tau,\sigma = \pi) = X^\mu_\pi(\tau),
\ee
where $X^\mu_0$ and $X^\mu_\pi$ are arbitrary but fixed functions.
These conditions are known as Dirichlet boundary conditions. While the position of the string endpoints are fixed in the case of Dirichlet boundary conditions, there is non-zero momentum flow off the end of the string. For this reason, the surfaces on which an open string with Dirichlet boundary conditions ends must themselves be dynamical objects in order to be able to absorb this momentum. These objects are known as Dirichlet branes or D-branes, and will play a very important role in the following chapters.

The second option is to take so-called von Neumann boundary conditions, given by
\be
\partial_{\sigma} X^\mu(\tau,\sigma = 0) = 0 , \qquad \partial_{\sigma} X^\mu(\tau,\sigma = \pi) = 0.
\ee
In contrast to the previous case, the momentum flow off the end of the string is zero in the case of von Neumann boundary conditions.
Note that for open strings it is not necessary to take the same type of boundary conditions for all of the $D$ coordinates $X^\mu$. In fact, even for just one fixed coordinate $X^\mu$ it is admissible to impose Dirichlet boundary conditions at one end of the string and von Neumann boundary conditions at the other end. 

The equations of motion (e.o.m.) obtained from the Polyakov action~\eqref{S_pol_bosonic} are the vanishing of the energy momentum tensor
\be
T_{\alpha\beta} = -\frac{2}{T} \frac{1}{\sqrt{-h}} \frac{\delta S_{P}}{\delta h^{\alpha\beta}} = 0
\label{eom_h_before_GF}
\ee
and
\be
h^{\alpha\beta} \partial_\alpha \partial_\beta X^\mu =0.
\label{eom_X_before_GF}
\ee
Apart from the usual diffeomorphism symmetry on the worldsheet, the Polyakov action enjoys a Weyl symmetry under a conformal rescaling of the world-sheet metric $h_{\alpha\beta} \rightarrow \exp(f(\tau,\sigma))\ h_{\alpha\beta}$. Note that this additional symmetry appears only for the case of strings, where the world-volume is 2-dimensional, and would be absent in a theory based on membranes or other  higher-dimensional objects. The local Weyl$\times$diffeomorphism symmetry is large enough to allow us to always find a gauge in which the world-sheet metric takes the form of a 2-dimensional Minkowski metric\footnote{Such a choice may not be possible globally on worldsheets of more complicated topology, but this potential subtlety will not be important for us.}. After moving to such a flat gauge, the e.o.m.~\eqref{eom_X_before_GF} simplifies into the 2-dimensional wave equation. The general solution of this equation of motion is a superposition of plane waves $e^{-2in(\tau-\sigma)}$, the precise form of which depends on the chosen boundary conditions.

To quantise the theory, the coefficients $\alpha_n^\mu$ of the waves $e^{-2in(\tau-\sigma)}$ appearing in the general solution of the e.o.m. are promoted into operators, and canonical commutation relations 
\be
\left[ \alpha_n^\mu, \alpha_m^\nu \right] = n \delta_{m+n,0} g^{\mu\nu}
\label{comm_rel_alphas}
\ee
are imposed. The operators $\alpha_n^\mu$ can then be interpreted as raising and lowering operators belonging to different oscillation modes of the string and can be used to create a Fock space of one-particle states in the usual manner. Note that as the local symmetries of the worldsheet were used to fix the form of the worldsheet metric $h$, the e.o.m associated with the field $h$, i.e. the vanishing of the energy momentum tensor~\eqref{eom_h_before_GF}, must now be imposed by hand. 

From~\eqref{eom_h_before_GF} the energy momentum tensor $T_{\alpha\beta}$ is obtained as a function of the fields $X^\mu$. Due to the symmetries of the theory and the boundary conditions $T_{\alpha\beta}$ actually contains only one independent component in the case of open strings, while there are two components for closed strings. For simplicity we will consider the open string case in the following, although the main ideas can be carried over straightforwardly to the case of closed strings. After plugging in the oscillator expansion of the $X^\mu$, one arrives at a similar expansion into plane waves for the independent component of the energy momentum tensor. The coefficients appearing in this expansion are known as the Virasoro operators and denoted by $L_n$. 

The Virasoro operators play a crucial role in the quantisation of the theory. Using their expression in terms of the $\alpha_n^\mu$ together with the relation~\eqref{comm_rel_alphas} one it is possible to show that they obey the Virasoro algebra
\be
\left[ L_m, L_n \right] = (m-n) L_{m+n} + \frac{c}{12}m(m^2-1) \delta_{m+n,0},
\label{Virasoro}
\ee
with central charge $c$ equal to the dimensionality $D$ of spacetime. The appearance of the Virasoro algebra is due to the fact that there is a residual symmetry which is left unfixed even after imposing the flat gauge condition on the worldsheet. This residual symmetry is given by conformal Killing transformations, i.e. diffeomorphisms whose effect on the worldsheet metric can be undone by a conformal Weyl transformation. The Virasoro operators are generators of precisely these residual gauge transformations, and the appearance of the central charge term in their algebra signals the fact that this symmetry suffers from an anomaly after quantisation. 

In the quantum theory the condition that the energy momentum tensor must vanish is rewritten in terms of the action of the Virasoro operators on physical states. Due to the appearance of the central charge in~\eqref{Virasoro}, not all of the operators can be chosen to annihilate the physical states. Instead, one imposes the condition $L_m - a\delta_{m,0}\left|\phi\right> = 0 ,\ m\geq 0$ for all physical states $\left|\phi\right>$, where the constant term $a$ for $m=0$ is an unknown normal ordering constant arising from the moving the operators in $L_0$ into the correct order.

Restricting the space of physical states by imposing the Virasoro constraints is not quite sufficient to lead to a consistent, unitary quantum theory for arbitrary values of the central charge $c$ (or spacetime dimension $D$). For example, it is possible to explicitly construct states fulfilling all Virasoro constraints but having negative norm if $a>1$, or if $a=1$ and $D>26$ (see e.g.~\cite{Green:87} for an explicit construction). It is possible to show that the free bosonic theory is free of negative norm states and is unitary up to one-loop level if and only if $a=1$ and $D=26$. In other words, (free) bosonic string theory would predict the existence of 26 dimensions of spacetime.

The underlying reason for the appearance of a critical number of dimensions in string theory is a quantum anomaly of the conformal symmetry on the worldsheet. This fact becomes much more clearly visible by considering BRST quantisation of the worldsheet theory instead of the canonical quantisation mentioned above. As part of the Faddeev-Popov gauge fixing procedure, a set of ghost fields is introduced to take into account the diffeomorphism$\times$Weyl symmetry of the worldsheet theory. By an explicit computation it is easy to check that the Faddeev-Popov ghosts contribute a value of $c_{g} = -26$ to the central charge of the total Virasoro algebra. It is thus clear that exactly 26 spacetime dimensions are needed to ensure that the central charge of the $X^\mu$ system cancels the ghost contribution, such that the total conformal anomaly vanishes.

Although it is consistent in $D=26$ dimensions, the bosonic string theory discussed so far is clearly not much use as a description of our world simply because it does not include fermionic degrees of freedom. In addition, in a flat background spacetime the ground state of the theory is tachyonic, signalling an instability of such backgrounds.
Both of these problems are rectified when moving to superstring theory. This amounts to introducing a set of worldsheet spinor fields\footnote{It is also possible to introduce a different number of worldsheet spinor fields, which leads to heterotic string theory. However, we focus here on the worldsheet action that leads to type I and II string theory.} $\psi^\mu$ to partner the bosonic fields $X^\mu$ and modifying the Polyakov action~\eqref{S_pol_bosonic} to\footnote{For strings propagating in more complicated backgrounds, further terms may be added to describe the interaction of the string with other background fields beside the metric $g$. Specifically, the fundamental string couples to the dilaton and Kalb-Ramond fields that are introduced in section~\ref{sec:action_IIB_bulk}, but we will neglect these potential additional terms here for simplicity.}
\be
S_{P} = -\frac{1}{2\pi} \int d^2 \sigma \sqrt{-\det h} \  g_{\mu\nu} \left[ \pi T \partial_\alpha X^\mu \partial^\alpha X^\nu - i\bar{\psi}^\mu \rho^\alpha \partial_\alpha \psi^\nu \right].
\label{S_pol}
\ee
Here $\rho^\alpha$ denote 2-dimensional Dirac matrices and we have followed the conventions of~\cite{Green:87}.

When choosing boundary conditions for the fermionic fields $\psi^\mu$ an additional sign ambiguity appears which is not present in the bosonic case. Depending on the choice of sign in the boundary conditions, one speaks of the Neveu-Schwarz (NS) or Ramond (R) sectors. In the case of closed strings the sign may be chosen independently for left- and right-moving modes, leading to 4 different sectors known as the R-R, R-NS, NS-R and NS-NS sectors. 

The superstring theory defined by~\eqref{S_pol} exhibits manifest worldsheet supersymmetry and can be quantised in a similar manner to the bosonic theory discussed previously. The worldsheet theory now admits a superconformal symmetry leading to additional ghost fields appearing in the Faddeev-Popov gauge fixing procedure. The contributions to the Virasoro central charge from the various ghost fields partially cancel, leaving an overall ghost contribution of -15 to the central charge. As each fermion field $\psi^\mu$ contributes 1/2 to the central charge, one obtains the famous result that superstring theory requires $D=10$ to guarantee absence of the superconformal anomaly.

In principle it would be possible to construct a large number of different superstring theories simply by projecting to different combinations of the various possible sectors of states mentioned above.
However, additional consistency conditions beyond the restriction to 10 spacetime dimensions must be fulfilled in order to obtain a consistent interacting theory, which rule out most of these would-be superstring theories. When calculating string scattering amplitudes, asymptotic in-going and out-going states are mapped to suitable operators known as vertex operators. The scattering amplitudes are then obtained by evaluating the path integral with the insertion of the vertex operators at different points on the intermediate string world-sheet connecting the participating asymptotic states. Due to the underlying superconformal symmetry of the theory these scattering amplitudes must be invariant under modular transformations. Furthermore, the products of vertex operators are not in all cases well-defined as the positions of the operators on the world-sheet are varied. 

In order to obtain a theory with consistent interactions it is necessary to project out parts of the spectrum of states and keep only certain combinations of sectors. This truncation of the spectrum is known as the Gliozzi-Sherk-Olive (GSO) projection~\cite{Gliozzi:1976qd}. A further important aspect of the GSO projection relates to the supersymmetry of the theory from the point of view of the 10-dimensional spacetime. Despite the fact that supersymmetry on the worldsheet is built in from the beginning, spacetime supersymmetry depends on the choice of sectors included in the theory. The GSO projection acts precisely in such a way that the resulting theory exhibits spacetime supersymmetry.

Evaluating the abovementioned conditions in detail shows that there are in fact only 4 consistent inequivalent theories of oriented closed strings which originate from the worldsheet action~\eqref{S_pol}~\cite{Polchinski:1998}. Two of these theories contain a tachyonic state in the spectrum and are thus unstable, leaving two stable theories known as type IIA and type IIB string theory because they lead to $\cN=2$ supersymmetry in 10 dimensions. When including both open and closed strings in a flat 10-dimensional spacetime without extra structure there is only one consistent theory, which is known as type I string theory and leads to $\cN=1$ spacetime supersymmetry. In addition, two further consistent theories in 10 dimensions may be constructed, which can be seen as hybrids between bosonic string theory and superstring theory and are known as heterotic string theories. Although these different string theories were initially thought to be independent, an intriguing web of dualities connecting the different theories has been discovered over the years. This suggests that they can all be thought of as different limiting cases of an as yet unknown underlying theory. Nevertheless, in this thesis we will focus almost exclusively on type IIB string theory and the related M- and F-theories, while the other string theories mentioned above will play at most minor roles.

Although string theory allows for the calculation of transition amplitudes between different quantum states just like any other quantum theory, the techniques involved are quite different from the field theoretic techniques generally used in the field of particle physics. In order to begin analysing string theory from a phenomenological point of view it is therefore advantageous to translate the dynamics encoded in the scattering amplitudes of string theory into the language of an effective field theory, which gives rise to the same scattering amplitudes. In fact, we will see that in particular in the case of type IIB string theory this reformulation of the theory in terms of an effective action does more than just make it easier to read off particle interactions and couplings. Namely, the effective action will make apparent an additional symmetry and eventually lead us towards a duality with M-theory via F-theory that was not visible from the fundamental formulation of the type IIB string theory.

\section{Type IIB SUGRA and orientifold compactifications}
\label{sec:IIBsugra_compactification}

\subsection{The type IIB effective action of the closed string sector}
\label{sec:action_IIB_bulk}
The starting point of our investigations will be the type IIB supergravity theory in 10 dimensions, which is well-known to describe the low energy dynamics of the massless degrees of freedom of closed type IIB string theory~\cite{Green:87, Polchinski:1998}. More precisely, we will focus on the bosonic part of the action for simplicity, which in principle is sufficient to reconstruct the full action by supersymmetry. The massless bosonic spectrum of closed type IIB superstrings from which the action is constructed can be split into the Neveu-Schwarz-Neveu-Schwarz (NS-NS) and Ramond-Ramond (R-R) sectors and is given in table~\ref{tab:IIBfields}. 
\begin{table}[ht]
\centering
\begin{tabular}{llcl}
\hline
&Field & Symbol & Spacetime transformation \\
\hline
\multirow{3}{*}{NS-NS sector} &
Graviton / metric & $g$ & Symmetric 2-tensor \\
& Dilaton & $\Phi$ & Scalar \\
& Kalb-Ramond $B$-field & $B_2$ & 2-form \\
\hline
\multirow{3}{*}{R-R sector} & \multirow{3}{*}{$p$-form potentials} & $C_0$ & Scalar \\ && $C_2$ & 2-form \\ && $C_4$ & Self-dual 4-form \\
\hline
\end{tabular}
\caption{Bosonic fields of type IIB supergravity.}
\label{tab:IIBfields}
\end{table}

Due to the presence of an infinite tower of massive states in the spectrum the exact effective action for the massless modes will include an infinite number of terms with more and more derivatives (see e.g. the discussion in~\cite{Green:87}). When calculating scattering events at an energy $E$, a term with $n$ such derivatives will give rise to a contribution suppressed by a factor of $\left(E/M_s\right)^n$, where $M_s = \ell_s^{-1}$. As we are interested in energies far below the mass scale of the lightest massive states of string theory, we may safely keep only the leading powers of derivatives in our action. Complementary to this grouping in terms of powers of $\alpha'$, the terms in the effective action may be arranged into a power series in the string coupling $g_s = e^{\Phi}$. In the type IIB setting we will always assume to be in the weak coupling limit and restrict ourselves to the leading order in $g_s$. Hence the effective actions presented in the following represent the leading terms in terms of powers of derivatives and of the string coupling.

Before proceeding let us point out that the self-duality of the Ramond-Ramond 4-form field makes a purely Lagrangian description of the dynamics of the massless modes of closed type IIB superstrings impossible. Instead, one can only define what is strictly speaking a pseudo-action in terms of an unconstrained 4-form field and then impose the self-duality constraint
by hand at the level of the equations of motion. Keeping this caveat in mind, we will nevertheless drop the linguistic distinction and speak simply of a type IIB supergravity action in the following. In terms of the fields listed in table~\ref{tab:IIBfields}, this action is given in the string frame by\footnote{See appendix~\ref{sec:conventions} for details on the differential form notation used and in particular for our conventions regarding the Hodge star operator $\ast$.}~\cite{Polchinski:1998} 
\be
\begin{aligned}
S_{IIB} =& \frac{1}{2\kappa_{10}^2} \int d^{10}x \sqrt{-\det g} \, e^{-2\Phi} R  + \frac{2}{\kappa_{10}^2} \int e^{-2\Phi} d \Phi\wedge\ast d\Phi \\ & 
- \frac{1}{4\kappa_{10}^2} \int \Big[ e^{-2\Phi} H_3 \wedge\ast H_3  + G_1\wedge \ast G_1 + G_3\wedge\ast G_3  \\ &  + \frac12 G_5\wedge\ast G_5 + \left(C_4-\frac12 B_2\wedge C_2 \right) \wedge H_3 \wedge F_3 \Big]. 
\end{aligned}
\label{S_IIB_nondemoc}
\ee
Here we have used the field strengths
\be
\begin{aligned}
H_3 =&\ d B_2 ,\\
F_{p} = &\ d C_{p-1} \ , \qquad p = 1,3,5, \\
G_p = &\ \genfrac{\{}{.}{0pt}{}{F_1 , }{F_p - C_{p-3}\wedge H_3,}\quad  \genfrac{}{}{0pt}{}{p=1,\ \ \ }{p=3,5.}
\end{aligned}
\ee
In this language the self-duality constraint that must be imposed at the level of the equations of motion reads
\be
\ast G_5 = G_5.
\label{duality_constr_non-democ}
\ee

Before proceeding, let us briefly consider the mass dimensions of the various fields in~\eqref{S_IIB_nondemoc}. Throughout this thesis, we choose conventions in which the spacetime coordinates $x^i$ as well as the metric components $g_{ij}$ and all $p$-form fields are dimensionless (see also appendix~\ref{sec:conventions}). This means that also the 10-dimensional gravitational coupling constant $\kappa_{10}$ is dimensionless and is simply given by
\be
\kappa_{10}^2 = \frac{1}{4\pi}.
\label{kappa_10}
\ee
These conventions are particularly convenient for the dimensional reduction of the action from 10 to 4 dimensions. To extract physical quantities like masses in the 4-dimensional theory it will then of course be necessary to rescale the spacetime coordinates and perform a suitable Weyl rescaling of the metric\footnote{Note that large parts of the string theory literature, such as e.g. references~\cite{Becker:2007zj, Blumenhagen:2006ci}, work directly with a metric related to our metric by such a Weyl rescaling such that $\kappa_{10}^2 = \ell_s^8/4\pi$.}. This rescaling reintroduces appropriate powers of $\alpha'$, or equivalently of $\ell_s$. This makes it clear that although no explicit factors of $\ell_s$ or $\alpha'$ appear in~\eqref{S_IIB_nondemoc} in our conventions, possible corrections which we have described as being of higher order in $\alpha'$ are actually of higher order in derivatives or in the metric in our conventions.

In many references, including the standard reference~\cite{Polchinski:1998}, the type IIB supergravity action is written not in terms of $C_4$ as above but rather in terms of a shifted field related to $C_4$ by
\be
\tilde{C}_4 = C_4 - \frac12 B_2\wedge C_2.
\ee
This choice would lead to a different definition of $G_5$ and would simplify the appearance of the Chern-Simons term in the IIB action. However, the advantage of using the field $C_4$ is that it is this field which appears in the standard expression for the Chern-Simons action of D-branes, see e.g. the discussion in~\cite{Myers:1999ps}. Furthermore, the democratic formulation of type IIB supergravity, which we will introduce momentarily, can be written more concisely in terms of $C_4$ rather than the shifted $\tilde{C}_4$. We will therefore stick to using $C_4$ throughout this thesis.

While the action~\eqref{S_IIB_nondemoc} supplemented with the self-duality constraint~\eqref{duality_constr_non-democ} is fine when considering only closed string effects, it is advantageous to use a slightly different reformulation of the theory when including D-branes. The basic idea stems from the fact that in 10 dimensions the theory of a $p$-form potential is dual to that of an $(8-p)$-form potential, where the roles of Bianchi identities and equations of motion of the potential are simply exchanged and the field strengths are related by
\be
\ast F_{p+1} = \pm F_{9 - p}.
\ee
This duality is the analog of the well-known electric-magnetic duality in 4 dimensions. In the absence of any sources the choice of which formulation to use is simply a question of preference, but this changes when sources are introduced in the form of D-branes. A $Dp$-brane couples electrically to a $p+1$-form potential while forming a magnetic source for a $(7-p)$-form potential. While the electric coupling can be easily described by a term of the form
\be
S \supset \int_{Dp\text{-brane}} C_{p+1}
\label{electric_coupling_schematic}
\ee
in the action, the magnetic coupling cannot be easily given in such a manner. Therefore the question of which of the equivalent formulations of the theory is best suited depends on the D-brane content of the model at hand. 

The most flexible and at the same time the most symmetric choice is the so-called democratic formulation~\cite{Fukuma:1999jt, Bergshoeff:2001pv}, which includes all $p$-form potentials $C_p,\ p=0,2,4,6,8$ on an equal footing in the pseudo-action. Of course, the doubling of the number of degrees of freedom in the pseudo-action must be compensated for by introducing additional duality relations similar to the one given in~\eqref{duality_constr_non-democ}. More precisely, the new pseudo-action is defined by
\be
\begin{aligned}
S_{IIB} =& \frac{1}{2\kappa_{10}^2} \int d^{10}x \sqrt{-\det g} \, e^{-2\Phi} R  + \frac{2}{\kappa_{10}^2} \int e^{-2\Phi} d \Phi\wedge\ast d\Phi \\ & 
- \frac{1}{4\kappa_{10}^2}  \int \left[ e^{-2\Phi} H_3 \wedge\ast H_3 + \frac12 \sum_{p=1,3,5,7,9} G_p\wedge \ast G_p \right], 
\end{aligned}
\label{S_IIB_democ}
\ee
with the field strengths $G_p$ given as before by
\be
\begin{aligned}
G_p = &\ \genfrac{\{}{.}{0pt}{}{F_1 , }{F_p - C_{p-3}\wedge H_3,}\quad  \genfrac{}{}{0pt}{}{p=1,}{p>1,}\\
F_{p} = &\ d C_{p-1} \ , \qquad p = 1,3,5,7,9.
\end{aligned}
\label{def_Gps}
\ee
The duality relations to be imposed after deriving the equations of motion from~\eqref{S_IIB_democ} read
\be
G_1 = \ast G_9, \qquad G_3 = - \ast G_7,\qquad \text{and } G_5=\ast G_5.
\label{duality_relations}
\ee

Note that the Chern-Simons coupling in the action~\eqref{S_IIB_nondemoc} is no longer present in the democratic action~\eqref{S_IIB_democ}. Nevertheless, it can easily be checked that the equations of motion derived from the democratic action with the field strengths defined by~\eqref{def_Gps} reduce precisely to the original equations obtained from~\eqref{S_IIB_nondemoc} when the relations~\eqref{duality_relations} are imposed. In particular, the elimination of the redundant degrees of freedom using the duality relations reproduces the effects of the Chern-Simons coupling.

Furthermore let us emphasise that in the democratic formulation we can consistently describe interactions between R-R fields and all types of D-branes using terms in the action of the form~\eqref{electric_coupling_schematic}. In other words, the magnetic interaction between a Ramond-Ramond potential and a D-brane can be replaced by the more easily treatable electric interaction of the brane with the dual potential. It is due to this advantage that we will use the democratic formulation throughout the majority of this thesis. 

Each R-R potential $C_p$ appearing in the IIB action is associated with an Abelian gauge symmetry~\cite{Bergshoeff:2001pv}. We denote the associated $p$-form gauge parameters by $\Lambda_p$. To write down the associated gauge transformation in a concise form it is convenient to combine the various parameters into a formal sum denoted by $\mathbf{\Lambda}=\sum_p \Lambda_p$. The gauge transformation can now be written as
\be
C_p \rightarrow C_p + \left[ \mathbf{\Lambda}\ e^{B_2} \right]_p,
\label{RR_gaugetrf_bulk}
\ee
where $\left[\ldots\right]_p$ denotes the $p$-form part of the formal sum of differential forms of various degrees contained in the brackets. It is straightforward to check that this transformation leaves the field strengths $G_p$ and therefore also both the action~\eqref{S_IIB_democ} and the duality relations~\eqref{duality_relations} invariant, provided only that the gauge parameters $\Lambda_p$ are closed forms. For the scalar field $C_0$ the gauge transformation~\eqref{RR_gaugetrf_bulk} is simply a shift by a constant, explaining why $C_0$ is sometimes referred to as the R-R axion field. Finally, we note that a shift of $B_2$ by an arbitrary closed 2-form is also a symmetry, because both the action and the duality relations involve $B_2$ only through the field strength $H_3$. As we will discuss in more detail in section~\ref{sec:SL(2,Z)}, in the quantum theory the shift symmetries of $C_0$ and $B_2$ are broken to discrete sub-symmetries due to the fact that these fields appear in the action of $D(-1)$ and string world-sheet instantons, respectively.

\subsection{Compactifications and Calabi-Yau orientifolds}
\label{sec:compactifications}
String theory exhibits many beautiful theoretical properties that make it probably the most promising candidate theory available today for unifying gravity and gauge theories into a common theoretical framework. Nevertheless, as with any physical theory it is of course essential to work out string theory's phenomenological implications at low energy scales, with the aim of eventually testing them experimentally.

Having obtained the effective action given in the previous subsection, it is natural to proceed in the usual manner by constructing a vacuum of the theory and calculating the scattering amplitudes of asymptotic states within this background. The canonical choice of vacuum to perform perturbation theory around would be simple Minkowski space. However, the fact that string theory requires a 10-dimensional spacetime for mathematical consistency while our visible universe has only four dimensions means that one has to instead consider vacua involving spacetimes of the general form
\be
M_{10} = M_{1,3} \times M_{6}.
\label{eq:spacetime_ansatz}
\ee
Here $M_{1,3}$ denotes 4-dimensional Minkowski spacetime (or more generally a maximally symmetric spacetime), while $M_6$ is a compact manifold of sufficiently small size to have evaded direct experimental detection up to now. 

Of course, the geometry of the internal manifold $M_6$ is restricted by the fact that the metric must form a solution of the equations of motion arising from~\eqref{S_IIB_nondemoc}. Nevertheless, there is still an enormous freedom in the choice of $M_6$. The situation may be simplified by considering solutions with vanishing internal field strengths $H_3, F_i = 0$ and a constant dilaton profile as well as imposing that the theory after compactification should exhibit some unbroken supersymmetry. Of course, what is meant here is supersymmetry at the compactification scale, and supersymmetry must still be broken by another mechanism at a lower scale. While of course not strictly necessary, looking to preserve supersymmetry during compactification has several advantages. From a phenomenological point of view it makes possible models with low-energy supersymmetry, which exhibit the usual advantages of such models such as alleviating the hierarchy problem. On a more technical level, it can be shown that a compactification preserving supersymmetry will be a solution of the equations of motion and hence a true vacuum of the theory~\cite{Candelas:1985en}, and the supersymmetry equations are generally much easier to solve than the more complicated equations of motion. Furthermore, supersymmetric compactifications are automatically stable while the vacuum may be destabilised e.g. by tachyons in the non-supersymmetric case. 

In the absence of fluxes the requirement of supersymmetry requires $M_6$ to be a K\"ahler manifold and to admit a covariantly constant spinor, which means that $M_6$ must have $SU(3)$ holonomy~\cite{Candelas:1985en}. In other words, $M_6$ must be taken to be a Calabi-Yau manifold. In the following we will usually use the notation $X_n$ or $Y_n$ to denote Calabi-Yau manifolds of the complex dimension $n$. The situation changes slightly if we consider compactifications involving either non-vanishing fluxes or sources for the closed string fields, such as D-branes. In this case an ansatz of the form~\eqref{eq:spacetime_ansatz} with a pure direct product metric is no longer possible and one must instead consider a warped product in which the metric of the external spacetime is scaled by a warp factor depending on the internal coordinates, as first noted in~\cite{Strominger:1986uh}. For $Dp$-branes with $p<7$ the warping can be seen as a local back-reaction effect in the vicinity of the brane, and vanishes asymptotically as one moves away from the source~\cite{Horowitz:1991cd}. In the so-called probe brane approach one may therefore consistently neglect warping effects as a first approximation. However, for D7-branes back-reaction effects become more important and we will return to this point in more detail in section~\ref{sec:ell_fibration_D-brane_backreaction}.

Compactifying type II string theories on a Calabi-Yau manifold leads to a 4-dimensional theory with $\cN = 2$ supersyetry, as both gravitinos present in the original 10-dimensional theory yield 4-dimensional gravitinos after dimensional reduction. This situation is of course disfavoured phenomenologically due to the restrictive nature of $\cN = 2$ supersymmetric theories, which do not allow for a chiral matter sector. To eliminate one of the supersymmetry generators and obtain a $\cN=1$ theory one therefore typically mods out the particle spectrum by a so-called orientifold projection. This projection involves a world-sheet parity operation and an isometric involution $\sigma$ of the underlying Calabi-Yau manifold. In this case one speaks of compactifying on a Calabi-Yau orientifold. In fact, partial supersymmetry breaking is not the only reason to consider Calabi-Yau orientifolds, because orientifold geometries are necessary for consistency when looking to add D-branes to the vacuum in order to introduce non-Abelian gauge theories into the model. Implementing an orientifold involution leads to the appearance of orientifold planes at the fixed locus of the involution $\sigma$. These orientifold planes can be shown to carry tension as well as being charged with respect to the Ramond-Ramond form fields. The charge and tension of orientifold planes have the opposite sign to the charge and tension of D-branes~\cite{Polchinski:1995mt}, which makes it possible to cancel the overall net charges in the model as is necessary for consistency due to the compactness of $M_6$. In type IIB orientifolds $\sigma$ must be taken to be holomorphic~\cite{Acharya:2002ag}, and to preserve Poincar\'e invariance it must act as the identity on the external spacetime. Therefore in the models considered the orientifold planes will be a product of the external spacetime $M_{1,3}$ with an even-dimensional submanifold within $M_6$. Depending on the choice of $\sigma$ it is possible to obtain models with either $O3$- and $O7$-planes or with $O5$- and $O9$-planes~\cite{Acharya:2002ag}\footnote{We use the standard notation $Dp$-brane denote to denote a D-brane with $(p+1)$-dimensional world-volume. Similarly an $Op$-plane is a $(p+1)$-dimensional object.}. As we will be primarily interested in type IIB models with $D7$-branes, we will stick to the case with $O3/O7$-planes for easier R-R tadpole cancellation.

As mentioned before, the full orientifold action is a combination of the geometric orientifold involution and a parity operation on the world-sheet. Each field in the type IIB spectrum given in table~\ref{tab:IIBfields} picks up a particular sign under the world-sheet parity operation. To obtain a well-defined state in the orientifold geometry the fields must exhibit the correct parity under pullback along the orientifold involution to cancel the sign caused by the world-sheet parity. For future convenience, we note that the parities of the various fields of table~\ref{tab:IIBfields} in compactifications with $O3/O7$-planes are given by~\cite{Grimm:2004uq}
\be
\sigma^* F = \genfrac{\{}{.}{0pt}{}{F,}{-F,} \quad \genfrac{}{}{0pt}{}{F = g, \Phi, C_0, C_4}{F = B_2, C_2 .\quad\quad \ }
\label{orientifold_parities}
\ee
In the democratic formulation the fields $C_6$ and $C_8$ carry the same orientifold parity as their respective dual fields $C_2$ and $C_0$. 

At this point we should mention a slight caveat that allows certain discrete positive parity $B$-field configurations to be switched on~\cite{Blumenhagen:2008zz}. To make the distinction between the parts with different orientifold parity clear we will in the following sometimes explicitly decompose $B_2 = B_+ + B_-$. The reason why non-zero $B_+$ is admissible is that, as noted at the end of section~\ref{sec:action_IIB_bulk}, the Kalb-Ramond field enjoys a discrete shift symmetry $B_2\rightarrow B_2 + \delta B$. For certain discrete values of $B_+$ it is possible to find such a gauge transformation which precisely transforms $B_+ = \sigma^* B_+$ into $-B_+$, so that~\eqref{orientifold_parities} is fulfilled up to a gauge transformation. However, because the gauge transformations are generated by closed forms $\delta B$ also the positive parity part $B_+$ of the Kalb-Ramond field must be closed. Therefore the field strength $H_3 = dB_2$ actually depends only on the negative parity part $B_-$.

While dimensional reduction on a compact manifold $M_6$ as described above is essential in order to extract 4-dimensional low energy physics from the underlying 10-dimensional string theory, the process does also dilutes the predictive power of string theory. This is because many features of the resulting low energy theory, including aspects such as the visible gauge group, particle content, number of generations or existence and magnitude of certain couplings, depend heavily on the geometric details of the compactification manifold. The situation is made even worse by the many different choices relating to the configuration of D-branes and fluxes for a given background manifold. The number of such configurations was famously estimated at $10^{\mathcal{O}(500)}$ in \cite{Douglas:2006es} for a certain class of type IIB models. Taking into account different types of string theories or even relaxing the Calabi-Yau condition will of course only exacerbate this issue. Of course, the estimates mentioned above are only heuristic in nature and do not fully take into account equivalences between various models. Nevertheless, they serve to show that the proposition of searching for individual concrete compactification models realising the Standard Model carries limited promise. This realisation has led to the development of statistical approaches, which attempt to analyse the distributions of values of various parameters and observables on the ensemble of string theory vacua and were initiated in \cite{Bousso:2000xa} and \cite{Douglas:2003qc}. In this thesis we will not pursue this direction further, but rather than constructing explicit compactifications we will mostly work with classes of Calabi-Yau manifolds and look for generic features of the resulting low energy theories within these classes of vacua.

\subsection{D-branes and their effective action}
\label{sec:Dbranes}
D-branes form an essential tool in phenomenological attempts to build string theory vacua which lead to realistic effective 4-dimensional theories. They owe their importance to the fact that they represent the only source of non-Abelian gauge symmetries in type II theories. D-branes are by definition objects on which open strings may end. In the presence of stacks of multiple identical D-branes it is therefore necessary to introduce so-called Chan-Paton labels to distinguish which of the branes a string is ending on. A more general open string state is then described by a Chan-Paton matrix $\lambda_{ij}$ whose entries give the number of strings stretching from brane $i$ to brane $j$~\cite{Paton:1969je}. For strings beginning and ending on a stack of $N$ D-branes this gives an $N^2$ multiplicity of all associated open string states, suggesting that the states transform in the adjoint representation of $U(N)$. This symmetry group is confirmed by an analysis of the open string scattering amplitudes, and the low energy effective theory on the brane worldsheet describing these string states is a $U(N)$ Yang-Mills theory\footnote{In an orientifold setting the gauge symmetry may be projected down to a symplectic or orthogonal group, depending on the precise orientation of the D-brane in relation to the orientifold planes, but we will ignore this subtlety for the moment.}. Similarly, strings stretching from one brane stack to another give rise to fields in the effective action which transform in the bifundamental representation of the gauge groups of the two brane stacks.

The dynamics of the open strings ending on a stack of $Dp$-branes can be described by an effective field theory on the $(p+1)$-dimensional brane world-volume. In the low energy limit we may again restrict ourselves to the massless fields as in the closed string sector. The massless field content of the low energy theory is given by a $(p+1)$-dimensional vector field $A$ and a set of $(9-p)$ scalar fields\footnote{While this number of massless scalars is correct for a D-brane in empty 10-dimensional Minkowski space, some of these fields can become massive in the presence of non-trivial fluxes or for D-branes on a non-trivial compactification manifold.} $s^i$, as well as their fermionic superpartners. All these world-volume fields transform in the adjoint representation of the gauge group on the stack of D-branes. In the Abelian case, the scalar fields mentioned above can be viewed as parametrising the transverse displacement of the brane in a tubular coordinate system around its rest position. More precisely, denoting coordinates on the brane world-volume by $(\xi^a),\  a=0,...,p $ and letting $X^m, \ m=0,...,9$ be coordinates of the 10-dimensional spacetime $M_{10}$, the embedding of the brane can be written in tubular coordinates as~\cite{Kerstan:2011dy}
\be
X^m (\xi) = \genfrac{\{}{.}{0pt}{}{ \delta^m_a \xi^a \ ,}{x^m_0 + s^{m-p}(\xi)\ ,} \qquad \genfrac{}{}{0pt}{}{0\leq m\leq p.}{p+1\leq m\leq 9.}
\label{embedding_brane}
\ee
Recall that we work in conventions in which the spacetime coordinates $X^m$ are dimensionless, so that the same holds true for the scalar fields $s^{m-p}$ and the world-volume coordinates $\xi^a$.
The intriguing appearance of non-commuting, matrix-valued spacetime coordinates when extending this to the non-Abelian case was first noted in~\cite{Witten:1995im} and much work has been carried out to clarify the interpretation of this effect and to understand its implications, but we will not pursue this direction further here. 

Fields arising from strings stretching between two different stacks of D-branes are generically massive and can be neglected in the low energy theory. An exception occurs at the intersection locus between two stacks of branes, where the strings stretching from one stack to the other can shrink to zero length. For these strings the mass of the lowest-lying string states, given by the product of the string tension and the string length, vanishes. In the language of the low energy theory one obtains additional massless fields supported on the intersection locus of the two brane stacks. Fields of this type will play an important role in our discussion of instanton effects in section~\ref{sec:fluxed_D3instantons}, but for now we will consider an isolated stack of D-branes for simplicity.

Similarly to the bulk action of equation~\eqref{S_IIB_nondemoc}, the effective action for the massless modes can be constructed by evaluating suitable open string scattering amplitudes and formulating a field theory action that reproduces these scattering amplitudes. As discussed in section~\ref{sec:action_IIB_bulk}, this effective action may be considered as a power series in the number of derivatives and the string coupling $g_s$. When considering the low energy theory we may restrict ourselves to the leading terms in the derivative expansion. As discussed around equation~\eqref{kappa_10}, the derivative expansion can also be viewed as an expansion in terms of $\alpha'$, although the factors of $\alpha'$ would not be directly visible in our conventions in which the coordinates and metric have been rescaled to make all fields dimensionless. We furthermore again assume to be in the weak coupling limit and restrict ourselves to the leading order in $g_s$, although as we will discuss more closely in section~\ref{sec:SL(2,Z)} there is a tension between the validity of the weak coupling assumption and the presence of $D7$-branes. In fact, this tension will be one of the key motivations to reformulating the theory in the language of F-theory. To summarise, the D-brane actions presented in the following are accurate to leading order in derivatives and in the string coupling, and may receive corrections at higher orders of $\alpha'$ or $g_s$.

The low energy effective theory on the world-volume of a D-brane can be split into two distinct parts. The first part, known as the Dirac-Born-Infeld or DBI action~\cite{Fradkin:1985qd, Leigh:1989jq}, contains the kinetic terms of the world-volume fields mentioned above and describes the couplings to the closed string fields of the NS-NS sector, i.e. the metric, the dilaton and the Kalb-Ramond 2-form $B_2$. We again restrict ourselves to the bosonic part of the DBI action, which for the Abelian case of a single $Dp$-brane with the world-volume $\cW_{p+1}$ is given in the string frame by
\be
S_{DBI} = - \mu_p \int_{\cW_{p+1}} d^{p+1} \xi \   e^{-\Phi} \sqrt{ - \det \left[ \iota^* g - \frac{1}{2\pi} \mathbf{F} \right] } .
\label{S_DBI}
\ee
In this expression we have introduced several new pieces of notation that require clarification. Firstly, $\iota: \cW_{p+1} \hookrightarrow M_{10}$ is shorthand for the embedding of the $Dp$-brane into the 10-dimensional spacetime as described in~\eqref{embedding_brane}. The brane tension is given in our conventions by
\be
\mu_p = 2\pi.
\ee
The field strength $F=dA$ of the brane gauge field has been combined with the $B$-field into the combination
\be
\frac{1}{2\pi} \mathbf{F} = \frac{1}{2\pi} F - \iota^* B_2.
\label{def_calF}
\ee
This combination is chosen in such a way that the invariance of the bulk action~\eqref{S_IIB_democ} under the symmetry transformation $B_2 \rightarrow B_2 + d \chi$ is preserved in the presence of D-branes if it is accompanied by a shift $A \rightarrow A + 2\pi \chi$ of the brane gauge field. Note that this is an additional Abelian gauge symmetry that is distinct from the Yang-Mills gauge symmetry on the D-brane. The determinant appearing in~\eqref{S_DBI} is taken over the world-volume indices of the 2-tensors $\iota^* g$ resp. $\mathbf{F}$.
Finally, let us note that although in our conventions both $A$ and $F=dA$ are dimensionless, many authors work instead in conventions where $A$ and $F$ have dimensions of $\ell_s^{-1}$ and $\ell_s^{-2}$, respectively. In that situation, the dimensionless combination entering the D-brane actions is $2\pi\alpha' \mathbf{F} = 2\pi\alpha' F - \iota^* B_2$~\cite{Blumenhagen:2006ci}. It is important to keep the relative normalisation factor of $\ell_s^2$ in mind when comparing the results of this thesis to the existing literature, in particular to the results of references~\cite{Grimm:2011dj, Grimm:2011tb} and~\cite{Kerstan:2012cy}, which will be heavily used in the later chapters of this thesis.

In general we will of course usually be more interested in the case with non-Abelian gauge symmetry, i.e. with more than one $Dp$-brane in the stack. In this case, the general form of the action is complicated to write down due to the appearance of non-commuting matrix-valued forms, such as the spacetime coordinates $X^m(\xi)$ appearing in the pullback formula
\be
(\iota^* g)_{ab} = g_{mn} \partial_a X^m \partial_b X^n.
\ee
Both these spacetime coordinates and the world-volume gauge field $A$ are represented as Hermitian $N\times N$ matrices for a stack of $N$ branes, and transform in the adjoint of the gauge group as mentioned before~\cite{Tseytlin:1997csa, Myers:1999ps}. In particular, a trace over the gauge indices has to be taken in a manner symmetrised over the various possible field orderings to remove ordering ambiguities~\cite{Tseytlin:1997csa}. We will refrain from writing out the full non-Abelian form of the action in a general background, which was worked out by Myers and can be found in~\cite{Myers:1999ps}. In the following we will always consider backgrounds with D-brane configurations which preserve supersymmetry. As noted by Witten~\cite{Witten:1995im}, the matrices $X^m$ are proportional to the unit matrix in gauge space for such supersymmetric configurations, so that in particular no additional effects from non-commutativity arise. Furthermore, we will focus almost exclusively on the diagonal $U(1)$ lying in the $U(N)=SU(N)\times U(1)$ gauge group of a stack of $N$ branes\footnote{We assume again that the branes are not placed in such a way as to develop symplectic or orthogonal gauge groups.}. The generator associated with this diagonal $U(1)$ is simply an $N\times N$ unit matrix. Hence as long as we consider supersymmetric backgrounds and restrict to switching on field strengths $F$ along the diagonal $U(1)$ we may work with the simpler form~\eqref{S_DBI}, and the only effect from the non-Abelian symmetry that we must keep in mind is an overall factor of $N$ arising from the trace.

The second part of the D-brane action is given by the Chern-Simons (CS) action~\cite{Douglas:1995bn}, which describes the coupling of the brane to the closed string fields of the Ramond-Ramond sector. Restricting as above to the bosonic part and the Abelian case for simplicity, the action is given by
\be
S_{CS} = - \mu_p \int_{\cW_{p+1}} \ch\left(\frac{1}{2\pi} \mathbf{F} \right) \wedge \sqrt{\frac{\hat{A}(R_T)}{\hat{A}(R_N)}} \wedge \sum_{q} C_q ~ ,
\label{S_CS}
\ee
where we follow the conventions of~\cite{Blumenhagen:2006ci}.
Here $\ch\left(1/2\pi \mathbf{F}\right)$ is the Chern character of the gauge bundle of the D-brane, whose curvature is given by the gauge invariant combination of $F$ and $B_2$ defined in~\eqref{def_calF}. $R_T$ and $R_N $ denote the curvature forms of the tangent and normal bundles to the brane world-volume, respectively, while $\hat{A}$ is the A-roof genus. Note that $R_T$ and $R_N$ are dimensionless in our conventions, as the metric and the world-volume coordinates are dimensionless. The characteristic classes used in this expression are defined in appendix~\ref{sec:char_classes}. The integrand in~\eqref{S_CS} is formally a power series involving forms of various degrees. It is of course understood that the integral picks out only the terms of total degree $(p+1)$ from this power series. 
Throughout this thesis we will work under the assumption that all curvature radii can be taken to be large compared to the string length, as is necessary to ensure validity of the supergravity assumption.

As in the DBI case, some modifications, including the introduction of a symmetrised trace over gauge indices, need to be made in order to obtain the full non-Abelian Chern-Simons action~\cite{Myers:1999ps}. However, as discussed above we will restrict ourselves to field strengths corresponding to the diagonal $U(1)\subset U(N)$ and to supersymmetric brane configurations in which the fields describing the embedding of the brane are proportional to the unit matrix. In this case the additional complications introduced by the non-commutativity of the fields in the non-Abelian case are largely absent and we may use the Abelian form~\eqref{S_CS} of the action, keeping in mind to include a factor of $N$ from the gauge trace when considering a stack of $N$ branes.

The introduction of $Dp$-branes breaks some or all of the original $\cN=2$ supersymmetry of type II string theory. Polchinski discovered in 1995~\cite{Polchinski:1995mt} that $Dp$-branes with odd $p$ in type IIB (or even $p$ in type IIA) are stable half-BPS states that leave half of the supersymmetry unbroken and lead to an effective theory with $\cN=1$ supersymmetry. The observation of half-BPS $Dp$-branes with odd $p$ of course matches the fact that precisely the $C_q$ R-R potentials with even $q$ exist in type IIB string theory, and in fact the existence of a set of gauge symmetries with respect to which the $Dp$-branes with odd $p$ carry a conserved charge can be seen as the reason for their stability. While $Dp$-branes with the ``wrong'' values of $p$ can in principle be introduced, they are unstable and break all of the supersymmetry, and such branes will not be considered in this thesis.

When including D-branes in compactifications on Calabi-Yau spaces (or compactifications with non-trivial $B_2$-field backgrounds), further conditions must be satisfied to avoid breaking the supersymmetry of the low energy theory completely. A D-brane in such a compactification will wrap a cycle $\Gamma$ within the compactification space $X_3$. The conditions for unbroken supersymmetry can be translated into a set of geometric calibration conditions for these cycles, which were worked out in~\cite{Becker:1995kb, Bergshoeff:1997kr, Marino:1999af} and depend on the dimension of the cycle $\Gamma$. In this thesis we will mainly be interested in spacetime-filling $D7$-branes and Euclidean $D3$-brane instantons, in which case the internal cycle $\Gamma$ is 4-dimensional. The calibration conditions then require $\Gamma$ to be a holomorphic cycle or in other words a divisor, and there must hold
\begin{align}
d^4 \xi \sqrt{\det \left[\iota^*(g) + \frac{1}{2\pi} F - \iota^*(B_2)\right]} = &\  \frac12 e^{-i\theta} \left[ \iota^*(J)+i\left(\frac{1}{2\pi} F - \iota^*(B_2)\right)\right]^2, \label{calibration_vol}\\
\left[\frac{1}{2\pi} F - \iota^*(B_2)\right]^{2,0} = & \ 0. \label{calibration_F}
\end{align}
In this expression $\iota^*$ denotes the pullback of forms from $X_3$ to the cycle $\Gamma$, and $J$ is the K\"ahler form of $X_3$. The parameter $\theta$ appearing in~\eqref{calibration_vol} controls which linear combination of the original 2 supersymmetry generators remains unbroken. The second condition~\eqref{calibration_F} implies that the internal part of the gauge invariant combination $\mathbf{F}$ defined in~\eqref{def_calF} must be a $(1,1)$-form (see appendix~\ref{sec:conventions}).

In a compactification on a Calabi-Yau orientifold the parameter $\theta$ in~\eqref{calibration_vol} can no longer be chosen freely.
This is due to the fact that the process of orientifolding itself already breaks half of the supersymmetry, so to maintain $\cN=1$ supersymmetry the D-brane placement and the orientifold projection must be correlated to leave precisely the same combination of supercharges unbroken. As discussed in~\cite{Jockers:2004yj}, this implies that $\theta=0$. As long as we fulfill the calibration conditions with $\theta=0$ in each case, we may of course also add multiple stacks of D-branes without breaking supersymmetry completely. The stability of such a configuration with multiple half-BPS D-branes is guaranteed by a cancellation between the mutual gravitational attraction and the repulsion mediated by the R-R form fields~\cite{Polchinski:1995mt}, such that the supersymmetric branes exert no net force upon one another.

As already mentioned in section~\ref{sec:compactifications}, if some of the D-branes are spacetime-filling the consistency of the theory requires compactifying on a Calabi-Yau orientifold rather than on a normal Calabi-Yau manifold. This is due to the fact that D-branes act as sources for the R-R form fields as discussed above, and the net R-R charge on a compact space must be zero. More precisely, the equation of motion for a Ramond-Ramond potential $C_{p+1}$ in the  presence of a single $Dp$-brane and no other sources takes the schematic form
\be
d \ast G_{p+2} = \delta_{Dp},
\label{RR_tadpole}
\ee
where $\delta_{Dp}$ stands for the form Poincar\'e-dual to the world-volume of the brane. If $\delta_{Dp}$ has indices purely in the internal compactification manifold $M_6$, we may integrate~\eqref{RR_tadpole} over a basis of the appropriate homology group of $M_6$. This yields the the condition that the part of $\delta_{Dp}$ that is non-trivial in the cohomology of $M_6$ must be cancelled by additional terms on the right hand side of~\eqref{RR_tadpole} coming from other D-branes and orientifold planes\footnote{As first noted in~\cite{Minasian:1997mm, Witten:1998cd}, the natural mathematical framework to describe D-brane charges and R-R fields is actually K-theory rather than the simpler cohomology of differential forms. Full consistency thus actually requires the cancellation of all R-R charges in K-theory~\cite{Uranga:2000xp}, which can in some cases render models inconsistent even though the R-R charges cancel in cohomology. Keeping this potential caveat in mind, we will stick to the cohomological description and only check for R-R charge cancellation at this level.}. However, if $\delta_{Dp}$ cannot be fully defined as a form on $M_6$, one would have to integrate over cycles lying partially in the non-compact part $M_{1,3}$ of spacetime. In this case Stokes' theorem cannot be applied to show that the left hand side vanishes, so no non-trivial condition is obtained. As $\delta_{Dp}$ has indices in the directions normal to the world-volume of the brane, this argumentation shows that only spacetime-filling D-branes contribute to potentially dangerous R-R tadpoles. In particular, non-trivial $Dp$-brane tadpoles which lead to potential inconsistencies exist only for $p\geq3$. 

The objects cancelling the R-R charge in a Calabi-Yau orientifold compactification are the orientifold planes located at the fixed point locus of the orientifold involution\footnote{Note that charge cancellation alone can also be achieved by including anti-branes, however this breaks supersymmetry completely and will not be considered here.}. The conditions arising from the requirement of vanishing net R-R charge are often referred to as tadpole cancellation conditions. Note that to obtain a well-defined theory after taking the orientifold quotient, D-brane stacks wrapped on cycles that are not pointwise invariant under the orientifold action can only appear together with an image stack such that the pair of stacks together is invariant under the orientifold action.

Although orientifold planes are not dynamical objects and do not give rise to additional world-volume fields as is the case for D-branes, they do couple to the fields of the closed string sector. These couplings can be described by a worldvolume action formally similar to that of the D-branes with world-volume fields set to zero, given for an $Op$-plane\footnote{The objects we consider here and throughout this thesis can be referred to more precisely as $Op^-$-planes. A slightly different choice for the orientifold action on the string degrees of freedom gives rise to $Op^+$ planes which have R-R charges of the same sign as $Dp$-branes~\cite{Denef:2008wq, Blumenhagen:2008zz}, but we will stick to $Op^-$-planes throughout.} with worldvolume $\cW_O$ by~\cite{Morales:1998ux, Scrucca:1999uz}
\begin{align}
S^{Op}_{DBI} = &\  2^{p-4}\  \mu_p \int_{\cW_O} d^{p+1}\xi \ e^{-\Phi} \sqrt{-\det\left[ \iota^*(g)\right]}, \\
S^{Op}_{CS} = &\ 2^{p-4}\  \mu_p \int_{\cW_O} \sqrt{\frac{L(R_T/4)}{L(R_N/4)}} \wedge \sum_q \iota^* C_q.  \label{S_CS_oplane} 
\end{align}
The notation is similar to that used in the actions~\eqref{S_DBI} and~\eqref{S_CS} except for the appearance of the Hirzebruch $L$-polynomial defined in appendix~\ref{sec:char_classes}.

The sum of forms $\sum_{q} \iota^* C_q$ appearing in the Chern-Simons actions of D-branes and O-planes in type IIB theory includes a non-dynamical 10-form $C_{10}$ in addition to the Ramond-Ramond forms of degree 2, 4, 6 and 8 introduced in section~\ref{sec:action_IIB_bulk}. Despite its non-dynamical nature, this additional form has an important consequence in the presence of $D9$-branes~\cite{Polchinski:1995mt}. To cancel the 10-form tadpole, the theory must include an O9-plane, which means that if no lower-dimensional branes are present it is none other than type I string theory. Furthermore, from the relative charge of the Chern-Simons actions we see that for tadpole cancellation the number of $D9$-branes must be exactly 32, which explains the appearance of the gauge group SO(32) in type I string theory. However, in this thesis we will focus on backgrounds without $D9$-branes, so that the 10-form will play no role.

In analogy with the Ramond-Ramond tadpoles discussed above, the theory may also include tadpoles for the NS-NS fields. However, the equations of motion are in a sense more forgiving toward the appearance of NS-NS tadpoles than towards the appearance of uncancelled R-R as discussed above~\cite{Blumenhagen:2006ci}. More precisely, while the equations of motion of the R-R fields were inconsistent in the presence of uncancelled D-brane tadpoles, the NS-NS equations can often still be solved in principle by altering the background field profile. For example, as mentioned in section~\ref{sec:compactifications}, the presence of $D7$-branes requires a warped background spacetime as well as a non-trivial dilaton profile. In this sense the possible NS-NS tadpoles encode the back-reaction of the D-branes, which we will discuss in more detail in section~\ref{sec:SL(2,Z)}.

When trying to build a vacuum satisfying the tadpole cancellation conditions it is important to note that a $Dp$-brane may act as a source not only for $C_{p+1}$, but also for the lower R-R forms\footnote{As shown in~\cite{Myers:1999ps}, additional couplings to the \emph{higher} R-R potentials $C_q,\ q>p+1$, appear in the case where the fields describing the embedding of the brane in the ambient spacetime do not commute. However, as mentioned before we will consider the supersymmetric case where the embedding fields can be taken to be diagonal, so that these additional couplings do not arise.}. As can be easily read off from the Chern-Simons action~\eqref{S_CS}, this occurs if the brane carries a topologically non-trivial flux of the (twisted) world-volume gauge field $\mathbf{F}$ defined in equation~\eqref{def_calF}. This means that the tadpole cancellation conditions will act as a restriction on the world-volume fluxes that can be included in the model. Conversely, switching on non-trivial fluxes may allow adding configurations of lower-dimensional branes which would otherwise be forbidden by the tadpole cancellation conditions. For example, a model with fluxed $D7$-branes may allow for the inclusion of $D5$-branes even without any $O5$-planes being present in the model.

\section{F-theory as a non-perturbative extension of type IIB}
\label{sec:FtheoryAsNonperturbIIB}

\subsection{The SL$(2,\bZ)$ symmetry of type IIB string theory}
\label{sec:SL(2,Z)}

The tadpole cancellation conditions discussed in the previous section are global conditions in the sense that they only require the overall net R-R charge to vanish. However, they allow for charges to be cancelled between orientifold planes and D-branes wrapped on topologically distinct cycles of the compactification manifold. This means that locally there may exist uncancelled R-R charges which affect the profile of the R-R potentials in the spacetime surrounding the branes. 

A particularly interesting and at first sight puzzling case is the profile of the field $C_0$ in the vicinity of a $D7$-brane. Let $\cW_8$ and $\cW'_8$ be the world-volumes of the $D7$-brane and its image brane, respectively. As mentioned in section~\ref{sec:action_IIB_bulk}, the interaction between $C_0$ and a $D7$-brane is most easily described not in terms of $C_0$ itself but in terms of its dual field $C_8$. Using the IIB bulk action~\eqref{S_IIB_democ} and the Chern-Simons action~\eqref{S_CS} of the 7-brane, the equation of motion of $C_8$ reads\footnote{As brane and image brane provide equivalent descriptions of the same physics in the orientifold geometry, the physical action is obtained by adding the actions of brane and image brane and then dividing the result by 2. An additional factor of 1/2 must be taken into account in the Chern-Simons action when deriving these equations of motion in the democratic formulation, see also the discussion in~\cite{Grimm:2011tb}.}
\be
 d\ast G_9 = \frac12 \left[\delta(\cW_8) + \delta(\cW'_8)\right] + \text{(other source terms)}.
 \label{eom_C8}
\ee
In the above $\delta(\cW_8)$ and $\delta(\cW'_8)$ describe the Poincar\'e-dual 2-forms which are sharply localised around the world-volumes of the brane and image brane, respectively. The other source terms in~\eqref{eom_C8} arise from other brane stacks and orientifold planes, at least some of which must be present due to tadpole cancellation. For simplicity we will assume for now that the $D7$-brane on $\cW_8$ is isolated, so that the forms describing the other sources have no support in the vicinity of the $D7$-brane. As long as we consider only a neighborhood of the chosen $D7$-brane, the other source terms may then be safely neglected.  

For the case of $D7$-branes, the transverse space to the brane world-volume is 2-dimensional, and the branes can be viewed as points in this transverse plane. Consider integrating $d C_0$ over a path $\partial S$ lying in this transverse plane and encircling the $D7$-brane once. In order to yield a well-defined path in the orientifold geometry, $\partial S$ must similarly encircle the image brane once. The notation has been chosen to suggest that $\partial S$ is the boundary of a region $S$ in the transverse plane, which by assumption includes both the positions of the brane and the image brane. Using the duality relation $\ast G_9 = G_1 = d C_0$ together with~\eqref{eom_C8} we obtain
\be
\int_{\partial S} d C_0 = \int_{\partial S} \ast G_9 = \int_S d\ast G_9 = \frac12 \int_S \left[\delta(\cW_8) + \delta(\cW'_8)\right] = 1.
\label{monodromy_C0_integrated}
\ee
In other words, the field $C_0$ develops a non-trivial monodromy around the $D7$-brane. This means that the potential $C_0$ can only be defined patchwise, and that the different values must be related by a symmetry transformation of the theory on the overlap of two patches to allow for a consistent interpretation of the background.

A closer inspection of the IIB supergravity action~\eqref{S_IIB_nondemoc} indeed shows that the action enjoys a classical SL($2, \bR$) symmetry which can explain the shift of $C_0$ required by~\eqref{monodromy_C0_integrated}. To make this symmetry manifest, it is helpful to make a change of variables and transform the action into the so-called Einstein frame in which the curvature term has the canonical form. Following~\cite{Blumenhagen:2006ci} we define
\be
\begin{aligned}
g_{E\mu\nu} =&\ e^{-\Phi/2} g_{\mu\nu},&\tau = &\ C_0 + i e^{-\Phi}, \\
\tilde{C}_4 = &\ C_4 - \frac12 B_2\wedge C_2&\quad \tilde{G}_3 =&\ F_3 - \tau H_3.
\end{aligned}
\label{trf_to_einstein_frame}
\ee
In terms of these variables the action~\eqref{S_IIB_nondemoc} takes the form
\be
\begin{aligned}
S_{IIB}& =\  \frac{1}{2\kappa_{10}^2} \int d^{10}x \sqrt{-\det g_E} \,  R_E  - \frac{1}{4\kappa_{10}^2} \int \frac{1}{(\mathrm{Im}\tau)^2} d \tau\wedge\ast_E d\overline{\tau} \\ & \
- \frac{1}{4\kappa_{10}^2} \int \left[\frac{1}{\mathrm{Im}\tau} \tilde{G}_3 \wedge\ast_E \overline{\tilde{G}_3}  +\frac12 G_5\wedge\ast_E G_5  - \frac{1}{2i\mathrm{Im}\tau} \tilde{C}_4 \wedge \tilde{G}_3 \wedge \overline{\tilde{G}_3} \right].
\end{aligned}
\label{S_IIB_Einstein}
\ee
We have explicitly added the subscript $E$ in~\eqref{S_IIB_Einstein} to signify that all metric-dependent quantities are evaluated using the Einstein frame metric $g_E$. An SL$(2,\bR)$ transformation can be parametrised by a matrix
\be
A = \left(\begin{matrix}p & q \\ r & s \end{matrix}\right), \ \text{ with } p,q,r,s \in \bR \text{ and } \det A = ps-qr=1.
\label{def_A_pq}
\ee
It is now straightforward to check that the action~\eqref{S_IIB_Einstein} is invariant under the transformation which is associated with such an SL$(2,\bR)$ matrix $A$ and acts on the Einstein frame fields by~\cite{Polchinski:1998}
\be
\begin{aligned}
\tau\rightarrow &\ \frac{s\tau+r}{q\tau + p}, & \qquad \left(\begin{matrix}B_2 \\ C_2 \end{matrix}\right) \rightarrow &\ A \left(\begin{matrix}B_2 \\ C_2 \end{matrix}\right), \\
g_E \rightarrow &\ g_E, & \tilde{C}_4 \rightarrow &\ \tilde{C}_4.
\end{aligned}
\label{SL2R_trf}
\ee
The Weyl rescaling of the metric performed in moving from the string frame to the Einstein frame does not change the action of the Hodge star operator~\eqref{Hodge_star} when acting on 5-forms. In particular, the form of the self-duality constraint is not changed when moving from one frame to the other
\be
G_5 = \ast G_5 = \ast_E G_5,
\ee
so that the self-duality constraint is clearly also invariant under the SL$(2,\bR)$ transformations.
Note that from~\eqref{SL2R_trf} the imaginary part of $\tau$ transforms as
\be
\mathrm{Im}\tau \rightarrow \frac{1}{\left| q\tau+p \right|^2} \mathrm{Im}\tau,
\label{trf_Im_tau}
\ee
so that the sign of $\mathrm{Im}\tau$ remains fixed even though the magnitude might change. This property is important because the imaginary part of $\tau$ is none other than the inverse of the string coupling $g_s = e^\Phi$, which must remain positive for consistency. 

Upon quantisation of the theory, the classical SL$(2,\bR)$ symmetry is broken non-perturbatively to the discrete subgroup SL$(2,\bZ)$. This can be seen in various ways. For example, the transformation~\eqref{SL2R_trf} takes a fundamental string into an object having $p$ units of NS-NS charge under $B_2$ and $q$ units of R-R charge with respect to $C_2$, which is conventionally referred to as a $(p,q)$-string. Due to the Dirac quantisation condition, the charges $p$ and $q$ must be integers\footnote{Note that the integers $p$ and $q$ (as well as $r$ and $s$) are relatively coprime due to the condition $ps-qr=1$ and B\'ezout's lemma.}~\cite{Polchinski:1998, Becker:2007zj}. Furthermore, D-brane instantons contribute to the low-energy action through terms involving factors of the schematic form $\exp(iS_{CS})$, which remain invariant under the SL$(2,\bZ)\subset\text{SL}(2,\bR)$ subgroup~\cite{Blumenhagen:2009qh}. The transformation necessary to explain the observed monodromy~\eqref{monodromy_C0_integrated} is simply the special case of the general SL$(2,\bZ)$ transformation~\eqref{SL2R_trf} obtained by taking $q=0$ and $p=r=s=1$, which yields
\be
C_0 \rightarrow C_0+1, \qquad C_2\rightarrow C_2+B_2.
\label{monodromy_C0}
\ee

As a side note let us point out that the transformation~\eqref{monodromy_C0} can also be understood as a Ramond-Ramond gauge transformation of the type~\eqref{RR_gaugetrf_bulk} with $\Lambda_0=1$ and $\Lambda_p=0,\ p>0$. At first sight it may seem that the inclusion of D-branes breaks the R-R gauge symmetry discussed in section~\ref{sec:action_IIB_bulk} due to the appearance of the non-derivative couplings in the Chern-Simons action~\eqref{S_CS}. However, a closer inspection reveals that in a Lorentz invariant and tadpole-free brane and flux configuration the R-R gauge symmetry is actually maintained due to a cancellation between the variations of the actions of different brane stacks. To see this, it is necessary to carefully distinguish between the positive and negative parity components of $B_2$ introduced in section~\ref{sec:compactifications}. As discussed there, any possible positive parity $B$-field contribution $B_+$ must be closed. These contributions may therefore be absorbed into the gauge parameters $\Lambda_p$ appearing in~\eqref{RR_gaugetrf_bulk}, so that the gauge transformation may be written as
\be
\delta C_p = \left[ \mathbf{\Lambda}\ e^{B_-} \right]_p.
\ee
Under this transformation, the Chern-Simons action~\eqref{S_CS} of a single $Dp$-brane shifts by
\be
\delta S_{CS} = -\mu_p \int_{\cW_{p+1}} \iota^* \mathbf{\Lambda} \wedge \exp\left[\frac{1}{2\pi} F - \iota^* B_+\right].
\label{gauge_shift_S_CS}
\ee
Because the negative parity form $B_-$ does not restrict to the world-volume of an orientifold plane one similarly obtains
\be
\delta S^{Op}_{CS} = 2^{p-4}\  \mu_p \int_{\cW_O} \iota^* \mathbf{\Lambda}.
\label{gauge_shift_S_CS_Op}
\ee

It is not obvious that the variation of the action vanishes, and indeed we will see that this relies on a cancellation between the contributions from several branes which only occurs in a Lorentz invariant and tadpole-free vacuum. First note that in order to preserve Lorentz invariance all D-branes must be spacetime-filling\footnote{The exception is given by instantonic D-branes lying purely in the internal manifold $M_6$. As before, the presence of such instantons non-perturbatively breaks the classical shift symmetries generated by forms $\Lambda_p$ with non-trivial pullback to $M_6$ to a discrete subgroup.}, such that they are all related by the tadpole cancellation conditions. By the same reasoning any fluxes $ F/2\pi - \iota^* B_+$ must be taken to be forms lying fully in the internal space $M_6$. This implies that only terms involving $\Lambda_p$, $p\geq4$ can contribute to a nonzero value in~\eqref{gauge_shift_S_CS}, as all dependence on the external coordinates of $M_{1,3}$ must come from the gauge parameters $\Lambda_p$.

After summing up the shifts~\eqref{gauge_shift_S_CS} and~\eqref{gauge_shift_S_CS_Op} of the actions of all the D-branes and orientifold planes present in the model, one finds that the overall coefficients of the $\Lambda_p$ terms with $p\geq4$ are precisely the combinations that are required to vanish due to the R-R tadpole cancellation conditions\footnote{More precisely, the coefficients of the $\Lambda_p$ only vanish in cohomology, however this is sufficient to ensure that the total associated gauge shift vanishes due to the fact that the $\Lambda_p$ are closed.}. In order to verify this result it is important to note that, as shown in~\cite{Blumenhagen:2008zz}, the contribution of $B_-$ drops out of the tadpole cancellation constraints so that they involve precisely the combination $ F/2\pi - \iota^* B_+$ that also appears in~\eqref{gauge_shift_S_CS}. 

While we have just seen that in a consistent compactification to a Lorentz invariant 4-dimensional theory the R-R gauge transformations survive as a symmetry and can explain the monodromy~\eqref{monodromy_C0}, consistency in an uncompactified 10-dimensional setting requires the SL($2,\bZ$) symmetry discussed above. This is one of the main reasons why, as argued in~\cite{Hull:1994ys, Schwarz:1995du, Vafa:1996xn}, the SL($2,\bZ$) symmetry is expected to survive as a symmetry of the full quantum theory and is not just viewed as an accidental symmetry of the IIB bulk action~\eqref{S_IIB_Einstein}. 

Note that equation~\eqref{trf_Im_tau} shows that for $q\neq 0$ the SL($2,\bZ$) duality relates backgrounds with large and small string couplings. As no intrinsically non-perturbative description of string theory is known, it is not possible to solve the strongly coupled theory explicitly to check the SL($2,\bZ$) symmetry directly. However, one prediction of the conjectured SL($2,\bZ$) duality which can be verified using perturbative means is the existence of ($p,q$)-strings with relatively co-prime $p$ and $q$. Such states were constructed in the weakly coupled theory in~\cite{Schwarz:1995dk, Witten:1995im}, where they can be seen as bound states between fundamental strings and D-strings. As these states are BPS states, they must continue to exist in the strongly coupled theory by supersymmetry (see e.g. the discussion in~\cite{Polchinski:1998}).

If we denote the coupling of a general 1-brane or string state to $B_2$ and $C_2$ by means of a charge vector $(q_1, q_2)^\mathrm{T}$, we may also interpret the SL($2,\bZ$) transformation~\eqref{SL2R_trf} as acting on the charge vector instead of on the fields $C_2$ and $B_2$. The transformation $A_{[p,q]}$ of equation~\eqref{def_A_pq}, which takes a fundamental string with charges $(1,0)$ into a $(p,q)$ string\footnote{Although specifying $p$ and $q$ does not fully fix the integers $r$ and $s$ appearing in the matrix $A_{p,q}$ of equation~\eqref{def_A_pq}, the charges $p$ and $q$ encode the full physical information and the ambiguity in $r$ and $s$ drops out of physically observable quantities\cite{Douglas:1996du}.}, acts on a general charge vector by
\be
\left(\begin{matrix}q_1 \\ q_2 \end{matrix}\right) \rightarrow A_{[p,q]}^\mathrm{T}\left(\begin{matrix}q_1 \\ q_2 \end{matrix}\right) = \left(\begin{matrix}p & r \\ q & s \end{matrix}\right)\left(\begin{matrix}q_1 \\ q_2 \end{matrix}\right).
\ee

As we have just seen, the SL($2,\bZ$) transformation of the fields $B_2$ and $C_2$ can be interpreted alternatively as a transformation of the charged objects these fields couple to\footnote{This becomes particularly clear if one views D-branes simply as solitonic configurations of the background fields that are being transformed\cite{Gaberdiel:1997ud}.}, leading to the appearance of a whole SL($2,\bZ$) family of 1-branes. By duality there exists a similar family of 5-branes, including as special cases the familiar $D5$- and $NS5$-branes. As the SL($2,\bZ$) transformation also acts on the field $\tau$ coupling naturally to $D(-1)$-branes, the same consideration shows that there must be families of $D(-1)$-branes and by duality also $D7$-branes labelled by integers ($p,q$). This can also be seen from the point of view of charge conservation: different $(p,q)$-strings carry different charges, which implies that the branes that they end on must have somewhat different world-volume theories to account for these charges~\cite{Strominger:1995ac, Douglas:1996du}. 7-branes on which $(p,q)$-strings can end are will be referred to as $[p,q]$ 7-branes in the following.

The monodromy~\eqref{monodromy_C0} caused by an ordinary $D7$-brane can be described by the monodromy matrix 
\be
M_{[1,0]} = \left( \begin{matrix} 1&1\\ 0&1\end{matrix} \right),
\ee
which acts on the fields in a similar manner as the matrix $A$ in equation~\eqref{SL2R_trf}.
The monodromy of a more general $[p,q]$ 7-brane can be evaluated most easily by first transforming into a frame in which the brane is an ordinary $D7$-brane using $\left(A_{[p,q]}^T\right)^{-1}$, applying the monodromy $M_{[1,0]}$, and then transforming back into the original SL(2$,\bZ$) frame~\cite{Douglas:1996du}. The resulting monodromy matrix is given by
\be
M_{[p,q]} = A_{[p,q]}^T \ M_{[1,0]} \  \left(A_{[p,q]}^T\right)^{-1} = \left( \begin{matrix} 1-pq & p^2 \\ -q^2 & 1+pq\end{matrix} \right).
\label{monodromy_pq}
\ee
In a model containing only a single type of $[p,q]$ brane, one may of course always transform into a frame where all the branes are ordinary $D7$-branes. However, this is clearly no longer possible branes of several different $[p,q]$-types are present. In this situation one speaks of mutually non-local 7-branes, and the non-trivial monodromy~\eqref{monodromy_pq} must be taken seriously. As discussed in~\cite{Gaberdiel:1997ud}, a consistent interpretation of these monodromies requires taking into account more exotic objects like string junctions or multi-pronged strings. Remarkably, this leads to the appearance of exceptional gauge groups, which are not visible in perturbative type IIB string theory involving only $[1,0]$ branes.
As we will see, monodromies of the form~\eqref{monodromy_pq} and exceptional gauge groups arise in a very elegant manner in the F-theory formulation.

\subsection{The axio-dilaton as the modular parameter of a torus}
\label{sec:axio_dilaton_from_torus}

The discussion of subsection~\ref{sec:SL(2,Z)} strongly suggests that the SL(2,$\bZ$) symmetry is not just an accidental perturbative symmetry of type IIB supergravity, but is instead a crucial ingredient of the full quantum theory. This naturally leads to the question of whether there might exist an alternative formulation of the theory in which the essential role of the SL(2,$\bZ$) symmetry is manifest. If so, it might be expected that this formulation might be more suited towards understanding the interesting non-perturbative phenomena associated with the symmetry.

By the mid 1990s, the study of supergravity theories obtained by compactifications of string theory had turned up a number of symmetries or dualities which could be connected to geometric properties of the compactification manifold. Examples of such dualities include T-duality, the S-duality of toroidally compactified heterotic strings or the type II analogue dubbed U-duality\footnote{The various symmetries are reviewed e.g. in~\cite{Giveon:1994fu, Hull:1994ys, Schwarz:1995du, Witten:1995ex, Polchinski:1998}.}. Guided by these observations, string theoreticians soon started trying to interpret the 10-dimensional type IIB SL(2,$\bZ$) symmetry as arising from geometric properties of a dimensional reduction of some higher-dimensional theory.

A clue towards such a geometrical reformulation lies in the fact that the group SL(2,$\bZ$) is identical to the symmetry group PSL(2,$\bZ$) of the 2-dimensional torus, up to an identification of matrices differing only by an overall minus sign\footnote{The extra generator present in SL(2,$\bZ$) is identified in the type IIB setting as being associated with a reversal of the string orientation.}. Due to its importance in the following, let us briefly review the key features of the torus geometry as discussed e.g. in~\cite{Font:2005td}. A flat 2-torus can be seen as a lattice in the complex plane, defined by two period vectors $\omega_1,\omega_2$ and endowed with the induced Euclidean metric. The lattice is left invariant by an SL(2,$\bZ$) transformation acting on the period vectors by
\be
\left(\begin{matrix}\omega_1 \\ \omega_2 \end{matrix}\right) \rightarrow \ A \left(\begin{matrix}\omega_1 \\ \omega_2 \end{matrix}\right), \qquad A = \left(\begin{matrix} p&q\\r&s \end{matrix}\right)\in \text{SL}(2,\bZ).
\label{trf_torus_period_vectors}
\ee
The torus geometry can be characterised by two parameters: the K\"ahler modulus $\rho$ giving the area of the lattice unit cell and the complex structure or modular parameter $\tau_T = \frac{\omega_2}{\omega_1}$, which can be interpreted as the ratio of the lengths of the two fundamental cycles on the torus. 

Using~\eqref{trf_torus_period_vectors} it is straightforward to check that $\tau_T$ transforms as
\be
\tau_T \rightarrow \ \frac{s\ \tau_T+r}{q\ \tau_T + p},
\ee
matching the transformation~\eqref{SL2R_trf} of the IIB axio-dilaton field $\tau$. It is therefore very tempting to identify the axio-dilaton with the complex structure $\tau_T$ of a suitable torus, and we will drop the subscript $T$ in the following. This identification suggests that type IIB string theory might be obtained by compactifying a suitable theory on a torus. 

At the level of low energy effective actions, there is not much choice for this candidate theory. In fact, as explained in~\cite{Polchinski:1998}, an analysis of the dimensionalities of spinor representations shows that a supergravity theory in $D$-dimensional Lorentzian space with $D>10$ can exist only for $D=11$. The 11-dimensional supergravity theory was explicitly constructed in~\cite{Cremmer:1978km}, and attempts were made to understand it as forming the long-wavelength limit of a theory of quantised membranes. These attempts were of course inspired by the familiar relationship between string theory and $D=10$ supergravity~\cite{Bergshoeff:1987cm}. The discovery of the various string dualities served as further evidence for the existence of a quantum theory dubbed M-theory~\cite{Witten:1995em} underlying $D=11$ supergravity, although the fundamental formulation of the theory remains elusive.

Indeed, as first studied in~\cite{Aspinwall:1995fw, Schwarz:1995dk}, it is possible to establish a correspondence between M-theory compactified on a 2-torus $T^2$ and type IIB string theory. More precisely, the compactification of M-theory on the torus first yields type IIA theory compactified on an $S^1$ to 9 dimensions~\cite{Witten:1995ex}, which in turn is T-dual to type IIB compactified on a dual $S^1$. This correspondence, which we will use extensively in the remainder of this thesis, gives a beautiful geometric explanation for the observed SL(2,$\bZ$) symmetry of the compactified IIB theory. 

A drawback of the correspondence as described above is that the original 10-dimensional IIB supergravity is only reached indirectly by decompactification in the limit in which the radius of the circle on which the IIB theory is compactified goes to infinity. By T-duality, this limit corresponds on the IIA side to the case where the radius of the compactification circle goes to zero. This in turn implies that the area of the torus $T^2$ of the M-theory compactification vanishes, so that strictly speaking the decompactification limit on the IIB side corresponds to a singular limit on the M-theory side. These observations make it plausible to consider the existence of a 12-dimensional theory which would be directly connected to IIB by dimensional reduction on a torus, just as M-theory and type IIA string theory are directly connected via compactification on a circle.

Some hints towards such an underlying 12-dimensional theory were uncovered in~\cite{Vafa:1996xn} by Vafa, who gave the theory the name F-theory.
By considering the world-volume theory of $[p,q]$-strings, Vafa was able to provide evidence~\cite{Vafa:1996xn} that F-theory might admit a reformulation in terms of objects with a 4-dimensional world-volume of signature (2,2) instead of strings with a world-sheet of signature (1,1). Such a theory would naturally be defined in a spacetime of signature (10,2)\footnote{Intriguingly, the appearance of a second time direction might offer a way around the 'no-go theorem' mentioned above restricting supergravity theories to $D\leq11$, because spacetimes of signature (10,2) admit 32-component Majorana-Weyl spinors~\cite{Polchinski:1998}. Nevertheless, no proper 12-dimensional supergravity theory has been formulated yet, although the possibility of a supersymmetric theory in (10,2) spacetime has been explored using the different ideas of 2-time physics in~\cite{Bars:2010zw}.}. Vafa furthermore argued that the BRST quantisation conditions would ensure that the spectrum of the theory matches the spectrum of 10-dimensional IIB string theory, and that the moduli space of theories that would be obtained by compactifying to (9,1) dimensions on a (1,1) space would be the same as if one had compactified on an Euclidean torus\footnote{More generally, on a manifold admitting a holomorphic elliptic fibration~\cite{Morrison:1996na, Morrison:1996pp}.}. Despite these hints, a more thorough understanding of the fundamental nature of F-theory remains elusive. For this reason, and the duality with M-theory compactifications outlined remains the most fruitful approach towards the construction of lower-dimensional F-theory vacua.

\subsection{D-brane back-reaction, varying dilaton profiles and elliptic fibrations}
\label{sec:ell_fibration_D-brane_backreaction}
As already noted by Vafa in~\cite{Vafa:1996xn}, the F-theory construction of IIB vacua carries more advantages than simply offering a natural explanation of the SL(2$,\bZ$) symmetry. One such advantage, which will play an important role in the later parts of this thesis, is that F-theory compactifications naturally give rise to exceptional gauge groups. As mentioned above, exceptional groups are visible in the perturbative IIB approach only through the complicated dynamics of $[p,q]$-string junctions. A closely related but even more immediate advantage is that the F-theory formulation is well-suited to constructing consistent compactifications with a varying axio-dilaton field. This is highly relevant in compactifications involving D-branes due to the back-reaction on the fields of the bulk or closed string sector. 

To illustrate this point, let us first consider the D-brane back-reaction from the original type IIB perspective, following the discussion in~\cite{Weigand:2010wm}. Consider a $Dp$-brane spanning the directions $x^0,\ldots,x^p$ in 10-dimensional Minkowski space. By symmetry, finding a solution for the bulk fields in the presence of this brane reduces to a problem in the $9-p$ spatial dimensions normal to the brane. In this normal space, the equations of motion of the bulk fields take the form of a Poisson equation with a source term at the origin. The solutions to the Poisson equation have significantly different dependence on the radial distance $r$ to the brane depending on the codimension $n=9-p$ of the brane. If the codimension $n$ is larger than 2, corresponding to the case $p<7$, the solutions of the Poisson equation vanish asymptotically away from the brane as some power of $1/r$. In fact, an explicit solution of the supergravity equations of motion describing a $Dp$-brane with $p<7$ was constructed in~\cite{Horowitz:1991cd}, in which it was shown that the spacetime metric as well as the dilaton and R-R fields depend on the radial coordinate only through the function
\be
H_p = 1+\frac{\text{const.}}{r^{7-p}}.
\label{def_Hp}
\ee 
The constant appearing in~\eqref{def_Hp} is related to the overall charge of the $Dp$-brane at the origin and the asymptotic value of the dilaton, and controls the size of the region in which the presence of the D-brane noticeably affects the surrounding geometry. Note that the non-trivial part of~\eqref{def_Hp} decays fast enough that asymptotically away from the brane the field configuration approaches flat Minkowski space with a constant dilaton field. This asymptotic value of the dilaton field is a modulus of the solution and can be chosen to be arbitrarily small. 

The behaviour of the dilaton field is particularly important, because as mentioned before it controls the magnitude of the string coupling $g_s = e^\Phi$. Clearly, if this coupling becomes too large, the perturbative description of string theory can no longer be trusted. In the case of $p<7$ discussed above, the coupling can be kept small everywhere except for at most a finite region in the vicinity of the brane. In this sense, it is possible to consistently include $Dp$-branes with $p<7$ in a perturbative setting, at least in compactifications in which the volume of the internal space is large enough that the back-reaction can be considered negligible on the majority of the compactification manifold.

This situation changes significantly in the case of $p=7$, in which case the space normal to the brane is only 2-dimensional. Essentially, the problem stems from the fact that solutions of the Poisson equation in 2 dimensions decay only logarithmically with the distance to the brane. If one introduces a complex coordinate $z=x^8+ix^9$ on the normal plane, it was shown in~\cite{Greene:1989ya} that supersymmetric solutions for the axio-dilaton $\tau = C_0 + i e^{-\Phi}$ which saturate a BPS bound for the energy are holomorphic\footnote{Anti-holomorphic solutions would carry the same energy, but would correspond to the opposite supersymmetry generator remaining unbroken. By convention the holomorphic solution for $\tau$ is associated with a $D7$-brane, while the anti-holomorphic solution describes an anti-$D7$-brane.} in $z$. The monodromy~\eqref{monodromy_C0_integrated} of the real part of $\tau$ around a $D7$-brane located at $z=z_0$ thus implies that the solution can be written locally in a neighborhood of the brane as
\be
\tau(z) = \tau_0 + \frac{1}{2\pi i} \log (z-z_0) + \mathcal{O}(z-z_0).
\label{expansion_tau}
\ee
The specific form of the solution and the monodromies that appear were analysed in much more detail in~\cite{Greene:1989ya, Gibbons:1995vg,  Bergshoeff:2006jj, Braun:2008ua}. In contrast to the case of $p<7$ considered previously, the appearance of non-trivial monodromies or branch cuts breaks the rotational invariance in the directions normal to the brane. Note that the string coupling $g_s = 1/\text{Im}(\tau)$ actually vanishes at the position of the brane, which implies that gravity does not couple to the energy density of the D-brane itself. However, the non-trivial $\tau$-field gives rise to an energy density which can be shown to give rise to a non-vanishing deficit angle in the limit $r = |z|\rightarrow \infty$, even though asymptotically far away from the brane the spacetime is locally flat. Furthermore, the asymptotic value of the string coupling in the presence of a single $D7$-brane\footnote{A small asymptotic string coupling can be achieved in more complicated configurations involving at least 6 D-branes of varying $[p,q]$-types~\cite{Bergshoeff:2006jj}.} is fixed to the non-perturbative value $g_s = \frac{2}{\sqrt{3}}$.

The discussion above means that it is in general not possible neglect the effect of $D7$-branes at large distances. However, there is instead a region in the immediate vicinity of the brane in which the string coupling is small and spacetime is approximately flat with no deficit angle~\cite{Braun:2008ua}. The size of this region is related to the exponential of the parameter $\tau_0$ appearing in~\eqref{expansion_tau}. If there is only a single $D7$-brane present, this parameter can be chosen so that the region of weak coupling is arbitrarily large. However, in a compactification with multiple branes and orientifold planes present it is clear that generically the profile of $\tau$ will vary non-trivially over the compactification manifold and there will be regions in which the string coupling takes on non-perturbative values. 

As we have seen, when constructing vacua of type IIB string theory including $D7$-branes it is necessary to take the back-reaction into account and in particular to consider solutions with varying axio-dilaton profile. Although such solutions of IIB supergravity can be constructed in 10 dimensions~\cite{Bergshoeff:2006jj}, the problem becomes even harder when working within a non-trivial compactification manifold. Fortunately, the F-theory construction via the duality with an M-theory compactification on a torus gives a much better handle on this problem. Recall from subsection~\ref{sec:axio_dilaton_from_torus} that in this approach the axio-dilaton of type IIB is identified with the modular parameter of the torus. Roughly speaking, IIB compactifications with the axio-dilaton varying over the internal space\footnote{For Lorentz invariance all D-branes are taken to be spacetime-filling, so that the normal space to the branes lies purely in the compactification manifold.} therefore correspond to M-theory compactifications on manifolds in which the shape of the torus varies as a function of the other coordinates. 

To make this correspondence more precise, let us follow~\cite{Denef:2008wq} and consider compactifying M-theory on a space formed by attaching a torus on to some 9-dimensional space $M_9$. This situation can be described by the metric ansatz
\be
ds^2_M= \frac{v}{\tau_2} \left((dx +\tau_1 dy)^2 + \tau_2^2 dy^2\right) + ds^2_9.
\ee
Here $v$ describes the overall volume while $\tau = \tau_1+i \tau_2$ is the complex structure of the torus parametrised by coordinates $x,y$. For a fixed point on $M_9$, the equations $y=0$ and $x=0$ define two 1-cycles on the torus. We denote these cycles as the A- resp. B-cycle and their lengths in units of $\ell_s$ by $L_A$ resp. $L_B$. By dimensional reduction along the A-cycle, this theory is dual to type IIA theory on a 10-dimensional space with the metric~\cite{Denef:2008wq}
\be
ds^2_{IIA} = \frac{\sqrt{v}}{L\sqrt{\tau_2}} \left( v\tau_2 dy^2 + ds^2_9 \right).
\ee
The overall scale of the IIA metric is controlled by the parameter $L$, which can be chosen freely in the reduction from M-theory to IIA but which also appears in the expression for the resulting type IIA string coupling
\be
g_{IIA} = L\left( \frac{v}{L^2\tau_2} \right)^{\frac34}.
\ee
In the reduction, the real part of the complex structure of the M-theory torus gives rise to the type IIA 1-form potential, $C_1 = \tau_1 dy$.

The next step is to perform a T-duality along the B-cycle. Under this duality, $C_0$ is obtained from the component of $C_1$ along the cycle direction, which is just $\tau_1$. Furthermore, the coupling constants are related by
\be
g_{IIB} = \frac{1}{L_B^{IIA}} g_{IIA} = \frac{\sqrt{L}}{v^{3/4} \tau_2^{1/4}} g_{IIA} = \frac{1}{\tau_2},
\ee
just as required to be able to identify the IIB axio-dilaton with the modular parameter of the M-theory torus. The form of the (string frame) metric on the IIB side can be deduced from the fact that the length $L_B^{IIB}$ of the B-cycle measured in the IIB frame is related to the corresponding length measured on the IIA side by $L_B^{IIB} = 1 / L_B^{IIA}$, which yields
\be
ds^2_{IIB} = \frac{\sqrt{v g_{IIB}}}{L} \left( \frac{L^2}{v^2} dy^2 + ds_9^2 \right).
\ee
Finally, after rescaling into the Einstein frame and choosing $L=\sqrt{v}$ this can be brought to the form
\be
ds^2_{IIB} = \frac{1}{v} dy^2 + ds_9^2. 
\ee
A striking feature of this result is that the $y$-direction, which originated from a torus cycle on the M-theory side, can be decompactified by taking the limit $v\rightarrow 0$. Moreover, this procedure works in a Poincar\'e invariant manner, in the sense that if $M_9 = B_d \times M_{1,8-d}$ is the product of some $d$-dimensional internal space $B_d$ with $(9-d)$-dimensional Minkowski space $M_{1,8-d}$ we obtain type IIB on $B_d \times M_{1,9-d}$ in the limit $v\rightarrow 0$. This observation implies that we can construct phenomenologically interesting IIB vacua involving 4-dimensional Minkowski space\footnote{More generally, as mentioned in subsection~\ref{sec:compactifications}, we must consider warped products of Minkowski space and the compactification manifold in the presence of non-trivial sources like fluxes or D-branes~\cite{Strominger:1986uh, Becker:1996gj, Dasgupta:1999ss}. Remarkably, the various warp factors cancel out in precisely the right manner that 4-dimensional Poincar\'e invariance is maintained even though the 4 dimensions do not share a common origin on the M-theory side~\cite{Denef:2008wq}.} by starting with a compactification of M-theory to 3 dimensions on an 8-dimensional compact space!

It is clear that the entire procedure outlined above goes through unchanged if the parameters $\tau$ and $v$ are allowed to vary as functions of the (local) coordinates of $M_9$, i.e. if we consider a torus fibration instead of a direct product. This means that as mentioned before type IIB vacua with a varying axio-dilaton profile can indeed be constructed via a fiberwise duality with the M-theory compactification on a non-trivial torus fibration. Furthermore, supersymmetry is preserved both under the reduction from M-theory to type IIA as well as under T-duality. More precisely, to obtain a 4-dimensional theory with $\cN=1$ supersymmetry the M-theory reduction must be chosen in such a way as to lead to an $\cN=2$ theory, as can be seen by a simple counting of the supercharges~\cite{Polchinski:1998}. A 3-dimensional theory with $\cN=2$ supersymmetry is obtained by compactifying on a Calabi-Yau fourfold~\cite{Witten:1996bn}, or a warped generalisation in the presence of fluxes~\cite{Becker:1996gj, Dasgupta:1999ss}. Under certain conditions this theory may be related to a 4-dimensional theory with $\cN=1$ supersymmetry~\cite{Witten:1996bn, Denef:2008wq, Grimm:2010ks}. As mentioned before, a necessary condition for  supersymmetry on the IIB side is that the axio-dilaton is a holomorphic function. In the M-theory language, the fact that the complex structure of the torus varies holomorphically over the base of the fibration means that we are actually considering an elliptic fibration in which the torus may be described as a holomorphic elliptic curve~\cite{Vafa:1996xn, Morrison:1996na, Morrison:1996pp}. In turn, this means that the powerful mathematical tools of algebraic geometry are available to construct and describe the compactification manifolds. It is this fact which encodes much of the power of the F-theory reformulation, as constructing elliptically fibered Calabi-Yau fourfolds is much simpler than attempting to directly construct consistent compactifications with varying dilation profiles on the type IIB side.

To illustrate the power of the F-theoretic description, let us briefly consider as an example the compactification of F-theory on an elliptically fibered K3 manifold. This model was first discussed in~\cite{Vafa:1996xn} and is closely related to the 4-dimensional cosmic string solution of~\cite{Greene:1989ya}. The torus fiber is described by an elliptic curve defined as the vanishing locus of cubic a polynomial of the form
\be
P = y^2 - x^3 - f x - g = 0.
\label{weierstrass_rough}
\ee
In this language, the shape of the torus is controlled by the parameters $f$ and $g$. A non-trivial fibration is obtained by letting $f$ and $g$ vary over the base manifold, which we take to be the projective plane $\bP^1$. The requirement that the total space of the fibration be Calabi-Yau (in other words, a K3 surface) can be shown to imply that $f$ and $g$ must be taken to be polynomials of degree 8 and 12 respectively in the coordinates of the base $\bP^1$~\cite{Denef:2008wq}. 

When analysing the behaviour of $\tau$ in section~\ref{sec:axio_dilaton_from_torus}, we found that in the vicinity of an ordinary $D7$-brane $\tau$ approaches $i\infty$. As $\tau$ describes the ratio of two periods of the torus, this corresponds to a location in which one of the 1-cycles of the torus shrinks to zero size. Such a degeneration of the torus described by the elliptic curve~\eqref{weierstrass_rough} occurs at the points on the base where the so-called discriminant
\be
\Delta = 4 f^3 + 27 g^2
\ee
vanishes~\cite{Denef:2008wq}. By the Calabi-Yau condition $\Delta$ is a polynomial of degree 24 on the base, so there are in general 24 such degeneration points. Locally around each such point one can use the SL(2,$\bZ$) invariance to go into a frame in which $\tau$ takes the form~\eqref{expansion_tau}. However, this is in general not possible globally, so that although any one of the zeroes of $\Delta$ can be chosen to describe the location of an ordinary $D7$-brane, the other zeroes will describe branes of different $[p,q]$ types~\cite{Vafa:1996xn}. A careful analysis of the monodromies shows that the interplay between these branes of different types is such that some combinations mimic the behaviour of IIB orientifold planes and one indeed obtains a consistent compactification, even though naively the presence of 24 7-branes in the compact K3 surface might have been expected to lead to inconsistencies~\cite{Denef:2008wq}. While such configurations are in general very difficult to describe in the type IIB language (see e.g.~\cite{Bershadsky:1995sp}), the F-theory description above gives a much more powerful handle and only requires specifying the polynomials $f$ and $g$! In the following sections, we will review some of the key features of elliptically fibered Calabi-Yau manifolds which elucidate the description of D-branes in this language and discuss how some of the configurations constructed in this manner may be understood from a perturbative IIB perspective.

\section{Introduction to constructing F-theory vacua}
\label{sec:intro_to_ftheory}

\subsection{Elliptically fibered Calabi-Yaus and the Weierstrass form}
\label{sec:elliptic_CalabiYaus}
Elliptic surfaces and algebraic geometry in general are extremely rich mathematical subjects with long histories, and in the following we will present only a selection of important facts following the discussion in~\cite{Esole:2011sm, Weigand:2010wm}. Some important definitions regarding Calabi-Yau manifolds constructed as hypersurfaces of toric ambient spaces can be found in appendix~\ref{sec:app_maths}, but we refer to text books such as~\cite{Hartshorne:1977, Griffiths:1978, Busam:2005} for proofs and a far more complete discussion of the subject.

An elliptic curve can be defined as an algebraic variety of genus one, which means that it can be represented as the vanishing locus of a suitable set of polynomials in a (weighted) projective space. While an elliptic curve can be represented in a variety of different ways, the most important for many of our purposes will be the so-called Weierstrass form. In this form, the elliptic curve is given by a degree 6 hypersurface in the weighted projective space $\bP_{2,3,1}$. The subscript indicates that the homogeneous coordinates $x,y,z$ transform with weights 2, 3 and 1, respectively, under the projective rescaling\footnote{Further details on the definitions and notation used can be found in appendix~\ref{sec:projective_varieties}.}. By suitable coordinate redefinitions, every polynomial of degree 6 in this space may be brought to the form
\be
P_W = y^2 - x^3 - f x z^4 - g z^6.
\label{Weierstrass_form}
\ee
As a consequence of the implicit function theorem, a surface defined as the vanishing locus of a function is singular at most at the points where the function and its first derivative vanish simultaneously\footnote{Note that the ambient space $\bP_{2,3,1}$ is itself singular, which could potentially lead to further singularities. However, the curve~\eqref{Weierstrass_form} misses these singularities of the ambient space regardless of the choices of $f$ and $g$, so no further complications arise~\cite{Braun:2010ff}.}. Evaluating the condition $P_W = d P_W = 0$ for the Weierstrass polynomial~\eqref{Weierstrass_form} shows that such a singular point exists if and only if the discriminant vanishes~\cite{Weigand:2010wm}, i.e. if
\be
\Delta = 4 f^3 + 27 g^2 = 0.
\label{discriminant_vanishing}
\ee

An elliptic curve defined in the manner above can be endowed with an addition operation in a canonical manner, which gives the curve the 
structure of an Abelian group. The distinguished zero element with respect to this addition is the point $z=0$. Note that the condition~\eqref{discriminant_vanishing} only tells us that the curve has a singularity, but not where on the curve the singularity lies. However, it is easy to see that the point $z=0=P_W$ is never a singular point, regardless of the choice of $f$ and $g$. As we will later be interested mainly in possible singularities of the elliptic curve, it will sometimes be convenient to work in a coordinate patch in which $z\neq0$, in which the scaling relation of $\bP_{2,3,1}$ can be used to set $z=1$.

Before proceeding to the physically more interesting case of elliptic fibrations, let us briefly describe how the abstract description of an elliptic curve given above relates to the more intuitive picture of a torus defined as the complex plane modulo a lattice. Associated to such a lattice $\Lambda$ is a particular doubly-periodic meromorphic function, the famous Weierstrass $\wp$ function, whose periods are given by the lattice vectors. This function obeys the differential equation~\cite{Busam:2005}
\be
\wp_{\tau}'(z)^2 = 4 \wp_{\tau}(z)^3 - g_{2,\tau} \wp_{\tau}(z) - g_{3,\tau},
\label{weierstrass_p_identity}
\ee
where the constants $g_{2,\tau}$ and $g_{3,\tau}$ are the so-called modular invariants of the lattice. Despite the name, neither the $\wp$ function nor the modular invariants are actually invariant under SL(2,$\bZ$) transformations, as emphasised by the subscript $\tau$. However, their transformation is covariant in the sense that if $\tau$ and $\tau'$ are related by an SL(2,$\bZ$) transformation, then~\eqref{weierstrass_p_identity} is fulfilled for all $z$ if and only if it is also fulfilled for all $z$ with $\tau$ replaced by $\tau'$~\cite{Busam:2005}. If one now defines a map from the torus $T^2 = \bC/\Lambda$ into $\bC^3$ by\footnote{The powers of 4 are implemented to take into account that the definition of the modular discriminant in the mathematical literature differs from the definition used in~\eqref{discriminant_vanishing} by an overall factor of -16.}
\be
z\mapsto \genfrac{\{}{.}{0pt}{}{(4^{2/3}\wp_\tau(z), 2 \wp'_\tau(z), 1), \quad }{(1,1,0),}\genfrac{}{}{0pt}{}{z\neq 0,}{z=0,}
\ee
the identity~\eqref{weierstrass_p_identity} together with the identifications
\be
f = - 4^{1/3} g_{2,\tau} ,\qquad g = - 4 g_{3,\tau},
\label{relation_fg_eisenstein}
\ee
shows that the image of this map is none other than the elliptic curve~\eqref{Weierstrass_form}. Note that this shows that under an SL(2,$\bZ$) transformation both the (inhomogeneous) coordinates $x, y$ and the parameters $f,g$ appearing in the Weierstrass equation~\eqref{Weierstrass_form} transform non-trivially, although the actual surface defined by the equation remains unchanged.

While the modular discriminant is not invariant under SL(2,$\bZ$) transformations, it transforms covariantly in such a way that the vanishing locus of the discriminant is well-defined without any SL(2,$\bZ$) ambiguity\footnote{In mathematical terms, $\Delta$ can be viewed as a modular form of weight 12~\cite{Busam:2005}.}. Nevertheless, it is often more convenient to characterise the lattice (or equivalently the elliptic curve) by a manifestly SL(2,$\bZ$)-invariant object, the so-called Jacobi $j$-invariant. For future convenience we note that the $j$-invariant is related to the parameters of the elliptic curve by~\cite{Weigand:2010wm}
\be
j(\tau) = \frac{4 (24 f)^3}{\Delta} = \frac{4 (24 f)^3}{4 f^3 + 27 g^2}.
\label{def_j}
\ee
The $j$-invariant can be viewed as an bijective map from the fundamental region of SL(2,$\bZ$) into the complex plane. While an explicit formula for the $j$-invariant can be given in terms of Jacobi theta functions, for the purposes of this thesis it will be sufficient to note that in the vicinity of $z\sim i\infty$ it may be expanded into a Laurent series whose leading terms are
\be
j(z) = e^{-2\pi i z} + 744 + e^{2\pi i z} + \mathcal{O}(e^{4\pi i z}).
\label{expansion_j}
\ee
A crucial fact is that the pole of $j$ at $z\sim i\infty$ is the only pole lying in the closure of the fundamental region of SL(2,$\bZ$), although of course there are further poles at all of its SL(2,$\bZ$) transforms.

As motivated previously, for the construction of F-theory vacua we are interested in compactification manifolds in which a torus or elliptic curve varies over a suitable base space, i.e. an elliptic fibration. A complex $n$-dimensional manifold $Y_n$ is elliptically fibered if it admits a holomorphic projection map $\pi$ to a base manifold $B_{n-1}$, whose generic fiber is a non-singular elliptic curve. In the less abstract language of equation~\eqref{Weierstrass_form}, an elliptic fibration can (locally) be built by allowing the parameters $f$ and $g$ of the elliptic curve to vary holomorphically over the base $B_{n-1}$.

Given a general elliptic fibration, one may ask whether the projection $\pi$ may be inverted in some sense in order to obtain an embedding of the base in the total space $Y_n$, or in other words whether the fibration admits a section. Note that if the fibration can be written at least locally in Weierstrass form~\eqref{Weierstrass_form}, such a section is always present irrespective of the details of the fibration. The canonical section is given by the so-called zero section $z=0$, which as discussed above will always avoid any possible singularities of the fibration and is thus manifestly well-defined and holomorphic. The importance of the Weierstrass form lies in the fact that in a sense the converse of the above statement is also true, in that any elliptic fibration with a section can be brought into the Weierstrass form via a birational transformation~\cite{Esole:2011sm, Weigand:2010wm}. As emphasised in~\cite{Esole:2011cn}, F-theory compactifications on birationally equivalent spaces may lead to inequivalent low energy theories, so that Weierstrass fibrations do not cover the full landscape of F-theory vacua. In this thesis we will nevertheless exclusively consider fibrations with section\footnote{F-theory models on elliptic surfaces without a section were considered in~\cite{Witten:1996bn} and correspond to IIB vacua with non-trivial configurations of $B_2$ and $C_2$ in 4 dimensions.} and work mostly with fibrations in the Weierstrass form or the closely related Tate form.

As discussed above, the elliptic fiber degenerates over the codimension 1 locus in the base where the discriminant $\Delta$ vanishes\footnote{Such a locus will exist unless the fibration is trivial and the functions $f$ and $g$ in~\eqref{Weierstrass_form} are constant, as we are considering closed and compact base manifolds. A holomorphic hypersurface in $B_{n-1}$, such as the locus defined by $\{\Delta = 0\}$ is often referred to as a divisor. See appendix~\ref{sec:char_classes} for further details.}. The loci where $f$ and $\Delta$ vanish simultaneously lead to a more complicated degeneration of the elliptic fiber. As we will discuss in the following subsections, these loci are responsible for interesting physical effects such as the appearance of additional massless states and enhanced gauge symmetries. For now, we focus on the case where $f$ and $\Delta$ intersect transversally and do not share a common factor, so that these more complicated degenerations can occur only at higher codimension in the base.
Fixing an arbitrary point $p$ on the discriminant locus, we can then generically assume that $f$ does not simultaneously vanish at $p$. Locally around $p$, we can parametrise the normal direction to the discriminant locus by a complex coordinate $z$, choosing $z=0$ to correspond to the discriminant locus at $p$. If the discriminant vanishes to order $n$ at $z=0$, equation~\eqref{def_j} shows that the leading term in a Laurent expansion of the $j$-invariant is of the form $z^{-n}$. The value of the $j$-invariant of course determines $\tau$ only up to an SL(2,$\bZ$) transformation. Locally we may take $\tau$ to lie within the fundamental domain of SL(2,$\bZ$), in which $j$ has only one pole. Around this pole it obeys the Laurent expansion~\eqref{expansion_j}, which can be inverted to obtain
\be
\tau(z) \simeq \frac{i}{2\pi} \log j \simeq \tau_0 + \frac{n}{2\pi i} \log z + \ldots.
\ee
Comparison with~\eqref{expansion_tau} shows that this is precisely the behaviour that is expected in the vicinity of a stack of $n$ $D7$-branes. This motivates the view that from a type IIB perspective the component of the discriminant locus under consideration should be identified with part of the world-volume of a stack of $n$ D-branes. Note that while we may locally always choose $\tau$ to lie in the fundamental domain of SL(2,$\bZ$), this is in general not possible globally, so that other parts of the discriminant will describe $[p,q]$-branes with $[p,q]\neq[1,0]$. 

While from a type IIB perspective it is natural to study the behaviour of the axio-dilaton, in the geometric F-theory perspective $\tau$ actually appears only rather indirectly by locally inverting the $j$-invariant. In this language, the presence of the $D7$-branes in the shape of the discriminant locus manifests itself more directly in the appearance of monodromies which act on the 1-cycles on the torus fiber. At the locus where the discriminant vanishes, one of the cycles of the torus shrinks to zero size. This can be seen e.g. by the fact that $\tau$ essentially describes the ratio of the lengths of two cycles, and that by an SL(2,$\bZ$) transformation we may always have $\tau$ going to infinity at the discriminant locus. With respect to a fixed local basis of 1-cycles, we may denote an arbitrary cycle by a pair of integers. Let $(p,q)$ be the coordinates of the vanishing cycle with respect to this basis\footnote{In the SL(2,$\bZ$) basis in which $\tau \rightarrow i\infty$ at discriminant locus, we would have $(p,q)=(1,0)$.}. As discussed in~\cite{Braun:2010ff}, upon transport around the discriminant locus an arbitrary cycle $(a,b)$ is transformed as
\be
\left(\begin{matrix} a\\b \end{matrix} \right) \rightarrow \left(\begin{matrix} 1-p q & p^2\\ - q^2 & 1+p q \end{matrix} \right) \left(\begin{matrix} a\\b \end{matrix} \right),
\label{monodromy_cycles}
\ee
which matches the monodromy of a $[p,q]$-brane derived in equation~\eqref{monodromy_pq}. Note that the cycle $(p,q)$ itself is invariant under the transformation~\eqref{monodromy_cycles}, so that the type of the vanishing cycle can be defined consistently at least in the vicinity of the corresponding brane. However, if we transport the $(p,q)$-cycle around a different part of the discriminant locus at which the $(p',q')$-cycle vanishes, the $(p,q)$-cycle will be transformed non-trivially unless~\cite{Braun:2010ff} $pq'-qp'=0$. This transformation acts on the monodromy matrix by conjugation such that $M_{[p,q]} \rightarrow M_{[p',q']}M_{[p,q]}M_{[p',q']}^{-1}$. In the case of branes with $pq'-qp'\neq 0$ one speaks of mutually non-local 7-branes, and the $(p,q)$ type of each brane can no longer be defined in a globally meaningful manner.

Up to now it might seem that we may construct F-theory models with $D7$-branes lying along arbitrary divisors of the base manifold. This might seem surprising from a type IIB perspective, where as discussed in section~\ref{sec:Dbranes} the branes are subject to a tadpole cancellation condition. It turns out that the locations of the 7-branes in a consistent F-theory compactification are restricted by the requirement that the compactification manifold is Calabi-Yau. The reason for this is that, as already noted by Kodaira~\cite{Kodaira:1963}, the singular locus of the fibration gives rise to a negative contribution to the curvature of the total space. More precisely, the first Chern class of the total space $Y_n$ can be related to the first Chern class of the base $c_1(B_{n-1})\equiv c_1(T_{B_{n-1}})$ and the cohomology class of the discriminant by
\be
c_1(Y_n) = \pi^* (c_1(B_{n-1}) - \frac{1}{12} [\Delta]).
\label{Kodaira_formula}
\ee
In particular, we see that in order for the total space to have vanishing first Chern class as required for a Calabi-Yau manifold the base $B_{n-1}$ must have a positive Chern class to cancel the contribution of $\Delta$. To make it clearer that this in some sense plays the role of a tadpole cancellation condition let us split the discriminant into its irreducible components $[\Delta] = \sum_i \delta_i [\Delta_i]$. Here the integers $\delta_i$ denote the vanishing orders of the discriminant along the components $\Delta_i$, or in the type IIB language the number of $D7$-branes on the divisor $\Delta_i$. The Calabi-Yau condition thus simplifies to\footnote{Strictly speaking, these formulae hold for the case where the discriminant locus has no complicated self-intersections rendering the total space $Y_n$ singular. In such a case, further corrections appear on the right hand sides of~\eqref{Kodaira_formula} and~\eqref{CY_condition_fibration} which encode the contributions of the higher codimension loci of singularity enhancement~\cite{Morrison:1996na}, but these corrections will not be important for us at the moment.}
\be
12 c_1(B_{n-1}) = \sum_i \delta_i [\Delta_i].
\label{CY_condition_fibration}
\ee
As we will discuss in more detail in the next subsection, it is possible to tune the parameters $f$ and $g$ of the fibration in such a manner that the string coupling becomes small over the entire base manifold. In this weak coupling limit the theory reduces to perturbative type IIB orientifold compactification such that $B_{n-1}$ can be viewed as the orientifold quotient of a suitable Calabi-Yau manifold. In these weak coupling limits $c_1(B_{n-1})$ can be related to the homology class of the orientifold plane, which in turn relates the condition~\eqref{CY_condition_fibration} to the type IIB $D7$ tadpole cancellation condition~\cite{Aluffi:2009tm}.

The discussion around equation~\eqref{relation_fg_eisenstein} already served to show that the modular discriminant as well as the quantities $f$ and $g$ transform non-trivially under a transformation of $\tau$. This implies that they cannot truly be regarded as functions on the base manifold $B_{n-1}$, but must instead be viewed as sections of appropriate line bundles. The relation~\eqref{Kodaira_formula} now allows us to identify the line bundles in question as powers of the canonical bundle $K_B$ of the base, which has first Chern class $c_1(K_B) = - c_1(B_{n-1})$. More precisely, by~\eqref{Kodaira_formula} $\Delta$ must be a section of $K_B^{-12}$, which together with the defining equation~\eqref{discriminant_vanishing} implies that $f$ and $g$ transform as sections of $K_B^{-4}$ and $K_B^{-6}$, respectively. Homogeneity of the Weierstrass equation~\eqref{Weierstrass_form} requires the coordinates $x$ and $y$ to also transform as sections of similar bundles. Putting this information together, we deduce that the Weierstrass equation is to be read as a degree (6,6) hypersurface in the weighted projective bundle~\cite{Esole:2011sm, Marsano:2011hv} $\bP_{2,3,1}(K_B^{-2}\oplus K_B^{-3}\oplus \cO)$. To each of the coordinates appearing in the Weierstrass form we may assign two integers denoting the weights with respect to the scaling relations associated with the $\bP_{2,3,1}$ and the anticanonical bundle $K_B^{-1}$, respectively. Written in this way, the weights are
\be
\begin{aligned}
x: (2,2), &\qquad y:(3,3) , &\qquad z: (1,0), \\
f: (0,4) , &\qquad g: (0,6).&
\end{aligned}
\label{weights_Weierstrass}
\ee
We have already encountered one application of these general relations, in the example of the elliptic K3 surface briefly considered at the end of section~\ref{sec:ell_fibration_D-brane_backreaction}. There, the base was just a $\bP^1$, whose anticanonical bundle $K_B^{-1}=\cO(-2)$ is the bundle whose sections are homogeneous polynomials of degree 2 in the coordinates of the $\bP^1$. The appearance of 24 D-branes in the shape of 24 zeroes of the discriminant in the K3 compactification then immediately follows from the general  rules~\eqref{weights_Weierstrass}.

\subsection{The Sen limit}
\label{sec:Sen_limit}

The discussion in the previous subsection shows that F-theory allows us to relatively easily construct consistent vacua involving $7$-branes and a varying axio-dilaton profile. Explicitly constructing such a model essentially reduces to choosing a base manifold whose anticanonical bundle is sufficiently easy to handle that we can construct the various sections given in~\eqref{weights_Weierstrass}. In particular, the analogue of the type IIB tadpole constraints are in a sense automatically satisfied in a consistent Calabi-Yau compactification of F-theory. Nevertheless, the perturbative type IIB formulation has the advantage of offering a more intuitive interpretation of the physical properties of the model in terms of collections of D-branes and orientifold planes with open strings stretched between them, and it would be nice to have a more precise interpretation of the F-theory vacua in this language. Conversely, a more precise relationship between the two pictures might allow us to better understand some aspects of perturbative IIB vacua from the vantage point of F-theory.

At first sight our previous discussion of F-theory vacua seems to suggest that they will typically involve non-local 7-branes and that there will always be regions in which the string coupling is large, regardless of the chosen SL(2,$\bZ$) frame. Of course, as we discussed in section~\ref{sec:SL(2,Z)} this is also the generic situation in the type IIB setting. On the other hand it is known that there do exist purely perturbative IIB vacua, the simplest examples being situations in which all branes coincide with the orientifold planes so as to cancel all tadpoles locally. This gives an additional motivation to study how such situations can arise in F-theory.

The question of how to formally relate the F-theory construction to the weakly coupled IIB picture was addressed by Sen in~\cite{Sen:1996vd, Sen:1997kw, Sen:1997gv}. Although the techniques are independent of the dimensionality of the compactification, for concreteness we will in the following focus on F-theory compactifications to 4 dimensions on a fourfold $Y_4$ fibered over a threefold base $B_3$. Sen's approach\footnote{For more recent discussions of weak coupling limits of models with non-Abelian gauge groups or of fibrations not in the Weierstrass form, see~\cite{Aluffi:2009tm, Esole:2012tf}.}, which we summarise in the following, is based on attempting to specialise the forms of $f$ and $g$ in such a way that a weak coupling limit can be obtained by tuning one of the parameters of the model to zero. Concretely, $f$ and $g$ may be rewritten as~\cite{Weigand:2010wm}
\be
f = -3 h^2 + \epsilon \eta , \qquad g = -2 h^3 + \epsilon h \eta - \frac{\epsilon^2}{12} \chi,
\label{ansatz_sen}
\ee
with $h,\eta,\chi$ and $\epsilon$ sections of appropriate powers of $K_B^{-1}$ as required by~\eqref{weights_Weierstrass}. In particular, if we choose $\epsilon$ to be a constant, this parametrisation is completely general in the sense that any $f$ and $g$ may be rewritten in this manner. The advantage of this reparametrisation is that the discriminant carries an overall prefactor of $\epsilon^2$~\cite{Sen:1997gv}
\be
\Delta = \epsilon^2 \left( \eta^2(4\epsilon \eta - 9 h^2) - \frac{18}{4} h(\epsilon \eta - 2h^2)\chi +\frac{3}{16} \epsilon^2 \chi^2    \right).
\label{discrim_general}
\ee

Together with equation~\eqref{def_j}, the form of the discriminant implies that by taking $\epsilon$ to be small it is possible to ensure that $|j|$ is arbitrarily large everywhere on the base manifold except at the locus where $f$ vanishes. If $|j|$ stays uniformly large as we move around the base, then $\tau$ stays in a neighborhood of an inverse image $\tau_0 = j^{-1}(\infty)$. Of course, the inverse image $\tau_0$ is defined only up to an SL(2,$\bZ$) transformation, but the key point is that $\tau$ stays within the same SL(2,$\bZ$)-patch\footnote{By this we mean a subset of the complex plane which can as a whole be mapped into the fundamental region of SL(2,$\bZ$) by the same SL(2,$\bZ$) transformation.} of the complex plane. Choosing the SL(2,$\bZ$) frame in which $\tau_0 = i\infty$, we see that the limit $\epsilon\rightarrow 0$ thus corresponds to the case of (almost) globally weak coupling\footnote{Different limits are possible in which the coupling is constant over the base, but in these cases it cannot be tuned to be perturbatively small~\cite{Dasgupta:1996ij}}. 

In the weak coupling limit, we expect a dual description as a perturbative IIB vacuum to be possible. Sen provided this interpretation by observing that the discriminant at leading order in $\epsilon$ factorises as
\be
\Delta = -9 \epsilon^2 h^2 (\eta^2 - h \chi) + \cO(\epsilon^3).
\ee
In other words, the discriminant locus splits into several components as $\epsilon\rightarrow 0$. By analysing the monodromies around the loci $h=0$ and $\eta^2 = h \chi$, Sen deduced that these equations should describe the locations of an $O7$-plane and a pair of $D7$-branes in the base, respectively. Note that locally around a point at which $h\chi\neq 0$, the portion $\eta^2 - h \chi$ of the discriminant looks like it describes two separate branes located at $\eta = \pm \sqrt{h \chi}$, but globally they are recombined into a single object~\cite{Sen:1997gv}.

Having identified the surface $h=0$ as a potential orientifold plane by the monodromy analysis, Sen confirmed this identification by explicitly constructing the double cover of $B_3$ and the orientifold involution which leaves the locus $h=0$ invariant. In a similar manner to the Weierstrass construction of elliptic fibrations discussed above, Sen attempted to define a double cover $X_3$ of $B_3$ as a hypersurface in the anticanonical bundle over $B_3$. In order to define a double cover of $B_3$ branched over $h=0$, it must be possible to choose coordinates in which the hypersurface equation takes the form
\be
X_3 : \quad \xi^2 - h = 0.
\label{def_X3}
\ee
The orientifold involution is then simply $\sigma: \xi \mapsto -\xi$. In the limit $\epsilon\rightarrow 0$, we expect that the F-theory model reduces to the type IIB compactification on $X_3$ modded out by the involution $\sigma$, with a $D7$-brane at the locus
\be
D7: \quad \eta^2 - \xi^2 \chi = 0.
\label{eq_D7_sen_limit}
\ee
In general, this is a single object invariant under the orientifold action, however for in the special situation where $\chi$ is a perfect square, $\chi = \psi^2$, the brane splits into a pair of branes located at $\eta = \pm \xi\psi$. Clearly, the two branes are precisely exchanged by the orientifold involution $\sigma$, so that they describe a brane-image brane pair.

As noted in~\cite{Braun:2008ua, Collinucci:2008pf}, the form of the equation~\eqref{eq_D7_sen_limit} describing the $D7$-brane is not entirely generic. Instead, it requires the brane to intersect the orientifold plane only in double points. Although this requirement follows directly from the F-theory construction, its origin from the point of view of perturbative type IIB is far from obvious. The authors of~\cite{Collinucci:2008pf} showed that it could be derived by Dirac charge quantisation arguments, but the double point intersection still showed that the F-theory construction can shed light on new and surprising aspects of perturbative IIB compactifications. Note that by specialising the choice of the sections $\eta$ and $\chi$, the equation~\eqref{eq_D7_sen_limit} can also be split into several factors describing individual $D7$-brane stacks. As the double point intersection property must hold for each individual brane stack~\cite{Collinucci:2008pf}, each one of the factors can be put into the form of equation~\eqref{eq_D7_sen_limit}.

The weak coupling limit of F-theory allows us to explore the interplay between tadpole cancellation and Calabi-Yau conditions when switching back and forth between the F-theory and IIB pictures. To illustrate these relationships, let us follow~\cite{Collinucci:2008pf} and consider a particularly simple example with $B_3 = \bP^3$. The sections of the anticanonical bundle $K_{\bP^3}^{-1}$ are simply homogeneous polynomials of degree 4 in the homogeneous coordinates of $\bP^3$ (see appendix~\ref{sec:projective_varieties} for further details). In order for the ansatz~\eqref{ansatz_sen} to yield the correct degree of $f$, $h$ must be a section of $K_B^{-2}$, i.e. a polynomial of degree 8. For the equation~\eqref{def_X3} defining the double cover $X_3$ to transform homogeneously, this in turn requires $\xi$ to be a polynomial of degree 4. The total space of the $K_B^{-1}$ bundle in which the equation~\eqref{def_X3} is defined can be viewed as a weighted projective space whose coordinates are the coordinates $u_i,\ i=1,...,4$ of $\bP^3$ and $\xi$. The sum of the weights of the coordinates is thus 8. As this matches the degree of the equation defining $X_3$, it follows from the adjunction formula that $X_3$ is a Calabi-Yau manifold~\cite{Denef:2008wq}. Furthermore, the Sen ansatz~\eqref{ansatz_sen} implies that $\eta$ and $\chi$ are of degree 16 and 24, respectively. This means that in terms of the hyperplane class $H = [{u_i = 0}]$ on $\bP^3$ the classes of the $D7$-brane and the orientifold plane are given by
\be
[D7] = [\eta^2] = 32 H = 4 [h] = 4[O7].
\ee
This is precisely the relation that is required by 7-brane tadpole cancellation~\cite{Blumenhagen:2008zz}. In other words, the Calabi-Yau condition for $Y_4$ on the F-theory side implies both the Calabi-Yau condition for $X_3$ as well as tadpole cancellation on the IIB side. It is straightforward to check that the converse is also true, i.e. tadpole cancellation together with the Calabi-Yau condition for $X_3$ fix the degrees of the polynomials and ensure that $Y_4$ is also Calabi-Yau.

A second consequence of the non-generic form~\eqref{eq_D7_sen_limit} of the $D7$-brane besides the double intersection with the orientifold plane was noted in~\cite{Collinucci:2008pf}. Namely, the brane world-volume generically has singularities that locally look like the so-called Whitney umbrella singularity. For this reason, a brane of the form~\eqref{eq_D7_sen_limit} is often referred to as a Whitney brane. The presence of these singularities complicates the calculation of the contribution of the $D7$-brane to the $D3$-brane tadpole, which in the smooth case involves calculating the Euler characteristic of the $D7$-brane cycle. The calculation of the contribution of the singularities to the $D3$-tadpole directly in type IIB is rather complicated~\cite{Collinucci:2008pf}, but it can be determined much more easily by comparison with the tadpole on the F-theory side. This serves as another example that the F-theory formulation can be advantageous even in calculating quantities that are well-defined in the perturbative IIB limit.

When moving away from the perturbative limit by allowing finite values of $\epsilon$, the discriminant~\eqref{discrim_general} no longer factorises for generic choices of the sections $h,\eta,\chi$. This means that the orientifold plane and the 7-branes that were visible in the perturbative limit recombine into a single object. This recombination is a truly non-perturbative effect, as we can see by estimating that for small $\epsilon$ we have $\epsilon \sim j(\tau)^{-1/2} \sim \exp ( i\pi \tau) \propto \exp ( - \pi/ g_s )$~\cite{Sen:1996vd}. Alternatively, this can be seen by noting that there are no suitable recombination modes in the perturbative spectrum, so that the recombination modes are presumably non-perturbative $(p,q)$-strings~\cite{Weigand:2010wm}. In general, at finite coupling the discriminant locus is a single irreducible object that has no simple interpretation in terms of orientifold planes and D-branes. Nevertheless a factorisation of the discriminant into objects that may admit such an interpretation can be achieved by suitably specialising the forms of the sections $f, g$ specifying the elliptic fibration. This is the key point that makes it possible to explicitly engineer fibrations with singularities leading to certain gauge groups as required e.g. in the construction of the F-theory GUT models. This point will be discussed slightly more detail in section~\ref{sec:FtheoryGUTs}. One example in which the discriminant factorises for finite $\epsilon$ was obtained by Sen in~\cite{Sen:1997gv} by setting $\chi=0$ in the ansatz~\eqref{ansatz_sen}. In this case the discriminant still describes a pair of coincident branes at $\eta=0$, but the erstwhile orientifold plane splits non-perturbatively into two [$p,q$]-branes located at $h = \pm \frac{2}{3} \sqrt{\epsilon \eta}$.

\subsection{Non-Abelian gauge groups in F-theory}
\label{sec:nonabelian_groups}
In our discussion of F-theory compactifications so far, we have essentially viewed the elliptic fibration as just an elegant bookkeeping device that allows us to keep track of the non-trivial behaviour of the axio-dilaton $\tau$. The singularities of the fibration (at codimension 1 in the base) were interpreted as 7-branes due to the appearance of SL(2,$\bZ$) monodromies around them. However, beyond simply acting as sources for bulk fields, D-branes in type IIB compactifications are associated with another essential feature for model building purposes in that they can give rise to non-Abelian gauge symmetries. This immediately raises the question of how such gauge symmetries appear in the F-theory description.

As 7-branes in F-theory are associated with singularities of the elliptic fibration, it is natural to expect that the gauge group is related in some way to the geometric structure of the singularity. Even before the advent of F-theory, the fact that some singularities of the compactification manifold can be tied to the appearance of enhanced gauge symmetries had been noted in the context of type II and heterotic strings~\cite{Witten:1995ex, Bershadsky:1995sp, Strominger:1995cz, Aspinwall:1995zi}. The key to understanding the relationship between gauge symmetry and geometry in the F-theory context lies in the fact that the ways in which an elliptic Calabi-Yau may degenerate are quite restricted. Kodaira~\cite{Kodaira:1963} studied the types of singularities and the structure of the space obtained by resolving the singularities of elliptic K3 surfaces\footnote{Not every singularity of the elliptic fiber actually renders the total space singular, to be precise the space only needs to be resolved if the vanishing order of the discriminant is at least 2~\cite{Weigand:2010wm}.}. He found that in the resolved geometry the previously singular fibers are replaced by a set of intersecting $\bP^1$'s, whose intersection structure can be depicted by a simply-laced affine Dynkin diagram (with possibly some nodes appearing with multiplicities). Such Dynkin diagrams are associated with the Lie algebras of the families $A_n, D_n$ and $E_n$~\cite{Witten:1995ex, Morrison:1996na}. The type of singular fiber which appears in an elliptic K3 surface in the Weierstrass form was related by Kodaira~\cite{Kodaira:1963} to the vanishing orders of the sections $f, g$ and $\Delta$ along the portion of the discriminant in question. This relationship is shown in table~\ref{tab:Kodaira_class}. If the vanishing orders of $f$ and $g$ are too large, specifically if ord($f)\geq 4$ and ord($g)\geq 6$, the singularities are called non-minimal. A non-minimal singularity at codimension 1 in the base signals that the singular space can no longer be resolved to a Calabi-Yau manifold~\cite{Bershadsky:1996nh}; such cases will not be considered in this thesis.

\begin{table}[t]
\centering
\begin{tabular}{cccccc}
\hline
\multirow{2}{*}{ord($f$)} $\ \ $ & \multirow{2}{*}{ord($g$)} $\ \ $ & \multirow{2}{*}{ord($\Delta$)} &$\ \ $ Fiber type $\ \ $& Number of $\ \ $& Intersection  \\
&&& (Kodaira) & components $\ $ & graph (ADE) \\ \hline
$\geq 0$ & $\geq 0$ & 0 & smooth & 1 & none \\ 
0&0& 1 & $I_1$ & 1 & none \\
$\geq 1$ & 1 & 2 & $II$ & 1 & none \\
1 & $\geq 2 $& 3 & $III$ & 2 & $A_1$ \\
$\geq 2$ & 2 & 4 & $IV$ & 3 & $A_2$ \\
0& 0& $n\geq 2$ & $I_n$ & $n$ & $A_{n-1}$ \\
2 & $\geq 3$ & $n+6$ & $I_n^*$ & 5+$n$ & $D_{n+4}$ \\
$\geq 2$ & 3 & $n+6$ & $I_n^*$ & 5+$n$ & $D_{n+4}$ \\
$\geq 3$ & 4 & 8 & $IV^*$ & 7 & $E_6$ \\
3 & $\geq 5$ & 9 & $ III^* $ & 8 & $E_7$ \\
$\geq 4$ & 5 & 10 & $ II^*$ & 9 & $E_8$ \\
\hline
\end{tabular}
\caption{Kodaira classification of singular fibers (based on~\cite{Bershadsky:1996nh, Morrison:1996na, Esole:2011cn}). The number of components given does not include possible multiplicities of some components. The fiber types $I_1$ and $II$ correspond to a nodal and a cuspidal curve, respectively. In the cases where the fiber consists of multiple components, the different intersection points are always distinct except for the type $III$ and $IV$ fibers.}
\label{tab:Kodaira_class}
\end{table}

Although many of the techniques employed in Kodaira's analysis carry over to the case of higher-dimensional Calabi-Yau spaces, the above enumeration of singular fibers is complete only for elliptic K3 surfaces. In higher dimensions, non-trivial monodromies can appear as one moves about the base, which can lead to some of the fiber $\bP^1$'s being identified~\cite{Bershadsky:1996nh}. As explained lucidly in~\cite{Braun:2010ff}, these monodromies act at the level of Dynkin diagrams by folding the diagram, leading to the appearance of double intersections between some fiber components. In this manner, Dynkin diagrams of all simple Lie groups including those of type $C_n$, $B_n$, $F_4$ and $G_2$ may arise~\cite{Bershadsky:1996nh}. Additional complications can arise for singularities at higher codimension in the base, especially for fibrations not in the Weierstrass form. As we will see in the following, singularities at higher codimension in the base play an important role in the construction of F-theory GUT models and their structure has been studied extensively in the recent literature, see e.g~\cite{Esole:2011sm, Krause:2011xj, Morrison:2011mb, Esole:2011cn, Tatar:2012tm, Mayrhofer:2012zy, Lawrie:2012gg, Borchmann:2013jwa}. At higher codimension, the fiber type can in some cases deviate from the Kodaira types, either through the intersection structure or the multiplicity of the nodes. Nevertheless, these potential subtleties have little effect on most of the topics covered in this thesis, and the physical intuition obtained from table~\ref{tab:Kodaira_class} will be sufficient in many cases.

For elliptically fibered Calabi-Yau $n$-folds with $n\geq 3$, the classification of table~\ref{tab:Kodaira_class} can be further refined to distinguish between different ways in which the monodromy group acts on the components of the fiber. The monodromies that may appear were analysed in detail in~\cite{Bershadsky:1996nh}, where in particular it was shown that singular fibers of type $C_n$ and $B_n$ can indeed be constructed if the sections $f,g$ are specialised in a particular manner. This means that the spectrum of singular fibers is rich enough to include fibers with intersection diagrams isomorphic to the Dynkin diagrams of $SU(n)$, $SO(n)$ and $Sp(n)$, i.e. all the gauge groups that occur in perturbative IIB compactifications. The new feature in table~\ref{tab:Kodaira_class}, which has no perturbative IIB analogue, is the appearance of the exceptional Lie algebras $E_6, E_7$ and $E_8$. Note that generically (more precisely, if $f$ and $g$ do not share a common factor) only singularities of type $I_n$ appear at codimension 1 in the base. However, enhancements to the higher Lie algebras can easily occur at codimension 2 and 3 even in a reasonably generic setting. As we will see, the appearance of loci of exceptional enhancement is one of the main reasons for the popularity of F-theory constructions for GUT model building, as it allows the construction of certain Yukawa couplings that are forbidden in perturbative type II compactifications.

The discussion above naturally suggests that a singularity of the compactification manifold leads to the appearance of the corresponding gauge symmetry in the low energy effective theory obtained by compactifying F-theory on the singular space. Note that due to the lack of a true fundamental formulation of F-theory, it is not easy to show this directly. Nevertheless, strong indirect evidence exists in the shape of striking dualities with other string theories in which the gauge symmetry is known, in particular with type IIB for the $SU(n)$, $SO(n)$ and $Sp(n)$ groups and with certain heterotic compactifications for the exceptional groups~\cite{Bershadsky:1996nh, Donagi:2008ca, Beasley:2008dc, Beasley:2008kw, Hayashi:2008ba}. 

Further evidence for the appearance of gauge symmetries matching the fiber singularities comes from the duality of F-theory and M-theory discussed in subsection~\ref{sec:ell_fibration_D-brane_backreaction}. This duality gives a natural interpretation for the gauge degrees of freedom in terms of $M2$-branes wrapping $\bP^1$'s in the elliptic fiber~\cite{Denef:2008wq, Weigand:2010wm, Maharana:2012tu}. Strictly speaking, the $\bP^1$'s are not directly visible in the singular geometry and become manifestly visible only after the singularities are resolved. In the resolved situation, the $\bP^1$'s have non-zero volume, so that the corresponding gauge bosons are massive\footnote{More precisely, only the Cartan generators of the gauge group remain massless, so that when compactifying on the resolved space we are in the Coulomb branch of the gauge group~\cite{Bershadsky:1996nh, Grimm:2010ks}.}. However, the mass vanishes in the F-theory limit in which the fiber volume is taken to zero~\cite{Grimm:2010ks}, leading to the appearance of additional massless states describing the gauge bosons of the non-Abelian gauge group. 

This picture is actually quite similar to the IIB picture in which non-Abelian gauge generators come from strings stretched from one brane to another, leading to gauge symmetry enhancement when some of these states become massless for stacks of coincident branes. In fact, the correspondence can be made even more precise using the language of $(p,q)$-strings and string junctions~\cite{Gaberdiel:1997ud}. In this language, a singularity with a non-Abelian gauge group is viewed as corresponding to several coincident 7-branes of different $[p,q]$-types. As discussed in subsection~\ref{sec:elliptic_CalabiYaus}, ($p,q$)-string states undergo non-trivial monodromies in the presence of such non-local 7-branes. The appearance of these monodromies and multi-pronged string junctions means that string states can no longer be labelled by just two Chern-Paton labels as in the perturbative IIB case. The analysis of~\cite{Gaberdiel:1997ud, Gaberdiel:1998mv, DeWolfe:1998zf} shows that this richer structure can indeed account for the exceptional groups not visible in the perturbative IIB setting. 

At the codimension 2 loci where two components of the discriminant meet, the singularity type worsens and additional massless states appear in the F-theory limit~\cite{Katz:1996xe}. For elliptic Calabi-Yau fourfolds, these loci are curves in the base known as matter curves\footnote{Intersections between two divisors carrying some gauge group are not the only way to obtain chiral matter in F-theory. Starting with a gauge group $G$ on a divisor one can break the group to a subgroup $G_1 \subset G$ by switching on suitable flux~\cite{Donagi:2008ca, Weigand:2010wm}. Chiral matter then arises from a decomposition of fields in the adjoint of $G$ into the representations of $G_1$. Such fields are sometimes referred to as 'bulk matter' as they propagate over the entire divisor and are not localised further on intersection curves.}. In the type IIB picture the corresponding phenomenon is the appearance of massless charged matter localised at the intersection of two brane stacks. Similarly, interactions between these charged matter states can arise at so-called Yukawa points where two matter curves meet~\cite{Donagi:2008ca, Beasley:2008dc, Beasley:2008kw}.

The results quoted above make it clear that the key to constructing phenomenologically viable F-theory models lies in being able to engineer and manipulate singularities of the fibration in a controlled manner. An explicit constructive algorithm to build fibrations with particular singularities was given by Tate in~\cite{Tate:1975}. The Weierstrass form~\eqref{Weierstrass_form} that we have used so far is actually not the best suited formulation to engineer singularities in this manner, and it is instead advantageous to rewrite the fibration in a related form known as the Tate form~\cite{Bershadsky:1996nh, Esole:2011sm}. Following the conventions of~\cite{Grimm:2009yu, Grimm:2010ez}, we define the fibration in the Tate form by the equation
\be
P_T = x^3 - y^2 + a_1 x y z + a_2 x^2 z^2 + a_3 y z^3 + a_4 x z^4 + a_6 z^6 = 0.
\label{P_Tate}
\ee
This equation is defined in the same space $\bP_{2,3,1}(K_B^{-2}\oplus K_B^{-3}\oplus \cO)$ as the original Weierstrass equation. Note that homogeneity of the equation requires the $a_i$ to transform as sections of $K_B^{-i}$. The Tate form can be related to the Weierstrass form by the coordinate shift~\cite{Cvetic:2013nia}
\be
\tilde{x} = x + \frac{1}{12} \left(4 a_2+ a_1^2\right) z^2 , \qquad \tilde{y} = y - \frac12 a_1 x z - \frac{1}{2} a_3 z^3.
\ee
This shift brings the Tate equation~\eqref{P_Tate} into the usual Weierstrass form
\be
P_T = \tilde{x}^3 - \tilde{y}^2 + f \tilde{x} z^4 + g z^6.
\ee
Here $f$ and $g$ related to the sections $a_i$ by~\cite{Bershadsky:1996nh, Grimm:2010ez}
\be
f = -\frac{1}{48}\left(\beta_2^2-24 \beta_4\right), \qquad g = -\frac{1}{864}\left(-\beta_2^3 + 36 \beta_2 \beta_4 - 216 \beta_6 \right),
\label{rel_weierstrass_tate}
\ee
with
\be
\beta_2 = a_1^2 + 4a_2, \qquad \beta_4 = a_1 a_3 + 2 a_4, \qquad
\beta_6 = a_3^2 + 4 a_6.
\ee
Plugging the expressions~\eqref{rel_weierstrass_tate} for $f$ and $g$ into the general formula~\eqref{discriminant_vanishing} yields the discriminant in terms of the Tate sections~\cite{Weigand:2010wm}
\be
\Delta = -\frac14 \beta_2^2(\beta_2\beta_6 - \beta_4^2) - 8\beta_4^3 - 27\beta_6^2 + 9 \beta_2\beta_4\beta_6.
\label{discriminant_tate}
\ee
This analysis shows that it is always possible to bring a fibration from the Tate form into the Weierstrass form. As discussed in~\cite{Katz:2011qp}, the converse is not always possible globally and for all gauge groups. However, the Tate form can be constructed at least locally for the gauge groups usually used in GUT model building, and for simplicity we will restrict our attention to fibrations that can be globally written in Tate form throughout this thesis.

The advantage of the Tate form from a model building perspective is that by specifying the vanishing orders of the sections $a_i$ we can fix\footnote{In a few specific cases, namely $SO(7)/SO(8)$, $SO(11)/SO(12)$ and $SO(4k+3)/SO(4k+4)$, the same vanishing orders of the $a_i$ can lead to different groups depending on the factorisation properties of the Tate polynomial, see table~\ref{tab:Tate_class} and the discussion in~\cite{Bershadsky:1996nh, Braun:2010ff}.} not only the vanishing orders of $f$ and $g$ (and therefore the original Kodaira fiber type), but also the monodromy acting on the fiber $\bP^1$'s~\cite{Bershadsky:1996nh}. The way in which this monodromy acts makes it possible to identify three distinct fiber types dubbed split, semi-split and non-split in~\cite{Bershadsky:1996nh}, where the split case is the case of trivial monodromies. The relationship between the vanishing orders of the $a_i$ and the singular fibers as obtained from Tate's algorithm is given in table~\ref{tab:Tate_class}.

\begin{table}[ht!]
\centering
\begin{tabular}{cccccc|cc}
\hline
&& Vanishing & order &&& Fiber type & Group \\
$\hspace{0.2cm} a_1 \hspace{0.2cm}$ & $\hspace{0.2cm} a_2 \hspace{0.2cm}$ & $\hspace{0.2cm} a_3\hspace{0.2cm} $ & $\hspace{0.2cm} a_4 \hspace{0.2cm} $ & $\hspace{0.2cm} a_6 \hspace{0.2cm} $ & \hspace{0.2cm} $\Delta \hspace{0.2cm}$ & & \\ \hline
0 & 0 & 0 & 0 & 0 & 0 & $I_0$ & none \\
0 & 0 & 1 & 1 & 1 & 1 & $I_1$ & none \\
0 & 0 & 1 & 1 & 2 & 2 & $I_2$ & $SU(2)$ \\
0 & 0 & 2 & 2 & 3 & 3 & $I_3^{ns}$ & $Sp(1)$ \\
0 & 1 & 1 & 2 & 3 & 3 & $I_3^s$ & $SU(3)$ \\
0 & 0 & $k$ & $k$ & $2k$ & $2k$ & $I_{2k}^{ns}$ & $Sp(k)$ \\
0 & 1 & $k$ & $k$ & $2k$ & $2k$ & $I_{2k}^s$ & $SU(2k)$ \\ 
0 & 0 & $k+1$ & $k+1$ & $2k+1$ & $2k+1$ & $I_{2k+1}^{ns}$ & $Sp(k)$ \\
0 & 1 & $k$ & $k+1$ & $2k+1$ & $2k+1$ & $I_{2k+1}^s$ & $SU(2k+1)$ \\ 
1 & 1 & 1 & 1 & 1 & 2 & $II$ & none \\
1 & 1 & 1 & 1 & 2 & 3 & $III$ & $SU(2)$ \\
1 & 1 & 1 & 2 & 2 & 4 & $IV^{ns}$ & $Sp(1)$ \\
1 & 1 & 1 & 2 & 3 & 4 & $IV^s$ & $SU(3)$ \\
1 & 1 & 2 & 2 & 3 & 6 & $I_0^{*ns}$ & $G_2$ \\
1 & 1 & 2 & 2 & 4 & 6 & $I_0^{*ss}$ & $SO(7)$ \\
1 & 1 & 2 & 2 & 4 & 6 & $I_0^{*s}$ & $SO(8)^*$ \\
1 & 1 & 2 & 3 & 4 & 7 & $I_1^{*ns}$ & $SO(9)$ \\
1 & 1 & 2 & 3 & 5 & 7 & $I_1^{*s}$ & $SO(10)$ \\
1 & 1 & 3 & 3 & 5 & 8 & $I_2^{*ns}$ & $SO(11)$ \\
1 & 1 & 3 & 3 & 5 & 8 & $I_2^{*s}$ & $SO(12)^*$ \\
1 & 1 & $k$ & $k+1$ & $2k$ & $2k+3$ & $I_{2k-3}^{*ns}$ & $SO(4k+1)$ \\
1 & 1 & $k$ & $k+1$ & $2k+1$ & $2k+3$ & $I_{2k-3}^{*s}$ & $SO(4k+2)$ \\
1 & 1 & $k+1$ & $k+1$ & $2k+1$ & $2k+4$ & $I_{2k-2}^{*ns}$ & $SO(4k+3)$ \\
1 & 1 & $k+1$ & $k+1$ & $2k+1$ & $2k+4$ & $I_{2k-2}^{*s}$ & $SO(4k+4)^*$ \\
1 & 2 & 2 & 3 & 4 & 8 & $IV^{*ns}$ & $F_4$ \\
1 & 2 & 2 & 3 & 5 & 8 & $IV^{*s}$ & $E_6$ \\
1 & 2 & 3 & 3 & 5 & 9 & $III^*$ & $E_7$ \\
1 & 2 & 3 & 4 & 5 & 10 & $II^*$ & $E_8$ \\
1 & 2 & 3 & 4 & 6 & 12 & non-min. & - \\
\hline
\end{tabular}
\caption{Classification of singular fibers at codimension 1 obtained from Tate's algorithm; taken from~\cite{Katz:2011qp}.}
\label{tab:Tate_class}
\end{table}

The discussion so far has centered on gauge groups that can be associated with codimension 1 divisors in the base, and the enhancements that take place when two or more such loci intersect. However, it is also possible that the fibration is singular along a locus of codimension 2 (or 3) which cannot be written as the intersection of two codimension 1 singularity loci. In the process of resolving the singular space, such singular loci also give rise to additional $\bP^1$ fibers, which in turn lead to additional gauge bosons in the low energy effective theory that are not associated with Cartan generators of some non-Abelian gauge group. In this manner, further Abelian gauge factors can arise in F-theory compactifications in addition to the groups listed in table~\ref{tab:Tate_class}~\cite{Grimm:2010ez, Mayrhofer:2012zy, Cvetic:2012xn, Borchmann:2013jwa, Cvetic:2013nia}. Such Abelian symmetries can have important consequences for the phenomenology of the low energy theory e.g. by forbidding certain couplings and interactions. The investigation of the construction, the mathematical description and the effects of Abelian gauge symmetries in F-theory compactifications will be a central topic of the following chapters.

\subsection{Aspects of F-theory GUT model building}
\label{sec:FtheoryGUTs}

From a phenomenological perspective, the main aim of studying string compactifications is to determine the properties of the resulting effective 4-dimensional theory. A very important part of this procedure is obtaining the low energy gauge group and the matter field spectrum including their interactions. These aspects can be addressed quite explicitly in type IIB orientifold compactifications and it is comparatively simple to build realistic effective theories in intersecting brane models~\cite{Blumenhagen:2005mu, Blumenhagen:2006ci, Maharana:2012tu}. In order to obtain chiral matter spectra, such constructions usually\footnote{Chiral spectra can also be achieved by including spacetime-filling $D3$-branes located at singular points of the compactification manifold, see e.g.~\cite{Maharana:2012tu}. However, trying to uplift IIB models based on singular manifolds to F-theory requires elliptic fibrations over singular bases, which are mathematiclly difficult to handle. Throughout this thesis we will stick to smooth bases and rely on fluxed intersecting $D7$-branes to generate chirality.} rely on intersecting $D7$-branes with world-volume fluxes~\cite{Maharana:2012tu, Blumenhagen:2008zz}. As we have seen in the previous sections, such 7-brane configurations generically lead to vacua with varying axio-dilaton profiles which are not easy to describe outside the F-theory framework. Nevertheless, from a phenomenological perspective it is tempting to use the comparatively simpler language of intersecting branes for model building without explicitly specifying the resulting complicated axio-dilaton background. The implicit expectation in such an approach is that these IIB models can in principle be embedded into a consistent F-theory compactification, but that the main phenomenological features can be determined already in the IIB language without having to make recourse to the more complicated mathematical framework of F-theory. Much of the recent phenomenological interest in F-theory constructions comes from the realisation that this expectation fails to hold in models with grand unification of the gauge group (GUT models)~\cite{Donagi:2008ca, Donagi:2008kj, Beasley:2008dc, Beasley:2008kw}. In particular, the structure of the Yukawa couplings in such models can only be fully understood by taking into account F-theoretic strong coupling effects in the shape of the appearance of exceptional symmetry groups. In the following we will briefly review the relevant reasoning for the case of $SU(5)$ GUTs, following the discussion of~\cite{Maharana:2012tu, Blumenhagen:2008zz}.

The main appeal of grand unified theories lies in the fact that the minimal supersymmetric extension of the Standard Model predicts the (approximate) unification of the three gauge couplings at a scale of around $10^{16}$ GeV~\cite{Blumenhagen:2008aw}. This suggests that the SM gauge groups might have a common origin in a single larger gauge group, the GUT group, which is broken at the unification scale. The simplest such GUT model is based on an $SU(5)$ gauge group and was considered already by Georgi and Glashow~\cite{Georgi:1974sy}. We will stick to the $SU(5)$ model throughout this thesis, although many statements can be easily adapted to other GUT groups as well. After embedding the Standard Model gauge group into the $SU(5)$, the actual breaking $SU(5) \rightarrow SU(3)\times SU(2) \times U(1)_Y$ can be achieved e.g. by turning on a flux along the hypercharge generator $U(1)_Y$. In this breaking process, the leptons of the Standard Model arise from the fundamental \textbf{5} and the antisymmetric \textbf{10} representation of $SU(5)$ according to the breaking pattern~\cite{Slansky:1981yr}
\begin{align}
\label{breaking_SU5_1}
\mathbf{5} & \rightarrow  (\mathbf{3},\mathbf{1})_{-1/3} + (\mathbf{1},\mathbf{2})_{1/2} \ , \\
\mathbf{10} & \rightarrow  (\mathbf{3},\mathbf{2})_{1/6} + (\bar{\mathbf{3}},\mathbf{1})_{-2/3} + (\mathbf{1},\mathbf{1})_1 \ .
\label{breaking_SU5_2}
\end{align}
In order to obtain all lepton generations of the Standard Model, the model must therefore include 3 multiplets in each of the representations $\bar{\mathbf{5}}$ and $\mathbf{10}$, while a further pair of $\mathbf{5}$ multiplets combine to describe the Higgs fields.

The natural starting point when attempting to realise an $SU(5)$ GUT model in a type IIB orientifold compactification\footnote{As before we focus on orientifolds with $O3/O7$-planes and without singularities.} is to begin with a stack of 5 $D7$-branes wrapped along a divisor $D_a \subset X_3$ in the compactification manifold $X_3$. This gives rise to a $U(5) = SU(5)\times U(1)_a$ low energy gauge group. To obtain matter in the $\mathbf{5}$ representation of $SU(5)$ it is necessary to introduce (at least) an additional brane on a different divisor $D_b\subset X_3$. The states in the $\mathbf{5}$ representation are then localised at the intersections between $D_a$ and $D_b$ (and their orientifold images), while the $\mathbf{10}$ representation arises from strings stretched between the GUT stack and its orientifold image~\cite{Blumenhagen:2008zz}. The gauge group of the IIB GUT model thus has (at least) two further Abelian factors $U(1)_a$ and $U(1)_b$ in additional to the actual GUT group. 

By a suitable choice of the divisors $D_a$ and $D_b$ both of the additional $U(1)$s can be made massive through the St\"uckelberg mechanism (see section~\ref{sec:U(1)inIIB}), so that they do not explicitly appear in the low energy gauge group~\cite{Blumenhagen:2008zz}. 
Nevertheless, the presence of the additional Abelian gauge symmetries still has a crucial effect because the $SU(5)$ multiplets appearing in~\eqref{breaking_SU5_1} and~\eqref{breaking_SU5_2} carry charges with respect to these $U(1)$s, which leads to selection rules on the possible couplings. In particular, the GUT origin of the top Yukawa coupling would be a coupling of the form~\cite{Maharana:2012tu}
\be
\mathbf{10}^{(2,0)} \mathbf{10}^{(2,0)} \mathbf{5}_H^{(1,-1)} ,
\label{top_Yukawa}
\ee 
where the superscripts denote the charges of the fields under $U(1)_a \times U(1)_b$. Clearly, this coupling is forbidden perturbatively by $U(1)$ gauge invariance. Of course, in more complicated models with more brane stacks further symmetries arise and the $U(1)$ charges of the fields need not be exactly as in equation~\eqref{top_Yukawa}. Nevertheless, this does not affect the fact that the top Yukawa coupling is forbidden. To see this, note that the $\mathbf{10}$ states arise from strings stretching between the GUT stack and its image, so that the only $U(1)$ symmetry they will be charged under is the diagonal $U(1)_a \subset U(5)$. The Higgs field in the $\mathbf{5}_H$ representation on the other hand must come from strings stretched between the GUT stack and some other brane, so that they will always be charged under the $U(1)$ associated to this other brane. This reasoning also explains why couplings involving two fields arising from strings between the GUT stack and a different brane stack, for example the down-type Yukawa coupling~\cite{Blumenhagen:2008zz}
\be
\mathbf{10}^{(2,0)} \bar{\mathbf{5}}^{(-1,-1)} \bar{\mathbf{5}}_H^{(-1,1)},
\ee 
may be gauge invariant.

Note that these observations do not imply that the top Yukawa coupling is completely absent in type IIB $SU(5)$ GUTs. Indeed, on a non-perturbative level the coupling can be generated e.g. by suitable Euclidean $D3$-instantons~\cite{Blumenhagen:2006xt, Blumenhagen:2007zk}. However, by definition such non-perturbative instanton contributions are expected to be relatively small in a perturbative type IIB setting~\cite{Maharana:2012tu}, so that some tuning of the parameters of the compactification is required to obtain the observed fermion mass hierarchies given the fact that the down-type (and lepton) Yukawa couplings are present perturbatively~\cite{Blumenhagen:2007zk, Blumenhagen:2008zz, Weigand:2010wm}.

Much of the recent phenomenological interest in F-theory stems from the realisation~\cite{Beasley:2008dc} that strong coupling effects in F-theory resolve the problem discussed above. In particular, F-theory models can allow all types of Yukawa couplings, including the top Yukawa coupling in $SU(5)$ GUTs, to appear on an equal footing. On a qualitative level this can be understood by noting that F-theory allows for the appearance of multi-pronged strings ending on multiple different brane stacks~\cite{Maharana:2012tu}. Roughly speaking, if the analogues of the $\mathbf{10}$ states in~\eqref{top_Yukawa} arise from multi-pronged strings rather than strings ending only on the GUT brane, they can simultaneously carry charges under different $U(1)$s. This can allow for charge cancellation to occur in a manner which is not possible in equation~\eqref{top_Yukawa}. To understand this more precisely, let us review how the various matter states and interactions arise in an F-theory $SU(5)$ GUT model, following the discussion in~\cite{Weigand:2010wm}.

Let $S\subset B_3$ be the divisor in the base of the elliptic fibration on which the GUT group should arise, and $w$ a (local) holomorphic coordinate parametrising the normal direction to $S$ in $B_3$. The analysis of~\cite{Katz:2011qp} shows that at least locally the fibration can be written in Tate form, and the vanishing orders of the sections $a_i$ near $S$ that are required in order to obtain an $SU(5)$ gauge group can be read off from table~\ref{tab:Tate_class}. To make the vanishing orders manifestly visible, it is convenient to rewrite
\be
a_1 = b_5, \quad a_2 = b_4 w, \quad a_3 = b_3 w^2, \quad a_4 = b_2 w^3, \quad a_6 = b_0 w^5,
\label{a_i_SU(5)}
\ee
where the sections $b_i$ do not vanish on the GUT brane $S$ located at $w=0$. Using the expression~\eqref{discriminant_tate} for the discriminant one obtains~\cite{Blumenhagen:2009yv}
\be
\Delta = -w^5\left( b_5^4 P + w b_5^2(8 b_4 P + b_5 R) + w^2 (16 b_3^2 b_4^2 + b_5 Q) + \cO(w^3) \right),
\label{discriminant_SU(5)}
\ee
where $P$, $Q$ and $R$ are certain polynomials in the $b_i$. We will only need the expression for $P$, which is given by~\cite{Blumenhagen:2009yv}
\be
P= b_3^2 b_4 - b_2 b_3 b_5 + b_0 b_5^2.
\label{def_5_curve}
\ee
Following the discussion of subsection~\ref{sec:nonabelian_groups} we expect matter fields at loci where the singularity enhances. Equation~\eqref{discriminant_SU(5)} makes it clear that there are two codimension 2 loci at which the vanishing order of $\Delta$ exceeds 5, which are located at $w = b_5 = 0$ and $w = P = 0$. Applying Tate's algorithm as in table~\ref{tab:Tate_class}, we expect that the states localised at these matter curves fill out the adjoint representations of $SO(10)$ and $SU(6)$, respectively. To determine the transformation properties of these fields with respect to the $SU(5)$ gauge group the adjoint representations of $SO(10)$ and $SU(6)$ are decomposed as~\cite{Weigand:2010wm, Maharana:2012tu}
\be
\begin{aligned}
SO(10)\rightarrow SU(5)\times U(1) &: \qquad \mathbf{45} \rightarrow \mathbf{24}_0 \oplus \mathbf{1}_0 \oplus \mathbf{10}_2 \oplus \overline{\mathbf{10}}_{-2}, \\
SU(6)\rightarrow SU(5)\times U(1) & : \qquad \mathbf{35} \rightarrow \mathbf{24}_0 \oplus \mathbf{1}_0 \oplus \mathbf{5}_1 \oplus \overline{\mathbf{5}}_{-1}.
\end{aligned}
\label{matter_decompositions}
\ee
These decompositions are the reason why the curve $P_{10}\equiv \{ w = b_5 = 0\}$ is often referred to as the $\mathbf{10}$-curve, while $P_5\equiv \{w=P=0\}$ is called the $\mathbf{5}$-curve. Note that the $U(1)$ group appearing in~\eqref{matter_decompositions} is at the moment only formally defined by the local enhancement to $SO(10)$ resp. $SU(6)$, and may or may not survive as a symmetry of the full low energy action~\cite{Maharana:2012tu}, depending on the details of the fibration. Additional GUT singlet states which describe true physical $U(1)$ gauge symmetries can arise from other singular loci at codimension two which do not lie fully within the GUT brane $S$~\cite{Grimm:2010ez, Weigand:2010wm}.

To determine the interactions of these matter states, we must similarly identify the higher singularity enhancements at codimension 3. Such enhancement points are expected to lie at the intersection of the matter curves. Equation~\eqref{def_5_curve} shows that $P$ factorises over the locus $b_5 = 0$, so that the matter curves intersect at two separate loci given by~\cite{Weigand:2010wm}
\be
\begin{aligned}
w=b_5 = b_3 & = 0 \quad \rightarrow \quad D_6 \Leftrightarrow SO(12) , \\
w = b_5 = b_4 & = 0 \quad \rightarrow \quad E_6.
\end{aligned}
\ee
Here we have already indicated the enhancement that is expected from Tate's algorithm. To identify the fate of the matter fields we again decompose the adjoint representations of the enhancement groups, which can be broken to $SU(5)\times U(1) \times U(1)$~\cite{Maharana:2012tu}
\be
\begin{aligned}
E_6:  \ \mathbf{78} \rightarrow  \mathbf{24}^{(0,0)} \oplus \left[ \big(  \right. &  \mathbf{1}^{(0,0)} \oplus \mathbf{1}^{(-5,-3)}  \\
& \left.\left.\ \ \oplus \, \mathbf{5}^{(-3,3)} \oplus \mathbf{10}^{(-1,-3)} \oplus \mathbf{10}^{(4,0)} \right) \oplus \text{ c.c.} \right],  \\
SO(12):  \ \mathbf{66} \rightarrow \mathbf{24}^{(0,0)} \oplus \left[ \big(  \right.& \left.\left. \mathbf{1}^{(0,0)} \oplus \mathbf{5}^{(-1,0)} \oplus \mathbf{5}^{(1,1)} \oplus \mathbf{10}^{(0,1)} \right) \oplus \text{ c.c.} \right].
\end{aligned}
\label{Yukawa_enhancement}
\ee
Again, the simple fact that the singularity enhances in this manner at the interaction points is not sufficient to determine whether the $U(1)$'s appearing here (and hence the associated selection rules) survive in the full low energy action or not. Regardless of this, the charges given in~\eqref{Yukawa_enhancement} show that the sought-after top Yukawa coupling
\be
	\mathbf{5}^{(-3,3)} \mathbf{10}^{(-1,-3)} \mathbf{10}^{(4,0)},
\ee
is indeed allowed to appear at the point of $E_6$ enhancement, while the point of enhancement to $SO(12)$ can yield the down type Yukawa. Finally, there may be additional enhancement points at $w=P=R = 0, (b_5,b_3)\neq (0,0)$, where the singularity is expected to enhance to $SU(7)$ in models with $b_0 \equiv 0$~\cite{Hayashi:2009ge, Andreas:2009uf, Blumenhagen:2009yv, Esole:2011sm}. This enhancement point is different in nature from the other $E_6$ and $SO(12)$ points in that it does not arise from the intersection of two matter curves. Instead, it can be viewed as the intersection of a codimension 2 curve carrying a GUT singlet with the GUT brane~\cite{Maharana:2012tu}. This enhancement point yields a $\mu$-term type coupling of the form $\mathbf{5}\overline{\mathbf{5}}\mathbf{1}$~\cite{Blumenhagen:2009yv}.

We should point out that Tate's algorithm, which was used above to motivate the appearance of matter and interaction loci, is strictly speaking applicable only at codimension 1 in the base. The structure of the fiber $\bP^1$s found after resolving the singularities was studied very explicitly in~\cite{Krause:2011xj} (see also~\cite{Blumenhagen:2009yv, Grimm:2009yu, Esole:2011sm, Marsano:2011hv, Morrison:2011mb, Esole:2011cn, Tatar:2012tm, Mayrhofer:2012zy, Lawrie:2012gg, Borchmann:2013jwa} for a variety of different approaches). The results show that while the structure of the fiber $\bP^1$s over the enhancement points does not exactly match the expectations of Tate's algorithm, the form of the Yukawa couplings is unaffected.

An interesting aspect of F-theory GUT models is that, as discussed above, the massless states charged under the visible gauge group and their interactions are localised on the GUT divisor $S$. This forms the motivation for the so-called local approach to F-theory model building, which focuses on the geometry of the fibration near $S$ and attempts to deduce as many aspects of the low-energy theory as possible without explicitly embedding $S$ in a compact Calabi-Yau fourfold~\cite{Andreas:2009uf}. An advantage of the local approach is that the particle spectrum and the coefficients of the couplings can be studied and in some cases calculated explicitly in local coordinate patches within $S$. For example, while the discussion above strictly speaking only shows that certain Yukawa couplings can appear, local models can allow the wavefunctions of the localised modes to be obtained explicitly by solving the associated Dirac equation~\cite{Donagi:2008ca, Beasley:2008kw}. This in turn makes it possible to explicitly calculate coefficients of Yukawa couplings and study flavour hierarchies and mixing matrices~\cite{Beasley:2008kw, Heckman:2008qa, Hayashi:2009ge, Heckman:2009mn, Font:2009gq, Heckman:2010bq}.

One approach to local F-theory model building which was pioneered in~\cite{Beasley:2008dc, Beasley:2008kw} is based on the fact that the structure of symmetry enhancements which can be obtained from an elliptic fibration in Tate form~\eqref{P_Tate} follows an $E_8$ breaking pattern~\cite{Grimm:2010ez}. The underlying reason for this is that an $E_8$ singularity can be described locally by the equation~\cite{Beasley:2008dc}
\be
y^2 = x^3 + z^5,
\ee
of which the Tate model~\eqref{P_Tate} describes a general unfolding\footnote{This can be generalised immediately to the cases of other ADE singularities corresponding to a different gauge group $G$, such as $G=E_6$ or $G=E_7$~\cite{Beasley:2008dc}. The resulting fibrations are correspondingly known as $G$-fibrations~\cite{Aluffi:2009tm}. For the moment we will stick to the usual Tate equation, i.e. the case $G=E_8$, although other forms play an important role in the study of Abelian gauge symmetries~\cite{Grimm:2010ez}.}. The authors of~\cite{Beasley:2008dc, Beasley:2008kw} argued that as the various singularities appearing over $S$ in the Tate model correspond to subgroups of $E_8$, they should admit an interpretation in terms of Higgsing of an underlying 8-dimensional $E_8$ gauge theory on $S$. More precisely, taking a Higgs bundle with structure group $H\subset E_8$ breaks the symmetry to the commutant $G= E_8/H$ over the bulk of $S$. In particular, the case $G=SU(5)$ is achieved by taking also $H=SU(5)$~\cite{Weigand:2010wm}. To distinguish it from the actual gauge group $G$, the commutant is often denoted by $H=SU(5)_\perp$. In the Higgs bundle picture the loci of enhanced symmetry are identified with the loci on $S$ on which particular components of the Higgs field vanish.

A different but closely related tool for local model building is the so-called spectral cover construction. The spectral cover can be defined by dropping all terms in the Tate polynomial of higher order in the coordinate $w$ normal to the brane, and keeping only the leading terms. This reduces the sections $b_i$ appearing in~\eqref{a_i_SU(5)} to sections of $K_S$. These sections can be used to define the so-called spectral surface, which for the case of an $SU(5)$ gauge group on $S$ is given by the equation~\cite{Donagi:2009ra, Weigand:2010wm, Maharana:2012tu} 
\be
\cC^{(5)}: \ b_0 s^5 + b_2 s^3 + b_3 s^2 + b_4 s + b_5 = 0,
\label{spectral_cover}
\ee
within a $K_S$ bundle with coordinate $s$ over $S$. This surface effectively models the base-dependent parts of the Tate equation~\eqref{P_Tate} and hence the behaviour of the discriminant in the vicinity of $S$. The spectral cover picture can be related to the Higgs bundle picture discussed above by taking~\eqref{spectral_cover} to be part of an equation describing a deformed $A_4$ singularity fibered over $S$~\cite{Maharana:2012tu}. The $SU(5)$ group associated with this $A_4$ singularity can be identified with the commutant group $H = SU(5)_\perp$. The fact that the $A_4$ singularity is fully deformed over generic points of $S$ corresponds to the fact that the $SU(5)_\perp$ subgroup of the overall $E_8$ symmetry is broken, leaving the unbroken gauge group $G=E_8/SU(5)_\perp = SU(5)$. Over certain loci in $S$ some of the roots of the equation~\eqref{spectral_cover} may coincide, corresponding to some of the $\bP^1$'s of the resolved $A_4$ shrinking to zero size~\cite{Maharana:2012tu}. This signals an enhancement of the group $G$ at these loci, identifying them as the matter curves and Yukawa points in the spectral cover language.

One advantage of the spectral cover approach is that it allows for an explicit construction of gauge fluxes and allows to calculate the chirality of the matter spectrum induced by the presence of these fluxes~\cite{Marsano:2009gv}. However, as the spectral cover construction explicitly holds only in the vicinity of $S$ care must be taken when extrapolating the results beyond the local limit~\cite{Marsano:2010ix, Marsano:2011hv}, and we will not pursue this approach to constructing fluxes in this thesis. A second advantage of the spectral cover approach is that (leaving aside questions regarding its global validity for a moment) it offers an explicit approach towards constructing models with additional Abelian gauge factors. To understand this, we follow the discussion in~\cite{Maharana:2012tu} and rewrite~\eqref{spectral_cover} (up to an overall factor) as a product of 5 factors of the form $(s+t_i), \ i=1,...,5$. The fact that $b_1=0$ in~\eqref{spectral_cover} is known as the tracelessness constraint as it translates to $\sum_i t_i = 0$. Each $t_i$ can be identified as corresponding to a $U(1)$ symmetry. More precisely, the 4 independent $U(1)$'s remaining after taking the tracelessness constraint into account are taken to describe the Cartan $U(1)$'s of $SU(5)_\perp$. 

A crucial point is that the $t_i$'s are not defined globally, but are instead related by monodromies as one moves around the GUT divisor $S$~\cite{Heckman:2009mn, Marsano:2009gv}. In the completely generic case all the $t_i$ are identified by the monodromies, such that no $U(1)$ symmetry remains after the tracelessness constraint is taken into account~\cite{Maharana:2012tu}. However, one can reduce the monodromy group by restricting to the case of a so-called split spectral cover, in which the equation~\eqref{spectral_cover} factorises globally on $S$. In this case the $t_i$ arising from the various factors are not mixed by monodromies, and some $U(1)$ symmetries may remain. The simplest case is the so-called $4+1$ split defined by~\cite{Marsano:2009gv, Maharana:2012tu}
\be
\cC^{(5)}: \ (c_0 s + c_1)(d_0 s^4 + d_1 s^3 + d_2 s^2 + d_3 s + d_4) = 0,
\ee
together with the tracelessness constraint $c_0 d_1 + c_1 d_0 = 0$. Such a configuration can be seen as reducing the commutant group from $H= SU(5)_\perp$ to $H=S[U(4)\times U(1)]$, leaving an unbroken symmetry group of $SU(5)\times U(1)$~\cite{Maharana:2012tu}. Cases with more $U(1)$'s are considered e.g. in~\cite{Dudas:2010zb}. The appearance of such additional Abelian factors is very interesting from a phenomenological point of view, because they lead to selection rules on the allowed couplings between the matter fields. As discussed more extensively in~\cite{Weigand:2010wm, Maharana:2012tu}, this gives hope that the $U(1)$ symmetries may be used to forbid phenomenologically dangerous operators leading e.g. to proton decay or to large neutrino masses. Nevertheless, the $U(1)$ symmetries and selection rules obtained from such a local spectral cover construction are vulnerable to being broken by global effects~\cite{Hayashi:2010zp, Grimm:2010ez}. This underlines the importance of studying the appearance and nature of Abelian gauge symmetries and the associated gauge fluxes in a global setting, which is one of the main topics of the remainder of this thesis.

The discussion above makes it clear that F-theory compactifications offer a framework well-suited to the construction and study of phenomenologically interesting GUT models. In fact, the F-theoretic construction offers several advantages over the study of GUTs in 4-dimensional field theories. An example of this is the fact that additional GUT breaking mechanisms are possible, such as the breaking via hypercharge flux in the compactification manifold~\cite{Weigand:2010wm}. A significant part of the recent literature has been devoted to studying the phenomenological implications of these scenarios, such as the possible appearance of light exotic particles or the effects on gauge coupling unification. A discussion of these results as well as a list of references can be found in the recent reviews~\cite{Weigand:2010wm, Heckman:2010bq, Maharana:2012tu}. It is to be expected that the combination of the global and local approaches to model building and a growing understanding of their respective strengths and limitations will lead to many further interesting results in the context of F-theory GUTS in the coming years.

\chapter{$U(1)$ symmetries in type IIB and F-theory}
\label{sec:u(1)inIIBandFtheory}
As discussed at the end of chapter~\ref{chapter:IIBtoFtheory}, the appearance of Abelian gauge symmetries in string compactifications can have important consequences for the phenomenology of the resulting low energy theory. Such Abelian gauge factors are ubiquitous in compactifications of type II string theory with intersecting branes, although the gauge bosons often obtain a mass through the St\"uckelberg mechanism so that the $U(1)$s affect the low energy theory only indirectly through selection rules on possible couplings. While Tate's algorithm yields a straightforward approach towards identifying and constructing non-Abelian gauge symmetries, the Abelian factors and the resulting selection rules are more difficult to treat in the F-theory framework. This chapter forms a summary of the investigations into the nature of massive $U(1)$s in F-theory which was presented in~\cite{Grimm:2011tb}. In section~\ref{sec:U(1)inIIB} we begin by reviewing the dimensional reduction of type IIB string theory on a Calabi-Yau orientifold with $D7$-branes, focusing on effects relating to massless and massive Abelian gauge factors. We continue in section~\ref{sec:U(1)inFtheory} by discussing how the F-theory analogues of these $U(1)$ symmetries can be described in an M-theory reduction on an elliptically fibered fourfold, and how massive $U(1)$s and the associated fluxes can be related to certain non-K\"ahler deformations of the fourfold. Sections~\ref{sec:MtheoryReduction} and~\ref{sec:FtheoryLimit} contain details of the M-theory reduction on such fourfolds, as well as a discussion of the F-theory limit and the match between the $U(1)$ effects in the F-theory and IIB settings. Finally, section~\ref{sec:U(1)geometry} contains a brief discussion of the geometry of U(1) symmetries in F-theory, including a summary of recent results on the construction of models with massless $U(1)$s using an approach based on constructing fibrations with multiple sections.

\section{$U(1)$ symmetries in type IIB compactifications with $D7$-branes}
\sectionmark{$U(1)$s from $D7$-branes in type IIB}
\label{sec:U(1)inIIB}

\subsection{Kaluza-Klein reduction of type IIB with $D7$-branes}
\label{sec:IIBreduction}

The 4-dimensional low energy theory of type IIB string theory compactified on a Calabi-Yau orientifold can be computed by a Kaluza-Klein reduction of 10-dimensional type IIB supergravity~\cite{Bodner:1989cg, Michelson:1996pn, Bohm:1999uk, Dall'Agata:2001zh, Louis:2002ny, Grimm:2004uq, Grimm:2005fa}. Spacetime-filling $D7$-branes can be treated in a similar manner, by integrating out the internal part of the brane world-volume in the Dirac-Born-Infeld and Chern-Simons actions. As already discussed in section~\ref{sec:action_IIB_bulk}, it is convenient to use the democratic formulation of type IIB supergravity when including D-branes. The dimensional reduction of the combination of the democratic type IIB bulk action and the action of a spacetime-filling $D7$-brane was first performed in reference~\cite{Jockers:2004yj}, whose conventions we largely follow.

The starting point for the compactification is a Calabi-Yau threefold $X_3$ endowed with a holomorphic involution $\sigma$. We choose the orientifold action in such a way that it leads to $O3/O7$-planes, as tadpole cancellation in models involving $D7$-branes can be most easily implemented in the presence of $O7$-planes. When modding out the field spectrum of the effective theory by the orientifold symmetry, only fields with certain parities under pullback along the involution $\sigma$ survive. Specifically, the parities of the fields of the democratic IIB bulk action given in equation~\eqref{S_IIB_democ} are~\cite{Grimm:2004uq, Jockers:2004yj}
\be
\sigma^* F = \genfrac{\{}{.}{0pt}{}{F,}{-F,} \quad \genfrac{}{}{0pt}{}{F = g, J, \Phi, C_0, C_4, C_8, \, }{F = \Omega_3, B_2, C_2, C_6 . \qquad }
\label{orientifold_parities_full}
\ee
To perform the Kaluza-Klein reduction, we assume that the 10-dimensional metric and hence also the Laplacian can be split into two parts corresponding to 4-dimensional Minkowski space and the 6-dimensional Calabi-Yau space $X_3$. Note that this is strictly speaking only an approximation valid in the case where the volume of $X_3$ is relatively large and the warp factor induced by the presence of $D7$-branes and orientifold planes varies slowly over $X_3$. We make a similar ansatz for the other fields of the low energy action and split them into 'external' pieces propagating on Minkowski space and 'internal' parts defined on $X_3$. After integrating out the internal space, the expectation value of the Laplacian evaluated on the part of the fields defined on $X_3$ yields a mass term for the corresponding 4-dimensional fields. The internal parts of the fields can be expanded into eigenfunctions of the Laplacian on $X_3$, yielding a tower of 4-dimensional fields with increasing masses. The mass scale of the first massive level is known as the Kaluza-Klein (KK) scale and can be straightforwardly estimated by\footnote{Strictly speaking,~\eqref{KK_scale_bulk} gives the KK scale relevant for fields propagating in the bulk of the Calabi-Yau $X_3$. As discussed e.g. in~\cite{Conlon:2008wa}, the KK scale may be different for fields propagating on D-branes wrapped on sub-manifolds of $X_3$, in which case it depends on the volume of the cycle wrapped by the brane rather than the volume $\cV$ of $X_3$.}
\be
M_{KK} = \frac{M_P}{\cV^{2/3}}.
\label{KK_scale_bulk}
\ee
Here $M_P \simeq 2.4 \times 10^{18}$ GeV is the 4-dimensional (reduced) Planck mass, and $\cV$ denotes the volume of the compactification manifold $X_3$ measured in units of the string length $\ell_s$
\be
\cV = \frac{1}{3!} \int_{X_3} J\wedge J \wedge J.
\label{dimless_volume} 
\ee
Recall that we work in conventions where the metric and hence also the K\"ahler form $J$ are dimensionless, which is why no explicit factors of $\ell_s$ appear in~\eqref{dimless_volume}.

Although we expect $\cV$ to be large in order to be able to trust the $\alpha'$ expansion of the action, $M_{KK}$ typically still lies far beyond the energy scales accessible with current experiments. For example, even in the so-called LARGE volume scenarios~\cite{Balasubramanian:2005zx}, a typical numerical value might be $\cV \simeq 10^{10}$. Such a volume would lead to the Kaluza-Klein mass scale $M_{KK}\simeq 10^{11} $ GeV. In specific cases, e.g. for highly anisotropic compactification manifolds, it may be possible to obtain certain lighter Kaluza-Klein states. Nevertheless, the above estimates show that in general the massive Kaluza-Klein modes may be safely neglected when determining the 4-dimensional theory at energy scales within or close to experimental reach. In the following we therefore restrict to the massless Kaluza-Klein modes in the dimensional reduction. 

The massless Kaluza-Klein modes are described by harmonic forms\footnote{A collection of the most important definitions and facts regarding to the calculus of differential forms as well as a summary of the conventions used in this thesis can be found in appendix~\ref{sec:conventions}. Specific applications to Calabi-Yau manifolds are summarised in appendix~\ref{sec:Calabi-Yau}.} on $X_3$, which in turn are in one-to-one correspondence with the cohomology groups of $X_3$. The non-trivial  independent cohomology groups on a Calabi-Yau threefold are $H^{1,1}(X_3)$ and $H^{2,1}(X_3)$. As the orientifold involution $\sigma$ squares to the identity, the cohomology groups may be split into parts with positive or negative parity under pullback along $\sigma$, so that e.g. $H^{1,1}(X_3) = H^{1,1}_+(X_3) \oplus H^{1,1}_-(X_3)$. We follow the conventions of~\cite{Jockers:2004yj} and denote bases of the spaces of positive and negative parity $(1,1)$-forms by
\be
\begin{aligned}
H^{1,1}_+(X_3): & \qquad \omega_\alpha , \quad \alpha = 1,\ldots,h^{1,1}_+  , \\ 
H^{1,1}_-(X_3): & \qquad \omega_a , \quad a = 1,\ldots,h^{1,1}_- .
\end{aligned}
\label{cohom_bases_1,1}
\ee

.


The volume form $\frac16 J^3$ has even parity with respect to $\sigma^*$, so the Hodge star induces isomorphisms $\ast: H^{1,1}_+(X_3) \rightarrow H^{2,2}_+(X_3)$ and $\ast: H^{1,1}_-(X_3) \rightarrow H^{2,2}_-(X_3)$. We may therefore choose bases $\{\tilde{\omega}^\alpha\}$ for $H^{2,2}_+$ and $\{\tilde{\omega}^a\}$ for $H^{2,2}_-$ which are dual to the bases of equations~\eqref{cohom_bases_1,1} in the sense that
\be
\int_{X_3} \omega_\alpha \wedge \tilde{\omega}^\beta =  \delta_\alpha^\beta , \qquad  \int_{X_3} \omega_a \wedge \tilde{\omega}^b = \delta^b_a.
\label{dual_cohom_base}
\ee
One could similarly choose bases for $H^{2,1}(X_3)_\pm$, but as we aim to focus on the Abelian gauge symmetries rather than reviewing the full reduction of the action, the cohomology groups of even degrees will suffice for our purposes.

The basis elements of $H^{1,1}(X_3)$ are normalised in such a way that they are Poincar\'e-dual to a basis of the divisor group\footnote{
The definition of a divisor as well as several examples can be found in appendices~\ref{sec:char_classes} and~\ref{sec:projective_varieties}.} of $X_3$. This in particular means that the integrals
\be
\cK_{\alpha\beta\gamma} =  \int_{X_3} \omega_\alpha\wedge\omega_\beta\wedge\omega_\gamma , \qquad\quad \cK_{\alpha a b} =   \int_{X_3} \omega_\alpha\wedge\omega_a\wedge\omega_b,
\label{triple_intersections}
\ee
describe the geometric intersection number of three divisors and are therefore integer-valued. In fact, as discussed in~\cite{Grimm:2011tb}, $\cK_{\alpha\beta\gamma}$ and $\cK_{\alpha a b}$ take values only in the subset of \emph{even} integers. This can be understood geometrically by noting that the divisors associated to the $\omega_\alpha$ and $\omega_a$ have a well-defined orientifold parity. In other words, the divisors are left invariant by $\sigma$ up to a possible reversal of orientation. This in turn means that all intersection points must appear in point-image-point pairs\footnote{An exception to this rule is given by possible intersection points that lie on an orientifold plane, however in this case the intersection is guaranteed to be a double point. This can be seen e.g. by focusing on a local neighborhood of the intersection point, in which the orientifold action reduces to a simple reflection along the orientifold plane. It is then clear that any divisor with well-defined orientifold parity must intersect the O-plane in double points.}, leading to even intersection numbers. Note that due to the orientifold parity the only non-vanishing triple intersection numbers take the form shown in~\eqref{triple_intersections} and involve either zero or two negative parity forms $\omega_a$.

The massless 4-dimensional fields from the closed string sector are obtained by expanding the 10-dimensional fields of type IIB supergravity into the cohomology groups of $X_3$, taking into account the orientifold parities as given in~\eqref{orientifold_parities_full}. Restricting to the expansion into forms on $X_3$ of even degree, this yields
\be
\begin{aligned}
 J &\ =   v^{\alpha} \omega_{\alpha}, \quad\quad  C_2 =  c^{a} \omega_a, \quad\quad B_2 \equiv   B_- + B_+ =  b^{a} \omega_a +  b^{\alpha} \omega_{\alpha}, \\
 C_4 &\ = c_{\alpha} \tilde{\omega}^{\alpha} +  c_2^{\alpha} \wedge \omega_{\alpha} + \ldots, \quad \quad
C_6 =  (\tilde c_2)_{a}  \wedge \tilde{\omega}^a + \ldots,\ \\
C_0 &\ = l , \quad \quad\quad\ \ \, C_8 = \tilde l^{(2)} \wedge \frac{ 1 }{\cV} \Omega_3 \wedge \bar{\Omega}_3.
\label{JBC_expand}
\end{aligned}
\ee
In the course of the Kaluza-Klein reduction, all the coefficient fields in these expansions are taken to depend only on the 4-dimensional coordinates. Note that $c_2^{\alpha}, \ (\tilde c_2)_{a}$ and $\tilde l^{(2)}$ are spacetime 2-forms, while all other expansion fields shown in~\eqref{JBC_expand} are scalars. As already discussed in section~\ref{sec:compactifications}, certain discrete $B$-field configurations with positive orientifold parity are allowed due to the symmetry of the theory under shifts of $B_2$ by integer quantised harmonic forms. To be precise, the coefficients $b^\alpha$ in the expansion of $B_+$ may take the values 0 or 1/2, so that $B_2$ and $-\sigma^* B_2$ can be identified up to a symmetry transformation.

The fixed point locus of the orientifold involution $\sigma$ defines a divisor\footnote{In addition to the fixed divisor $D_{O7}$, $\sigma$ may have isolated fixed points in $X_3$, which give rise to $O3$-planes.} within $X_3$, which is wrapped by a spacetime-filling $O7$-plane and denoted by $D_{O7}$. We use the same notation to describe a divisor in $X_3$ and its Poincar\'e-dual 2-form, while we write $[D_{O7}]$ for the corresponding class in (co-)homology. The $O7$-plane introduces a tadpole for the Ramond-Ramond form $C_8$ which must be cancelled by a suitable configuration of $D7$-branes. In a supersymmetric configuration, the $D7$-brane stacks also wrap divisors in $X_3$~\cite{Becker:1995kb, Bergshoeff:1997kr, Marino:1999af}, which we denote by $D_A$. Here the index $A$ labels the different stacks of $N_A$ $D7$-branes each, each of which is accompanied by an image stack wrapped on the divisor $D_A'=\sigma(D_A)$. The $D7$-brane tadpole cancellation condition then reads~\cite{Blumenhagen:2006ci}
\be
\sum_A N_A\left( [D_A]+[D_A'] \right) = 8 [D_{O7}].
\ee
It is often convenient to combine the brane divisors $D_A$ and their orientifold images into the combinations $D_A^\pm = D_A \cup (\pm D_A')$, where the minus sign signifies a reversal of orientation. The Poincar\'e-dual cohomology classes are again denoted by $[D_A^\pm]$. In the case where the brane divisor is invariant under the orientifold action we include an additional factor of 1/2 in the definition of $D_A^+$, such that $[D_A^+] = [D_A]$. As the classes $[D_A^\pm]$ have a well-defined orientifold parity they may be expanded into the bases of $H^{1,1}_\pm$ introduced above as
\be
[D_A^+] = C_A^\alpha\ \omega_\alpha, \qquad\quad [D_A^-] =  C_A^a\ \omega_a.
\label{expansion_D_A}
\ee
Using~\eqref{dual_cohom_base}, the (dimensionless) wrapping numbers $C^\alpha_A$ and $C^a_A$ may also be calculated as
\be
C^\alpha_A =  \int_{D_A^+} \tilde{\omega}^\alpha, \qquad \quad C^a_A = \int_{D_A^-} \tilde{\omega}^a.
\ee

The addition of spacetime-filling $D7$-branes leads to the appearance of additional massless fields propagating on the world-volume of the brane. As already discussed in more detail in section~\ref{sec:Dbranes}, each stack gives rise to a set of scalar fields describing the deformations of the brane divisors and a one-form gauge field $A^A$. These fields transform in the adjoint representation of the gauge group associated to the brane stack and are represented by suitable matrices in the world-volume DBI and CS actions of the D-brane~\cite{Tseytlin:1997csa}. The world-volume gauge group depends on the precise location of the divisors $D_A$ in relation to the orientifold plane. Throughout this thesis we will focus on the cases in which the gauge group is $U(N_A)$. This occurs either if $[D_A]\neq[D_A']$, or if the brane lies in the same homology group as the orientifold plane without lying completely on top of the divisor $D_{O7}$. In the other cases, i.e. if either $D_A = D_A'$ pointwise or if $[D_A]=[D_A']\neq[D_{O7}]$, the gauge group is either $SO(2N_A)$ or $Sp(N_A)$~\cite{Blumenhagen:2006ci}.

In the course of the Kaluza-Klein reduction, the massless 4-dimensional fields which arise from the brane gauge and deformation fields can again be obtained by expanding them into harmonic forms on the brane world-volume. This expansion is performed in detail in~\cite{Jockers:2004yj}. In the following we will focus on the expansion of the gauge field $A^A$ and the associated field strength $\hat{F}^A$, following the conventions of~\cite{Grimm:2011tb}. As already mentioned in section~\ref{sec:Dbranes}, it is helpful to combine the field strength with the pullback of the Kalb-Ramond field into a combined field $\mathbf{F}^A_{D7} = \hat{F}^A - 2\pi\iota^* B_2$, as it is this combination which appears in the world-volume action of the D-branes. $\mathbf{F}^A_{D7}$ can be expanded in terms of the generators of $U(N_A)$ as
\be
\frac{1}{2\pi} \mathbf{F}^A_{D7} = \left(\frac{1}{2\pi} \hat{F}_0^A - \iota^* B_2\right) T^0_A + \frac{1}{2\pi} \sum_{i=1}^{N_A-1} \hat{F}_i^A T_A^i + \frac{1}{2\pi} \sum_{j=N_A}^{N_A^2-1} \hat{F}_j^A T_A^j.
\label{expansion_F_generators}
\ee
The first term describes the diagonal $U(1)_A \subset U(N_A)$, whose generator $T_A^0$ is just the $N_A\times N_A$ unit matrix, while the second stands for the Cartan subsector of $SU(N_A)$ with generators $T_A^i$. In the following we will concentrate on the Abelian subsector and in particular on the diagonal $U(1)_A$. This is because we aim to eventually match the IIB results with F-theory compactifications, in which only the Cartan of the non-Abelian gauge algebra remains visible after resolving the compactification manifold and moving to the Coulomb phase of the gauge theory. We use the capital index $I$ running from $I=0$ to $I=N_A-1$ to enumerate the set of generators $(T^0_A, T^i_A)$.

A crucial role in the following analysis will be played by topologically non-trivial background configurations of the brane gauge field. To make these background fluxes explicitly visible, we write
\be
\hat{F}_I^A = d A_I^A + \cF^A_I.
\label{field_strength_and_flux}
\ee
In order to preserve Lorentz invariance of the 4-dimensional vacuum, the fluxes must correspond to 2-forms living completely on the internal part of the brane world-volume, i.e. on the divisor $D_A\subset X_3$. In a supersymmetric configuration, they are further restricted to lie in $H^{1,1}(D_A) \subset H^2(D_A)$~\cite{Marino:1999af}. Part of $H^{1,1}(D_A)$ is spanned by the pullback of the forms $\omega_\alpha$ and $\omega_a$, which form a basis of $H^{1,1}(X_3)$. It is possible to split $H^{1,1}(D_A)$ into a direct sum of the subspace spanned by these pullbacks, and a second subspace which is orthogonal to the pullbacks with respect to the wedge product~\cite{Jockers:2004yj, Grimm:2011dj}. Flux lying in this orthogonal subspace is often referred to as variable flux, and denoted $\cF^A_{I, \tv}$. In summary, the gauge flux can be expanded as
\be
\cF_I^A = \cF_I^{A, \alpha} \omega_\alpha + \cF_I^{A,a} \omega_a + \cF^A_{I, \tv},
\label{flux_expansion_twoforms}
\ee
where we have suppressed the explicit pullbacks to the brane world-volume. 
 
While most of the expansion coefficients of the fields appearing in~\eqref{JBC_expand} are 4-dimensional fields, the coefficients in the expansion of the gauge flux~\eqref{flux_expansion_twoforms} are discrete constants which form a part of the specification of the chosen background. This is because the gauge flux is restricted by the Freed-Witten quantisation condition~\cite{Blumenhagen:2008zz, Freed:1999vc}
\be
 \frac{1}{2\pi} \hat{F}^A + \frac12 c_1(K_{D_A}) T^0_A \in H^2(D_A, \bZ)_{N_A\times N_A}.
 \label{FW_quant}
\ee
Here $c_1(K_{D_A})$ is the first Chern class of the canonical bundle of $D_A$, which is defined in appendix~\ref{sec:char_classes}. The Freed-Witten quantisation condition can be derived by considering the path integral of an open string ending on the brane, whose endpoints couple to the brane gauge field $A^A$. As the boundary of the string worldsheet is taken around a closed loop in $D_A$ the path integral picks up a phase related to the integral of $\hat{F}^A/2\pi$ over the two-cycle traced out by the loop. The condition that this phase vanishes then translates into the quantisation condition~\eqref{FW_quant} for the gauge flux. For future convenience, we follow~\cite{Grimm:2011tb} and combine the flux along the diagonal $U(1)$ with the other discrete quantity introduced above, the positive parity component $B_+$ of the Kalb-Ramond field, into the discrete quantity
\be
\tilde{\cF}_0^A = \frac{1}{2\pi} \cF_0^A - \iota^* B_+ \equiv \tilde{\cF}_0^{A,\alpha} \omega_\alpha + \tilde{\cF}^{A,a}_0 \omega_a +  \cF^A_{0, \tv}.
\label{def_tilde_cF}
\ee

If the brane divisors are not orientifold-invariant, the involution $\sigma$ cannot be directly restricted to $D_A$. Therefore the gauge field strength does not in general have a well-defined orientifold parity like the other fields in~\eqref{orientifold_parities_full}. Instead, the orientifold projection relates the gauge invariant combination of gauge and $B_2$-field on $D_A$ to the corresponding combination on the image stack $D_A'$ by
~\cite{Grimm:2011tb}
\be
\frac{1}{2\pi}\hat{F}_I^{A'}-\iota_{A'}^* B_2 = - \sigma^* \left(\frac{1}{2\pi}\hat{F}_I^A - \iota_A^* B_2 \right).
\label{fluxes_on_image_stack}
\ee

The second part of the field strengths in~\eqref{field_strength_and_flux} besides the fluxes comes from the exterior derivative of the gauge field $A_I^A$. In the Kaluza-Klein reduction, the 10-dimensional gauge field $A_I^A$ gives rise to a 4-dimensional gauge field $(A_I^A)_\mu dx^\mu$. In the presence of non-trivial 1-cycles on $D_A$, $A_I^A$ additionally yields a set of Wilson line moduli counted by $b^1_-(D_A^+)$~\cite{Jockers:2004yj}. However, these Wilson line moduli will play no role in our analysis and for simplicity we assume that $b^1_-(D_A^+)=0$, so that $F_I^A \equiv d A_I^A$ is the 4-dimensional gauge field strength.

The 4-dimensional effective action can now be computed by inserting the expansions of the various fields into the 10-dimensional supergravity action of equation~\eqref{S_IIB_democ} and the $D7$-brane DBI and CS actions, and proceeding to integrate out the internal space. After this dimensional reduction, the redundant degrees of freedom of the democratic formulation of type IIB supergravity have to be eliminated using the duality relations~\eqref{duality_relations}. 
As the brane and flux configuration is assumed to satisfy the supersymmetry conditions of refs.~\cite{Becker:1995kb, Bergshoeff:1997kr, Marino:1999af}, the effective theory is supersymmetric and can thus be cast into the general form of an $\cN=1$ supergravity\footnote{The signs of the various terms are governed by our conventions for the Hodge star given in appendix~\ref{sec:conventions}. Note in particular $*_4 1 = - \sqrt{|\det g|} d^4x$ and $F^A\wedge*_4 F^B = -\frac12\sqrt{|\det g|} F^A_{\mu\nu}F^{B\mu\nu}d^4x$.}~\cite{wess1992supersymmetry, Jockers:2004yj}
\begin{multline} \label{eq:N_1}
   \mathcal{S}^{(4)}_{\mathcal{N}=1}=\frac{1}{2\kappa_4^2}
      \int \Big[ - R\:*_41 - 2\:K_{M\bar N}\nabla M^M\wedge*_4\nabla\bar M^{\bar N} \\
      + \mathrm{Re} f_{AB}\, F^A\wedge*_4 F^B
      +  \mathrm{Im} f_{AB}\, F^A\wedge F^B
      +\left(V_{\rm F}+V_{\rm D}\right)\:*_41 \Big] \ .
\end{multline}
In this expression, $M^M$ enumerates the complex chiral fields and $F^A$ the field strengths of the gauge fields of the theory, while $K_{M\bar N} = \partial_{M^M} \partial_{M^{\bar{N}}} K$ are the components of a K\"ahler metric. The gauge kinetic function $f_{AB}$ is a holomorphic function of the chiral fields. The F-term potential $V_{\rm F}$ can be computed from the superpotential $W$ with the help of the K\"ahler metric and the K\"ahler covariant derivatives $D_M W = \partial_{M^M}W + (\partial_{M^M} K) W$. It is given by
\be
\label{eq:V_F}
   V_{\rm F}=e^K\left(K^{M\bar N}D_MW
      D_{\bar N}\bar W-3\left|W\right|^2\right).
\ee
The D-term potential $V_{\rm D}$ involves the inverse of the gauge kinetic function as well as the so-called Killing potentials $D_A$. These Killing potentials or D-terms are in one-to-one correspondence with the gauged isometries of the target space manifold parametrised by the chiral fields, and can be determined from the Killing vectors $X^M_A$ of the gauged isometries by
\be \label{genD}
     i \partial_{M} D_A = K_{M \bar N} \bar X^N_A\ .
\ee
The D-term potential is then given by
\be \label{eq:V_D}
   V_{\rm D}=\frac{1}{2}\left(\mathrm{Re}\, f^{-1}\right)^{AB}D_A D_B \ .
\ee
Finally, the 4-dimensional gravitational coupling constant $\kappa_4$ is the inverse of the 4-dimensional Planck mass, $\kappa_4 = M_P^{-1}$. In our conventions, it is equal to the 10-dimensional coupling constant introduced in section~\ref{sec:action_IIB_bulk}, such that $\kappa_4^2 = \kappa_{10}^2 = 1 / 4\pi$. 

Jockers and Louis~\cite{Jockers:2004yj} were able to determine the characteristic $\cN=1$ data of the effective theory obtained from type IIB string theory with $D7$-branes on a Calabi-Yau orientifold by working out the dimensional reduction in detail and comparing the result with the general form~\eqref{eq:N_1}. They found that a set of chiral fields of the $\cN=1$ theory are given in terms of the scalar fields appearing in the reduction~\eqref{JBC_expand} by
\begin{align}
G^a = & \ c^a - \tau b^a \label{def_Ga}\\
\label{def_Talpha}
T_{\alpha} =& \ \frac{1}{2}  \cK_{\alpha \beta \gamma} v^\beta v^\gamma + i \left(c_{\alpha} - \frac12  \cK_{\alpha bc} c^b {b}^c \right)  + \frac{i}{2\left(\tau-\bar{\tau}\right)} \cK_{\alpha bc} G^b \left( G^{c} - \overline{G}^c\right)  \nonumber \\
=& \  \frac{1}{2}\cK_{\alpha \beta \gamma} v^\beta v^\gamma+ i \left(c_{\alpha} - \cK_{\alpha bc} c^b {b}^c \right) + \frac{i}{2} \tau \cK_{\alpha b c}  b^b   \, b^c \;.
\end{align}
Here $\tau = l + i e^{-\Phi} $ is the axio-dilaton, which combines with the deformation moduli of the $D7$-branes into another chiral field denoted $S$. The massless chiral spectrum is completed by the complex structure moduli which can be obtained from a variation of the holomorphic (3,0)-form $\Omega_3$. In the presence of Wilson line moduli the fields $T_\alpha$ are corrected by an additional term, but this would not affect the present analysis and so we assume for simplicity that no Wilson lines are present.

The K\"ahler potential $K$ of the effective theory can be written as~\cite{Jockers:2004yj}
\be
\label{KpotIIB}
   K = - \log \Big[-i \int_{X_3} \Omega_3 \wedge \bar \Omega_3 \Big] - \log\big[-i(\tau - \bar \tau)\big] - 2\log[\cV]\ .
\ee
In order to evaluate the components of the K\"ahler metric, $\Omega_3$, $\tau$ and $\cV$ have to be viewed as functions of the chiral fields. In the following, we will only require the derivatives of $K$ with respect to $G^a$ and $T_\alpha$ and the component $K_{G^a \bar{G}^b}$ of the K\"ahler metric, which are given by~\cite{Grimm:2004uq, Grimm:2005fa}
\begin{align}
\label{derivs_kaehler_pot}
   \partial_{G^a} K = & \ - \frac{i}{2 \cV }  \int_{X_3} \omega_a \wedge J \wedge B_2  \ , \qquad \partial_{T_\alpha} K = -\frac{v^\alpha}{2 \cV} , \\
   K_{G^a \bar{G}^b} = & \ - \frac14 e^\Phi \frac{\cK_{\alpha a b} v^\alpha}{\cV}  + \frac{1}{16 \cV^2} G^{\alpha\beta} \cK_{\alpha a c} b^c \cK_{\beta b d} b^d  . \label{Kaehler_metric_GG}
\end{align}
The matrix $G^{\alpha\beta}$ appearing in~\eqref{Kaehler_metric_GG} is the inverse of
\be
G_{\alpha\beta} = -\frac14\left( \frac{\cK_{\alpha\beta\gamma}v^\gamma}{\cV} - \frac14 \frac{\cK_{\alpha\gamma\delta}v^\gamma v^\delta \cK_{\beta \epsilon \zeta} v^\epsilon v^\zeta}{\cV^2} \right).
\label{g_alphabeta}
\ee

The index $A$ in~\eqref{eq:N_1} runs over the field strengths $F^A$ of all the gauge bosons of the theory. Each stack\footnote{As before we assume for ease of notation that all brane stacks carry a world-volume gauge group of $U(N_A)$.} of $D7$-branes in the model contributes the $N_A^2$ gauge fields appearing in the expansion~\eqref{expansion_F_generators}, which are associated to the generator of the diagonal $U(1)_A$ and the generators of $SU(N_A)$. The real and imaginary parts of the gauge kinetic function corresponding to these fields can in principle be obtained by dimensionally reducing the DBI and CS actions, respectively, although this may be complicated in the general case of multiple brane stacks with non-Abelian gauge groups. In the following we will require only the simplest case of a single stack of $N_A$ $D7$-branes in the absence of fluxes, in which case the gauge kinetic function for the diagonal $U(1)_A$ was determined in~\cite{Jockers:2004yj}
\be
\label{gaugeKinIIB}
f_{AA} = - \frac{1}{16\pi^2} N_A  C^\alpha_A \, T_\alpha.
\ee

It should be mentioned at this point that the $D7$-branes do not form the only source of gauge bosons in the low energy theory. An additional set of $U(1)$ gauge symmetries arises from the expansion of the Ramond-Ramond form $C_4$ into harmonic 3-forms on $X_3$. To distinguish them from the brane $U(1)$s, we refer to these additional $U(1)$s as R-R $U(1)$s or bulk $U(1)$s. In general, kinetic mixing between the bulk and brane $U(1)$s is possible, depending on the pullback of the harmonic 3-forms from $X_3$ to the brane world-volume. However, as discussed in more detail in~\cite{Jockers:2004yj}, in this case the entire derivation of the chiral $\cN=1$ coordinates becomes more complicated. Throughout this thesis we will assume that kinetic mixing of this form is absent, so that the gauge kinetic matrix $f_{AB}$ splits into distinct blocks corresponding to bulk and brane gauge fields, respectively. In this case, none of the scalar $\cN=1$ fields are charged under any of the bulk $U(1)$s. Furthermore, as we will briefly discuss in section~\ref{sec:KKreduxFtheory}, the bulk $U(1)$s have a very straightforward and clear uplift to F-theory. The discussion of section~\ref{sec:nonabelian_groups} already shows that the case of the brane gauge fields, and in particular the diagonal $U(1)_A$, is more involved and interesting because at first sight the F-theoretic framework seems to account for only the $SU(N_A)$ factor of the IIB gauge group. In addition, the diagonal $U(1)_A$ may participate in gauging a shift symmetry of the scalar fields and become massive via the St\"uckelberg mechanism, as we will review in the next section. For these reasons, we will largely ignore the bulk $U(1)$s in this thesis and focus instead on the brane gauge fields, with particular emphasis on the diagonal $U(1)_A$.

\subsection{The St\"uckelberg mechanism and massive $U(1)$s from $D7$-branes}
\label{sec:Stueckelberg}

The primary motivation behind studying type IIB compactifications involving intersecting $D7$-branes with non-trivial gauge flux is that they lead to a chiral massless 4-dimensional spectrum. The chiral index which measures the net chirality of the matter localised at the intersection between the brane stacks $A$ and $B$ is given by the topological integral~\cite{Blumenhagen:2008zz}
\be
I_{AB} = - \int_{D_A \cap D_B} \left[\tilde{\cF}^A - \tilde{\cF}^B \right] = - \int_{X_3} [D_A] \wedge [D_B] \wedge \left[\tilde{\cF}^A - \tilde{\cF}^B \right].
\label{chiral_index_AB}
\ee
When applying this formula to an intersection where the brane $B$ is an image brane, it is important to take into account the fact that the fluxes on image brane stacks are given by~\eqref{fluxes_on_image_stack}, so that
\be
I_{AB'} = -  \int_{D_A \cap D'_B} \left[\tilde{\cF}^A - \tilde{\cF}^{B'} \right] = - \int_{X_3} [D_A] \wedge [D_B'] \wedge \left[\tilde{\cF}^A + \sigma^* \tilde{\cF}^B \right].
\ee

In the following, we will focus on the contribution from flux $\cF_0$ in the diagonal $U(1)$, although as discussed e.g. in~\cite{Blumenhagen:2008zz} analogous formulae apply for fluxes in the non-Abelian part of the gauge group.
The dependence of the chiral indices on the flux quanta and the wrapping numbers of the D-branes can be made explicitly visible by using the expansions~\eqref{def_tilde_cF} and~\eqref{expansion_D_A} together with the intersection numbers~\eqref{triple_intersections}. These expansions will be helpful when comparing the flux-induced chirality calculated in the F-theory setting with the IIB result. One obtains~\cite{Grimm:2011tb}
\begin{align}
I_{AB} =& - \frac{1}{4 }  \Big( \cK_{\alpha \beta\gamma}C^\beta_A C^\gamma_B + \cK_{\alpha a b} C^a_A C^b_B \Big) \Big( \tilde{\cF}_0^{A,\alpha} - \tilde{\cF}_0^{B,\alpha} \Big) \nonumber \\
& - \frac{1}{4 } \Big( \cK_{\alpha a b}C^\alpha_A C^a_B + \cK_{\alpha a b} C^a_A C^\alpha_B \Big) \Big( \tilde{\cF}_0^{A,b} - \tilde{\cF}_0^{B,b} \Big) \ , \\
I_{AB'} =& - \frac{1}{4 }  \Big( \cK_{\alpha \beta\gamma}C^\beta_A C^\gamma_B - \cK_{\alpha a b} C^a_A C^b_B \Big) \Big( \tilde{\cF}_0^{A,\alpha} + \tilde{\cF}_0^{B,\alpha} \Big) \nonumber \\
& - \frac{1}{4 }  \Big( \cK_{\alpha a b}C^\alpha_B C^a_A - \cK_{\alpha a b} C^a_B C^\alpha_A \Big) \Big( \tilde{\cF}_0^{A,b} - \tilde{\cF}_0^{B,b} \Big) \ .  
\end{align}
In particular, the chiral index of the matter localised at the intersection of a brane and its own orientifold image reduces to
\be
I_{AA'} = -\frac{1}{2} \Big( \cK_{\alpha \beta\gamma}C^\beta_A C^\gamma_A - \cK_{\alpha a b} C^a_A C^b_A \Big) \tilde{\cF}_0^{A,\alpha}.
\ee

The chiral matter at the intersection of two brane stacks, which is counted by the indices above, is charged under the gauge groups of both stacks. More precisely, it transforms in the bifundamental representation of the two gauge groups~\cite{Blumenhagen:2008zz}. In the presence of such charged chiral matter, it is in principle possible for gauge anomalies to appear which would render the theory inconsistent. In principle these anomalies could affect both the Abelian and non-Abelian parts of the gauge group. However, the anomalies involving the gauge bosons of the diagonal $U(1)_A$ are automatically cancelled in string compactifications with intersecting branes~\cite{Blumenhagen:2006ci}. The mechanism responsible for this non-trivial cancellation is known either as the Green-Schwarz or the St\"uckelberg mechanism, and we will use both names interchangeably.
 
The anomalies involving $U(1)_A$ can be grouped into different types depending on whether the other bosons participating in the anomalous interaction are also Abelian gauge bosons, non-Abelian gauge bosons of $SU(N_A)$, or gravitons. If the different anomalies do not cancel amongst one another, the overall anomaly may be cancelled by additional terms in the Lagrangian which transform under $U(1)_A$ gauge transformations in a specific manner. One way to achieve this is to include a 2-form field $c_2$ in the 4-dimensional action, which is coupled to the field strength $F_0^A$ of $U(1)_A$ via a so-called St\"uckelberg coupling of the form 
\be
S_{St.} \supset \int_{M^{1,3}} c_2 \wedge F_0^A.
\label{stueckelberg_schematic}
\ee 
Anomaly cancellation via couplings of the St\"uckelberg form is also known as the Green-Schwarz mechanism. As we will explicitly review below, the dimensional reduction of the Chern-Simons action of $D7$-branes contributes couplings of the form~\eqref{stueckelberg_schematic}. Both the coefficients of the St\"uckelberg coupling and the anomaly coefficients are determined by the fluxes and geometric data of the D-brane configuration. As discussed more extensively in~\cite{Blumenhagen:2006ci}, this dependence is such that the anomalies cancel exactly. 

In the following, we will not attempt to demonstrate this anomaly cancellation in detail and will focus instead on an important side-effect of the St\"uckelberg mechanism. Namely, the $U(1)_A$ gauge boson acquires a mass in the course of the St\"uckelberg mechanism by absorbing the scalar field which is dual to the 2-form $c_2$ appearing in~\eqref{stueckelberg_schematic} as a longitudinal mode\footnote{While anomalous $U(1)$s always obtain a mass from the St\"uckelberg mechanism, the converse is not always true and it is also possible for a non-anomalous $U(1)$ to become massive~\cite{Blumenhagen:2006ci}.}. Due to this mass, the gauge field is no longer directly visible in the low energy effective action. Crucially, however, the selection rules on 4-dimensional couplings resulting from the original $U(1)_A$ gauge symmetry remain intact. Therefore, the $U(1)_A$ symmetry survives as an accidental global symmetry of the effective 4-dimensional low energy theory even after becoming massive. When the theory is re-expressed in the language of F-theory, the effective selection rules survive although the underlying cause in the shape of the massive $U(1)_A$ gauge symmetry is not directly visible in the F-theory construction. In reference~\cite{Grimm:2011tb}, we proposed a method of making the underlying symmetry visible by including a certain class of non-harmonic forms in the Kaluza-Klein reduction of the F-theory model. A review of this construction and the evidence for it will form a central part of the present chapter. To establish the connection between the proposed F-theory construction and the type IIB results, it is necessary to carefully analyse the St\"uckelberg couplings and their dependence on the type IIB compactification geometry, to which we now turn.

To derive the form of the 4-dimensional St\"uckelberg couplings which appear in a type IIB orientifold compactification with $D7$-branes, we dimensionally reduce the Chern-Simons action~\eqref{S_CS}. In the orientifold setting, a brane stack and its corresponding image stack are viewed as providing equivalent descriptions of the same physical object. The overall action of such an orientifold pair of brane stacks is thus obtained by adding the DBI and CS actions of the individual stacks and dividing by 2~\cite{Grimm:2011tb}. The actual dimensional reduction is accomplished by inserting the expansions of the 10-dimensional fields given in equations~\eqref{JBC_expand} and~\eqref{field_strength_and_flux} into the action and integrating out the internal part of the brane world-volume. Using the various intersection numbers defined in section~\ref{sec:IIBreduction}, it is straightforward to derive that the 4-dimensional couplings linear in the field strength $F_A^0$ of the diagonal $U(1)$ can be written as
\be
\label{St-coupl}
S_{St.} = - \frac{1}{4 \kappa_4^2 } \sum_A \Big(    Q_A^a \int_{{M}^{1,3}}  F_{0}^A \wedge \big(\tilde c_{2\, a}  - \cK_{\alpha a c} b^c c_2^\alpha \big)  - Q_{A\alpha}  \int_{{M}^{1,3}}  F_{0}^A \wedge  c_{2}^\alpha  \Big).
\ee
The charges $Q_{A\alpha}$ and $Q_A^a$ appearing in this expression are given by\footnote{Note that by definition the variable fluxes in~\eqref{flux_expansion_twoforms} do not couple to the bulk forms, and thus don't contribute to the charges.}~\cite{Jockers:2004yj, Grimm:2011tb}
\begin{align}
Q_A^a =&  \frac{1}{2\pi} N_A C^a_A \;, \label{Qsodd} \\
Q_{A \alpha} =&  - \frac{1}{2\pi} N_A \Bigl( \cK_{\alpha \beta \gamma}  \, \tilde \cF^{A,\beta}_0 \, C^{\gamma}_A  +  \cK_{\alpha b c }  \, \tilde \cF^{A,b}_0   \, C^{c}_A \Bigr) .\label{Qseven}   
\end{align}
The factors of $N_A$ appearing here originate from the trace over the generator $T_A^0$ of the diagonal $U(1)_A$, which is just the $N_A\times N_A$ unit matrix. By inserting the expansions~\eqref{JBC_expand}, the duality conditions~\eqref{duality_relations} can be used to relate the 4-dimensional 2-forms $\tilde{c}_{2\, a}$ and $c_2^\alpha$ to the scalar fields $c_\alpha$ and $c^a$, respectively. After eliminating the 2-forms in favor of the dual scalars, the St\"uckelberg couplings~\eqref{St-coupl} lead to a gauging of the axionic shift symmetries of $c_\alpha$ and $c^a$. More precisely, the chiral fields $G^a$ and $T_\alpha$ of the 4-dimensional effective $\cN=1$ action~\eqref{eq:N_1}, which are related to $c^a$ and $c_\alpha$ by their definitions~\eqref{def_Ga} and~\eqref{def_Talpha}, couple to the $U(1)_A$ gauge potentials through the covariant derivatives appearing in their kinetic terms. These covariant derivatives take the form~\cite{Grimm:2011tb}
\begin{align}
\nabla G^a =& \  d G^a - Q_A^a A^A,  \label{gauging1} \\
\nabla T_\alpha =&\ dT_\alpha - i Q_{A \alpha} A^A . \label{gauging2}
\end{align}

The fact that the shift symmetries of $G^a$ and $T_\alpha$ are gauged in this manner immediately leads to a mass term for the $U(1)_A$ gauge potentials in the 4-dimensional effective action, which schematically takes the form
\be
\mathcal{S}^{(4)}_{\mathcal{N}=1} \supset -\frac{1}{\kappa_4^2} \int_{M^{1,3}} K_{M\bar{N}} X^M_A \bar{X}^N_B A^A\wedge\ast A^B.
\ee
As before, the indices $M, \ \bar{N}$ run over the chiral fields of the theory, which include $G^a$ and $T_\alpha$. A crucial fact is that in compactifications admitting forms of negative orientifold parity a non-vanishing mass term arises even in the complete absence of fluxes, as the charges $Q_A^a$ are independent of the fluxes. We will refer to the gauging~\eqref{gauging1} and the $U(1)$ mass terms induced by $Q_A^a$ as geometric gauging and geometric mass terms, respectively, because the $Q_A^a$ depend only on the geometry of the $D7$-brane configuration through the wrapping numbers $C_A^a$.

To obtain the physical masses of the gauge fields it is necessary to diagonalise these mass terms and then rescale the metric and the gauge fields to get both the Einstein-Hilbert term and the kinetic terms into canonical form. In the general case involving multiple brane stacks, this is non-trivial and kinetic mixing between the different $U(1)$s may appear if the kinetic and mass terms cannot be diagonalised simultaneously. In this thesis we will not attempt to derive the general expression for the masses. To get an idea of the mass scales involved, we will instead stick to the simplest case of a single $D7$-brane stack without fluxes. The mass of the $U(1)_A$ gauge boson in this situation was worked out in\footnote{Our numerical prefactors differ slightly from those in~\cite{Grimm:2011tb} due to our choice of different conventions regarding the mass dimensions of the fields.}~\cite{Grimm:2011tb}
\begin{align}
\label{massIIB}
m^2 =&\  \frac{M_P^2}{2\pi  \mathrm{Re} f_{AA}} K_{G^a \bar G^b} N_A^2 C^a_A C^b_A  \nonumber \\
=& \  4\pi M_P^2 N_A C^a_A C^b_A  \Big( C^\alpha_A \cK_\alpha - e^\phi C^\alpha_A \cK_{\alpha b c} b^b b^c \Big)^{-1} \nonumber \\
&\ \times \Big[ \frac{e^\phi}{\cV} \cK_{a b} -\frac{1}{2\cV^2} \cK_{ac}\cK_{bd} b^c b^d +\frac{1}{\cV} \cK^{\alpha \beta}\cK_{\alpha a c}\cK_{\beta b d} b^c b^d \Big].
\end{align}
Here we have used the K\"ahler metric~\eqref{Kaehler_metric_GG} and the gauge kinetic function~\eqref{gaugeKinIIB}. The quantities $\cK_{ab}$, $\cK_{\alpha\beta}$ and $\cK_\alpha$ are defined following~\cite{Jockers:2004yj} by contracting the intersection numbers of equation~\eqref{triple_intersections} with suitable numbers of the K\"ahler moduli fields $v^\alpha$. This means that in an isotropic compactification $\cK_{ab}$ and $\cK_{\alpha\beta}$ scale as $\cV^{1/3}$, while $\cK_\alpha \sim \cV^{2/3}$. $\cK^{\alpha\beta}$ is defined as the inverse matrix of $\cK_{\alpha\beta}$ and therefore scales as $\cK^{\alpha\beta} \sim \cV^{-1/3}$.

In the absence of fluxes the $U(1)$ mass~\eqref{massIIB} contains several contributions which depend on the moduli fields $b^a$ as well as one term which depends only on the geometric and volume moduli. In comparison with the geometric mass term, the contributions involving $b^a$ involve an additional factor of $\cV^{-2/3}$, so that they are subleading in a geometric large volume regime~\cite{Grimm:2011tb}. By comparison with~\eqref{KK_scale_bulk}, we see that the purely geometric mass is suppressed by an additional factor of $g_s$ compared to the (bulk) Kaluza-Klein mass scale. As already discussed in section~\ref{sec:ell_fibration_D-brane_backreaction}, the presence of $D7$-branes prevents $g_s$ from being arbitrarily small over the entire compactification manifold\footnote{The dilaton tadpole cannot be cancelled locally in the case we are considering, as that would require the $D7$-branes to be placed on top of the $O7$-plane. In that case, the wrapping numbers $C^a_A$ would vanish and the gauge group would not be $U(N_A)$.}. Therefore, we expect the mass scale of the geometrically massive $U(1)$s to be roughly comparable to the Kaluza-Klein scale, so that the massive $U(1)$s can be neglected in the low energy theory. Note that the estimates above are valid for an approximately isotropic compactification manifold, and the mass may lie significantly lower e.g. if the volume of the $D7$-brane divisor is much smaller than $\cV^{2/3}$~\cite{Conlon:2008wa}. However, we will assume a generic isotropic compactification throughout this thesis for simplicity.

Let us emphasise once again that geometric mass terms of the form discussed above will only appear for the diagonal $U(1)$s associated with brane stacks whose divisors $D_A$ are not invariant under the orientifold action. In particular, if $X_3$ has no non-trivial orientifold-odd divisors\footnote{This is the case e.g. for toric orientifolds built using an involution $\sigma$ which can be written as a reflection $x\rightarrow -x$ of a homogeneous coordinate.} and $h^{1,1}_- (X_3) = 0$, no geometric gauging can occur. Even if $h^{1,1}_- (X_3) > 0$, it is possible to obtain $U(N_A)$ gauge symmetries with a massless diagonal factor in the low energy action. This situation arises in the presence of $D7$-branes which are placed in the same homology class as the orientifold plane without lying completely on top of it. Let us further note that a compactification which includes $n$ brane stacks $D_{A_1},\ldots,D_{A_n}$ with $[D_{A_i}] \neq [D_{A_i}']$ does not necessarily lead to a low energy theory with $n$ massive $U(1)$s. This is because the physical $U(1)$s of the low energy action are superpositions of the diagonal $U(1)$s of several brane stacks and are obtained by diagonalising the $U(1)$ mass matrix. The number of massive $U(1)$ combinations is given by the rank of the matrix $C_A^a$, which may be smaller than $n$. On the other hand, every stack of $D7$-branes which hosts a $U(N_A)$ gauge theory and obeys $[D_A] = [D_A']$ immediately contributes a massless diagonal $U(1)$ to the low energy theory.

\subsection{D-terms and flux-induced tadpoles in type IIB}
\label{sec:IIB_tadpoles}

The form of the covariant derivatives~\eqref{gauging1} and~\eqref{gauging2} means that the shift symmetries of $G^a$ and $T_\alpha$ are gauged by the non-linearly realised $U(1)_A$ gauge symmetries. The non-linear action of the gauge symmetry makes it easy to read off the Killing vectors corresponding to these gauged isometries, which are given by~\cite{Grimm:2011tb}
\be
 \label{Killing_sum}
   X_A^a = - Q_A^a \ , \qquad  X_{A \alpha} = - i Q_{A \alpha}\ .
\ee
As these Killing vectors are constants and have no dependence on the moduli fields of the theory, equation~\eqref{genD} may be integrated straightforwardly to yield the D-term corresponding to the $U(1)_A$ gauge symmetry
\begin{align}
 D_A = -i K_{\bar{M}} \bar{Q}_A^M = & \ \frac{N_A}{2 \pi \cV}  \int_{D_A} J \wedge ( \frac{1}{2\pi} \cF_A - \iota^* B_2 ) \nonumber \\
  = & \ \frac{v^\alpha}{4 \pi \cV} N_A  \big(    \cK_{\alpha \beta \gamma} \tilde \cF^{A,\beta}_0 C_A^\gamma  +     \cK_{\alpha ac} ( \tilde \cF^{A,c}_0 - b^c ) C_A^a   \big) .
  \label{DtermIIB}
\end{align}
Even in the absence of fluxes, a Fayet-Iliopoulos D-term depending on the moduli $b^a$ can thus arise, corresponding to the geometric gauging of the shift symmetry of $G^a$. This flux-independent D-term appears only in the case where the brane stack is not orientifold-invariant and the wrapping numbers $C_A^a$ do not vanish. This is consistent with the fact that the associated $U(1)_A$ is geometrically massive\footnote{Strictly speaking, in the case with several brane stacks the $U(1)_A$ may be a superposition of massive and massless physical $U(1)$s if the mass matrix (or, equivalently, the charge matrix $Q_A^a \propto C_A^a$) does not have full rank.} if $C_A^a \neq 0$.

As we have seen above, gauge fluxes along the diagonal $U(1)_A$ of the $U(N_A)$ gauge theory which is present on suitable stacks of $D7$-branes can form a powerful tool from a model building perspective. This is because such fluxes have many important consequences for the low energy theory through their contributions to the $U(1)$ mass matrix, the D-terms and the chiral indices of charged matter.
This makes it crucial to understand any restrictions that might constrain the fluxes which can be switched on. For one, fluxes must obey the Freed-Witten quantisation condition~\eqref{FW_quant}, which restrict the fluxes to integer or half-integer values depending on the form of the canonical bundle of the brane divisor. A second set of constraints arises from the requirement of $D5$- and $D3$-brane tadpole cancellation, which yield conditions relating fluxes from different $D7$-brane stacks.

The relevant tadpole cancellation conditions in the presence of spacetime-filling D-branes can be derived from the equations of motion of the Ramond-Ramond form fields $C_4$, $C_6$ and $C_8$. These equations of motion are computed from the total action obtained by adding up the IIB bulk action~\eqref{S_IIB_democ} and the actions of all brane stacks and orientifold planes in the model. As discussed in detail in~\cite{Grimm:2011tb}, two important additional factors of 1/2 must be taken into account when computing these equations of motion. The first factor of 1/2 stems from the orientifold geometry and relates to the fact that, as already mentioned above, brane and image brane stacks are indistinguishable from a physical perspective. Therefore, the physical action of a brane-image-brane pair is obtained by adding the individual brane actions and dividing the result by 2. The second factor is required to take into account the doubling of the degrees of freedom which occurs in the democratic formulation of type IIB. This doubling made it possible to include both electric and magnetic couplings on an equal footing in the Chern-Simons actions of D-branes and O-planes. As the two types of couplings are physically dual to one another, only one should be taken into account when deriving the tadpole cancellation conditions. In~\cite{Grimm:2011tb} it was argued that this can be achieved by including an additional factor of 1/2 with the CS actions of D-brane and O-plane stacks.

Taking the factors of 1/2 mentioned above into account, the equation of motion of $C_8$ reads~\cite{Grimm:2011tb}
\be
d \ast G_9 = \frac12 \sum_A N_A \delta(D_A^+) - 4 \delta(D_{O7}).
\ee
Here $\delta(D_A^+)$ and $\delta(D_{O7})$ are specific representatives of the cohomology classes $[D_A^+]$ and $[D_{O7}]$ which are sharply peaked around the corresponding divisors. Using the duality relations~\eqref{duality_relations}, this can be translated into a Bianchi identity for $F_1$
\be
d F_1 = \frac12 \sum_A N_A \delta(D_A^+) - 4 \delta(D_{O7}).
\label{Bianchi_id_F1}
\ee
As a consistency check for the factor of 1/2 appearing on the right hand side of this equation, we follow~\cite{Grimm:2011tb, Collinucci:2008pf} and consider computing the monodromy of $C_0$ along a path which encircles a $D7$-brane stack in the orientifold quotient $B_3 = X_3/\sigma$. By definition, a single $D7$-brane should carry a unit of $C_0$ charge and so this monodromy should be 1. In the double cover $X_3$, the path lifts to a path which encircles both brane and image brane stacks. This implies that the factor of 1/2 in~\eqref{Bianchi_id_F1} is precisely as required to still obtain a unit monodromy along this path.

To preserve Lorentz invariance of the vacuum, all branes (and the $O7$-plane) are assumed to be spacetime-filling. This implies that the forms on the right hand side of~\eqref{Bianchi_id_F1} are forms lying completely in the compact space $X_3$. By integrating~\eqref{Bianchi_id_F1} over a basis of $H_2(X_3)$ and noting that the left hand side vanishes by Stokes' theorem, we thus obtain the $D7$-brane tadpole cancellation condition
\be
\label{D7-tadpole_IIB}
\sum_A N_A ([D_A] + [D_A']) = 8 [D_{O7}].
\ee

The $D5$- and $D3$-brane tadpoles can be derived in an analogous manner by considering the equations of motion of $C_6$ and $C_4$, respectively. The only qualitative difference is that in the presence of non-trivial gauge fluxes $C_6$ and $C_4$ couple to more than one type of D-brane. As before, we focus on $D7$-branes with gauge group $U(N_A)$ and specifically on flux in the diagonal $U(1)_A$, whose contribution to the $D5$-brane tadpole is encoded in the Bianchi identity
\be
\begin{aligned}
dF_3 =& \  - \frac12  \sum_A N_A\, \left[ \left(\frac{1}{2\pi} {\cal F}^{A}_0 - \iota^* B_2\right) \wedge \delta(D_A) \right. \\ & \ \left. -  \sigma^*\left( \frac{1}{2\pi}  {\cal F}_0^{A} - \iota^* B_2\right) \wedge  \delta(D_A' )  \right]   \\ \label{Bianchi_id_F3}
& \ + 4 \iota^* B_2^- \wedge \delta(D_{O7} )  - \frac12 (\delta(D_5) + \delta(D_5')) \;.
\end{aligned}
\ee
For completeness we have indicated the form of a possible contribution from $D5$-branes in the final term, although we will assume for simplicity that no $D5$-branes are present in the following. Note that due to the negative orientifold parity of $C_6$, the coupling to the $O7$-plane involves only $B_2^-$ and not $B_2^+$. Upon integrating~\eqref{Bianchi_id_F3} over a basis of $H_4(X_3)$, the $D7$ tadpole cancellation condition can be used to show that the contribution of $B_2^-$ actually drops out of the $D5$ tadpole cancellation condition~\cite{Blumenhagen:2008zz, Grimm:2011tb}
\be
\label{D5tadInt}
  \sum_A N_A \big( \cK_{\alpha b c} \tilde  \cF_0^{A, \alpha} C^b_A+ \cK_{a \beta c} \tilde  \cF_0^{A,a}  C^\beta_A  \big)= 0\ .
\ee
This was of course to be expected, because the $D5$ tadpole describes a net overall $D5$-brane charge, which obeys a Dirac quantisation condition and should thus not depend on the continuous moduli $b^a$.

The $D3$-brane tadpole potentially obtains contributions not only from $D3$-branes and $O3$-planes, but also from fluxed $D7$- or $D5$-branes. We will follow~\cite{Grimm:2011tb} and focus on the contribution of the diagonal $U(1)_A$ flux $\cF_0^A$ along $D7$-branes, and in particular we assume for simplicity that no $D5$-branes are present. The (integrated) $D3$-brane tadpole cancellation condition may then be written as~\cite{Blumenhagen:2008zz}
\be
N_{D3} + N_{\rm flux} = \frac{N_{03}}{4} + \sum_A N_A \frac{\chi_0(D_A)}{24} + \frac{\chi(D_{07})}{12}.
\label{D3tadpole}
\ee
Here $N_{D3}$ and $N_{O3}$ are the numbers of spacetime-filling $D3$ brane-image-brane pairs and $O3$-planes present in the model. The last two terms on the right hand side involving the Euler characteristics of the $D7$-branes and the $O7$-plane originate from the $\hat{A}$-genus in the CS actions~\eqref{S_CS} and~\eqref{S_CS_oplane}. The index $0$ in $\chi_0(D_A)$ indicates that in the presence of singularities on the brane divisor the Euler characteristic must be slightly modified as discussed in~\cite{Collinucci:2008pf}. Finally, the term $N_{\rm flux}$ includes possible contributions from both $D7$-brane gauge fluxes and from bulk fluxes $H_3 = dB_2$ and $F_3 = dC_2$. The contribution of the diagonal $U(1)_A$ flux $\cF_0^A$ to $N_{\rm flux}$ is given by~\cite{Grimm:2011tb}
\be
\begin{aligned}
N_{\rm flux}^{(0)} =& - \frac{1}{4 }  \, \sum_A \, N_A \Big( \int_{D_A}   \tilde {\cal F}^A_0 \wedge \tilde {\cal F}^A_0  + \int_{D_A'} \tilde {\cal F'}^A_0 \wedge \tilde {\cal F'}^A_0 \Big)   \\
        =& - \frac{1}{4} \sum_A N_A \Big( \cK_{\alpha \beta \gamma} C^\alpha_A \tilde \cF_0^{A, \beta} \tilde \cF_0^{A, \gamma} +   \cK_{\alpha b c} C^\alpha_A  \tilde \cF_0^{A,b}  \tilde \cF_0^{A, c} +  \\
        &  2    \cK_{a b \gamma} C^a_A \tilde \cF_0^{A, b}  \tilde \cF_0^{A, \gamma}             \Big). \label{d3iibNgauge}
\end{aligned}
\ee
In the second line of~\eqref{d3iibNgauge} we have restricted ourselves to fluxes which can be written as pullbacks of forms from $X_3$ to the brane world-volume for ease of notation, although variable fluxes also contribute to the $D3$-brane tadpole~\cite{Blumenhagen:2008zz}. Furthermore, we have used the $D5$-brane tadpole cancellation condition to eliminate the continuous $B_-$ moduli~\cite{Blumenhagen:2008zz}. In the following sections, we will discuss how fluxes along the diagonal $U(1)_A$ can be uplifted into an F-theory compactification. In particular, we will see how the expressions for the flux-dependent chiral indices, $U(1)$ masses and tadpoles derived above are reproduced in the F-theoretic setting.

\section{$U(1)$s in the effective action of F-theory compactifications}
\sectionmark{$U(1)$s in the F-theory effective action}
\label{sec:U(1)inFtheory}
In the previous sections we have seen that type IIB compactifications with $D7$-branes generically lead to low energy theories with $U(N)$ gauge group factors, and that in particular the diagonal $U(1)\subset U(N)$ and its associated fluxes have many important consequences for the effective theory. In contrast to $U(1)$s originating from the Cartan subsector of a non-Abelian gauge group, which were already discussed briefly in section~\ref{sec:nonabelian_groups}, the diagonal $U(1)$s are somewhat hidden from view in F-theory compactifications. In the following subsections, we will outline a proposal which was presented in~\cite{Grimm:2011tb} and allows both massive and massless diagonal $U(1)$s and their fluxes to be explicitly included in the F-theory reduction. We then present additional evidence for the proposed construction by showing that it allows the type IIB effects related to the diagonal $U(1)$s, which were discussed in section~\ref{sec:U(1)inIIB}, to be reproduced consistently in F-theory.

\subsection{Kaluza-Klein reduction of F-theory at the massless level}
\label{sec:KKreduxFtheory}
The starting point when constructing the 4-dimensional effective action of F-theory is to use the correspondence with an M-theory compactification on an elliptically fibered fourfold, as discussed in section~\ref{sec:ell_fibration_D-brane_backreaction}. The M-theory compactification initially leads to an $\cN=2$ supersymmetric theory in 3 dimensions, which is uplifted to a 4-dimensional $\cN=1$ theory in the F-theory limit of vanishing fiber volume. The dimensional reduction and derivation of the F-theory effective action at the massless level was carried out in~\cite{Grimm:2010ks}. In the following, we will not attempt to present the full details of this reduction, and will instead focus on aspects relating to $U(1)$ gauge symmetries and their fluxes.

The elliptically fibered Calabi-Yau\footnote{Further details on the mathematical description of such fibrations in terms of Weierstrass models, which will be used throughout the present chapter, can be found in section~\ref{sec:elliptic_CalabiYaus}.} fourfold which forms the basis for the dimensional reduction of F-theory will be denoted by $Y_4$. $Y_4$ is endowed with the structure of a fibration by a suitable projection map $\pi_Y : Y_4 \rightarrow B_3$ to a (complex) 3-dimensional base manifold $B_3$. As we will rely quite heavily on the type IIB results for inspiration, we focus on F-theory setups which have a well-defined type IIB limit. This means that $B_3$ is identified with the orientifold quotient $X_3/\sigma$ of a Calabi-Yau threefold $X_3$ which defines a IIB compactification.
As in the type IIB case, the M-theory compactification is performed by carrying out a Kaluza-Klein reduction of the fields appearing in the low energy effective action of M-theory and then integrating out the internal space $Y_4$. Strictly speaking, in an F-theory compactification involving non-Abelian gauge groups the dimensional reduction cannot be directly carried out on $Y_4$ itself, because $Y_4$ develops $ADE$ type singularities as discussed in section~\ref{sec:nonabelian_groups}. Instead, it is necessary to first resolve these singularities to obtain a smooth Calabi-Yau manifold $\hat Y_4$, on which objects like cohomology groups and topological intersection numbers may be defined in a meaningful manner. 

The low energy limit of M-theory can be described by the unique\footnote{Here, of course, we mean unique up to duality transformations which lead to equivalent theories. We consider a dual democratic formulation of 11-dimensional supergravity in appendix~\ref{sec:app_democ_Mtheory}.} 11-dimensional supergravity, whose bosonic field content is given by the 11-dimensional metric $g$ and a 3-form potential $C_3$. Up to terms of higher order in the curvature scalar, the bosonic part of the action is given by~\cite{Denef:2008wq, Becker:2007zj}
\be
\label{S11normal}
S_{11} = \frac{1}{2\kappa_{11}^2} \int d^{11} x \sqrt{-g} R - \frac{1}{2\kappa_{11}^2} \int \left( \frac12 G_4 \wedge \ast G_4 + \frac{1}{6} C_3 \wedge G_4\wedge G_4 \right) .
\ee
Here $G_4 = d C_3$ is the 4-form field strength associated to $C_3$. As in the IIB setting, we work in conventions where the 11-dimensional spacetime coordinates as well as the fields $C_3$, $g$ are dimensionless. In these conventions, the 11-dimensional gravitational coupling constant simply reduces to\footnote{Our action can be related to the action of~\cite{Denef:2008wq, Becker:2007zj} by a rescaling $x^i\rightarrow x^i/\ell_M$ of the coordinates followed by a Weyl rescaling $g_{ij} \rightarrow g_{ij}/\ell_M^2$, where $\ell_M$ is the characteristic M-theory length scale. Under these rescalings the 11-dimensional gravitational coupling changes to $\kappa_{11}^2 = \ell_M^9/4\pi$.} $\kappa_{11}^2 = 1/4\pi$.


Just as in the type IIB reduction discussed in section~\ref{sec:IIBreduction}, the multiplicities of the massless 3-dimensional fields appearing in the Kaluza-Klein reduction of the 11-dimensional fields are given by the Hodge numbers of $\hat{Y_4}$. Our notation for the cohomology groups of $\hat Y_4$ will be based upon the notation used in refs~\cite{Grimm:2010ks, Grimm:2011tb, Kerstan:2012cy}. The variation of the 11-dimensional metric gives rise to $h^{1,1}(\hat Y_4)$ real scalar fields $v^\Lambda$ and a set of $h^{3,1}(\hat Y_4)$ complex scalars $z^M$~\cite{Haack:1999zv, Haack:2001jz}. While the $z^M$ can be related to variations of the complex structure, the fields $v^\Lambda$ arise from the expansion
\be
\label{kaehler_exp_fourfold}
J = v^\Lambda \omega_\Lambda
\ee
of the K\"ahler form $J$ of $\hat Y_4$ into a basis $\{\omega_\Lambda\}$ of the cohomology group $H^{1,1}(\hat Y_4, \bZ)$. For the Kaluza-Klein reduction of $C_3$, we additionally require a basis of $H^3(\hat Y_4, \bZ)$. In the following, we will often write simply $H^p(\hat Y_4)$ instead of $H^p(\hat Y_4, \bZ)$ for the sake of brevity, although all the basis forms used in this thesis are normalised to have integer intersections. As $\hat{Y}_4$ is Calabi-Yau, there holds $h^{3,0}(\hat Y_4)=0$ and therefore $H^3(\hat Y_4) = H^{2,1}(\hat Y_4) \oplus H^{1,2}(\hat Y_4)$. For future convenience, we choose a basis of $H^3(\hat Y_4, \bZ)$ which can be divided into two sets $\{\alpha_\kappa\},\ \{\beta^\kappa\}$ of real 3-forms which obey
\be
\int_{\hat Y_4} \omega_\Lambda \wedge \alpha_\kappa \wedge \alpha_\lambda = \int_{\hat Y_4} \omega_\Lambda \wedge \beta^\kappa \wedge \beta^\lambda = 0.
\label{int_alphabeta_harmonic}
\ee
In terms of the basis forms introduced above, the harmonic Kaluza-Klein reduction of $C_3$ reads~\cite{Grimm:2010ks}
\be
C_3 = A^\Lambda \wedge \omega_\Lambda + c^\kappa \alpha_\kappa + b_\kappa \beta^\kappa.
\label{expansion_C3}
\ee
Here the $A^\Lambda$ are 3-dimensional vector fields which pair up with the $v^\Lambda$ to form 3-dimensional vector multiplets, while $c^\kappa$ and $b_\kappa$ are real scalar fields. 

In the type IIB limit, the massless 3-dimensional fields appearing in these expansions lift to the 4-dimensional massless fields obtained in the Kaluza-Klein reduction of type IIB on $X_3$. The precise identification was worked out in detail in~\cite{Grimm:2010ks} and will be briefly reviewed in the following. At first sight it may seem tempting to assume that e.g. the vectors $A^\Lambda$ must uplift to 4-dimensional vectors and scalars $c^\kappa, b_\kappa$ lead to 4-dimensional scalars. However, this naive assumption fails because scalar and vector fields are dual in 3 dimensions. Instead, to find the correct correspondence between 3- and 4-dimensional fields it is necessary to analyse the relationships between the cohomology groups of $\hat Y_4$ and $B_3$ (or $X_3$) which are induced by the structure of the elliptic fibration. 

We focus on the cohomology groups which are relevant for the description of Abelian gauge symmetries in the low energy action, namely $H^{1,1}(\hat Y_4)$ and $H^{2,1}(\hat Y_4)$. The fact that $\hat Y_4$ is elliptically fibered over $B_2$ and arises by resolution from the space $Y_4$ makes it possible to divide the forms $\omega_\Lambda \in H^{1,1}(\hat Y_4)$ into 3 distinct categories~\cite{Grimm:2010ks}.
First, $H^{1,1}(\hat Y_4)$ includes a form $\omega_0$ which is Poincar\'e dual
 to the homology class of the base $B_3$ in $\hat Y_4$. By this we actually mean that $\omega_0$ is dual to the divisor in $\hat Y_4$ defined by the vanishing of the zero section\footnote{Throughout this thesis we are assuming that a section exists, so that $\hat{Y_4}$ can be written in Weierstrass form. For the definition of the zero section of a Weierstrass fibration, see section~\ref{sec:elliptic_CalabiYaus}.} of the fibration. $\omega_0$ can be used to pull back integrals of differential forms from $B_3$ to $\hat Y_4$. 
More precisely, given a $p$-form $\eta_p$ on $B_3$ and a $p$-dimensional submanifold $\Gamma_p$ of $B_3$ there holds
\be
 \int_{\Gamma_p \subset B_3} \eta_p = \int_{\pi_{\hat Y}^{-1}(\Gamma_p) \subset \hat Y_4} \omega_0 \wedge \pi_{\hat Y}^* \eta_p.
\label{integrals_B3_and_Y4}
\ee 
In the case where there are additional sections beyond the canonical zero section, each independent section contributes a further harmonic (1,1)-form~\cite{Morrison:1996pp, Morrison:2012ei, Cvetic:2013nia}. This fact will be crucial for the discussion of massless $U(1)$s in section~\ref{sec:sections_and_U(1)s}, but for the moment we assume for simplicity that the fibration admits only the zero section.

In addition to $\omega_0$, one obtains a set of forms $\omega_\alpha$ which arise by pullback of $H^{1,1}(B_3)$ along the projection $\pi_{\hat{Y}}$. As the notation suggests, these forms are in one-to-one correspondence with the basis of $H^{1,1}_+(X_3)$, which was introduced in section~\ref{sec:IIBreduction} and which survives the orientifold projection to $B_3$. 
We will use the same notation for the forms $\omega_\alpha$ defined on $\hat Y_4$, $B_3$ and $X_3$, trusting that it will always be clear from the context which specific form is being considered. 
The $\omega_\alpha$ furnish Poincar\'e duals of the so-called vertical divisors $D_\alpha \equiv \pi_{\hat{Y}}^{-1}(D^b_\alpha)$, which are obtained by fibering the complete elliptic fiber over the divisors $D_\alpha^b \subset B_3$ of the base.

In the case without non-Abelian gauge symmetries, where $Y_4$ is smooth, the forms discussed above span $H^{1,1}(Y_4)$~\cite{Witten:1996bn}. However, in the general case the resolution $\pi_R:\hat Y_4 \rightarrow Y_4$ introduces additional resolution divisors with associated (1,1)-forms. More precisely, each component $D_A^b$ of the discriminant locus $\Delta \subset B_3$ which hosts a non-Abelian singularity gives rise to a set of forms in $H^{1,1}(\hat Y_4, \bZ)$ which correspond to Cartan generators of the gauge group $G_A$ on $D_A^b$~\cite{Grimm:2010ks}. Following~\cite{Grimm:2011tb}, we denote these forms by $\tw_{i A}, \ i = 1,\ldots, \text{rk}(G_A)$. The divisors dual to the $\tw_{iA}$ are formed by fibering the different fiber $\bP^1$'s which appear after resolution over the base divisor $D_A^b$~\cite{Krause:2011xj}. As already discussed in section~\ref{sec:nonabelian_groups}, the resolution $\bP^1$'s intersect according to the Cartan matrix $\cC^A_{ij}$ of the gauge group $G_A$. More precisely, the forms $\tw_{iA}$ obey~\cite{Krause:2011xj, Grimm:2011tb}
\be
\int_{\hat Y_4} \pi_{\hat Y}^* (\gamma )\wedge \tw_{iA} \wedge \tw_{jB} = - \delta_{AB} \cC^A_{ij} \int_{D_A^b} \gamma  \qquad \forall \gamma \in H^4(B_3). 
\label{cartan_intersection_harmonic}
\ee
The cohomology classes associated with the divisors $D_A^b$, or equivalently their pullbacks $D_A = \pi_{\hat Y}^{-1}(D_A^b)$, can be expanded into the forms $\omega_\alpha$ on $B_3$ resp. $\hat Y_4$. As in section~\ref{sec:IIBreduction}, 
we denote the expansion coefficients by $C_A^\alpha$, so that
\be
[D_A] =  C_A^\alpha \ \omega_\alpha.
\label{poincare_dual_DA}
\ee

After dualisation, the 3-dimensional vector multiplets involving $v^\alpha$ and $A^\alpha$ yield a set of $h^{1,1}(B_3)$ complex scalars. It is natural to expect that in the F-theory uplift these scalars correspond to the $h^{1,1}(B_3)$ scalars $T_\alpha$ which occurred in the type IIB compactification. The general discussion of section~\ref{sec:nonabelian_groups} showed that a component $D_A^b$ of the discriminant locus corresponds to a stack of $D7$-branes in the IIB limit, whose gauge group $G_A$ matches the ADE type of the singularity above $D_A^b$. This immediately suggests that the 3-dimensional vector multiplets\footnote{Strictly speaking, the vector multiplets involve fields $\xi^\Lambda$ which are related to $v^\Lambda$ by a rescaling involving the volume of $\hat Y_4$~\cite{Grimm:2010ks, Grimm:2011tb}, but we will ignore this slight distinction for the moment.} $(v^{iA}, A^{iA})$ should lift to 4-dimensional vectors describing the gauge bosons in the Cartan of $G_A$. These expectations were confirmed by explicitly performing the F-theory uplift in~\cite{Grimm:2010ks}. The same analysis showed that the leftover vector multiplet $(v^0, A^0)$ becomes a part of the 4-dimensional metric in the uplift from 3 to 4 dimensions. In the case where additional sections are present, the associated 3-dimensional vector multiplets lift to 4-dimensional gauge bosons. 

The fact that $\hat Y_4$ is elliptically fibered can also be used to split $H^{2,1}(\hat Y_4)$ into two\footnote{A third contribution to $H^{2,1}(\hat Y_4)$ can arise in the presence of Wilson lines on the $D7$-brane divisors~\cite{Grimm:2010ks}. Wilson lines play no role in the present analysis, and as in section~\ref{sec:U(1)inIIB} we assume that no Wilson lines are present in the model to simplify the notation.} distinct subspaces which are related to different cohomology groups of $B_3$. One part is simply given by the pullback of $H^{2,1}(B_3)$ to $\hat Y_4$. The reduction of $C_3$ along this part of $H^{2,1}(\hat Y_4)$ yields a set of 3-dimensional scalars, which are related in the F-theory uplift to the bulk $U(1)$s obtained by reducing $C_4$ on $H^{2,1}_+(X_3)$ in the type IIB compactification~\cite{Grimm:2010ks}. The second part of $H^{2,1}(\hat Y_4)$ arises from an uplift of $H^{1,1}_- (X_3)$, which can be heuristically understood as follows~\cite{Denef:2008wq, Blumenhagen:2010ja}. Due to their negative orientifold parity, the cohomology classes $\omega_a \in H^{1,1}_- (X_3)$ do not survive the orientifold projection to $B_3$. This means that they cannot be directly uplifted to $\hat Y_4$ as was possible with the positive parity forms $\omega_\alpha$. However, it turns out to be possible to create well-defined 3-forms on $\hat Y_4$ by combining the $\omega_a$ with the 1-forms $dx,\ dy$ on the elliptic fiber. The underlying reason is that the monodromy around an orientifold plane acts on $dx$ and $dy$ by sign reversal~\cite{Denef:2008wq}, so that $dx$ and $dy$ effectively have negative orientifold parity just like the $\omega_a$. As the negative parities of the two factors cancel, the schematic combinations
\be
\alpha_a \sim \omega_a \wedge dy , \qquad \beta^a \sim \delta^{ab}\omega_b \wedge dx
\label{uplift_omega}
\ee
can survive the orientifold projection and lead to well-defined 3-forms on $\hat Y_4$. The existence of the forms $\alpha_a$, $\beta^a$ can be confirmed by an explicit computation of the F-theory limit, which shows that the scalars obtained by reducing $C_3$ on $\alpha_a$ and $\beta^a$ are uplifted to the moduli fields $c^a$ and $b^a$ of the type IIB theory~\cite{Grimm:2011tb}.

Although they will not be needed in the following, let us note that the complex structure moduli $z^M$ of $\hat Y_4$ can be related to the complex structure moduli of $X_3$ and the deformation moduli of the $D7$-branes~\cite{Denef:2008wq}. In summary, the number of complex scalar fields in the 4-dimensional effective action of F-theory is given by~\cite{Grimm:2010ks}
\be
n_{s} = h^{3,1}(\hat{Y}_4) + h^{1,1}(B_3) + h^{2,1}(\hat Y_4) - h^{2,1}(B_3).
\ee
The remaining $h^{1,1}(\hat Y_4) + h^{2,1}(B_3) - h^{1,1}(B_3) - 1 $ massless 3-dimensional vector multiplets obtained in the Kaluza-Klein reduction on $\hat{Y}_4$ uplift to 4-dimensional gauge bosons. This number includes the multiplets $(v^{iA}, A^{iA})$ which are associated with the Cartan subalgebra of a non-Abelian part of the gauge group. As noted in~\cite{Grimm:2010ks}, these gauge bosons are absorbed into the non-Abelian gauge groups which re-appear in the F-theory limit of vanishing fiber volume. Hence the true number of massless Abelian gauge factors in the 4-dimensional F-theory gauge group is given by
\be
n_{U(1)} = h^{1,1}(\hat Y_4) + h^{2,1}(B_3) - h^{1,1}(B_3) - 1 - \sum_A \text{rk} (G_A).
\ee
Here we have included the $h^{2,1}(B_3)$ bulk $U(1)$ gauge bosons for completeness, although as in section~\ref{sec:U(1)inIIB} we will assume that $h^{2,1}(B_3)=0$ in the following to simplify the required notation.

\subsection{Geometrically massive $U(1)$s from non-harmonic forms}
\label{sec:massiveU(1)sFtheory}

The discussion above shows a beautiful match between the effective theories obtained from F-theory and type IIB compactifications, at least at the level of the massless degrees of freedom appearing in the effective action. The expansions of the 11-dimensional metric and of $C_3$ along the various cohomology groups described above have in particular allowed us to identify the F-theory analogues of the type IIB K\"ahler and complex structure moduli, as well as of the bulk $U(1)$s and the deformation moduli of $D7$-branes. Under the assumption that the uplift~\eqref{uplift_omega} is well-defined in cohomology, we have also found analogues of the scalars $c^a$, $b^a$ appearing in the reduction of the IIB fields $C_2$ and $B_2$ given in~\eqref{JBC_expand}. After including the gauge bosons corresponding to off-diagonal generators, which arise from $M2$-branes wrapped on $\bP^1$'s in the resolved fiber and become massless in the F-theory limit, we have also been able to account for the non-Abelian part of the $D7$-brane gauge groups. From a type IIB perspective, this leaves only the fate of the diagonal $U(1)_A$ gauge bosons\footnote{As before, we focus on the case in which the $D7$-brane gauge is $U(N_A)$.} discussed in section~\ref{sec:U(1)inIIB} unclear. A natural guess is that the diagonal $U(1)$s are accounted for by the additional vector multiplets which appear in F-theory compactifications with multiple sections. Such models have been studied recently e.g. in~\cite{Mayrhofer:2012zy, Borchmann:2013jwa, Cvetic:2013nia, Cvetic:2013uta}, although to the best of our knowledge the explicit relation to the type IIB picture has not been worked out in full detail. Nevertheless, it is plausible that additional sections can account for all the type IIB diagonal $U(1)$s which remain massless in the absence of fluxes. 

However, the F-theory reduction presented so far does not account for diagonal $U(1)$s which in type IIB obtain a mass via the geometric St\"uckelberg mechanism discussed in section~\ref{sec:Stueckelberg}. This was to be expected, because F-theory does not have a direct analogue of the type IIB geometric St\"uckelberg mechanism\footnote{On the other hand, flux-dependent St\"uckelberg masses can appear in F-theory, as we will see in section~\ref{sec:FtheoryLimit}.}~\cite{Grimm:2010ks, Grimm:2011tb}. In order for the effective theories obtained by the IIB and F-theory compactifications to match at the massless level, the geometrically massive $U(1)_A$ gauge bosons should therefore be absent from the beginning in the massless Kaluza-Klein reduction of F-theory. In a sense, we expect the effects of the geometric St\"uckelberg mechanism to be built in automatically into the geometry of the F-theory compactification. In particular, the F-theory setup should automatically guarantee the absence of the anomalies which in type IIB are cancelled by the Green-Schwarz mechanism involving geometrically massive $U(1)$s. This was explicitly checked and confirmed in~\cite{Cvetic:2012xn}.

Although the massless F-theory reduction yields a consistent and anomaly-free theory, it is worthwhile exploring the role of the geometrically massive $U(1)$s in F-theory further. The main motivation for this from a phenomenological point of view is that massive $U(1)$s contribute to selection rules on allowed field couplings in the low energy effective theory. As will be discussed further in chapter~\ref{sec:chap_instantons}, the $U(1)$ charges also restrict the shape of the superpotential contributions due to certain instanton configurations, which can have important consequences for moduli stabilisation. In the type IIB description, it is obvious that these selection rules remain intact even when the $U(1)$ gauge boson becomes massive. Furthermore, the selection rules are expected to survive in F-theory models which have a smooth type IIB limit. In~\cite{Grimm:2011tb}, a prescription was given which can be used to make the geometrically massive $U(1)$ symmetries, and hence the underlying reason for the low energy selection rules, explicitly visible in the F-theory reduction. 

To describe the geometrically massive $U(1)_A$ gauge bosons, we follow~\cite{Grimm:2011tb} and include an additional set of non-closed 2-forms $\tw_{0A}$ in the dimensional reduction of $C_3$
\be
\label{C3_expand}
  C_3 =  A^0 \wedge \omega_0 + A^\alpha \wedge \omega_\alpha + A^{i A} \wedge \tw_{iA} + A^{0A}\wedge \tw_{0A} + c^a \alpha_a + b_a \beta^a \equiv A^\Lambda \wedge \omega_\Lambda.
\ee
Here and in the following we will extend the notation $\omega_\Lambda$ to run over all the (1,1)-forms appearing in the Kaluza-Klein reduction, including $\tw_{0A}$, even though the index $\Lambda$ had originally been introduced in section~\ref{sec:KKreduxFtheory} to label the harmonic (1,1)-forms only.
The 3-dimensional vector $A^{0A}$ will be identified with the diagonal $U(1)_A$ of the brane stack on $D_A$ in the IIB limit, up to a normalisation factor which will be worked out in detail in section~\ref{sec:FtheoryLimit}. 

The non-closedness of $\tw_{0A}$ implies that the 4-form field strength includes a term of the form $G_4 \supset A^{0A}\wedge (d \tw_{0A})$. Upon dimensional reduction of the $\int G_4 \wedge \ast G_4$ term in the 11-dimensional supergravity action~\eqref{S11normal}, this yields a mass term for the gauge boson $A^{0A}$.
Allowing for non-closed forms in the Kaluza-Klein reduction is well-known as a method of generating masses in the effective theory, and was discussed e.g. in~\cite{Gurrieri:2002wz, Grimm:2008ed, Camara:2011jg} in the type II context. Formally, the inclusion of the non-closed forms $\tw_{0A}$ is very similar to the process of including fields from higher Kaluza-Klein mass levels in the dimensional reduction. Indeed, recalling the IIB mass formula~\eqref{massIIB} for the geometrically massive $U(1)$s shows that the $U(1)$ masses approach the Kaluza-Klein scale as we move away from the perturbative IIB limit and allow for larger values of the string coupling $g_s$ in F-theory~\cite{Grimm:2011tb}. This suggests that, in order to obtain a fully consistent F-theory effective action involving the massive gauge bosons $A^{0A}$ as dynamic degrees of freedom, other Kaluza-Klein modes of a comparable mass should also be included along with $A^{0A}$. However, our focus lies on effects relating to the geometrically massive $U(1)$ symmetries, and a full description of the F-theory effective action including all massive Kaluza-Klein modes is beyond the scope of this thesis. In the following we will therefore include only the non-closed forms required to describe the geometrically massive $U(1)$ symmetries and their associated fluxes, without making any claim that this corresponds to a consistent truncation of the massive field spectrum of the theory. 

The discussion of the geometric St\"uckelberg mechanism in section~\ref{sec:Stueckelberg} can be used to gain additional intuition about the relationship between $\tw_{0A}$ and the other forms of the dimensional reduction. A crucial point is that the diagonal $U(1)_A$ gauge boson is not the only field which is removed from the massless spectrum in the course of the geometric St\"uckelberg mechanism. Instead, for each massive gauge boson a certain linear combination of the would-be moduli $c^a$ is absorbed to form the longitudinal component of the massive vector field. At the same time, some of the fields $b^a$ obtain a mass because they appear in the D-term potential~\eqref{DtermIIB}. Therefore, following our previous logic not all of the $h^{1,1}_-(B_3)$ type IIB fields $c^a$ and $b^a$ should be uplifted to massless moduli fields in F-theory. In other words, not all of the forms $\alpha_a$ and $\beta^a$ appearing in~\eqref{C3_expand} should be harmonic. In fact, this can be intuitively understood as follows in the naive picture of the geometric origin of $\alpha_a$ and $\beta^a$ captured by equation~\eqref{uplift_omega}. The fields $c^a$, $b^a$ which participate in the geometric gauging are associated with forms $\omega_a \in H^{1,1}_-(X_3)$ which have non-vanishing support on a divisor $D_A$ wrapped by a stack of $D7$-branes, such that the wrapping numbers $C_A^a$ are non-zero. In the F-theory uplift, the elliptic fiber degenerates over the $D7$-branes, so that the 1-forms $dx$, $dy$ appearing in~\eqref{uplift_omega} are not well-defined over this locus. In other words, the naive uplift~\eqref{uplift_omega} may fail exactly if $\omega_a$ has non-zero restriction to a $D7$-brane divisor and the fields $c^a$, $b^a$ take part in the geometric St\"uckelberg mechanism. More precisely, we expect that an uplift of the forms $\omega_a$ to 3-forms on $\hat Y_4$ in the manner of~\eqref{uplift_omega} continues to exist, but that the resulting 3-forms fail to be harmonic in the case where the $\omega_a$ restrict non-trivially to a $D7$-brane. This expectation is based on the fact that the $b^a$ continue to exist in the IIB effective action, albeit with a non-zero mass.

As motivated above, we expect that both $\tw_{0A}$ and some of the forms $\alpha_a,\ \beta^a$ are non-harmonic in the presence of geometric gauging, and that the non-harmonicity is related in some way to the geometry of the $D7$-brane divisors $D_A$ in the IIB limit. To match the effective theory of the IIB compactification, we finally note that the $c^a$ appearing in the expansion $C_3 = c^a\alpha_a +\ldots$ should be absorbed in the process by which the vectors $A^{0A}$ become massive. In other words, we expect the geometric gauging~\eqref{gauging1} to be mirrored on the F-theory side after the non-harmonic forms are taken into account. Such a gauging appears naturally as a consequence of the expansion~\eqref{C3_expand} of $C_3$ if the forms $\tw_{0A}$ and $\alpha_a$ satisfy a relation of the form $ d\tw_{0A} = M_A^a \alpha_a$ with a constant matrix $M_A^a$~\cite{Grimm:2011tb}.
It was shown in~\cite{Grimm:2011tb} that the matrix $M_A^a$ can be related to the IIB quantities introduced in section~\ref{sec:U(1)inIIB} by
\be
\label{dtw_0}
d\tw_{0A} =  N_A C_A^a \alpha_a.
\ee
This relation captures the essence of the general discussion above in that the non-closedness of $\tw_{0A}$ and the subset of forms $\alpha_a$ which fail to be harmonic is determined by the $D7$-brane wrapping numbers $C_A^a$. It also immediately implies that the kinetic terms of the $c^a$ obtained by dimensionally reducing the field strength $G_4$ involve the covariant derivatives
\be
G_4 = dC_3 = (d c^a - N_A C_A^a  A^{0A}) \wedge \alpha_a + \ldots,
\ee
in full analogy with the IIB expression~\eqref{gauging1}.
The exact prefactors appearing in~\eqref{dtw_0} will be confirmed in section~\ref{sec:FtheoryLimit} by matching the F-theory effective action in the type IIB limit to the IIB results of section~\ref{sec:U(1)inIIB}.

To perform the dimensional reduction of the action, we will need to evaluate various integrals over $\hat Y_4$ involving both the harmonic forms introduced in section~\ref{sec:KKreduxFtheory} as well as the non-harmonic forms $\alpha_a, \beta^a$ and $\tw_{0A}$. In the following we give the values of the required integrals, which were worked out in~\cite{Grimm:2011tb} by performing the match with the IIB effective action. It will sometimes be convenient to group the non-harmonic form $\tw_{0A}$ together with the forms $\tw_{iA}$ describing the Cartan resolution divisors into the combined set
\be
\label{setI}
  \tw_{IA} \ ,\ I=0,i, \quad i=1,\ldots,\text{rank}(G_A) \ , \quad A = 1,\ldots, n_{D7}, 
\ee
with $n_{D7}$ the number of independent $D7$-brane stacks in the model.
 The intersection number of three base divisors is denoted by
\be
 \int_{B_3} \omega_\alpha \wedge\omega_\beta\wedge\omega_\gamma = \frac{1}{2} \cK_{\alpha\beta\gamma}.
\ee
The elliptic fibration structure and the fact that $X_3$ is a double cover of $B_3$ can be used to relate these intersection numbers to integrals on $X_3$ and $\hat Y_4$
\be
\label{def_Kalphabetagamma}
\cK_{\alpha\beta\gamma} =  \int_{X_3} \omega_\alpha \wedge\omega_\beta\wedge\omega_\gamma = 2 \int_{\hat Y_4} \omega_0 \wedge \omega_\alpha \wedge\omega_\beta\wedge\omega_\gamma,
\ee
where we have used the relation~\eqref{integrals_B3_and_Y4}. This shows that the $\cK_{\alpha\beta\gamma}$ are identical with the intersection numbers~\eqref{triple_intersections} appearing in the IIB compactification. The structure of the elliptic fibration further implies that intersections of 4 vertical divisors vanish~\cite{Grimm:2010ks}
\be
\int_{\hat Y_4} \omega_\alpha \wedge\omega_\beta\wedge\omega_\gamma \wedge \omega_\delta =   0.
\ee 
Furthermore, to obtain the correct IIB limit from the M-theory reduction the integrals involving both the non-harmonic form $\tw_{0A}$ and the Cartan forms $\tw_{iA}$ should vanish~\cite{Grimm:2011tb}. The generalisation of~\eqref{cartan_intersection_harmonic} to the full set $\tw_{IA}$ therefore takes the form
\be \label{twtw_Cartan}
  \int_{\hat Y_4} \tw_{IA} \wedge \tw_{JB} \wedge \omega_\alpha \wedge \omega_\beta
    = - \frac12 \delta_{AB}\, \cC^B_{IJ}\, C^\gamma_A \,  \cK_{\alpha \beta \gamma},   
\ee
where the matrix $\cC^B_{IJ}$ obeys $\cC^B_{0i} = 0$, $\cC^B_{00} = N_B$ and reduces to the Cartan matrix $\cC^B_{ij}$ of the gauge group $G_B$ for the other values of the indices. Finally, the integrals involving one form $\tw_{IA}$ and 3 forms out of the set $\{\omega_0,\omega_\alpha\}$ are expected to vanish, just as in the harmonic case analysed in~\cite{Grimm:2010ks, Krause:2011xj}.

The form of the integrals involving $\alpha_a$ and $\beta^a$ can be motivated by the schematic formula~\eqref{uplift_omega}. This picture makes it clear that all integrals involving $\alpha_a\wedge \alpha_b$ or $\beta^a \wedge \beta^b$ are expected to vanish just as in the harmonic case~\eqref{int_alphabeta_harmonic}. Furthermore, the forms $\alpha_a\wedge\beta^b$ already have two legs along the elliptic fiber, so that the only 2-forms which will yield a non-vanishing integral over $\hat Y_4$ after taking the wedge product with $\alpha_a\wedge\beta^b$ are the forms $\omega_\alpha$ originating from $B_3$. In this case, the forms $dx$, $dy$ in~\eqref{uplift_omega} are essentially integrated to unity over the elliptic fiber, so that the overall integral reduces to the corresponding integral over $X_3$ involving the $\omega_a$ rather than $\alpha_a$ and $\beta^a$. In summary, we expect~\cite{Grimm:2011tb}
\begin{align}
\int_{\hat Y_4} \omega_\alpha \wedge \alpha_a \wedge \beta^b  =& \ \frac{1}{2 }\int_{X_3} \omega_\alpha\wedge\omega_a\wedge\omega_c \delta^{cb} = \frac12 \cK_{\alpha ac} \delta^{cb}  , \label{int_omega_alpha_beta}\\
\int_{\hat Y_4} \omega_{IA} \wedge \alpha_a \wedge \beta^b  =&\  \int_{\hat Y_4} \omega_0 \wedge \alpha_a \wedge \beta^b = 0, \label{int_alpha_beta_tw}
\end{align}
with the constants $\cK_{\alpha a b}$ appearing in~\eqref{int_omega_alpha_beta} being the same as those introduced in section~\ref{sec:IIBreduction}. The factor of $1/2$ takes into account the fact that integrals involving forms of well-defined orientifold parity on $X_3$ are twice as large as on $B_3$, as $X_3$ is the double cover of $B_3$. Therefore, the $\cK_{\alpha a b}$ are even and the  intersections~\eqref{int_omega_alpha_beta} are integer despite the factor of 1/2. The integral relations involving $\alpha_a$ and $\beta^b$ will again be confirmed by matching the F-theory effective action to the IIB result in section~\ref{sec:FtheoryLimit}.

Note that consistency of the integral relation~\eqref{int_omega_alpha_beta} with the exactness of $C^a_A \alpha_a$ conjectured in~\eqref{dtw_0} prevents $\beta^a$ from being closed with respect to exterior differentiation. To describe $d\beta^a$, we follow~\cite{Grimm:2011tb} in introducing an additional set of non-harmonic 4-forms $\tilde{\tw}^{bA}$ which are related to $\beta^a$ by
\be
d \beta^a = - \delta^{ac} \frac12 \cK_{\alpha cb} \, N_A \, C^\alpha_A \, \tilde{\tw}^{bA} .
\label{d_beta^a}
\ee
The prefactors in this relation are chosen for later convenience in the dimensional reduction. Consistency between the various relations~\eqref{dtw_0},~\eqref{int_omega_alpha_beta},~\eqref{int_alpha_beta_tw} and~\eqref{d_beta^a} can now be achieved by imposing~\cite{Grimm:2011tb}
\begin{align}
\label{tw0-tildetw-omega}
 C^\alpha_A \cK_{\alpha a b} \int \omega_\beta \wedge \tw_{0 B} \wedge \tilde{\tw}^{b A} = & \  \delta^A_B \cK_{\beta a b} C^b_A , \\  C^\alpha_C \cK_{\alpha a b} \int \tw_{IA}\wedge \tw_{0 B} \wedge \tilde{\tw}^{b C} = & \ C^\alpha_C \cK_{\alpha a b} \int \omega_0 \wedge \tw_{0 B} \wedge \tilde{\tw}^{b C} = 0.
\end{align}
Finally, we follow~\cite{Grimm:2011tb} in introducing a second set of 4-forms $\tilde \tw_{aA}$, which are dual to the $\tilde{\tw}^{bA}$ in the sense that
\be
\label{tildetw_tildetw}
 \int_{\hat Y_4} \tilde \tw_{aA} \wedge \tilde \tw^{bB} = \delta_A^B \delta_a^b.
\ee
The forms $\tilde \tw_{aA}$ will prove useful for the description of the F-theory uplift of fluxes along the diagonal $U(1)_A$ discussed in the next subsection. Consistency with the various integral relations introduced above requires
\be
\label{dtwodd}
d\tilde \tw_{aA} =  N_A C^\alpha_A \omega_\alpha \wedge \alpha_a .
\ee
Furthermore, the forms $\tilde \tw_{aA}$ are taken to fulfill the integral relations~\cite{Grimm:2011tb}
\begin{align}
 \int_{\hat Y_4} \tilde \tw_{aA} \wedge \tilde \tw_{bB} = & \ - \frac12 \delta_{AB} \cK_{\alpha a b} N_A  C^\alpha_A, \\
\label{2tw0-tildetw}
 \int_{\hat Y_4} \tilde \tw_{aA} \wedge \tw_{IB} \wedge \omega_\alpha = & \ - \frac12 \delta_{AB} \delta_{I 0} \cK_{\alpha a b} N_A C^b_A  .
\end{align}
As we will see in section~\ref{sec:G4fluxes}, the first relation is required in order to obtain a match between the $M2$-brane tadpole induced by $G_4$ fluxes along $\tilde \tw_{aA}$ and the corresponding type IIB $D3$-brane tadpole. Similarly, the second relation~\eqref{2tw0-tildetw} can be deduced by requiring that the $G_4$-induced chiral indices proposed in section~\ref{sec:G4fluxes} correctly reproduce the known type IIB results.

A brief comment is in order regarding the proposed forms of the differential and integral relations involving $\beta^a$, $\tilde{\tw}^{aA}$ and $\tilde{\tw}_{aA}$. At first sight, it seems that~\eqref{d_beta^a} and the following relations can lead to the appearance of non-harmonic forms even in the case where $C_A^a=0$ and no geometric gauging occurs in the IIB picture. However, let us emphasise that the entire non-harmonic structure introduced above is aimed at describing the F-theory uplift of $D7$-brane configurations with gauge group $U(N_A)$ in the IIB picture. In other words, the brane index $A$ in equation~\eqref{d_beta^a} should run only over the stacks of branes with unitary gauge group. If $C_A^a = 0$, the gauge group $U(N_A)$ can appear only if the $D7$-brane divisor $D_A$ is in the same homology class as the $O7$-plane~\cite{Grimm:2011tb}. The fact that forms of negative orientifold parity have vanishing pullback to the orientifold plane then implies
\be
C_A^\alpha \cK_{\alpha a b} = \int_{D_A \subset X_3} \omega_a\wedge\omega_b =  \int_{D_{O7}} \omega_a\wedge\omega_b = 0.
\ee
Therefore the forms $\beta^a$ are actually closed in this case, and as expected no non-harmonic structure appears in the absence of geometric gauging.

Before proceeding with the description of the uplift of fluxes along the diagonal $U(1)_A$ to F-theory, a brief comment is in order about the implications of the appearance of $\tw_{0A}$ for the expansion of the K\"ahler form. In order to be able to match the 3-dimensional effective action obtained from the F-theory reduction with the $\cN=1$ effective action of type IIB, the 3-dimensional effective action must also be supersymmetric. This in turn requires the massive vectors $A^{0A}$ to combine with suitable scalar fields $v^{0A}$ into complete 3-dimensional vector multiplets. The scalars $v^{0A}$ are obtained by including $\tw_{0A}$ in the expansion of the K\"ahler form
\be
\label{Jexp_nonkaehler}
  J =  v^0 \omega_0  + v^\alpha \omega_\alpha + v^{i A}  \tw_{i A} + v^{0A} \tw_{0A} \equiv v^\Lambda \omega_\Lambda.
\ee

At first sight, this seems to be in contradiction with the usual statement that low energy supersymmetry requires a compactification on a K\"ahler manifold, because~\eqref{Jexp_nonkaehler} implies
\be
dJ =   v^{0A} N_A C_A^a \alpha_a. 
\ee
However, as discussed in more detail in~\cite{Grimm:2011tb, Becker:1996gj, Grana:2005jc}, a compactification on a non-K\"ahler space with a globally defined (1,1)-form\footnote{Despite the fact that $J$ is no longer a true K\"ahler form when $dJ\neq 0$, we will continue to refer to $J$ as `the K\"ahler form' in the following.} $J$ is actually not in contradiction with the existence of a supersymmetric effective theory. Instead, K\"ahlerity is only required for a supersymmetric vacuum, so that we should require a vanishing vacuum expectation value for $d J$~\cite{Grimm:2011tb}
\be
\label{vac_value_J}
\left< d J \right> \propto C_A^a \left< v^{0A} \right> = 0.
\ee
As the $v^{0A}$ become components of 4-dimensional vectors in the F-theory limit, their expectation value is anyway required to vanish in order to preserve 4-dimensional Lorentz invariance. In the following, we will therefore always assume that~\eqref{vac_value_J} is fulfilled.
Despite their vanishing vacuum expectation value, the fields $v^{0A}$ can be formally viewed as describing massive fluctuations of $\hat Y_4$ into an auxiliary non-K\"ahler space $\hat Z_4$~\cite{Grimm:2011tb}. Strictly speaking, the dimensional reduction involving the massive fields $v^{0A}, A^{0A}$ should be carried out on this auxiliary space $\hat Z_4$. However, all the topological integrals and differential relations introduced above do not rely on the form of the metric and hold on both $\hat Y_4$ and $\hat Z_4$. Therefore, we will ignore the subtle distinction between the two spaces in the following and continue to denote the non-K\"ahler space on which the derivation of the F-theory effective action is carried out by $\hat Y_4$.

\subsection{$G_4$ fluxes and the induced tadpoles and chiralities}
\label{sec:G4fluxes}

The discussion presented above shows that the inclusion of specific non-harmonic forms in the Kaluza-Klein reduction can be used to implement a geometric gauging scenario and geometrically massive $U(1)$ symmetries in F-theory compactifications. Although a precise match with the type IIB picture including all prefactors requires a careful discussion of the F-theory limit and is postponed to section~\ref{sec:FtheoryLimit}, it is evident that the massive gauge fields $A^{0A}$ introduced above provide an F-theoretic description of the diagonal $U(1)$s in type IIB. As we had seen in section~\ref{sec:U(1)inIIB}, many interesting features of the low energy action are controlled by gauge fluxes along these diagonal $U(1)$s, and we now turn to the description of this type of flux in the F-theory reduction. Several flux-dependent quantities, such as the flux-induced chiralities and tadpoles derived in the IIB setting in sections~\ref{sec:Stueckelberg} and~\ref{sec:IIB_tadpoles}, are actually independent of the K\"ahler moduli. We therefore expect that these quantities are not changed in the F-theory limit of vanishing elliptic fiber volume, so that the type IIB results should be visible directly in the M-theory picture. In the following, we will introduce the uplift of fluxes in the diagonal $U(1)$ to M-theory $G_4$ fluxes that was proposed in~\cite{Grimm:2011tb}, and show how the IIB expressions for the chiralities and tadpoles are reproduced in this proposal.

In section~\ref{sec:KKreduxFtheory}, the relationships between certain cohomology groups on $\hat Y_4$ and $X_3$ were used to derive correspondences between fields appearing in the M-theory and IIB reductions. As we will see, these correspondences can also be used to obtain an intuition for the uplift of the IIB fluxes which can be expanded along forms on the brane divisors that arise by pullback\footnote{Throughout this chapter we will assume for simplicity that the brane divisors do not admit any of the so-called variable fluxes defined in~\eqref{flux_expansion_twoforms}.} from the Calabi-Yau $X_3$. For the convenience of the reader, let us recall the expansion~\eqref{def_tilde_cF} of the flux along the diagonal $U(1)_A$ of the brane stack on $D_A \subset X_3$ in terms of the bases of the cohomology groups $H^{1,1}_\pm(X_3)$
\be
\tilde{\cF}_0^A = \tilde{\cF}_0^{A,\alpha} \omega_\alpha + \tilde{\cF}^{A,a}_0 \omega_a.
\label{def_tilde_cF_repeat}
\ee
In the absence of variable fluxes, the gauge flux is thus completely specified by the dimensionless flux quanta\footnote{The flux quanta are restricted to integer or half-integer values by the Freed-Witten quantisation condition~\eqref{FW_quant}. The fluxes $\tilde{\cF}_0^{A,\alpha}$ of positive orientifold parity include a contribution from the discrete $B_+$ field.}
$\tilde{\cF}_{0}^{A,\alpha}$ and $\tilde{\cF}^{A,a}_0$, and it is natural to expect that these flux quanta will also appear in the uplift of $\tilde{\cF}_0^A$ to an M-theory $G_4$ flux.

Let us consider first the $D7$-brane fluxes along the positive parity forms $\omega_\alpha$. As the $\omega_\alpha$ survive the uplift to $\hat Y_4$, it is natural to expect that the corresponding $G_4$ flux can be written as a wedge product of the IIB flux with a suitable 2-form on $\hat Y_4$. In fact, a very similar form is observed for the M-theory description of positive parity fluxes in the Cartan of $SU(N_A)$, which in the IIB language are given by $\tilde{\cF}_i^A = \tilde{\cF}_i^{A,\alpha} \omega_\alpha$. The corresponding $G_4$ flux in M-theory involves the forms $\tw_{iA}$ that are dual to the resolution divisors of the singularity over $D_A$, and can be written as~\cite{Krause:2011xj, Grimm:2010ks, Grimm:2011tb, Weigand:2010wm}
\be
G_4 =  -  \sum_i \tilde{\cF}_i^{A, \alpha} \omega_\alpha \wedge \tw_{iA}.
\label{G_4_Cartan}
\ee

It is natural to expect that the positive parity fluxes along the diagonal $U(1)_A$ can be described by the direct analogue of this expression involving the additional form $\tw_{0A}$, which describes the diagonal $U(1)_A$ in the M-theory lift. Following~\cite{Grimm:2011tb}, we therefore make the ansatz
\be
G_4 = -  \tilde{\cF}_{0}^{A,\alpha} \omega_\alpha \wedge \tw_{0A}
\ee
to describe the $G_4$ fluxes of this type.

The uplift of the negative parity fluxes $\tilde \cF^{A, a}_0$ is somewhat more involved as the uplift of the forms $\omega_a$ to $\hat Y_4$ is more complicated. Note that non-zero fluxes $\tilde \cF^{A, a}_0$ along the diagonal $U(1)_A$ exist only in the presence of geometric gauging, because as discussed in section~\ref{sec:massiveU(1)sFtheory} negative parity forms have vanishing pullback to branes with gauge group $U(N_A)$ and $[D_A]=[D_A']$. This suggests that their uplift to F-theory is connected to the non-harmonic structure associated with the 3-forms $\alpha_a$, $\beta^a$, which describe the uplift of the forms $\omega_a$ on $X_3$. As we will see momentarily, the forms $\tilde{\tw}_{aA}$ introduced in~\eqref{tildetw_tildetw} indeed have the required properties to describe this uplift. In summary, we use the form
\be
G_4 = - \tilde{\cF}_0^{A, \alpha} \omega_{\alpha} \wedge \tw_{0A}  - \tilde{\cF}_0^{A, a} \tilde \tw_{aA},
\label{newfluxes}
\ee 
which was originally proposed in~\cite{Grimm:2011tb}, for the uplift of fluxes along the diagonal $U(1)_A$ to F-theory.

Of course, fluxes along the diagonal $U(1)$s as in~\eqref{newfluxes} are not the only type of $G_4$ fluxes that can appear in an F-theory compactification. For example, $G_4$ may include other fluxes which originate from the open string sector in the IIB picture, such as the Cartan fluxes appearing in~\eqref{G_4_Cartan} or variable fluxes which are trivial from the perspective of the ambient bulk space but non-trivial on the $D7$-brane world-volume. In addition, $G_4$ is well-known to include the uplift of closed string fluxes, which in the IIB language are described by the field strengths $F_3$ and $H_3$ corresponding to the potentials $C_2$ and $B_2$, respectively~\cite{Weigand:2010wm}. This uplift can be understood as locally arising in a similar manner to the uplift of the odd forms $\omega_a$, such that~\cite{Denef:2008wq}
\be
(F_3, H_3) \rightarrow G_4 \sim H_3 \wedge dx + F_3\wedge dy. 
\label{uplift_F3_H3}
\ee 

In the type IIB setting we had seen in section~\ref{sec:IIB_tadpoles} that the presence of $D7$-brane flux along the diagonal $U(1)$ contributes to the effective $D5$-brane tadpole, which manifests itself in the modified Bianchi identity~\eqref{Bianchi_id_F3} for $F_3$. In the spirit of~\eqref{uplift_F3_H3}, it is natural to expect that the fluxes defined in~\eqref{newfluxes} lead to a similarly modified Bianchi identity for $G_4$~\cite{Grimm:2011tb}. Indeed, directly evaluating $dG_4$ by inserting the expansion~\eqref{newfluxes} and using the non-harmonicity of $\tw_{0A}$ and $\tilde{\tw}_{aA}$ described in equations~\eqref{dtw_0} and~\eqref{dtwodd} of the previous subsection leads to
\be
  dG_4 =  -  N_A \Big(\tilde{\cF}_0^{A, a} C^\alpha_A + \tilde{\cF}_0^{A, \alpha} C^a_A \Big) \, \omega_\alpha \wedge \alpha_a\,. 
  \label{G4_tadpole}
\ee
Keeping in mind that the $\alpha_a$ effectively describe the uplift of the forms $\omega_a$ to $\hat Y_4$, this indeed seems to exhibit the correct structure to describe the F-theory uplift of the IIB expression~\eqref{Bianchi_id_F3}. To demonstrate the exact match, it is necessary to keep in mind that the intersection forms of $\hat Y_4$ and $X_3$ differ by a factor of 2. To take these relative factors into account, it is advantageous to consider the uplift of the integrated tadpole contributions~\cite{Grimm:2011tb}
\be
\label{upliftD5tad}
\int_{X_3} d F_3 \wedge \omega_b \rightarrow  \int_{\hat Y_4} d G_4 \wedge \beta^b,
\ee
rather than directly comparing the Bianchi identities of $G_4$ and $F_3$.
Inserting the explicit expression~\eqref{G4_tadpole} for $dG_4$ and using the integrals~\eqref{int_omega_alpha_beta}, one obtains the result that the contribution to the $D5$-brane tadpole induced by a flux $\tilde{\cF}_0^A$ is
\be
\label{M5tad}
  \delta_{ab}  \int_{\hat Y_4} dG_4 \wedge \beta^b =  - \frac12 N_A \Big( \tilde{\cF}_0^{A, c} C^\alpha_A + \tilde{\cF}_0^{A, \alpha} C^c_A  \Big) \cK_{\alpha ac}\;.
\ee
This exactly matches the corresponding contribution~\eqref{D5tadInt} obtained in the IIB compactification. In particular, the restrictions on the fluxes which are obtained in the IIB and F-theory pictures from the condition that the tadpoles must vanish in a consistent compactification are identical.
An important aspect of the result~\eqref{M5tad} is that the fact that the continuous moduli $b_a$ do not appear in the integrated $D5$ tadpole is in a sense automatic in F-theory. Recall from section~\ref{sec:IIB_tadpoles} that the corresponding result in the IIB setting relied on the $D7$-brane tadpole cancellation condition~\eqref{D7-tadpole_IIB}. This observation can be viewed as another manifestation of the fact that, as already remarked in section~\ref{sec:elliptic_CalabiYaus}, the analogue of the IIB $D7$-brane tadpole cancellation condition is automatically fulfilled in a consistent F-theory compactification.

A second check on the proposed form~\eqref{newfluxes} of the F-theory fluxes is obtained by considering the analogue of the IIB $D3$-brane tadpole. As discussed in section~\ref{sec:IIB_tadpoles}, the $D3$-brane tadpole can be derived in the IIB setting by considering the equation of motion of the field $C_4$ which couples to the $D3$-branes. In the M-theory lift, the $D3$-branes correspond to $M2$-branes, which in turn couple directly to the M-theory 3-form potential $C_3$~\cite{Denef:2008wq}. Just as in the type IIB setting, the consistency condition corresponding to cancellation of net charge in the compact space, which in the M-theory setting is referred to as absence of the $M2$-brane tadpole, can be derived from the equation of motion of $C_3$~\cite{Denef:2008wq}. In addition to $M_2$ branes, two other types of sources contribute to the $M2$-brane tadpole. A curvature-related source term is induced by an additional term in the 11-dimensional supergravity involving higher orders of the curvature scalar, which we have not displayed in~\eqref{S11normal}. The curvature-induced contribution to the integrated $M2$-brane tadpole is proportional to the Euler characteristic $\chi$ of the fourfold used in the M-theory compactification to 3 dimensions, as first shown by Sethi, Vafa and Witten in~\cite{Sethi:1996es}. For a smooth elliptically fibered fourfold the Euler characteristic is determined fully in terms of the Chern classes of the base~\cite{Sethi:1996es}. However, in an F-theory compactification with non-Abelian singularities it is important to use the Euler characteristic $\chi(\hat Y_4)$ of the resolved space~\cite{Andreas:2009uf, Weigand:2010wm}, and a naive application of the Sethi-Vafa-Witten formula for smooth spaces can lead to significant deviations from the true result~\cite{Blumenhagen:2009up}.

A further source term for $C_3$ appears in the presence of non-trivial $G_4$-flux, as is immediately evident from the coupling proportional to $\int C_3 \wedge G_4\wedge G_4$ in the 11-dimensional action~\eqref{S11normal}. Putting the various contributions together, one obtains the integrated $M2$-brane tadpole cancellation condition~\cite{Weigand:2010wm, Denef:2008wq}
\be
N_{M2} +  \int_{\hat Y_4} G_4\wedge G_4 = \frac{\chi(\hat Y_4)}{24}.
\label{M2tadpole_overall}
\ee
In the type IIB limit, the spacetime-filling $M2$-branes counted by $N_{M2}$ are dual to spacetime-filling $D3$-branes, so that in relating~\eqref{M2tadpole_overall} to the IIB expression~\eqref{D3tadpole} there holds $N_{M2} = N_{D3}$. Furthermore, the term involving $\chi(\hat Y_4)$ accounts for the terms involving the modified Euler characteristics of the $O$-planes and $D7$-branes appearing on the right hand side of~\eqref{D3tadpole}~\cite{Collinucci:2008pf}. This leaves the term involving $\int G_4 \wedge G_4$ to account for the entire flux-dependent contribution $N_{\rm flux}$ in~\eqref{D3tadpole}, which includes contributions from both bulk and brane fluxes in the IIB language. In the following we focus on the contribution to the $M2/D3$-tadpole which is induced by gauge flux along the diagonal $U(1)$. In the type IIB setting this contribution was given in equation~\eqref{d3iibNgauge}. The contribution to the M-theory tadpole induced by the proposed $G_4$ fluxes~\eqref{newfluxes} can be computed using the integral relations presented in section~\ref{sec:massiveU(1)sFtheory}~\cite{Grimm:2011tb}
\be
\begin{aligned}
\label{M2_tadpole_flux}
\int_{\hat Y_4} G_4 \wedge G_4 =  - \frac{1}{4} \sum_A N_A \Big( & \ \cK_{\alpha \beta \gamma} C^\alpha_A \tilde \cF_0^{A, \beta} \tilde \cF_0^{A, \gamma} +   \cK_{\alpha b c} C^\alpha_A  \tilde \cF_0^{A,b}  \tilde \cF_0^{A, c}  \\ 
 & \ + 2    \cK_{a b \gamma} C^a_A \tilde \cF_0^{A, b}  \tilde \cF_0^{A, \gamma}             \Big).
\end{aligned}
\ee
Here we have used the $D5$-brane tadpole cancellation condition.
This purely flux-dependent contribution again precisely matches the type IIB result~\eqref{d3iibNgauge}. In fact, this match between the IIB and F-theory pictures for the $M2/D3$-tadpole extends also to the role of the continuous $b_a$ moduli. As discussed in section~\ref{sec:IIB_tadpoles}, the naive $D3$-tadpole actually contains contributions linear and quadratic in the continuous moduli. The quadratic contribution can be shown to vanish as long as the $D7$-brane tadpole is zero, while the absence of the linear contribution further requires the $D5$-brane tadpole to be cancelled~\cite{Blumenhagen:2008zz}. In the corresponding M-theory expression~\eqref{M2_tadpole_flux}, the quadratic term is automatically absent, matching the previously discussed expectation that the analogue of the IIB $D7$-brane tadpole cancellation is automatically encoded in the geometry of the M-theory compactification. However,~\eqref{M2_tadpole_flux} includes a contribution of the form
\be
\hspace{-0.2cm}\int \left[\tilde{\cF}_0^{A, a} \tilde \tw_{aA} +  \tilde{\cF}_0^{A, \alpha} \omega_{\alpha} \wedge \tw_{0A} \right] \wedge b_b d \beta^b \propto   \cK_{\alpha a c} N_A \delta^{cb} b_b \left( \tilde{\cF}_0^{A, a} C^\alpha_A + \tilde{\cF}_0^{A, \alpha} C^a_A \right).
\ee
Comparison with~\eqref{M5tad} confirms that this term is absent precisely if the $D5$-brane tadpole vanishes. This exactly mirrors the situation that was observed in the type IIB framework. The match between the $M2/D3$ tadpoles in the F-theory and IIB pictures yields a second non-trivial check for the proposed non-harmonic structure and the $G_4$ flux~\eqref{newfluxes}.

As already mentioned in section~\ref{sec:Stueckelberg}, the main motivation behind the inclusion of non-trivial fluxes in compactifications of string or F-theory is that their presence can lead to a chiral matter spectrum in the low energy theory. In particular, the chiral index measuring the chirality of charged matter localised at the intersection of 2 $D7$-branes in type IIB involves an integral of the relative $D7$-brane gauge flux over the intersection curve. The corresponding matter states in F-theory can be viewed as arising from $M2$-branes wrapped on $\bP^1$'s in the elliptic fiber which become visible after the enhanced singularity over the intersection curve is resolved~\cite{Beasley:2008dc, Weigand:2010wm}. Fibering these $\bP^1$s over the intersection curve leads to 4-dimensional surfaces known as matter surfaces. The index measuring the net chirality induced by $G_4$ flux in F-theory is now given by the integral of $G_4$ over these matter surfaces~\cite{Donagi:2008ca, Hayashi:2008ba}.  

Recent progress in the explicit construction of $G_4$ fluxes and matter surfaces, which was initiated in~\cite{Krause:2011xj, Braun:2011zm}, has made it possible to check these F-theoretic chirality formulae in concrete examples. In particular, the detailed study of the resolution of the so-called $U(1)$-restricted $SU(5)$ Tate model performed in~\cite{Krause:2011xj} made it possible to identify the harmonic 2-forms corresponding to the Cartan $U(1)$s and the additional massless $U(1)$. These forms were used to construct harmonic $G_4$ fluxes of the form~\eqref{G_4_Cartan}, whose integrals over the matter surfaces were indeed shown to correctly reproduce the known type IIB chiral indices.

The type IIB results presented in section~\ref{sec:Stueckelberg} show that also fluxes along massive $U(1)$ symmetries influence the chiral spectrum. As the chiral spectrum is protected in the uplift from IIB to F-theory~\cite{Grimm:2011tb}, this makes it clear that also $G_4$ flux corresponding to IIB fluxes along massive $U(1)$s must be included in the F-theory description in order to gain a complete understanding of the chiral matter spectrum. A first step towards this goal was taken in reference~\cite{Krause:2012yh}, which showed that a certain harmonic $G_4$ flux in the $SU(5)\times U(1)$ model could be identified with a type IIB flux configuration along massive $U(1)$ symmetries which is free of $D5$-brane tadpoles. As discussed in~\cite{Krause:2012yh}, this suggests that two complementary approaches are possible towards constructing $G_4$ fluxes corresponding to diagonal $U(1)$s in F-theory. The description~\eqref{newfluxes} given above in a sense provides an off-shell description in the spirit of the IIB approach, in which the vanishing of the $D5$ tadpole~\eqref{M5tad} must be imposed by hand. Although the individual forms $\tw_{0A}$ and $\tilde{\tw}_{aA}$ appearing in~\eqref{newfluxes} are in general non-closed, the overall tadpole-free combination is expected to correspond to a harmonic $G_4$. In other words, the observation of~\cite{Krause:2012yh} suggests that the $D5$ tadpole cancellation condition can be automatically accounted for in F-theory by harmonicity of $G_4$ flux. Note that this is in some ways similar to the fact that, as discussed in section~\ref{sec:elliptic_CalabiYaus}, the F-theory analogue of the type IIB $D7$-brane tadpole cancellation condition is automatically satisfied in a consistent compactification due to the geometric properties of the compactification manifold.

Although the discussion above suggests that the contributions of fluxes along both massive and massless $U(1)$s to the chiral indices can in some situations be computed by evaluating integrals of harmonic $G_4$ fluxes over matter surfaces, there are several reasons why it is worthwhile exploring how the chiral indices can be expressed in the language of non-harmonic forms used in~\eqref{newfluxes}. For one, the analysis of~\cite{Krause:2012yh} was carried out in the specific framework of an $SU(5)\times U(1)$ model. It therefore strictly speaking does not rule out the possibility that in other setups there may exist tadpole-free combinations of fluxes which do not admit a dual description in terms of a harmonic form. Furthermore, in some situations it may be simpler from a computational perspective to work with the non-harmonic forms in~\eqref{newfluxes} and solve the tadpole cancellation condition explicitly than to construct the harmonic $G_4$ fluxes directly. Finally, the analysis presented in~\cite{Krause:2012yh} focused on flux which could be viewed as the uplift of orientifold-even flux from IIB. It is not immediately clear how to extend this analysis to fluxes of negative orientifold parity, whereas such fluxes are explicitly included in~\eqref{newfluxes}. A proposal for the F-theoretic chirality formulae written directly in the language of non-harmonic forms was given in~\cite{Grimm:2011tb} and will be reviewed in the following. 

As a guiding principle, the proposed chirality formulae are required to reproduce the known chiralities of section~\ref{sec:Stueckelberg} in the type IIB limit. Furthermore, from the discussion above we expect the formulae for the chirality $I_{AB}$ of matter at the intersection of two brane stacks $A$ and $B$ to involve the integral of the wedge product of $G_4$ with a suitable 4-form. This additional 4-form should depend on the geometry of the two brane divisors $D_A$ and $D_B$, and effectively provides a description of the Poincar\'e duals of the matter surfaces in terms of the non-harmonic forms introduced in section~\ref{sec:massiveU(1)sFtheory}. Note that the intersection between $D_A$ and $D_B$ generally gives rise to two different types of matter surfaces corresponding to matter fields which have either the same or the opposite charge\footnote{The overall sign is a matter of convention. In the type IIB language changing the overall sign corresponds to reversing the roles of all branes and image branes in the model.} with respect to the $U(1)$ gauge symmetries of the two brane stacks~\cite{Grimm:2011tb}. These two types of matter states correspond in the type IIB language to matter at the intersection between the two branes or the intersection between a brane and an image brane. Correspondingly, we expect two types of chiral indices to arise measuring the net chiralities of the matter fields at the two types matter curves. These indices will be labeled $I_{AB}$ and $I_{AB'}$, in accord with the intuitive type IIB interpretation. Finally, we expect that as in the IIB picture the net chirality should depend on the orientation of the intersection in the sense that $I_{AB}= - I_{BA}$. 

The chirality formulae proposed in reference~\cite{Grimm:2011tb} on the basis of the requirements mentioned above can be conveniently written in terms of the forms $\tilde{\tw}^{aA}$ of equation~\eqref{d_beta^a} and the combinations
\be
\label{tildeDAnew}
[\tilde D_A] =  [D_A] - \sum_{I=0}^{{\mathrm rk}(G_A)} a_{IA} \, \tw_{IA}, \quad\quad\quad 
[\tilde D_{A'}] =  [D_A] - \sum_{I=0}^{{\mathrm rk}(G_A)} a_{IA'} \, \tw_{IA}.
\ee
The coefficients $a_{iA} = a_{iA'}$, $i=1, \ldots, {\rm rk}(G)$ are the Dynkin labels of the Cartan elements of the gauge group $G_A$, while $a_{0A}= \frac{1}{N_A} = - a_{0A'}$. Up to the term involving $a_{0A}$,~\eqref{tildeDAnew} describes the proper transform of the brane divisors $D_A^b$ under resolution of the singularities above $D_A^b$. As before, we focus on the case of unitary gauge groups and therefore take $a_{iA}=1$. 

In terms of the quantities introduced above, the proposed expressions for the chiral indices in terms of the non-harmonic forms used to describe massive $U(1)$s are~\cite{Grimm:2011tb}
\be
\begin{aligned}
\label{index1New}
I_{AB} = \frac{1}{4}  \int_{\hat Y_4} \Big( & \  ( [\tilde D_A] \wedge [\tilde D_{B'}]- [\tilde D_{A'}] \wedge [\tilde D_{B}])  \\ & \ + \cK_{\alpha a b} ( C^\alpha_A C^a_B \tilde{\tw}^{b A} - C^\alpha_B C^a_A  \tilde{\tw}^{b B}) \Big) \wedge G_4, 
\end{aligned}
\ee
\be
\begin{aligned}
I_{AB'} =  \frac{1}{4  }  \int_{\hat Y_4} \Big( & \ ([\tilde D_A] \wedge [\tilde D_{B}]- [\tilde D_{A'}] \wedge [\tilde D_{B'}] )  \\ & \ - \cK_{\alpha a b} ( C^\alpha_A C^a_B  \tilde{\tw}^{b A} + C^\alpha_B C^a_A \tilde{\tw}^{b B} ) \Big) \wedge G_4. \label{index2new}
\end{aligned}
\ee

The chiral indices in the presence of $G_4$ flux of the form~\eqref{newfluxes} can be explicitly evaluated using the various integral relations presented in section~\ref{sec:massiveU(1)sFtheory}. In this way, one obtains
\be
\begin{aligned}
\label{index1eval}
I_{AB} =& \  - \frac{1}{4 }  \Big( \cK_{\alpha \beta\gamma}C^\beta_A C^\gamma_B + \cK_{\alpha a b} C^a_A C^b_B \Big) \Big( \tilde{\cF}_0^{A,\alpha} - \tilde{\cF}_0^{B,\alpha} \Big) \\
& - \frac{1}{4 }  \Big( \cK_{\alpha a b}C^\alpha_A C^a_B + \cK_{\alpha a b} C^a_A C^\alpha_B \Big) \Big( \tilde{\cF}_0^{A,b} - \tilde{\cF}_0^{B,b} \Big) ,
\end{aligned}
\ee
\be
\begin{aligned}
\label{index2eval}
I_{AB'} =&\ - \frac{1}{4}  \Big( \cK_{\alpha \beta\gamma}C^\beta_A C^\gamma_B - \cK_{\alpha a b} C^a_A C^b_B \Big) \Big( \tilde{\cF}_0^{A,\alpha} + \tilde{\cF}_0^{B,\alpha} \Big) \\
&\ - \frac{1}{4}  \Big( \cK_{\alpha a b}C^\alpha_B C^a_A - \cK_{\alpha a b} C^a_B C^\alpha_A \Big) \Big( \tilde{\cF}_0^{A,b} - \tilde{\cF}_0^{B,b} \Big) , 
\end{aligned}
\ee
in agreement with the IIB results of section~\ref{sec:Stueckelberg}. This confirms that the proposed expressions~\eqref{index1New} and~\eqref{index2new} fulfill all the requirements to describe the desired chiral indices in the presence of $G_4$ flux along the diagonal $U(1)$. It would be interesting to extend the analysis of~\cite{Krause:2012yh} to the general case, and investigate whether the given expressions can always be rewritten in the standard M-theory form for chirality formulae involving the integral of a harmonic $G_4$ flux over suitable matter surfaces. However, such an analysis is beyond the scope of this thesis.

The discussion above shows that the proposed non-harmonic structure on $\hat Y_4$ yields a framework for implementing $G_4$ fluxes corresponding to the diagonal $U(1)$ which formally allows us to exactly reproduce the known type IIB results. Of course, in matching the F-theory and IIB results, we have always implicitly assumed that the flux quanta $\tilde{\cF}_0^{A, \alpha}$, $\tilde{\cF}_0^{A, a}$ appearing in the definition of $G_4$ correspond to the IIB flux quanta of equation~\eqref{def_tilde_cF}. In particular, the flux quanta must be restricted to integer or half-integer values in accord with the type IIB Freed-Witten quantisation condition~\eqref{FW_quant}. While this quantisation is obvious from the point of view of the duality with IIB, it must be possible to derive it directly in the F-theoretic setting without making recourse to the type IIB limit. It is natural to expect that the relevant conditions arise from a suitable generalisation of the standard M-theoretic quantisation condition for harmonic $G_4$ fluxes on a Calabi-Yau fourfold~\cite{Witten:1996md, Collinucci:2010gz},
\be
\tilde G_4 = G_4 - \frac12 c_2(\hat Y_4) \in H^4(\hat Y_4, \mathbb Z)\
\label{FW_quant_G4}
\ee
to the case involving non-harmonic forms. 

To include possible non-harmonic fluxes, it is helpful to rephrase~\eqref{FW_quant_G4} not in terms of cohomology groups but in terms of integrals of $\tilde G_4$ over suitable 4-cycles~\cite{Grimm:2011tb}. More precisely, given a basis $\Gamma_n$ of $H_4(\hat Y_4)$, the condition~\eqref{FW_quant_G4} is equivalent to requiring that
\be
 \int_{\Gamma_n} \tilde{G_4} \in \bZ
\ee
for all $n$. It is natural to assume that the generalisation to the case including non-closed $G_4$ involves evaluating similar integrals over certain 4-cycles which are trivial from the point of view of the homology of $\hat Y_4$~\cite{Grimm:2011tb}. Furthermore, the non-K\"ahler deformation of $\hat Y_4$ to $\hat Z_4$ might lead to a modification of the geometric part of~\eqref{FW_quant_G4}~\cite{Grimm:2011tb}. Although it would be interesting to investigate this further, the recent analysis of~\cite{Collinucci:2010gz, Collinucci:2012as} shows that the construction of the surfaces $\Gamma_n$ used to test for integrality of $\tilde{G}_4$ is a highly non-trivial task even in the harmonic case. For this reason, we will not pursue this direction further in this thesis.

\subsection{The dimensional reduction of M-theory on non-K\"ahler fourfolds}
\label{sec:MtheoryReduction}

In this subsection we discuss the 3-dimensional effective theory obtained when including the non-harmonic forms introduced in section~\ref{sec:massiveU(1)sFtheory} in the M-theory dimensional reduction. In particular, we take the massive fluctuations $\delta J = v^{0A}\tw_{0A}$ of the K\"ahler form into account, which formally describe a fluctuation of $\hat Y_4$ into a non-K\"ahler space. We will not attempt to include all details of the harmonic dimensional reduction, which is performed e.g. in~\cite{Haack:1999zv, Haack:2001jz, Berg:2002es}. Instead, we follow~\cite{Grimm:2011tb} and focus on the new aspects which arise upon inclusion of the non-harmonic forms related to the geometrically massive $U(1)$ symmetries. In section~\ref{sec:FtheoryLimit} we will discuss the F-theory limit and uplift the 3-dimensional action obtained from the M-theory compactification to 4 dimensions. As we will see, this 4-dimensional action will include shift symmetries gauged by the massive $U(1)$s, gauge boson masses and D-terms in agreement with the IIB results of section~\ref{sec:U(1)inIIB}. This serves as another check on the non-harmonic framework proposed in section~\ref{sec:massiveU(1)sFtheory}, which is independent from the match between the flux-induced tadpoles and chiralities found in section~\ref{sec:G4fluxes}.

The 3-dimensional effective action of M-theory may be computed by inserting the Kaluza-Klein expansions of the various fields into the 11-dimensional supergravity action~\eqref{S11normal} and integrating out the internal fourfold\footnote{In appendix~\ref{sec:app_democ_Mtheory}, we consider the dimensional reduction of a democratic version of the 11-dimensional supergravity theory. In the absence of branes, the standard dimensional reduction considered here leads to results that are fully equivalent to the results of the reduction of the democratic action, while being slightly simpler from a computational perspective. However, it will be advantageous to use the democratic formulation when including $M5$-branes in chapter~\ref{sec:chap_instantons}.}. In this way one obtains a theory given in terms of the real scalar and vector fields appearing in the expansions~\eqref{C3_expand} and~\eqref{Jexp_nonkaehler} of $C_3$ and $J$. In accord with $\cN=2$ supergravity in 3 dimensions, we expect that it must be possible to combine the vectors $A^\Lambda$ with the associated scalar fields $v^\Lambda$ into 3-dimensional vector multiplets. Similarly, the other real scalars should combine into complex chiral multiplets. To isolate the characteristic features of the effective action, it is useful to put it into a standard form of an $\cN=2$ supersymmetric theory in 3-dimensions. This is in full analogy to the approach taken in~\eqref{eq:N_1} for the 4-dimensional effective action of the IIB compactification. 

In 3 dimensions a slight additional ambiguity appears because the form of the characteristic supersymmetric $\cN=2$ action is not unique due to the duality between vector and scalar multiplets. In particular, it is possible to work with an action where all dynamical degrees of freedom belong to chiral scalar multiplets. In the case where some of the isometries of the target space are gauged, this action nevertheless includes a set of non-dynamical vector fields and is given by~\cite{Berg:2002es}
\be\label{kinetic_lin_gen}
  S^{(3)}_{\cN=2} = \frac{1}{\kappa_3^2} \int  \frac{1}{2} R_3 *1 -
  K_{A \bar B }\, \nabla N^A \wedge * \nabla \bar N^{\bar B}  
     - \frac{1}{2} \Theta_{AB} A^{A} \wedge F^{B} + V * 1 .
\ee
In our conventions, where all fields and spacetime coordinates are dimensionless, the 3-dimensional gravitational coupling constant $\kappa_3$ is once again simply given by $\kappa_3^2 = 1/4\pi$.
The $N^A$ denote the scalar fields of the theory, while the $A^A$ are vectors with field strength $F^A = dA^A$. The constant symmetric matrix $\Theta_{AB}$ is known as the embedding tensor. Together with the Killing vectors $\tilde{X}^{A B}$, it controls the form of the gauging through the covariant derivatives~\cite{Berg:2002es}
\be \label{DMThetaapp}
   \nabla N^A = d N^A + \Theta_{BC}\tilde{X}^{A B} \, A^{C}.
\ee
$K_{A\bar B}$ denote the components of the 3-dimensional K\"ahler metric, while $V$ is a scalar potential. Note that the action~\eqref{kinetic_lin_gen} includes no kinetic term for the auxiliary vector fields $A^A$, which are instead governed by the equation of motion
\be
\label{eomA}
 * F^{A} = 2 \R \big( K_{B\bar{C}} \tilde{X}^{\bar{C} A} \nabla N^{B} \big).
\ee

From the expansion~\eqref{C3_expand} of $C_3$, it is clear that the 3-dimensional action obtained directly from the M-theory reduction is in a slightly different form as it includes the dynamical vectors $A^\Lambda, \ \Lambda = 0, \alpha, IA$. It would be possible to extract the 3-dimensional characteristic $\cN=2$ data by dualising the vectors $A^\Lambda$ into scalars and comparing the result with the form~\eqref{kinetic_lin_gen}. The other option is to work out what the general $\cN=2$ action~\eqref{kinetic_lin_gen} looks like when some of the gauged scalars are dualised into dynamical vector multiplets, and compare the result to the action obtained from the M-theory reduction. We will follow reference~\cite{Grimm:2011tb} and use the second approach. 

After dualising some of the scalar multiplets from~\eqref{kinetic_lin_gen} into vectors, the field content of the theory includes a set $(\xi^\Lambda, A^\Lambda)$ of dynamical 3-dimensional vector multiplets. The remaining chiral multiplets will be denoted by $M^I$ following~\cite{Grimm:2011tb}. As worked out in detail in~\cite{Grimm:2011tb}, the general 3-dimensional $\cN=2$ action after dualisation takes the form
\be
\begin{aligned}
\label{kinetic_lin_gen_1}
  S^{(3)}_{\cN=2} = \frac{1}{\kappa_3^2} \int \Big[ &\ \frac{1}{2}R_3 *1  
 - \tilde K_{I \bar J }\, \nabla M^I \wedge * \nabla \bar M^{J}
  + \frac{1}{4} \tilde K_{\Lambda \Sigma}\, 
  d\xi^{\Lambda}\wedge * d\xi^{\Sigma} \\ 
  \phantom{\int \Big[} & \ + \frac{1}{4} \tilde K_{\Lambda \Sigma}\, F^{\Lambda} \wedge * F^{\Sigma}
     + \,  F^{\Lambda} \wedge \I (\tilde K_{\Lambda I} \, \nabla M^I) 
      \\ 
     &\ + \frac12 \Theta_{\Lambda \Sigma} A^{\Lambda} \wedge F^{\Sigma} - (V_\cT + V_{\mathrm{F}}) * 1 \Big]. 
\end{aligned}
\ee
In the following we will specify the various pieces of notation used in~\eqref{kinetic_lin_gen_1}, and discuss how the fields and couplings are related to the quantities appearing in the original action~\eqref{kinetic_lin_gen}. We stick closely to the discussion of~\cite{Grimm:2011tb}.
While the covariant derivatives
\be \label{DMTheta}
   \nabla M^I =d M^I + \Theta_{\bar J \Lambda}\tilde{X}^{I \bar J} A^\Lambda \equiv  d M^I + X^I_\Lambda \, A^{\Lambda}
\ee
remain unchanged from~\eqref{DMThetaapp}, the kinetic potential $\tilde K (M, \bar M, \xi)$ is related to the K\"ahler potential $K$ appearing in~\eqref{kinetic_lin_gen} by a Legendre transformation~\cite{Grimm:2011tb}
\be
\qquad K(t,\bar t,M,\bar M) = \tilde K - \frac12 (t_\Sigma + \bar t_\Sigma).
\ee
Here $t_\Sigma$ is the complex scalar dual to the vector multiplet $(\xi^\Sigma, A^\Sigma)$. The kinetic potential $\tilde{K}$ is restricted by $\R \, t_\Sigma = \tilde K_\Sigma$. As before, subscripts $I, \bar{I}$ and $\Lambda$ applied to the kinetic potential $\tilde K$ signify a partial derivative with respect to the fields $M^I, \bar M^I$ and $\xi^\Lambda$, respectively. Similarly, we write $\cT_I$, $\cT_{\bar I}$ and $\cT_\Lambda$ for the derivatives of the function $\cT$ which appears in
\be
\label{V_ct}
V_\cT = \tilde K^{I \bar J} \cT_I \cT_{\bar J} - \tilde K^{\Lambda \Sigma} \cT_\Lambda \cT_\Sigma - \cT^2.
\ee
$\tilde K^{I \bar J}$ and $\tilde K^{\Lambda \Sigma}$ are the inverses of $\tilde K_{I\bar J}$ and $\tilde K_{\Lambda\Sigma}$, respectively. 

As we will see in section~\ref{sec:FtheoryLimit}, $V_\cT$ can be viewed as the 3-dimensional analogue of the 4-dimensional D-term potential~\eqref{eq:V_D}. Just as in the 4-dimensional case, the first step to computing the potential is to evaluate the Killing potentials $D_\Sigma$, which are defined by
\be \label{3dDterm}
   i \partial_{M^I} D_\Sigma = \tilde K_{I \bar J} X^{\bar J}_\Sigma .
\ee
Together with the embedding tensor $\Theta$, the Killing potentials specify the function $\cT$ of equation~\eqref{V_ct} according to
\be \label{explicitcT}
   \cT(M, \bar M, \xi) = -  \frac12 \xi^\Sigma \Theta_{\Lambda \Sigma}  \xi^\Lambda +  \xi^\Sigma D_\Sigma.
\ee
In the case of interest to us, the Killing vectors of the gauged isometries will be field-independent constants. In this case,~\eqref{3dDterm} can be solved explicitly, and after inserting the results into~\eqref{explicitcT} and using~\eqref{V_ct} one obtains~\cite{Grimm:2011tb}
 \be
\begin{aligned}
 \label{generalVcT}
  V_\cT =&\ -(\tilde K_{I \bar J}-\tilde K_{\Gamma I}\tilde K^{\Gamma \Delta}\tilde K_{\Delta \bar{J}}) X^I_\Sigma  X^{\bar J}_\Lambda \xi^\Sigma \xi^\Lambda - \tilde K^{\Lambda \Sigma} D_\Lambda D_\Sigma \\
   &\ -  \Theta_{\Sigma \Gamma} \tilde K^{\Gamma \Delta} \Theta_{\Delta \Lambda} \xi^\Sigma \xi^\Lambda + 2 \tilde K^{\Lambda \Sigma} D_\Lambda \Theta_{\Sigma \Gamma} \xi^\Gamma  + 2i\tilde K^{\Lambda\Sigma}\tilde K_{\Lambda I} X^{I}_{\Delta} \Theta_{\Sigma \Gamma} \xi^\Delta \xi^\Gamma  \\&\ 
  - 2i\tilde K^{\Lambda\Sigma}\tilde K_{\Lambda I} X^{I}_{\Delta} D_{\Sigma}\xi^\Delta - \cT^2.
\end{aligned}
\ee

Although we will not be needing it in the following, let us for completeness also give the definition of the remaining potential $V_F$, which is the analogue of the 4-dimensional F-term potential~\eqref{eq:V_F}. In terms of the superpotential $W$ and the K\"ahler covariant derivatives $D_I W = \partial_{M^I} W + (\partial_{M^I} \tilde K) W$, $V_F$ is given by~\cite{Grimm:2011tb}
\be
V_{\rm F} =  e^K (\tilde K^{I \bar J} D_I W \overline{D_J W} - (4 + \xi^\Sigma \xi^\Lambda \tilde K_{\Sigma \Lambda} ) |W|^2) .
\ee 

This completes the specification of the general 3-dimensional $\cN=2$ action~\eqref{kinetic_lin_gen_1} involving both vector and gauged scalar multiplets. 
In the following, we discuss how the effective action obtained by compactifying M-theory on the non-K\"ahler space $\hat Y_4$ can be cast in the standard form~\eqref{kinetic_lin_gen_1}. We will not discuss the full details of the calculations, which are presented in~\cite{Grimm:2010ks} for the case of a harmonic Kaluza-Klein reduction. Instead, we follow the approach of~\cite{Grimm:2011tb} and focus on presenting the most important results and the modifications that arise due to the inclusion of non-harmonic forms in the compactification.

The first step is to identify the chiral and vector multiplets in terms of the fields introduced in the Kaluza-Klein reduction of sections~\ref{sec:KKreduxFtheory} and~\ref{sec:massiveU(1)sFtheory}. One set of chiral multiplets is obtained from the complex structure moduli $z^M$ which appear in the reduction of the 11-dimensional metric. A second set, which will be denoted by $N^a$ as in~\cite{Grimm:2011tb}, can be built from linear combinations of the fields $c^a$ and $b^a$ appearing in the reduction~\eqref{C3_expand} of $C_3$. Without loss of generality, the $N^a$ can be written as
\be \label{def-Na}
  N^a = c^a - i f^{ab}\, b_b
\ee
with a suitable matrix $f^{ab}$. The combinations $N^a$ can be viewed as arising from an expansion of $C_3$ into a new set of complex forms $\Psi_a$ and their conjugates $\bar \Psi_a$. These complex forms can in turn be related to the original basis $\alpha_a$, $\beta^a$ by requiring that $N^a\Psi_a + \bar{N}^a \bar{\Psi}_a$ reproduces the $c^a$ and $b^a$-dependence of the original reduction~\eqref{C3_expand}. In this way one obtains~\cite{Grimm:2011tb}
\be
\label{Psi_exp}
\Psi_a =  \frac{i}{2} \R f_{ba} (\beta^b  - i   \bar f^{cb} \alpha_c) \ , \qquad \Psi_a + \bar \Psi_a = \alpha_a ,
\ee
where $\R f_{ab}$ is the inverse of $\R f^{ab}$. As discussed in~\cite{Grimm:2010ks, Grimm:2011tb}, the functions $f^{ab}$ must be chosen in such a way that the $\Psi_a$ are in fact (2,1)-forms, so that the $f^{ab}$ depend on the complex structure moduli $z^M$. The vectors $A^\Lambda$ appearing in the 3-dimensional vector multiplets $(\xi^\Lambda, A^\Lambda)$ are identical with the vectors appearing in the Kaluza-Klein reduction of $C_3$. On the other hand, their scalar partners $\xi^\Lambda$ are related to the K\"ahler moduli $v^\Lambda$ of~\eqref{Jexp_nonkaehler} by~\cite{Grimm:2011tb}
\be \label{RLxi_def}
   \xi^0 = \frac{v^0}{\cV} \ , \qquad \xi^\alpha = \frac{v^\alpha}{\cV} \ , \qquad \xi^{IA} = \frac{v^{IA}}{\cV} ,
\ee
with $\cV$ being the (dimensionless) volume of $\hat Y_4$
\be
\cV = \frac{1}{4! } \int_{\hat Y_4} J^4.
\ee

An immediate consequence of the non-closedness of the forms $\tw_{0A}$ described by equation~\eqref{dtw_0} is that the shift symmetry of the real scalars $c^a$ is gauged. This fact can be directly read off from the expression for the 4-form field strength
\be
G_4 \supset dC_3 = F^\Lambda \wedge \omega_\Lambda + \left(d c^a -  N_A C^a_A A^{0A} \right) \wedge \alpha_a 
  + db_a \wedge \beta^a + b_a \, d\beta^a .
  \label{field_strength_gauging}
\ee
As the $c^a$ appear directly in the definition~\eqref{def-Na} of the chiral fields $N^a$, it is clear that the kinetic terms of these chiral fields will involve the covariant derivatives
\be \label{nablaN}
   \nabla N^a = d N^a - N_A C^a_A A^{0A} .
\ee
In other words, the 3-dimensional Killing vectors~\eqref{DMTheta} which describe the gauging of the target space isometries are given by
\be
\label{Killing_3D}
X^a_{0A} = - N_A C^a_A ,\qquad \quad X^I_\Lambda = 0, \text{ for } I\neq a \text{ or } \Lambda \neq 0A.
\ee

Let us emphasise that even in the presence of fluxes the geometric gauging~\eqref{Killing_3D} is the only gauging that is directly visible in the M-theory reduction. At first sight, this may seem puzzling in light of the flux-induced gauging of the IIB K\"ahler moduli $T_\alpha$ discussed in section~\ref{sec:Stueckelberg}. As shown in~\cite{Grimm:2011tb}, this gauging appears in the M-theory formulation upon dualisation of some of the vector multiplets $(\xi^\Lambda,A^\Lambda)$ into scalars $t_\Lambda$. More precisely, the gauging involves the embedding tensor $\Theta_{\Sigma\Lambda}$ and takes the form~\cite{Grimm:2011tb}
\be
  \nabla t_\Sigma = dt_\Sigma - 2 i \Theta_{\Sigma \Lambda} A^\Lambda.
  \label{flux_induced_gauging_Mtheory}
\ee
As we will see in the following, a non-zero embedding tensor indeed appears in the presence of non-trivial $G_4$ fluxes. In other words,~\eqref{flux_induced_gauging_Mtheory} describes the M-theoretic realisation of flux-induced gauging. By working out the precise relationship between the $t_\Sigma$ and the IIB chiral fields $T_\alpha$, we will confirm in section~\ref{sec:FtheoryLimit} that the IIB flux-induced gauging of~\eqref{gauging2} is indeed reproduced in the presence of $G_4$ fluxes of the form~\eqref{newfluxes}.

Before continuing with the discussion of flux-induced effects, let us briefly consider the kinetic potential $\tilde K$. The derivatives of $\tilde K$ can be obtained by comparing the dimensional reduction of the 11-dimensional action~\eqref{S11normal} to the general form~\eqref{kinetic_lin_gen_1}. The function $\tilde K$ which leads to the required derivatives was derived in the context of the purely harmonic Kaluza-Klein reduction and in the large volume limit in~\cite{Grimm:2010ks}. This kinetic potential admits a natural extension to the case including the possibly non-harmonic forms $\tw_{0A}$, $\alpha_a$ and $\beta^a$. Namely, it is possible to formally replace the relevant intersection numbers of harmonic forms with the corresponding intersection numbers introduced in section~\ref{sec:massiveU(1)sFtheory}. This leads to~\cite{Grimm:2011tb}
\be
\label{kinPot}
\tilde{K}(M,\bar M,\xi) = -3\log \cV + \frac{i}{4 }\xi^{\Lambda} (N^a - \bar{N}^a)(N^b- \bar{N}^{b})\int_{\hat Y_4}\omega_{\Lambda}\wedge \Psi_a \wedge\bar{\Psi}_{\bar{b}} + {K}_{CS} .
\ee

The part $K_{CS}$, which yields the kinetic terms of the complex structure moduli $z^M$, is given in terms of the holomorphic (4,0)-form $\Omega_4$ on $\hat Y_4$ by~\cite{Grimm:2010ks}
\be
K_{CS} = -\log \left[  \int \Omega_4 \wedge \bar \Omega_4 \right].
\ee
The form of $K_{CS}$ can in principle be confirmed by an explicit dimensional reduction of the 11-dimensional curvature scalar, but this will not be required in the following.
The second term in~\eqref{kinPot} can be verified by checking that it correctly reproduces the term $F^{\Lambda} \wedge \I (\tilde K_{\Lambda I} \, \nabla M^I)$ in~\eqref{kinetic_lin_gen_1}, which originates from the reduction of the 11-dimensional Chern-Simons coupling $\int C_3\wedge G_4\wedge G_4$. As a second consistency check, note that~\eqref{kinPot} involves only the imaginary parts of the $N^a$, in accordance with the fact that the real parts enjoy the gauged shift symmetry described by~\eqref{nablaN}.
It is somewhat less straightforward to confirm that~\eqref{kinPot} correctly reproduces the kinetic terms of the $N^a$, $\xi^\Lambda$ and $A^\Lambda$ obtained from the dimensional reduction. In the Calabi-Yau reduction of~\cite{Grimm:2010ks}, the match is obtained with the help of certain relations involving the Hodge star of the forms used in the Kaluza-Klein reduction, such as $\ast \Psi_a \sim \Psi_a \wedge J$. To confirm~\eqref{kinPot} in a rigorous manner, it has to be shown that these relations continue to hold when including the various non-harmonic forms introduced in section~\ref{sec:massiveU(1)sFtheory}. Similar problems were considered in the context of compactifications on non-Calabi-Yau threefolds e.g. in~\cite{House:2004pm, House:2005yc, Grana:2005ny, Benmachiche:2006df}. For the purpose of this thesis, we will not attempt to carry out the calculations explicitly. Instead, we will follow~\cite{Grimm:2011tb} in assuming that the relevant threefold techniques can in principle be generalised to the fourfold case and used to confirm~\eqref{kinPot}.

The final ingredients needed to complete the specification of the 3-dimensional action are the embedding tensor $\Theta_{\Lambda\Sigma}$ and the superpotential $W$. The superpotential only appears in the F-term potential $V_F$, which we will not require in the following. We therefore focus our attentions on the embedding tensor, which can be derived from the dimensional reduction of the 11-dimensional Chern-Simons term
\begin{align}
S_{CS}^{(11)} = &\ -\frac{1}{12\kappa_{11}^2} \int C_3 \wedge G_4 \wedge G_4 =  \frac{1}{2\kappa_3^2} \int_{M^{1,2}} \Theta_{\Lambda\Sigma} A^\Lambda \wedge F^\Sigma +\ldots .
\label{embedding_general}
\end{align}
It will be helpful to distinguish explicitly between topologically trivial and non-trivial parts of the 4-form field strength. We will therefore denote non-trivial fluxes by $G_4^f$ in the following, so that $G_4 = dC_3 + G_4^f$. In the context of the harmonic dimensional reduction of~\cite{Grimm:2010ks}, it is obvious that $\Theta_{\Lambda\Sigma}$ depends only on $G_4^f$ and not on $dC_3$. However, in the presence of the non-closed forms $\beta^a$ and $\tw_{0A}$, the reduction of the Chern-Simons term includes a term of the schematic form 
\be
\int_{M^{1,2}} b_a A^{\Lambda} \wedge F^\Sigma \int_{\hat Y_4} \omega_\Lambda \wedge \omega_\Sigma \wedge d\beta^a.
\label{b_a_dep_interaction}
\ee
At first sight, this seems to introduce a dependence on the moduli $b_a$ into the embedding tensor. However,~\eqref{b_a_dep_interaction} should actually be viewed as a contribution to the term of the form $F^{\Lambda} \wedge \I (\tilde K_{\Lambda I} \, \nabla M^I)$ in the 3-dimensional action~\eqref{kinetic_lin_gen_1}, and not to $\Theta_{\Lambda\Sigma}$. Therefore, even in the case including non-harmonic forms in $C_3$ we obtain\footnote{In determining the precise numerical prefactor, it is necessary to take into account that the Chern-Simons coupling $\int C_3 \wedge G_4 \wedge G_4$ is strictly speaking a shorthand notation for an interaction of the form $\int G_4^3$ over a suitable auxiliary 12-dimensional space. This effectively leads to a factor of 3 appearing in the dimensional reduction, as discussed in detail in~\cite{Witten:1996md}.}
\be
\Theta_{\Lambda\Sigma} =   - \frac{1}{2 } \int_{\hat Y_4} \omega_\Lambda\wedge\omega_\Sigma \wedge G^{f}_4 .
\ee
This shows that the embedding tensor is constant in the sense that it does not depend on any of the scalar fields of the theory. As discussed in detail in~\cite{Grimm:2011tb}, this property is crucial in order to allow for the vectors multiplets $(\xi^\Lambda, A^\Lambda)$ to be dualised into complex scalars $t_\Lambda$ consistently.

In a generic M-theory compactification, the form of the embedding tensor would not be further restricted. However, the completely general 3-dimensional action~\eqref{kinetic_lin_gen_1} does not necessarily admit an uplift to 4 dimensions. Such an uplift is of course necessary in order to interpret is as furnishing a description of an F-theory vacuum. The most general 3-dimensional action which can be obtained by a dimensional reduction from a 4-dimensional $\cN=1$ action was discussed in reference~\cite{Grimm:2011sk}, where it was shown that actions of this type must obey~\cite{Grimm:2011tb}
\be \label{Theta2}
  \Theta_{\alpha \beta} = 0 = \Theta_{rs} ,\qquad \text{with } r,s=0, (IA).
\ee
These conditions can therefore be viewed as a further non-trivial restriction on the fluxes $G^f_4$ which may be switched on in an F-theory compactification\footnote{Note that strictly speaking it is not necessary for~\eqref{Theta2} to hold in the full M-theory reduction, as the components $\Theta_{\alpha\beta}$ and $\Theta_{(IA)(JB)}$ are only required to vanish in the F-theory limit of vanishing fiber volume~\cite{Grimm:2011tb}.}, in addition to the tadpole cancellation conditions of section~\ref{sec:G4fluxes}. To evaluate the conditions in practice, as carried out e.g. in the context of an $SU(5)\times U(1)$ F-theory model in~\cite{Krause:2012yh}, a detailed understanding of the intersection properties of the basis of forms used in the dimensional reduction is required. In the following, we will focus on fluxes of the form~\eqref{newfluxes} and assume as in~\cite{Grimm:2011tb} that the basis $\omega_\Lambda$ used in the dimensional reduction is such that~\eqref{Theta2} yields no further non-trivial restrictions on the flux quanta $\tilde{\cF}_0^{A, \alpha}$ and  $\tilde{\cF}_0^{A, a}$.

From the discussion above, the possible non-trivial elements of the embedding tensor which can appear in the presence of fluxes along the diagonal $U(1)$ or the Cartan $U(1)$s are the components $\Theta_{(IA) \alpha}$. These components can be evaluated explicitly using the expansions~\eqref{G_4_Cartan} and~\eqref{newfluxes} of the fluxes and the integral relations of section~\ref{sec:massiveU(1)sFtheory}. Using the given intersections of the non-harmonic forms $\tw_{0A}$ and $\tilde{\tw}_{a,A}$, it is in particular easy to see that the components $\Theta_{(iA) \alpha}$ depend only on Cartan fluxes~\eqref{G_4_Cartan}, while in turn $\Theta_{(0A)\alpha}$ depends only on the fluxes along the diagonal $U(1)_A$. This shows that $\Theta_{(iA)\alpha}$, and therefore also the gauging~\eqref{flux_induced_gauging_Mtheory} induced by Cartan fluxes, is not altered by the introduction of the non-harmonic forms. As Cartan fluxes were already discussed in the harmonic dimensional reduction of~\cite{Grimm:2010ks}, we focus in the following on the new components
\be
\label{theta_alpha_0A}
\Theta_{(0A) \alpha} = -  \frac14 \cK_{\alpha \beta \gamma} N_A C_A^\gamma \tilde{\cF}_0^{A, \beta}    -  \frac14 \cK_{\alpha ab} N_A C^b_A \tilde{\cF}_0^{A, a}.
\ee

Through the embedding tensor, the fluxes contribute to the function $\cT$ of~\eqref{explicitcT} and hence to the 3-dimensional D-term potential $V_{\cT}$. Due to the simple form of the constant Killing vectors in~\eqref{Killing_3D}, the only non-vanishing Killing potentials are $D_{0A}$ and~\eqref{3dDterm} may be solved easily by
\be
D_{0A} =  i N_A C^a_A\left( \partial_{\bar{N}^a}\tilde{K}\right) .
\ee
It is straightforward to evaluate the derivatives of the kinetic potential directly using the explicit expression~\eqref{kinPot}. Combining the various components, one obtains the following expression for the function $\cT$~\cite{Grimm:2011tb}
\be 
\begin{aligned}
\cT  =& \  - \xi^{IA} \Theta_{(IA)\alpha} \xi^{\alpha} +  i \xi^{0A} N_A C^a_{A}\tilde K_{\bar{N}^a}  \\
    = & \ \frac{1}{ 4 \cV^{2}} \int_{\hat Y_4} J \wedge J \wedge G_4 . 
\label{evaluatedcT}
\end{aligned}
\ee
Note that this expression involves the full field strength $G_4$, and thus includes contributions from both $dC_3$ and from non-trivial fluxes $G_4^f$. As we will see in the following section, even in the absence of fluxes the purely geometric part in $\cT$ involving $dC_3$ yields a mass term for the scalars $\xi^{0A}$ in the D-term potential $V_{\cT}$~\cite{Grimm:2011tb}. This is of course to be expected, as the $\xi^{0A}$ combine with the vectors $A^{0A}$ into 3-dimensional multiplets and the $A^{0A}$ obtain a geometric mass term due to the gauging~\eqref{nablaN}.

\subsection{The F-theory limit}
\label{sec:FtheoryLimit}
The M-theory reduction on $\hat Y_4$ carried out in the previous section is the first step towards obtaining the 4-dimensional effective action of F-theory compactified on $\hat Y_4$. The second step is to uplift the 3-dimensional action~\eqref{kinetic_lin_gen_1} to 4 dimensions. Following the general logic outlined in section~\ref{sec:ell_fibration_D-brane_backreaction}, the M-theory effective action at finite fiber volume is expected to be T-dual to a compactification of a 4-dimensional $\cN=1$ supersymmetric action on a suitable circle. The 4-dimensional action underlying this circle compactification is what is referred to as the effective action of F-theory on $\hat Y_4$. It can be interpreted as describing a type IIB compactification on the base $B_3$ of the elliptically fibered fourfold $\hat Y_4$, in which a non-trivially varying dilaton profile and associated non-perturbative effects have been taken into account. The limit of vanishing fiber volume on the M-theory side corresponds to the limit in which the radius of the circle which defines the 4d$\rightarrow$3d reduction goes to infinity, such that full 4-dimensional Poincar\'e invariance is restored.

This discussion motivates the following strategy towards obtaining the F-theory uplift, which was developed in~\cite{Grimm:2010ks, Grimm:2011tb}. Starting with a general 4-dimensional $\cN=1$ action of the form~\eqref{eq:N_1}, we dimensionally reduce this on a circle. This leads to a 3-dimensional action which is specified in terms of the 4-dimensional characteristic $\cN=1$ data such as the 4-dimensional K\"ahler potential, gauge kinetic function etc.. The 3-dimensional action obtained via this circle reduction can then be compared to the action that was derived in the dimensional reduction of M-theory on $\hat Y_4$. This comparison determines the 4-dimensional characteristic $\cN=1$ data in terms of the geometry of $\hat Y_4$. Finally, we perform the 4-dimensional decompactification limit by taking the volume of the fiber of $\hat Y_4$ to zero.

The generic F-theory effective action at the geometrically massless level was computed using these techniques in~\cite{Grimm:2010ks}. Rather than reviewing the full discussion, we will follow~\cite{Grimm:2011tb} and focus on the effects related to the non-harmonic forms in the M-theory reduction. As extensively motivated above, these non-harmonic forms describe effects which in the IIB picture are associated with diagonal $U(1)$s on $D7$-brane stacks and the corresponding gaugings and D-terms. The relevant part of the 4-dimensional $\cN=1$ action is therefore
\be \label{S4gauge}
  S^{(4)}_{\rm gauge} = \frac{1}{\kappa_4^2} \int_{M^{1,3}} - K_{M \bar N} \nabla M^M \wedge *_4  \nabla \bar M^{\bar N} + \frac14 \R f^{(IA) (JB)} D_{(IA)} D_{(JB)} *_4 1 . 
\ee
Here we have used the general 4-dimensional $\cN=1$ notation introduced in section~\ref{sec:IIBreduction}. At this stage, the indices $(IA)$, $(JB)$ are actually general indices running over all the gauge symmetries present in the theory, corresponding to the indices $A$, $B$ in~\eqref{eq:N_1}. Of course, the 4-dimensional gauge bosons appearing in the gaugings and D-terms are expected correspond to the 3-dimensional gauge bosons describing the Cartan and diagonal $U(1)$s, which is why we use the same indices $(IA)$ that were introduced in the M-theory Kaluza-Klein reduction of section~\ref{sec:massiveU(1)sFtheory}.
We will focus on the case where the shift symmetries of the chiral fields $M^M$ are gauged in a non-linear manner, such that the associated Killing vectors can be directly read off from the covariant derivatives
\be
\nabla M^N = d M^N + X_{IA}^N A_{(4)}^{IA}.
\ee
Furthermore, we take the Killing vectors to be field-independent constants, which is the case of interest relating to the IIB and M-theory reductions discussed above.

In the following we will denote 4-dimensional spacetime indices by $\mu,\nu= 0,1,2,3$, while corresponding 3-dimensional indices are denoted by $r,s=0,1,2$ as in~\cite{Grimm:2011tb}. The compactification circle is taken to lie in the $x^3$ direction. 
In the course of the Kaluza-Klein reduction of~\eqref{S4gauge}, all fields are taken to be independent of $x^3$. Clearly, the 4-dimensional scalars then descend to 3-dimensional scalar fields, which we continue to denote by $M^M$. Each of the 4-dimensional vectors $A_{(4)}^{IA}$ gives rise to a 3-dimensional vector $\tilde{A}^{IA}$ and an additional scalar $\tilde{\xi}^{IA}$, such that $A_{(4)}^{IA} = (\tilde A^{IA},\ \tilde \xi^{IA})$. Finally, an additional 3-dimensional scalar $\xi^0$ and vector $A^0$ arise from the degrees of freedom associated with the components $g_{r 3}^{(4)}$ and $g_{33}^{(4)}$ of the 4-dimensional metric $g^{(4)}_{\mu\nu}$. The precise relation between $g^{(4)}_{\mu\nu}$ and the 3-dimensional metric $g^{(3)}_{rs}$ can be worked out by requiring that the reduction of the 4-dimensional curvature tensor yields a 3-dimensional Einstein-Hilbert term and standard kinetic terms for $A^0$ and $\xi^0$. In this way, one obtains~\cite{Ferrara:1989ik, Grimm:2010ks, Grimm:2011tb}
\be 
\begin{aligned}
\label{metric_red}
  g^{(4)}_{\mu \nu} = &\ \left(\begin{array}{cc} g^{(3)}_{rs} + (\xi^0)^{-1} A^0_r A^0_s & (\xi^0)^{-1} A^0_r \\ 
                                               (\xi^0)^{-1} A^0_r & (\xi^0)^{-1} \end{array} \right), \\  
  g_{(4)}^{\mu \nu} = & \ \left(\begin{array}{cc} g_{(3)}^{rs} & - g_{(3)}^{rs} A^0_s  \\ 
                                                - g_{(3)}^{rs} A^{0}_s  & \xi^0 + g_{(3)}^{rs} A^0_r A^0_s \end{array} \right).
\end{aligned}
\ee 
The field $\xi^0$ is therefore related to the circumference $r$ of the compactification circle along $x^3$ by~\cite{Grimm:2011tb}
\be
\xi^0 = \frac{1}{r^2}.
\ee

The dimensional reduction of~\eqref{S4gauge} to 3 dimensions can now be carried out by inserting the expansions of the metric and the vectors $A^{IA}_{(4)}$ and integrating out the circle direction. To put the kinetic terms into the standard form, it is helpful to write the resulting action in terms of fields $A^{IA}$, $\xi^{IA}$ which are related to the fields $\tilde A^{IA},\ \tilde \xi^{IA}$ introduced above by~\cite{Ferrara:1989ik, Grimm:2010ks}
\be \label{AAred}
  A_{(4)}^{IA} = (\tilde A^{IA},\ \tilde \xi^{IA}) =  (A^{IA}  + (\xi^0)^{-1} \xi^{IA} A^0,\ (\xi^0)^{-1} \xi^{IA}). 
\ee
After a suitable Weyl rescaling $g_3\rightarrow  \xi^0 g_3$ of the 3-dimensional metric, which is required to obtain a canonical Einstein-Hilbert term,~\eqref{S4gauge} reduces to\footnote{The seemingly different sign of the kinetic term is due to the Hodge star identity~\eqref{hodge_star_inner_prod}.}~\cite{Grimm:2011tb}
\begin{align} \label{S3gauge}
  S^{(3)}_{\rm gauge} =& \ \frac{1}{\kappa_3^2} \int  K_{M \bar N} \nabla M^M \wedge *_3  \nabla \bar M^{\bar N} + V_{\rm gauge} *_3 1 ,\\
  V_{\rm gauge} =&\  K_{M \bar N} X_{IA}^M X_{JB}^{\bar N}\xi^{IA} \xi^{JB}  + \frac14  \xi^0 \ \R f^{(IA) (JB)} D_{IA} D_{JB} .
  \label{V_gauge}
\end{align}
Note that the three-dimensional covariant derivatives
\be \label{3dgauge-der}
  \nabla M^M = d M^M + X_{IA}^M A^{IA}
\ee
involve only $A^{IA}$, while the term involving $A^0$ in~\eqref{AAred} cancels against contributions coming from the reduction of the metric~\eqref{metric_red}. 

The full 3-dimensional action obtained by reduction from 4 to 3 dimensions can be put into the standard form of~\eqref{kinetic_lin_gen_1} by using the kinetic potential~\cite{Grimm:2011tb}
\be \label{Kinetic_after_red}
   \tilde K_{4\rightarrow3}(M,\bar M,\xi) = \log \xi^0 + K(M,\bar M) + \frac{2}{\xi^0} \R f_{(IA)(JB)} \xi^{IA} \xi^{JB}.
\ee
This expression can be confirmed explicitly by checking that it correctly accounts for the kinetic terms and the D-term potential obtained after reduction from 4 to 3 dimensions.
Firstly, it is obvious that the kinetic terms of the $M^M$ in~\eqref{S3gauge} are correctly reproduced due to the term in~\eqref{Kinetic_after_red} involving the 4-dimensional K\"ahler potential $K(M, \bar M)$. Although we have not displayed the kinetic terms of the vector multiplets $(\xi^0, A^0_3),\ (\xi^{IA}, A^{IA}_3)$ obtained after reduction, it can similarly be checked directly that they are also described by~\eqref{Kinetic_after_red}~\cite{Haack:1999zv, Grimm:2011tb}.

To understand the form~\eqref{V_gauge} of the effective D-term potential, it is crucial to note that no 3-dimensional Chern-Simons terms appear in the reduction from 4 to 3 dimensions. In other words, the embedding tensor vanishes and
\be
\Theta^{(4\rightarrow3)}_{(IA)(JB)} = \Theta^{(4\rightarrow3)}_{(IA)0} = \Theta^{(4\rightarrow3)}_{00} = 0.
\ee
The 3-dimensional D-term potential of a theory obtained by reduction from 4 dimensions therefore simplifies from the general form~\eqref{generalVcT} to
\be
\label{Vct_4to3}
V^{(4\rightarrow3)}_{\cT} = \tilde K_{4\rightarrow3}^{M \bar N} \cT_M \cT_{\bar N} - \tilde K_{4\rightarrow3}^{(IA) (JB)} \cT_{IA} \cT_{JB} - \cT^2.
\ee
The function $\cT$ here is given simply by $\cT = D^{(4\rightarrow 3)}_{IA} \xi^{IA}$. The 3-dimensional D-terms $D^{(4\rightarrow 3)}_{IA}$ are determined by the kinetic potential~\eqref{Kinetic_after_red} and the constant Killing vectors using equation~\eqref{3dDterm}. Due to the fact that $\tilde K_{4\rightarrow3}$ involves the chiral multiplets $M^M$ only through the 4-dimensional K\"ahler potential $K$, they are identical to the 4-dimensional D-terms
\be
D^{(4\rightarrow 3)}_{IA} = -i (\partial_{\bar{M}^{\bar{N}}} \tilde K_{4\rightarrow3} )\, \bar{X}^{\bar N}_{IA} = -i (\partial_{\bar{M}^{\bar{N}}}  K )\, \bar{X}^{\bar N}_{IA} = D_{IA}.
\ee
 The inverse of the kinetic metric obtained from the kinetic potential~\eqref{Kinetic_after_red} can be directly evaluated, with the result~\cite{Grimm:2011tb}
\be
\begin{aligned} \label{tKinverse}
   \tilde K_{4\rightarrow3}^{00} = &\ -(\xi^0)^2 , \qquad\quad
   \tilde K_{4\rightarrow3}^{(IA)0} = -  \xi^0 \xi^{IA}  , \\
   \tilde K_{4\rightarrow3}^{(IA)(JB)} = &\   \frac14 \xi^0 \, \R f^{(IA)(JB)} - \xi^{IA} \xi^{JB}.
\end{aligned}
\ee
Plugging these expressions into the general formula~\eqref{Vct_4to3} leads precisely to the potential $V_{\rm gauge}$ obtained by direct dimensional reduction in~\eqref{V_gauge}. In particular, the form of the kinetic potential guarantees that the potential $V^{(4\rightarrow3)}_{\cT}$ is positive definite, which is not obvious from~\eqref{Vct_4to3} but is necessary for consistency in 4 dimensions. This completes the necessary checks to confirm that the kinetic potential of the 3-dimensional theory obtained by a reduction from 4 to 3 dimensions is indeed given in terms of the 4-dimensional K\"ahler potential $K$ and gauge kinetic function $f_{(IA)(JB)}$ by equation~\eqref{Kinetic_after_red}.

As our goal is to uplift the 3-dimensional action obtained from the M-theory reduction to 4 dimensions, we would like to essentially run the above reasoning in reverse. More precisely, we would like to derive the 4-dimensional Killing vectors, D-terms and K\"ahler potential from the 3-dimensional quantities using the relations~\eqref{3dgauge-der} and~\eqref{Kinetic_after_red}. However, as pointed out in~\cite{Grimm:2011tb}, it is not possible to directly compare~\eqref{Kinetic_after_red} to the kinetic potential~\eqref{kinPot} obtained from the M-theory reduction in the previous section. One way to see this is to note that the general action obtained by reduction from 4 to 3 dimensions has no embedding tensor, while the M-theory reduction of section~\ref{sec:MtheoryReduction} led to $\Theta_{\alpha (IA)} \neq 0$. To compare the different actions, it is therefore necessary to first dualise the vector multiplets $(\xi^\alpha, A^\alpha)$ in the M-theory reduction into complex scalars $t_\alpha$. In the course of this dualisation, the embedding tensor elements $\Theta_{\alpha (IA)}$ essentially become elements of a Killing vector describing the gauging of a shift symmetry of $t_\alpha$ by the vectors $A^{IA}$, as described by~\eqref{flux_induced_gauging_Mtheory}.

The set of scalars that can be identified as arising from 4-dimensional scalars is thus $M^N = (t_\alpha, N^a, z^M)$. On the other hand, the vector multiplets $ A^{IA}$, $A^0$ of section~\ref{sec:MtheoryReduction} can be identified with the vectors $A^{IA}$, $A^0$, which arise from the 4-dimensional vectors and the 4-dimensional metric according to~\eqref{AAred} and~\eqref{metric_red}, respectively. 
In the course of the Legendre transformation which realises the dualisation of the would-be vectors $(\xi^\alpha, A^\alpha)$ into the $t_\alpha$, the Kinetic potential $\tilde K$ of section~\ref{sec:MtheoryReduction} is transformed according to~\cite{Grimm:2011tb}
\begin{align}\label{legendreTrf}
 \tilde K_{4\rightarrow3}(z^M,N^a,t_\alpha,\xi^{IA},\xi^0) = &\ \tilde{K}(z^M,N^a, \xi^{\Lambda}) - \frac12 (t_{\alpha}+\bar{t}_{\alpha})\xi^{\alpha},
 \\ 
 \R\ t_\alpha = &\ \partial_{\xi^\alpha} \tilde K(z^M,N^a, \xi^{\Lambda}) .
 \label{def_re_t_alpha}
\end{align}
In principle, these relations suffice to uplift the action obtained from the M-theory reduction to 4 dimensions, and determine the characteristic quantities of the F-theory effective action in terms of the geometry of $\hat Y_4$.

Unfortunately, the complicated form of the kinetic potential $\tilde K$ given in equation~\eqref{kinPot} is such that the relation~\eqref{def_re_t_alpha} cannot be explicitly inverted in general~\cite{Grimm:2011tb}. This means that $\tilde K_{4\rightarrow3}$, and with it the 4-dimensional K\"ahler potential, will only be determined implicitly by the relation~\eqref{legendreTrf}. Note that this mirrors the situation in the type IIB compactification discussed in section~\ref{sec:IIBreduction}, where the K\"ahler potential also involves the volume of the Calabi-Yau, which cannot explicitly be written in terms of the IIB chiral fields $T_\alpha$. In the following, we will discuss the most important features of the F-theory K\"ahler potential and gauge kinetic function and demonstrate the match with the corresponding IIB expressions as in~\cite{Grimm:2011tb}.

To simplify the evaluation of the Legendre transform~\eqref{def_re_t_alpha}, it is helpful to expand $\tilde K$ in terms of the fields $\xi^\Lambda$ and focus on the leading terms which are dominant in the F-theory limit $v^0 \rightarrow 0$. The precise scaling behaviour of the various fields in the F-theory limit was worked out in~\cite{Grimm:2010ks} for the massless fields and extended in~\cite{Grimm:2011tb} to include the fields of the non-harmonic Kaluza-Klein reduction. In terms of a parameter $\epsilon$, which describes the F-theory limit by $\epsilon\rightarrow 0$, the scaling is given by
\be
\label{eps_scaling_v}
v^0 \sim \epsilon , \qquad v^\alpha \sim \frac{1}{\sqrt{\epsilon}} , \qquad v^{IA} \sim \epsilon^{3/2}, \qquad \cV \sim \frac{1}{\sqrt{\epsilon}} + \cO(\epsilon),
\ee
or equivalently
\be
\xi^0 \sim \epsilon^{3/2} , \qquad \xi^{\alpha} \text{ finite,} \qquad \xi^{IA} \sim \epsilon^2.
\ee
These scaling properties may be used to expand the quantum volume $\cV$, which appears in the large volume expression~\eqref{kinPot} for the kinetic potential, into a power series in $\epsilon$. Due to the fact that the intersection of 4 vertical divisors vanishes, this expansion starts with
\be
\cV = \int_{\hat Y_4} \frac{J^4}{4!} = \frac12 \frac{1}{3!} \cV^4 \cK_{\alpha \beta \gamma} \xi^\alpha \xi^\beta \xi^\gamma \xi^0 + \cO(\epsilon),
\ee
where we have used the intersection numbers~\eqref{def_Kalphabetagamma}. Note that due to the intersection properties of the forms $\tw_{IA}$ the expansion does not include a term linear in the $\xi^{IA}$. In order to capture the leading terms involving $\xi^{IA}$, we will therefore have to expand $\cV$ up to order $\epsilon^2$, for which we need the intersection numbers~\eqref{twtw_Cartan} as well as
\be
\cK_{00\alpha\beta} \equiv  \int_{\hat Y_4} \omega_0 \wedge\omega_0\wedge\omega_\alpha\wedge\omega_\beta .
\ee

Inserting the expansion of $\cV$ into the large volume kinetic potential~\eqref{kinPot} and using the fact that $\int_{\hat Y_4} \tw_{IA} \wedge \Psi_a \wedge\bar{\Psi}_{\bar{b}} = \int_{\hat Y_4} \omega_0 \wedge \Psi_a \wedge\bar{\Psi}_{\bar{b}} = 0 $ due to~\eqref{int_alpha_beta_tw} leads to~\cite{Grimm:2011tb}
\be
\begin{aligned}
 \label{tKexpansion}
    \tilde{K} =&\ \log \xi^0 + \log \left[\frac12\frac{1}{3!} \cK_{\alpha \beta \gamma} \xi^\alpha \xi^\beta \xi^\gamma + \frac14 \cK_{00\alpha\beta} \xi^0 \xi^\alpha\xi^\beta  \right.
    \\
    & \left. - \frac{1}{4 \xi^0} \cC^A_{IJ} C_A^\lambda \cK_{\lambda \alpha \beta} \xi^\alpha \xi^\beta \xi^{IA} \xi^{JA}  + \cO(\epsilon^3) \right]  \\
         &+  \frac{i}{4 } \xi^{\alpha} (N^a - \bar{N}^a)(N^b- \bar{N}^{b})\int_{\hat Y_4}\omega_{\alpha}\wedge \Psi_a \wedge\bar{\Psi}_{\bar{b}} + {K}_{CS}.
\end{aligned}
\ee
By expanding the logarithm and evaluating the derivative with respect to $\xi^\alpha$ we immediately obtain
\be\label{t_alpha_expansion}
\R \ t_\alpha = \frac{1}{4 \hat{\cK}}\cK_{\alpha \beta \gamma} \xi^\beta \xi^\gamma + \frac{i}{4} d_\alpha + \cO(\epsilon^{3/2}),
\ee
where we have introduced the abbreviations
\begin{align}
\label{def_hat_cK}
\hat{\cK} = &\ \frac12\frac{1}{3!} \cK_{\alpha \beta \gamma} \xi^\alpha \xi^\beta \xi^\gamma, \\ 
d_\alpha = &\ (N^a - \bar{N}^a)(N^b- \bar{N}^{b})\int_{\hat Y_4}\omega_{\alpha}\wedge \Psi_a \wedge\bar{\Psi}_{\bar{b}}.
\label{def_d_alpha}
\end{align}
On the other hand, comparing~\eqref{tKexpansion} to the general form~\eqref{Kinetic_after_red} suggests
\be
\R \ f_{(IA)(JB)} = - \frac{1}{8\hat{\cK}} \delta_{AB} \cC^A_{IJ} C_A^\alpha \cK_{\alpha \beta \gamma} \xi^\beta \xi^\gamma.
\ee

As the gauge coupling function must be a holomorphic function of the chiral fields, this suggests~\cite{Grimm:2010ks, Grimm:2011tb} that a term of the form 
\be
\label{missing_term_Kinpot}
-\frac{i}{4 \xi^0} \cC^A_{IJ} C_A^\alpha d_\alpha \xi^{IA} \xi^{JB}
\ee 
is missing in~\eqref{tKexpansion}, which is needed to match the second term in~\eqref{t_alpha_expansion}. Note that this missing term involves 5 of the fields $\xi^\Lambda, N^a$, while it is clear that the reduction of the M-theory action~\eqref{S11normal} can only lead to terms involving at most 3 fields through the Chern-Simons interaction $\int C_3\wedge G_4 \wedge G_4$. Therefore it seems natural to guess that the missing term~\eqref{missing_term_Kinpot} may be accounted for by higher order terms in the 11-dimensional action, which were not included in~\eqref{S11normal}. However, we will not try to investigate the possible origin of the term~\eqref{missing_term_Kinpot} any further and will simply assume that it arises in a dimensional reduction of the full 11-dimensional action. Including this term, one obtains the gauge kinetic function
\be
\label{gaugekin_Ftheory}
 f_{(IA)(JB)} = - \frac12 \delta_{AB}\cC^A_{IJ} C_A^\alpha t_\alpha.
\ee

The form of the gauge kinetic function obtained above is very reminiscent of the type IIB result~\eqref{gaugeKinIIB}. To demonstrate the precise match, we will require the correspondence between the fields on the M-theory and IIB sides which was worked out in~\cite{Grimm:2011tb}. The fields $N^a$ can be directly identified with the IIB fields $G^a$ defined in~\eqref{def_Ga}, while the vector fields on the M-theory and IIB sides are related by
\be
\label{relation_AA}
A^{0A} \leftrightarrow \frac{1}{2\pi} A^A.
\ee
Taking this scaling factor into account, it is clear that the gauging~\eqref{nablaN} of the $N^a$ precisely reproduces the IIB gauging~\eqref{gauging1} in the uplift from 3 to 4 dimensions. 
Furthermore, the K\"ahler moduli $v^\alpha$ arising from the expansion of $J$ into forms $\omega_\alpha$ are expected to be related to the K\"ahler moduli $v^\alpha_{IIB}$, which appear in the type IIB reduction on $X_3$, by
\be
v^\alpha_{IIB} = \sqrt{v^0}\,  v^\alpha + \cO(\epsilon). 
\label{kaehler_mod_IIB_ftheory}
\ee
The factor of $\sqrt{v^0}$, which was not included in~\cite{Grimm:2011tb}, is necessary in order to obtain finite K\"ahler moduli in the IIB limit in the face of the scaling properties of the fields given in~\eqref{eps_scaling_v}. As we will see in section~\ref{sec:partition_fct_match}, this factor can also be deduced by comparing the actions of vertical $M5$-instantons and $E3$-instantons. At leading order in $\epsilon$ we thus obtain
\be
\label{relation_xi^alpha}
\xi^\alpha \sim \frac{12 v^\alpha}{\cK_{\beta\gamma\delta}v^0 v^\beta v^\gamma v^\delta} + \cO(\epsilon) \rightarrow \frac{12 v_{IIB}^\alpha}{\cK_{\beta\gamma\delta}v_{IIB}^\beta v_{IIB}^\gamma v_{IIB}^\delta} = 2 \frac{v^\alpha_{IIB}}{\cV_{IIB}},
\ee
where the volume of the Calabi-Yau manifold $X_3$ used in the type IIB reduction is 
\be
\cV_{IIB} = \frac{1}{3!} \int_{X_3} J^3 = \frac{1}{3!} \cK_{\alpha\beta\gamma}v_{IIB}^\alpha v_{IIB}^\beta v_{IIB}^\gamma.
\ee

By plugging the relation~\eqref{relation_xi^alpha} into the definition~\eqref{t_alpha_expansion} of the $t_\alpha$ we see that in the F-theory limit there holds
\be
\label{rel_t_T}
t_\alpha = 3 \frac{\cK_{\alpha \beta \gamma} \xi^\beta \xi^\gamma }{\cK_{\alpha' \beta' \gamma'}\xi^{\alpha'} \xi^{\beta'} \xi^{\gamma'} } + \ldots \rightarrow \frac{1}{4} \cK_{\alpha \beta \gamma} v_{IIB}^\beta v_{IIB}^\gamma + \ldots = \frac12 T_\alpha,
\ee
with $T_\alpha$ the chiral field of the IIB effective action defined in~\eqref{def_Talpha}. The same result was obtained in a slightly different way in~\cite{Grimm:2011tb}. Taking into account the different normalisation of the gauge fields as given in~\eqref{relation_AA}, we thus obtain an exact match between the gauge kinetic function $f_{(0A)(0B)}$ of the diagonal $U(1)_A$ in the F-theory limit and the IIB result~\eqref{gaugeKinIIB}. In particular, this justifies the intersection numbers~\eqref{twtw_Cartan} with $\cC^A_{00} = N_A$ for the non-harmonic forms $\tw_{0A}$.

Having derived the precise relation between the fields $t_\alpha$ and $T_\alpha$, we are now in a position to verify that our treatment of $G_4$ fluxes along the diagonal $U(1)$ reproduces the flux-induced gauging of the $T_\alpha$ encoded in~\eqref{gauging2}. As shown explicitly in~\cite{Grimm:2011tb}, if the embedding tensor $\Theta_{\alpha (IJ)}$ is non-vanishing the Legendre transformation~\eqref{legendreTrf} results in a gauging of the shift symmetries of the $t_\alpha$. The covariant derivatives take the form~\cite{Grimm:2011tb}
\be
\nabla t_\alpha \equiv d t_\alpha + X_{\alpha (IA)}A^{IA} = d t_\alpha -2i\Theta_{\alpha (IA)} A^{IA}.
\label{gauging_t_alpha}
\ee
The element of the embedding tensor relevant for the gauging with respect to the diagonal $U(1)$ $A^{0A}$ was given in~\eqref{theta_alpha_0A}. After taking into account the factors which were shown above to appear in the relations between $t_\alpha$, $A^{0A}$ and their IIB counterparts $T_\alpha$, $A^A$, this exactly reproduces the gauging~\eqref{gauging2} obtained in the type IIB setting. This match serves as another piece of evidence supporting the proposed uplift~\eqref{newfluxes} of the fluxes along the diagonal $U(1)$ to F-theory.

In a similar manner, it is possible to see that the F-theory reduction correctly reproduces the D-terms~\eqref{DtermIIB} associated with the diagonal $U(1)_A$ in the IIB setting. To verify this, we require the fact that the Legendre transformation implies~\cite{Grimm:2011tb} $\partial_{t_{\alpha}} K = \partial_{t_{\alpha}} \tilde{\cK} = -\frac12 \xi^\alpha$ and $\partial_{N^a} K = \partial_{N^a} \tilde{K}$. The discussion above furthermore shows that the Killing vectors are unchanged in the uplift from 3 to 4 dimensions. Plugging these results into the general expression $D_{0A} = -i K_{\bar N^a}\bar{X}^a_{0A} -i K_{\bar t_\alpha}\bar{X}_{\alpha(0A)}$ for the 4-dimensional D-terms corresponding to $A^{0A}$ leads to~\cite{Grimm:2011tb}
\be
D_{0A} =  i\tilde{K}_{\bar{N}^a} N_A C^a_A + 2 K_{\bar t_\alpha} \Theta_{\alpha 0A} = i\tilde{K}_{\bar{N}^a} N_A C^a_A - \Theta_{\alpha 0A} \xi^\alpha.
\ee
The derivative of $\tilde K$ with respect to the $N^a$ can be directly evaluated from~\eqref{kinPot}. In order to facilitate the comparison with the type IIB results, it is helpful to use the definitions~\eqref{def-Na} and~\eqref{Psi_exp} of $N^a$ and $\Psi_a$ together with the integral relation~\eqref{int_omega_alpha_beta} to evaluate
\be
\begin{aligned}
\tilde{K}_{\bar{N}^a} = &\ - \frac{i}{4}\xi^\alpha (N^b-\bar{N}^b) \left[ \int_{\hat Y_4} \omega_\alpha\wedge \Psi_a \wedge \bar{\Psi}_b + \int_{\hat Y_4} \omega_\alpha \wedge \Psi_b \wedge \bar{\Psi}_a \right] \\
=&\  \frac{i}{4} \xi^\alpha \cK_{\alpha a c} \delta^{cb} b_b.
\label{K_barNa}
\end{aligned}
\ee
Using the form of $\Theta_{\alpha(0A)}$ given in~\eqref{theta_alpha_0A} we thus obtain
\be
D_{0A} = \frac{1}{4} \xi^\alpha N_A \Big( \cK_{\alpha\beta\gamma} \tilde{\cF}^{A,\beta} C^\gamma_A + \cK_{\alpha a b} (\tilde{\cF}^{A, a}-\delta^{ac}b_c) C^b_A \Big).
\ee
After taking into account~\eqref{relation_xi^alpha} and the relation between $A^{0A}$ and $A^A$, this again perfectly matches the type IIB result~\eqref{DtermIIB}.

The only aspect of the type IIB reduction relating to the geometrically massive $U(1)$s which remains to be discussed in the F-theory picture is the actual form of the geometric mass~\eqref{massIIB}. We have already established that the IIB expressions for the Killing vectors and the gauge kinetic function, which enters the mass~\eqref{massIIB} after rescaling the gauge bosons to put the kinetic terms into canonical form, are reproduced correctly in the F-theory compactification on $\hat Y_4$. It therefore remains to consider the elements $K_{N^a \bar{N}^b}$ of the 4-dimensional K\"ahler metric. These derivatives can be obtained by differentiating equation~\eqref{legendreTrf}, which describes the Legendre transformation of the kinetic potential $\tilde{K}$. In evaluating these partial derivatives, the $\xi^\alpha$ are to be considered as functions of the $N^a$ and $t_\alpha$ according to~\eqref{def_re_t_alpha}. In this way one obtains~\cite{Grimm:2011tb}
\be
K_{N^a\bar{N}^{b}} = \tilde K_{N^a\bar{N}^{b}} - \tilde{K}_{N^a\xi^{\alpha}} \tilde{K}^{\xi^{\alpha}\xi^{\beta}} \tilde{K}_{\xi^{\beta}\bar{N}^{b}}
\ee
Here $\tilde{K}^{\xi^{\alpha}\xi^{\beta}}$ is the inverse of the subblock $\tilde{K}_{\xi^{\alpha}\xi^{\beta}}$ of the kinetic metric $\tilde K$ given in~\eqref{kinPot}. Up to corrections of order $\cO(\epsilon^{3/2})$ one obtains
\be
\tilde{K}_{\xi^{\alpha}\xi^{\beta}} = \partial_{\xi^\alpha}\partial_{\xi^\beta} \log\left[ \cV^{-3} \right] =  \frac{1}{2 \hat \cK} \cK_{\alpha\beta\gamma} \xi^\gamma - \frac{1}{16\hat{\cK}^2}\cK_{\alpha\gamma\delta}\xi^\gamma\xi^\delta\cK_{\beta\gamma'\delta'}\xi^{\gamma'}\xi^{\delta'}.
\ee
Using relation~\eqref{relation_xi^alpha} to relate the $\xi^\alpha$ to the type IIB K\"ahler moduli $v^\alpha_{IIB}$ and taking the limit of $\epsilon\rightarrow 0$ this yields
\be
\tilde{K}_{\xi^{\alpha}\xi^{\beta}} \rightarrow \frac14 \cV_{IIB} \cK_{\alpha\beta}-\frac{1}{16} \cK_\alpha \cK_\beta.
\ee
Here $\cK_\alpha$ and $\cK_{\alpha\beta}$ are contractions of the intersection numbers $\cK_{\alpha\beta\gamma}$ with the IIB K\"ahler moduli $v^\alpha_{IIB}$, as in section~\ref{sec:Stueckelberg}. Therefore $\tilde{K}_{\xi^{\alpha}\xi^{\beta}}$ coincides with the matrix $G_{\alpha\beta}$ defined in~\eqref{g_alphabeta} up to an overall factor of $-1/\cV_{IIB}^2$. The mixed derivatives involving $\xi^\alpha$ and $N^a$ are easily obtained from~\eqref{K_barNa} and the fact that $\tilde{K}$ is real
\be
\tilde{K}_{N^a\xi^{\alpha}} = -\frac{i}{4}\cK_{\alpha a c} \delta^{cd} b_d , \qquad \tilde{K}_{\xi^{\beta}\bar{N}^{b}} = \frac{i}{4}\cK_{\alpha b c} \delta^{cd} b_d.
\ee
Furthermore,~\eqref{K_barNa} yields
\be
\tilde K_{N^a\bar{N}^{b}} = - \frac{1}{8}\xi^\alpha \cK_{\alpha a c} \delta^{cd}\R f_{db},
\ee
where $f^{bd}$ is the function appearing in the definition~\eqref{Psi_exp} of the forms $\Psi_a$. Putting the various pieces together allows us to compare $K_{N^a\bar{N}^{b}}$ to the corresponding IIB expression $K_{G^a \bar{G}^b}$ of~\eqref{Kaehler_metric_GG}. As expected, we find an exact match provided that in the F-theory limit we have
\be
\R f_{bd} \rightarrow \delta_{bd} e^{\Phi} = \frac{1}{\R(-i\tau)}\delta_{bd},
\ee
where $\tau$ is the axio-dilaton. This suggests that 
\be
f^{ab} = -i\tau \delta^{ab}.
\label{relation_f_tau}
\ee
Note that~\eqref{relation_f_tau} is just as expected from the identification of the $N^a$ with the $G^a$ if the fields $c^a$ and $b_a$ appearing in the two definitions can be identified. The relation~\eqref{relation_f_tau} will also be confirmed independently by the match of the partition functions of $E3$- and $M5$-instantons found in section~\ref{sec:partition_fct_match}, and by the dimensional reduction of the democratic M-theory supergravity carried out in appendix~\ref{sec:app_democ_Mtheory}.

The F-theory/IIB match found above completes our discussion of the effective description of massive $U(1)$ gauge symmetries and their associated fluxes in F-theory with the help of suitable non-harmonic forms in the dimensional reduction. However, before we proceed let us briefly comment on the special form of $f^{ab}$ in~\eqref{relation_f_tau}. Note that of course the function $f$ can be changed by choosing a different basis $\{\alpha_a, \beta^b\}$ of $H^{2,1}(\hat Y_4)$. The basis we have chosen is distinguished by the fact that $\alpha_a$ and $\beta_a$ are both constructed from the same form $\omega_a \in H^{1,1}(X_3)$. This is the rationale behind the expectation that the intersection numbers~\eqref{int_omega_alpha_beta} and the fields $c^a, b_a$ in~\eqref{C3_expand} can be identified with the corresponding IIB quantities. Now of course~\eqref{relation_f_tau} is only expected to be valid in this chosen basis and should not be assumed in a general F-theory model with no direct IIB connection. In order for this picture to be consistent, the relation~\eqref{relation_f_tau} should hold independently of the basis $\omega_a$ chosen on the IIB side. To check this, let us consider a general basis change $\omega_a \rightarrow \omega_b'= R_{ab}\omega_b$. Under this transformation we also have $\beta^a \rightarrow {\beta^a}' = R_{ab} \beta^b$, $\alpha_a \rightarrow \alpha_a' = R_{ab}\alpha_b$. Therefore, the intersection numbers~\eqref{int_omega_alpha_beta} of $\alpha_a'$ and ${\beta^b}'$ will still match the corresponding type IIB intersection numbers $\cK_{\alpha a b}'$ evaluated using $\omega_a'$, although of course $\cK_{\alpha a b}' \neq \cK_{\alpha a b}$. The complex forms $\Psi_a$ in~\eqref{Psi_exp} transform into $\Psi_a' = R_{ab}\Psi_b$, which is an admissible basis change because the $\Psi_a'$ will still be $(2,1)$-forms provided this was the case for the $\Psi_a$. It is easy to check that the new forms $\Psi_a'$ can still be written as a linear combination of the $\alpha_a'$, ${\beta^b}'$ as in~\eqref{Psi_exp} using a function ${f'}^{ab}$ related to $f^{ab}$ by
\be
f' = (R^\rT)^{-1}f R^\rT.
\label{change_f_basis}
\ee
Here we have used matrix notation and written $R^\rT$ for the transpose of the basis change matrix. Clearly, if $f$ is proportional to the unit matrix as in~\eqref{relation_f_tau} then $f'=f$. This is consistent with our expectation that~\eqref{relation_f_tau} holds in the F-theory limit in models with a IIB dual, independently of the basis chosen on the IIB side. However, let us emphasise that a general basis change on the M-theory side, in which the $\alpha_a$ and $\beta^b$ are rotated by different matrices, will not leave $f$ unchanged.

\section{Geometric interpretation of $U(1)$s in F-theory}
\label{sec:U(1)geometry}

The discussion of the previous section shows that geometrically massive $U(1)$ symmetries can be made explicitly visible in F-theory compactifications by including certain non-harmonic forms in the dimensional reduction. At the level of the effective supergravity all the effects due to these massive $U(1)$s that are known from the type IIB setting can be reproduced in F-theory in this way. The various differential and integral properties of the required non-harmonic forms given in section~\ref{sec:massiveU(1)sFtheory} were motivated based on expectations from the type IIB picture. However, from a purely F-theoretic point of view the introduction of these forms is somewhat \textit{ad hoc}, and in the framework of the Kaluza-Klein reduction it is not clear what distinguishes the chosen forms from the multitude of other non-harmonic forms present on the compactification manifold. It is natural to expect that the properties of the forms relating to the massive $U(1)$s are somehow encoded in the geometry of the elliptic fiber above the $D7$-brane locus in the base. Let us therefore now turn to a discussion of how $U(1)$ symmetries appear from a geometric point of view in the F-theory compactification. We begin by reviewing how certain cycles corresponding to Cartan $U(1)$ gauge symmetries appear upon resolution of a non-Abelian $A_n$ singularity, following~\cite{Katz:1996xe, Grimm:2011tb}. We then discuss the proposal given in~\cite{Grimm:2011tb} for how a cycle corresponding to the diagonal $U(1)$ can be identified in this framework. Finally, in section~\ref{sec:sections_and_U(1)s} we review recent progress in the description of massless non-Cartan $U(1)$ symmetries using a different approach based on constructing additional sections of the elliptic fibration.

\subsection{Cartan $U(1)$s from resolution of $SU(N)$ singularities}
\label{sec:geom_cartans}
In this subsection we consider Cartan $U(1)$s, which can by definition be viewed as arising from a larger non-Abelian gauge group. As reviewed in section~\ref{sec:nonabelian_groups}, such non-Abelian gauge groups are described in the F-theory construction by ADE singularities in the elliptic fiber of the compactification manifold. The group theory details of the associated gauge group are encoded in the geometric structure of the singularity, which is hidden in the singular phase but becomes visible upon resolution. In particular, the resolution introduces a set of holomorphic curves which are associated with the Cartan $U(1)$s of the gauge group. Let us review explicitly how these Cartan curves arise, following the presentation of~\cite{Katz:1996xe, katz1992gorenstein}. We focus on F-theory models with non-Abelian gauge group $SU(N)$. These models are of course expected to correspond to type IIB configurations with gauge group $U(N)$ and thus furnish the correct setting to study the uplift of the diagonal $U(1)$ in the next section. 

According to the Tate classification of table~\ref{tab:Tate_class}, the gauge group $SU(N)$ arises in the presence of singularities of the type $A_{N-1}$. Such a singularity can be locally described by the equation
\be
-XY + Z^N = 0
\label{A_n-1_sing}
\ee
in the weighted projective space $\bP_{N,\, N,\, 2}$~\cite{katz1992gorenstein}. In other words, we expect that the Tate polynomial~\eqref{P_Tate} which describes the elliptically fibered fourfold can be put into this form in a vicinity of the $SU(N)$ component of the discriminant locus by suitable coordinate redefinitions. Choosing local coordinates in the base in which the discriminant is described by $w=0$, it is possible to check that the Tate polynomial can be rewritten in this form if the sections $a_i$ appearing in~\eqref{P_Tate} have certain vanishing orders in $w$. These vanishing orders are determined by Tate's algorithm and given in table~\ref{tab:Tate_class}. For our purposes, it will not be necessary to include the full details of the Tate polynomial and we instead work directly with the standard form~\eqref{A_n-1_sing}. Nevertheless, it should be kept in mind that after embedding the singularity in an elliptically fibered fourfold the coordinate redefinitions necessary to bring the Tate polynomial into the form~\eqref{A_n-1_sing} ensure that $X, Y$ and $Z$ depend on the coordinates of the base of the fibration.  

The singularity of the hypersurface described by~\eqref{A_n-1_sing} is located at the origin $X=Y=Z=0$. To resolve the singularity, we first extend the ambient space by introducing $N-1$ additional $\bP^1$'s with homogeneous coordinates $(u_j, v_j),\ j=1,...,N-1$. The resolved surface is given by the intersection of the equations~\cite{katz1992gorenstein}
\be
\begin{aligned}
XY =& \ Z^N , \\
X v_j = &\ u_j Z^j , \qquad 1\leq j \leq N-1 ,\\
Z^{j-k} u_j v_k = & \ u_k v_j, \qquad 1\leq k < j \leq N-1,
\label{small_res_A_N}
\end{aligned}
\ee
inside the extended ambient space.
Note that away from the original singular point $X=Y=Z=0$ these equations suffice to fix the coordinates of the $\bP^1$'s completely. Therefore the component of the fiber away from the singularity, which we denote by $\tilde{t}_0$, remains essentially unaffected by the resolution. Additional fiber components appear after resolution in the so-called exceptional set given by the preimage of the singular point $X=Y=Z=0$. Specifically, one obtains $N-1$ independent curves (or 2-cycles) $\tilde{t}_i$, which can be specified by~\cite{katz1992gorenstein}
\be
\tilde{t}_i: X=Y=Z=u_1=...=u_{i-1} = v_{i+1}=...=v_{N-1} = 0.
\ee
We use lower case indices $i,j,k$ to run over the values 1 to $N-1$, while capital letters $I,J$ will be used in the following for indices running over the range 1 to $N$.

The fact that the fiber splits into the $N$ different components $\tilde{t}_0,\ldots,\tilde{t}_{N-1}$ can be encoded in the homological relation 
\be
[\tilde{t}_0] + \sum_{i=1}^{N-1}[\tilde{t}_i] = [e],
\label{hom_rel_fiber}
\ee
where $[e]$ denotes the homology class of the smooth elliptic fiber~\cite{Grimm:2011tb}. It is useful to formally define an additional curve by $\tilde{t}_N \equiv \tilde{t}_0 - [e]$. Then the relation~\eqref{hom_rel_fiber} can also be written simply as 
\be
\sum_{I=1}^{N} [\tilde{t}_I] = 0.
\label{hom_rel_fiber_weights}
\ee

The geometry of the resolved fiber can be related to the group theory of the gauge group by identifying the combinations $\tilde{v}_i \equiv \tilde{t}_i - \tilde{t}_{i+1}$ with the simple roots of $SU(N)$~\cite{Grimm:2011tb}. In other words they correspond to the nodes of the Dynkin diagram of $SU(N)$. By including the additional node $\tilde{v}_0 = \tilde{t}_0 - \tilde{t}_1$, one obtains the extended Dynkin diagram. The correspondence between the nodes of the Dynkin diagram and the 2-cycles $\tilde{t}_I$ is particularly clear in the case of F-theory compactifications on $K_3$. In that case, 2-cycles naturally intersect in points, and it is possible to verify explicitly that the 2-cycles intersect according to the extended Cartan matrix of $SU(N)$. In the physically more relevant case of F-theory compactifications on fourfolds, the natural intersection structure pairs 2-cycles and 6-cycles. To describe the Cartan intersection structure in this case, it is helpful to introduce 6-cycles $\tilde{t}^{I}$ which are defined to be dual to the 2-cycles $\tilde{t}_I$ in the sense that~\cite{Grimm:2011tb}
\be
\tilde{t}_I\cdot \tilde{t}^J = - \delta_I^J.
\label{dual_2cycle_6cycle}
\ee
The divisors $D_i\equiv \tilde{t}^i - \tilde{t}^{i+1}$ then clearly intersect the 2-cycles $\tilde{v}_j$ according to the Cartan matrix $\cC_{ij}$ of $SU(N)$, i.e.
\be
\tilde{v}_i \cdot D_j = \cC_{ij}.
\label{cartan_intersection_cycle_divisor}
\ee
The 2-forms $\tw_{i}$ which are Poincar\'e-dual to the divisors $D_i$ can therefore be identified with the 2-forms which were introduced in section~\ref{sec:KKreduxFtheory} to describe the Cartan $U(1)$ gauge symmetries in the M-theory Kaluza-Klein reduction.

The discussion so far was quite general in the sense that it involved only on the local geometry around the singular locus. In situations where the fibration can be written globally in the Tate form appropriate for an $A_{N-1}$ singularity, it is actually possible to directly resolve the codimension one singularities at the global level without first bringing the Tate equation into the standard form~\eqref{A_n-1_sing}. This global resolution process was described in detail in~\cite{Krause:2011xj} in the context of an $SU(5)\times U(1)$ GUT model, although the generalisation of the process to other special unitary gauge groups is clear. The resolution techniques used in~\cite{Krause:2011xj} are somewhat different from those used above. In equation~\eqref{small_res_A_N}, we had explicitly pasted in a set of $\bP^1$'s locally above the singular locus by first enlarging the ambient space of the fiber and then describing the resolved fiber by the intersection of several equations in this space. A resolution carried out in this manner is sometimes also referred to as a small resolution~\cite{Esole:2011sm}. In contrast, the approach of~\cite{Krause:2011xj} is to add a set of additional coordinates $e_i$ together with suitable scaling relations which act on the $e_i$ and the original coordinates $x,y,z$ of the elliptic fiber. This of course changes the fiber ambient space, but does not enlarge its dimension. Therefore the resolved fiber is still specified by a single equation, which is a modified form of the original Tate equation, in the new ambient space. The advantage of this approach is that it is easier to construct the exceptional divisors $D_i$ corresponding to the Cartan $U(1)$s. Whereas in the small resolution approach described above the Cartan divisors were defined only implicitly as 6-cycle duals to the curves $\tilde{t}_I$, the $D_i$ can be explicitly constructed in the approach of~\cite{Krause:2011xj} by intersecting the equations $e_i=0$ with the modified Tate equation. This makes it possible to verify the intersection properties of the divisors $D_i$ or their dual 2-forms $\tw_i$, by direct calculation~\cite{Krause:2011xj}. Note that the global resolution approach is not restricted to Cartan $U(1)$s arising from non-Abelian singularities at codimension 1 in the base. Indeed, the resolution of a so-called $U(1)$-restricted global $SU(5)$ Tate model, which exhibits an additional codimension 2 singularity with gauge group $U(1)$ in addition to the $SU(5)$ singularity, was discussed in~\cite{Krause:2011xj}. In complete analogy with the Cartan case, the resolution of the additional singularity introduces an additional coordinate with associated exceptional divisor. The harmonic 2-form associated with this additional resolution divisor then leads to a massless $U(1)$ gauge boson in the Kaluza-Klein reduction of $C_3$, which describes the extra Abelian factor in the low energy gauge group.

Before proceeding with the discussion of how the diagonal $U(1)\subset U(N)$ is expected to arise in the geometric resolution picture, let us note that the resolution processes outlined above are not the only way to remove the original singularity. A different approach is to simply deform the original equation~\eqref{A_n-1_sing} in such a way that the singularity is absent. Just as in the resolution picture discussed above, it is possible to construct specific cycles related to the roots of the gauge group corresponding to the singularity in the deformation picture\footnote{In fact, in compactifications on $K_3$ deformations and resolutions can be viewed as equivalent as they are related by mirror symmetry~\cite{Grimm:2011tb, Katz:1997eq}.}. In the following, we will briefly review the construction of these cycles in the geometry obtained by deforming the $A_{N-1}$ singularity, sticking closely to the presentation of~\cite{Grimm:2011tb}. Although from a physical perspective we are more interested in the the case where the singularity is removed by resolution, the deformation picture will provide useful intuition for our discussion of the diagonal $U(1)$s in the next subsection.

By suitable coordinate redefinitions, the most general deformation of the $A_{N-1}$ equation~\eqref{A_n-1_sing} can be put into the so-called preferred versal form~\cite{katz1992gorenstein}
\be
-XY + Z^N + \sum_{I=2}^{N-1} b_I Z^{N-I} = 0.
\label{pref_versal_deform}
\ee
The deforming polynomial can be factorised, which allows the deformed equation to be rewritten as
\be
-XY + \prod_{I=1}^N (Z+t_I) = 0,
\label{AN-1_total_deform}
\ee
with suitable parameters $t_I$. The parameters $b_I$ and $t_I$ can be chosen real without loss of generality~\cite{Braun:2010ff}. The fact that there is no term corresponding to $b_1$ in~\eqref{pref_versal_deform} translates into the constraint $\sum_I t_I = 0$. This relation is reminiscent of the homological relation~\eqref{hom_rel_fiber_weights}, or the analogous relation obeyed by the 6-cycles $\tilde{t}^I$ in the resolution picture. Indeed, the $t_I$ can be identified with the weights of the fundamental representation of $SU(N)$~\cite{Grimm:2011tb, Katz:1996xe}. Therefore, they can be viewed as playing analogous roles to the $\tilde{t}^I$ in the resolution picture\footnote{The fact that the weights should be associated to the 6-cycles $\tilde{t}^I$ rather than the 2-cycles $\tilde{t}_I$ can be deduced from the fact that the weights naturally live in the dual space to the simple roots, which in the resolution picture were given by the 2-cycles $\tilde{v}_i$~\cite{Katz:1996xe}.}.

In type IIB language, deforming the $A_{N-1}$ singularity can be viewed as moving apart the $N$ $D7$-branes which originally formed a stack with gauge group $U(N)$. Roughly speaking, the parameters $t_I$ then correspond to the positions of the $D7$-branes. It is clear that~\eqref{AN-1_total_deform} describes a non-singular surface in the completely deformed case in which all $t_I$ are pairwise distinct, such that all $N$ branes are completely separated from one another. If on the other hand $N'<N$ of the $t_I$ are equal to some common value $t$, we obtain an enhancement to an $A_{N'-1}$ singularity at the point $Z=t$. The full original $A_{N-1}$ singularity is recovered if all of the $t_I$ are equal, in which case they are required to vanish by the tracelessness constraint $\sum_I t_I = 0$.

We had already argued in section~\ref{sec:ell_fibration_D-brane_backreaction} that one of the 1-cycles of the elliptic fiber is expected to shrink to zero size over the location of a $D7$-brane. This vanishing cycle can be seen explicitly in the deformation picture~\cite{Grimm:2011tb}. To make the shrinking $S^1$ manifest, it is helpful to change coordinates in~\eqref{AN-1_total_deform} to $X \equiv \tilde{X}+i\tilde{Y} , \ Y\equiv \tilde{X}-i\tilde{Y}$, and to project the equation to plane spanned by the imaginary parts of $\tilde{X}$ and $\tilde{Y}$. This yields
\be
\left(\I\, \tilde{Y}\right)^2 + \left(\I\, \tilde{X}\right)^2 = -\prod_{I=1}^N \left(Z+t_I\right),
\ee
which shows that the circle in the $\I \, \tilde{X}-\I\, \tilde{Y}$ plane shrinks exactly at the points where $Z=t_I$ for some $I$. The collapsing circle can be viewed as the A-cycle of the elliptic fiber~\cite{Grimm:2011tb}. Because the A-cycle shrinks over the brane locations $t_I$, it is possible to construct closed 2-cycles by fibering the A-cycle over (real) lines in the base connecting two of the branes. Following~\cite{Grimm:2011tb}, we denote such a line running from $t_J$ to $t_I$ by $t_I - t_J$. 

Of particular importance are the 2-cycles $v_i,\ i=1,...,N-1$, which are constructed by fibering the A-cycle over the lines $t_i - t_{i+1}$. This is because these 2-cycles can be seen to intersect\footnote{Strictly speaking, a pair of 2-cycles can be directly intersected only on $K_3$. In the fourfold case, we mean that the intersections of the 2-cycles with dual 6-cycles yield the Cartan matrix as in~\eqref{cartan_intersection_cycle_divisor}.} according to the Cartan matrix of $SU(N)$ and can therefore be identified with the simple roots~\cite{Grimm:2011tb, Katz:1996xe, katz1992gorenstein}. More general 2-cycles $v_{IJ}$ fibered over $t_I-t_J$ can be viewed as sums or differences of the simple root cycles $v_i$. In the deformation picture, the gauge bosons of the full non-Abelian gauge group are viewed as arising from $M2$-branes wrapped on these 2-cycles $v_{IJ}$~\cite{Katz:1996xe}. It is obvious that the gauge boson state associated to the 2-cycle $v_{IJ}$ is massless if and only if $t_I = t_J$ and the cycle shrinks to zero size. Note that for elliptic Calabi-Yaus of complex dimension larger than 2, all gauge boson states including the Cartan gauge bosons corresponding to the cycles $v_i$ are massive in the fully deformed picture. This is in contrast to the case in which the singularity was removed by resolution, where the Cartan gauge bosons remained massless and only the gauge bosons corresponding to off-diagonal generators of the gauge group became massive.

\subsection{Geometry of the diagonal $U(1)$}
\label{sec:geom_diag_U(1)}

The discussion of the local resolution of an $A_{N-1}$ singularity in the previous subsection shows that the singular fiber naturally splits into $N$ components in the resolution process. The singularity structure is related to the $SU(N)$ gauge theory by identifying the $\mathrm{rk}(SU(N)) = N-1$ homologically independent resolution spheres in the fiber with the Cartan $U(1)$s. However, this identification leaves one component of the resolved fiber unaccounted for. It is natural to expect that this additional component is associated with an additional $U(1)$ symmetry~\cite{Grimm:2011tb}, which would match the type IIB picture where the gauge group of a stack of $N$ branes is $U(N)= SU(N)\times U(1)$.

To be more precise, let us follow~\cite{Grimm:2011tb} and denote the 2-forms Poincar\'e-dual to the 6-cycles $\tilde{t}^I$ by $\omega_I$. The intersection property~\eqref{dual_2cycle_6cycle} can then be written as
\be
\int_{\tilde{t}_I} \omega_J = - \delta_J^I.
\label{dual_2cycle_2form}
\ee
Note that the forms $\omega_I$ are individually non-closed. This can be easily shown using an argument given in~\cite{Grimm:2011tb}, which is based on the fact that the sum of the 2-cycles $\tilde{t}_I$ is homologically trivial. To begin with, we rewrite equation~\eqref{hom_rel_fiber_weights} as
\be
\sum_{I=1}^{N} \tilde{t}_I = \partial \cC,
\label{triviality_2-cycles}
\ee
where $\cC$ is a suitable 3-cycle. After summing over the index $I$ in~\eqref{dual_2cycle_2form} one then obtains
\be
-1 = \int_{\sum_I \tilde{t}_I} \omega_J = \int_\cC d\omega_J.
\label{nonclosed_omega_J}
\ee

The fact that the individual $\omega_J$ are non-closed is not in contradiction with the well-known fact that the Cartan $U(1)$s remain massless upon resolution~\cite{Grimm:2011tb}. This is because the Cartan $U(1)$s are associated to the 2-forms $\tw_i = \omega_i - \omega_{i+1}$, which makes it possible for the non-harmonic components of the $\omega_I$ to cancel out in the differences $\omega_i - \omega_{i+1}$. On the other hand, as discussed in section~\ref{sec:massiveU(1)sFtheory} the diagonal $U(1)$ should be described by a form $\tw_0$ which is in general non-closed. In~\cite{Grimm:2011tb} it was suggested that this form $\tw_0$ is described by the combination
\be
\tw_0 = \sum_{I=1}^N \omega_I.
\label{expansion_tw_0}
\ee

The proposed form~\eqref{expansion_tw_0} can be motivated by the following observations. Firstly, note that the exterior derivatives $d\omega_I$ must be independent of $I$ in order for the Cartan forms $\tw_i$ to be closed. Therefore~\eqref{expansion_tw_0} yields
\be
d \tw_0 = N d\omega_1,
\ee
so that the factor of $N$ in the differential relation~\eqref{dtw_0} appears naturally as a consequence of the form of $\tw_0$. Furthermore, the $\omega_I$ are expected to be mutually orthogonal in the sense that
\be
\int_{\hat Y_4} \omega_I \wedge\omega_J \wedge \omega_\alpha\wedge\omega_\beta \propto -\delta_{IJ},
\label{intersection_omega_I}
\ee
where $\omega_\alpha$ and $\omega_\beta$ are 2-forms arising by pullback from the base $B_3$ as in section~\ref{sec:U(1)inFtheory}. This relation is necessary in order for the Cartan forms $\tw_i$ to intersect according to the Cartan matrix of $SU(N)$ as in~\eqref{cartan_intersection_harmonic}. As argued in section~\ref{sec:U(1)inFtheory}, the form describing the diagonal $U(1)$ is expected to be orthogonal to all the Cartan forms $\tw_i$ in the sense that
\be
\int_{\hat Y_4} \tw_0 \wedge\tw_i \wedge \omega_\alpha\wedge\omega_\beta = 0,
\ee
see equation~\eqref{twtw_Cartan}. In light of~\eqref{intersection_omega_I}, the only combination of the $\omega_I$ which satisfies this relation is precisely the combination~\eqref{expansion_tw_0}. Finally, we note that~\eqref{intersection_omega_I} implies
\be
\int_{\hat Y_4} \tw_0 \wedge\tw_0 \wedge \omega_\alpha\wedge\omega_\beta \propto - N ,
\ee
in agreement with the intersection numbers given in~\eqref{twtw_Cartan}.

The arguments presented above show that the form $\tw_0$ related to the 2-cycles appearing in the resolution of the $A_{N-1}$ singularity by~\eqref{expansion_tw_0} has the right properties to be identified with the form $\tw_{0A}$ introduced in section~\ref{sec:massiveU(1)sFtheory}. As established in section~\ref{sec:U(1)inFtheory}, this form describes the diagonal $U(1)$ which is expected to be present from the IIB perspective in the case where this $U(1)$ is geometrically massive. It remains to discuss what happens in the F-theory uplift of IIB configurations in which the diagonal $U(1)$ remains massless at the geometric level. At first sight, the expression~\eqref{expansion_tw_0} seems to be problematic in this case because the $\omega_I$ are non-closed according to~\eqref{nonclosed_omega_J}. In~\cite{Grimm:2011tb} it was argued that this apparent contradiction can be resolved by taking into account that the homological relation~\eqref{triviality_2-cycles}, which arises in the local resolution of the $A_{N-1}$ singularity, fails to hold in the global picture. Let us briefly review the argumentation of~\cite{Grimm:2011tb} in the following.

In order to gain intuition about the F-theory geometry which may give rise to a massless diagonal $U(1)$, let us consider the corresponding type IIB picture. As outlined in section~\ref{sec:IIBreduction}, a $U(N)$ gauge symmetry with massless diagonal $U(1)$ arises in the case where the brane divisor $D$ is in the same homology class as the orientifold divisor $D_{O7}$. In this sense, the masslessness of the diagonal $U(1)$ is a non-local effect even in the IIB setting, because it depends on the full details of the orientifold involution and not only on the local geometry around the brane divisor $D$~\cite{Grimm:2011tb}. The key to understanding the relevant F-theory geometry is to note that in the IIB picture the brane configuration can be smoothly deformed into a configuration where the branes lie on top of the orientifold plane, which would lead to the gauge group $SO(2N)$. According to tables~\ref{tab:Kodaira_class} and~\ref{tab:Tate_class}, an $SO(2N)$ gauge group is described in the F-theory picture by singularity of type $D_N$. This suggests that a massless diagonal $U(1)$ is described by an F-theory geometry in which a $D_N$ singularity is deformed into an $A_{N-1}$ singularity~\cite{Grimm:2011tb}.

A $D_N$ singularity can be described by the equation~\cite{katz1992gorenstein}
\be
-X^2 - Y^2 Z + Z^{N-1} = 0
\ee
in the weighted projective space $\bP_{N-1,\, N-2,\, 2}$. The most general deformation of this equation can again be described by $N$ real deformation parameters $t_I$. Without loss of generality, the deformed equation can be taken to be in the preferred versal form
\be
-X^2 - Y^2 Z + Z^{N-1} - \sum_{i=1}^{N-1} \delta_{2i} Z^{N-i-1} + 2\gamma_N Y = 0,
\label{pref_versal_D_N}
\ee
where the $\delta_{2i}$ are elementary symmetric polynomials in the $t_I$ and $\gamma_N = \prod_I t_I$~\cite{Grimm:2011tb, katz1992gorenstein}. Note that in contrast to the case of the $A_{N-1}$ singularity, the $t_I$ do not have to satisfy a tracelessness constraint of the form~\eqref{hom_rel_fiber_weights}.

To construct the 2-cycles which shrink to zero size in the singular limit $t_I\rightarrow 0$, we proceed as in the case of the $A_{N-1}$ singularity by fibering a cycle in the fiber over a line in the base connecting two points where the fiber cycle collapses. We follow the discussion of~\cite{Grimm:2011tb}. The first step is to identify the collapsing fiber cycle, which is facilitated by changing coordinates to
\be
X' = \sqrt{Z}X, \qquad Y' = YZ + \prod_{I=1}^N t_I.
\label{coord_shift_D_N}
\ee
In terms of these coordinates, the deformed equation~\eqref{pref_versal_D_N} takes the form
\be
(X')^2 + (Y')^2 = \prod_{I=1}^N (Z+t_I^2).
\label{collaps_cycle_D_N}
\ee
Restricting to the real $X', Y'$ plane yields the equation of a circle which collapses to zero at the points $Z=-t_I^2$. It would be tempting to attempt to construct 2-cycles by simply fibering the collapsing cycle over lines in the $Z$-plane connecting the points $t_I$. However, in comparison with the case of the $A_{N-1}$ singularity a new subtlety appears due to the fact that the coordinate transformation~\eqref{coord_shift_D_N} introduces a branch cut for the coordinate $X'$~\cite{Grimm:2011tb}. 

To avoid the difficulty associated with the branch cut in the $Z$ plane, it is helpful to consider a local double cover of the base parametrised by $Z' = \sqrt{Z}$. Effectively, we now have two sets of degeneration points located at $Z'=\pm i t_I$. At the same time, the number of paths over which the collapsing cycle can be fibered appears doubled in the double cover picture, as the point $i t_I$ can be connected with either $it_J$ or $-it_J$. Note that after the coordinate change to $Z'$, equation~\eqref{collaps_cycle_D_N} essentially looks like a pair of deformed $A_{N-1}$ singularities with deformation parameters $it_I$ and $-it_I$, respectively~\cite{Grimm:2011tb}. In the IIB language, the two $A_{N-1}$ singularities can be interpreted as describing the homologous brane and image brane stacks. The 2-cycles constructed from fibrations over paths connecting points $it_I$ and $-it_J$ then correspond to strings stretched between the brane stack and its orientifold image in this picture. 

If a line running from one brane at $it_I$ to a different brane in the same stack at $it_J$ is denoted by $t_I - t_J$ as before, then the lines between $it_I$ and a brane in the image stack at $-it_J$ corresponds effectively to $t_I + t_J$~\cite{Grimm:2011tb}. In the IIB picture, the plus sign can be viewed as an effect of the change of orientation that is experienced in crossing the orientifold plane when moving from the brane stack to the image stack. Let us emphasise that the new type of lines $t_I + t_J$ and the associated 2-cycles do not have a direct analogue in the discussion of a general $A_{N-1}$ singularity as in section~\ref{sec:geom_cartans}. In this sense the appearance of a new type of 2-cycle in the singularity deformation can be really seen as intrinsic to the $D_N$ geometry. Of course, the appearance of the new cycles can also be seen in the single cover picture without making the coordinate change from $Z$ to $Z'$. Here the two types of paths are distinguished according to whether the path passes through the branch cut of $X'$ or not~\cite{Grimm:2011tb}.

As in the $A_{N-1}$ case, the 2-cycles $v_i$ constructed from fibrations over $t_i-t_{i+1}$ can be identified with the simple roots of the gauge group~\cite{Grimm:2011tb, katz1992gorenstein}. The novelty in the case of the $D_N$ singularity is that an additional, homologically independent simple root of the form $v_n = t_{N-1} + t_N$ appears. Clearly, all positive roots corresponding to fibrations over $t_i \pm t_{i+1}$ with $i<j$ can then be written as positive combinations of the $v_I$. The $D_N$ singularity can be deformed into an $A_{N-1}$ singularity by taking all $t_I$ to be equal, $t_I \equiv t$, which yields an $A_{N-1}$ singularity located at $Z=t^2$. The explicit resolution of the of the singularity performed in~\cite{katz1992gorenstein} shows that the $A_{N-1}$ contains the vanishing cycles $v_i$ just as in section~\eqref{sec:geom_cartans}. However, due to the underlying $D_N$ structure we have an additional non-trivial cycle $v_n$, which collapses as $t\rightarrow 0$. The fact that this cycle is linearly independent from the $v_i$ means that in the dual resolution picture the resolution 2-cycles no longer obey a relation of the form~\eqref{triviality_2-cycles}~\cite{Grimm:2011tb}. In particular, the 2-form $\tw_0$ constructed by the analogue of~\eqref{expansion_tw_0} can be harmonic, matching the fact that the diagonal $U(1)$ is should remain massless in this setting.

\subsection{Massless $U(1)$s and rational sections}
\label{sec:sections_and_U(1)s}

In the previous sections we have discussed how the appearance of massive or massless $U(1)$ gauge factors beyond the non-Abelian gauge groups appearing in Tate's classification of table~\ref{tab:Tate_class} can be understood in F-theory from the local geometry of the resolved fibers. While this approach is very explicit and in principle allows for a direct construction of the resolution curves, it is also technically rather involved and can in general only be carried out patchwise. For model building purposes a more direct approach to detecting the presence of $U(1)$ symmetries and computing the charges of the matter fields with respect to these $U(1)$s from the geometry of the compactification manifold would thus be desirable. Recently, significant progress has been made in this direction using techniques based on the study of the group of sections of the elliptic fibration~\cite{Morrison:2012ei, Mayrhofer:2012zy, Braun:2013yti, Borchmann:2013jwa, Cvetic:2013nia, Braun:2013nqa, Cvetic:2013uta}, whose relation to $U(1)$ symmetries in the low energy effective action had already been noted in~\cite{Morrison:1996pp}. In the following, we briefly review the concepts used in this approach. Our presentation in this section will be based mainly on refs.~\cite{Mayrhofer:2012zy, Braun:2013yti, Cvetic:2013nia}. Let us emphasise from the beginning that studying sections of the elliptic fibration directly yields information only on massless $U(1)$ symmetries. In particular, this approach does not help in understanding the geometric origin of the non-closed 2-forms describing massive $U(1)$ gauge symmetries, for which purpose we have to use the local resolution picture as in section~\ref{sec:geom_diag_U(1)}.

As was already mentioned in section~\eqref{sec:elliptic_CalabiYaus}, a section $s$ of an elliptically fibered fourfold  $Y_4$ can be viewed as an embedding of the base $B_3$ into  $Y_4$ which is compatible with the fibration structure in the sense that $\pi\circ s = \mathrm{id}_{B_3}$. If the elliptic fibration is holomorphic, which is assumed to be the case throughout this thesis, then such a globally defined section is also automatically holomorphic~\cite{Braun:2013yti}. As before, we focus on elliptic fibrations in the Weierstrass form, which are given by an equation
\be
P_W = y^2 - x^3 - f x z^4 - g z^6 = 0
\label{Weierstrass_form_2}
\ee
in the weighted projective space $\bP_{2,3,1}$ with homogeneous coordinates $x,y,z$. The geometry of the fibration is specified by the parameters $f$ and $g$, which are holomorphic sections of the bundles $K_B^{-4}$ and $K_B^{-6}$, respectively. A section can essentially be viewed as a map 
\be
s': \, u\in B_3 \mapsto [A(u),B(u),C(u)]
\label{section_explicit}
\ee 
from $B_3$ into $\bP_{2,3,1}$, in which $[A(u),B(u),C(u)]$ lies on the surface given by the Weierstrass equation~\eqref{Weierstrass_form_2} for every point $u$ in the base. Clearly, this generally requires $A, B$ and $C$ to depend on $f$ and $g$. However, the Weierstrass form always admits the holomorphic section $[1,1,0]$, which is also referred to as the zero section. Depending on the precise form of $f$ and $g$, it may or may not be possible to find additional holomorphic sections. This construction can be immediately generalised to the case where $A, B$ and $C$ are rational functions on $B_3$ instead of being holomorphic. In this case one speaks of rational sections~\cite{Braun:2013yti}. A rational section can fail to be well-defined on a codimension 2 locus in the base where two of the three functions $A, B$ and $C$ simultaneously develop poles~\cite{Braun:2013yti, Cvetic:2013nia}. After resolving the singularities of $Y_4$ and moving to the resolved fourfold $\hat Y_4$, a rational section $s'$ describes a birational morphism from $B_3$ to a surface $S'\subset \hat{Y}_4$. Over the points where $s'$ fails to be well-defined, $S'$ wraps an entire component of the resolved elliptic fiber~\cite{Braun:2013yti}. 
In the following, we will not distinguish between holomorphic and rational sections and will instead simply refer to both types of objects as sections.

Algebraically, the Weierstrass polynomial in~\eqref{Weierstrass_form_2} can be viewed as a polynomial with coefficients $f$ and $g$ in the function field on $B_3$~\cite{Morrison:2012ei, Cvetic:2013nia}. A point on an elliptic curve whose homogeneous coordinates $[x_0,y_0,z_0]$ lie in the field over which the curve is defined is known as a rational point~\cite{Cvetic:2013nia}. Clearly, for each point in the base in which it is well-defined, a section $[A(u),B(u),C(u)]$ yields such a rational point. As was already mentioned in section~\ref{sec:elliptic_CalabiYaus}, it is possible to canonically define an addition operation on the set of rational points of an elliptic curve. This addition gives this set the structure of an Abelian group, which is known as the Mordell-Weil group of the elliptic curve. When applied to an elliptic fibration rather than a single elliptic curve, the group structure extends directly to the rational sections and one speaks of the Mordell-Weil group of the elliptic fibration. A crucial mathematical result is that this group is finitely generated~\cite{Morrison:2012ei, Cvetic:2013nia}. In other words, there can be only finitely many linearly independent rational sections.

Let us now turn to the relation between sections of the fibration and $U(1)$ gauge symmetries. The key point is that each section $s'$ yields an associated divisor $S'$ in the resolved fourfold $\hat Y_4$. By the Shioda-Tate-Wazir theorem~\cite{wazir2004arithmetic}, these divisors are linearly independent from the other divisors that we have discussed previously, namely the vertical and Cartan divisors and the divisor associated to the zero section. In other words, each independent section $s'$ leads to an additional harmonic 2-form $\tw_{s'}$ which appears in the Kaluza-Klein reduction~\eqref{expansion_C3} of $C_3$ and yields an additional gauge field $A^{s'}$. 

The analysis of~\cite{Mayrhofer:2012zy} shows that the new divisors corresponding to additional sections can be seen as originating from the resolution of certain singularities\footnote{In this sense they are broadly similar to the Cartan divisors, which were discussed in section~\ref{sec:geom_cartans} and arise from singularities at codimension 1 in the base. However, the explicit resolution processes for singularities at different codimensions in the base differ somewhat at the technically level~\cite{Mayrhofer:2012zy, Braun:2011zm}.} in $Y_4$ which arise at codimension 2 in the base. To see this explicitly, it is convenient to work with the Tate form\footnote{For definiteness let us consider $SU(5)$ GUT models, in which it is possible to rewrite the fibration in Tate form at least locally around the GUT divisor~\cite{Mayrhofer:2012zy}. Although globally certain obstructions can in principle arise~\cite{Katz:2011qp}, this subtlety will not be important for the following and we assume for simplicity that the Tate form is globally defined.}
\be
P_T = x^3 - y^2 + a_1 x y z + a_2 x^2 z^2 + a_3 y z^3 + a_4 x z^4 + a_6 z^6 = 0,
\label{P_Tate_2}
\ee
rather than the Weierstrass form~\eqref{Weierstrass_form_2}. Let us follow~\cite{Mayrhofer:2012zy} and focus on sections $[A(u),B(u),C(u)]$ which fulfill $A^3 = B^2$. In order for such a section to lie inside the fourfold $Y_4$, the remaining part of the Tate polynomial~\eqref{P_Tate_2} must factorise at $x=A,$ $y=B$ as\footnote{As we are working in a projective setting the overall scale is irrelevant, so $x=A,$ $y=B$ is actually equivalent to the single condition $x^3 = y^2$.}
\be
P_T|_{x^3=y^2} = a_1 A B z + a_2 A^2 z^2 + a_3 B z^3 + a_4 A z^4 + a_6 z^6 = z(z-C)R[z;u].
\label{factorisation_at_x3y2}
\ee
The remainder $R[z;u]$ is some polynomial in $z$ and also depends on the coordinates $u$ of the base. The overall factor of $z$ describes the zero section, while the factor $z-C$ is present as we assume $[A(u),B(u),C(u)]$ to be a section. This implies that the overall Tate polynomial $P_T$ can be written as
\be
P_T = (x^3-y^2)Q + z(z-C)R[z;u],
\label{factorised_tate_section}
\ee
with a suitable function $Q$~\cite{Mayrhofer:2012zy}. An equation of this type is said to be in binomial form.

The significance of the fact that the Tate equation factorises into the binomial form~\eqref{factorised_tate_section} is that the resulting fourfold is manifestly singular at the locus given by the intersection of the 4 equations~\cite{Mayrhofer:2012zy, Braun:2011zm}
\be
x^3-y^2 = 0 \cap z-C =0 \cap Q=0 \cap R =0
\label{singular_locus_section}
\ee
in the ambient 5-fold.
The first two equations together with the rescaling relation fix the coordinates of the $\bP_{2,3,1}$ fiber, so that~\eqref{singular_locus_section} describes a curve of singularities at codimension 2 in the base. This singularity can be eliminated by means of a small resolution as in section~\ref{sec:geom_cartans}. In this process, two additional homogeneous coordinates $\lambda_1,\ \lambda_2$ of a $\bP^1$ are introduced and the resolved space is given by the intersection of the equations~\cite{Mayrhofer:2012zy, Braun:2011zm, Esole:2011sm}
\be
(x^3-y^2)\lambda_1 = R \lambda_2 \ \cap \ Q \lambda_2 = (z-C)\lambda_1
\ee
in the extended ambient space.
Note that over the original singular locus~\eqref{singular_locus_section} the additional coordinates $\lambda_i$ are unconstrained, so that the small resolution acts by pasting in an additional $\bP^1$ over the singularity. After resolution, the additional 6-cycle $S'$ associated to the section is given by~\cite{Mayrhofer:2012zy}
\be
\lambda_1 = (x^3-y^2) = z-C = 0,
\ee
away from the original singular locus~\eqref{singular_locus_section}, while it wraps the entire resolution $\bP^1$ over the erstwhile singularity.

From the discussion above it is clear that the appearance and form of additional sections is determined by the factorisation properties of the Tate polynomial. Focusing on sections of the form $x^3=y^2$, the number of $U(1)$'s in the low energy theory is equal to the number of different ways in which we can achieve a factorisation of the remainder of the Tate polynomial as in~\eqref{factorisation_at_x3y2}. The different inequivalent types of sections that can be obtained can be related to the different ways in which $P_T|_{x^3=y^2}$ can be factorised holomorphically~\cite{Mayrhofer:2012zy, Braun:2013yti}. Let us focus in particular on factorisations into at most two pieces. After extracting the overall factor of $z$ corresponding to the zero section, $P_T|_{x^3=y^2}$ describes a polynomial of order 5 in $z$. Different types of factorisations can then be labelled by the degrees in $z$ of the individual factors. At the level of two factors the possibilities are (5,0), (4,1) and (3,2). Let us emphasise that we explicitly allow for rational sections, so that in particular $C(u)$ in~\eqref{section_explicit} is allowed to be a rational function. This means that a rational section can be present even in the case labelled above as the (5,0) split, in which it is not possible to factorise $P_T|_{x^3=y^2}$ holomorphically in a non-trivial manner~\cite{Braun:2013yti}. 

At this point, let us briefly discuss the relationship between the explicit description of a section as in~\eqref{section_explicit} and the description used in~\cite{Mayrhofer:2012zy}. The authors of~\cite{Mayrhofer:2012zy} considered factorisations of the Tate polynomial of the form
\be
P_T = (x^3-y^2)Q + z \prod_i^n Y_i,
\ee
with suitable holomorphic factors $Y_i$. A section was then defined implicitly by the intersection of two equations $(x^3=y^2) \cap (Y_i = 0)$. In the case of a (4,1) split in a factorisation into two factors, it is clear that the factor linear in $z$ can be identified with the factor $z-C$ in~\eqref{factorisation_at_x3y2}. In the case of a (3,2) split,~\cite{Mayrhofer:2012zy} defined the section as the torus sum of the two roots of the quadratic factor. Even though these roots can vary holomorphically over the base if the corresponding equation $Y_1$ is holomorphic, the operation of adding points on an elliptic curves explicitly introduces ratios of the functions defining the individual roots~\cite{Morrison:2012ei}. Therefore this case is described in our language by a factorisation of the form~\eqref{factorisation_at_x3y2} with a rational function $C$.

The relationship between $U(1)$ symmetries and sections of the elliptic fibration discussed in this subsection is particularly interesting from a phenomenological point of view. This is because it offers an explicit approach towards geometrically engineering Abelian gauge factors in the low energy theory by tuning the parameters of the fibration in such a way that the Tate (or Weierstrass) polynomial factorises. In phenomenologically interesting models, the $U(1)$ gauge factors should of course be combined with a suitable non-Abelian gauge group. In this case, it is necessary to study the factorisation properties of the transformed Tate polynomial which is obtained after the non-Abelian singularities at codimension 1 have been resolved. The resolution geometry of fibrations with an $SU(5)$ GUT gauge group and one or more additional $U(1)$s was studied explicitly in~\cite{Mayrhofer:2012zy, Braun:2013yti, Borchmann:2013jwa, Cvetic:2013nia, Braun:2013nqa, Cvetic:2013uta}. The explicit description of the divisors corresponding to the $U(1)$ gauge symmetries makes it possible to directly calculate the intersections with the matter curves above the $SU(5)$ locus. These intersections encode the $U(1)$ charges of the various matter fields, and therefore yield the selection rules for the possible couplings in the low energy theory which result from the massless $U(1)$s.

\chapter{Fluxed $D3$- and $M5$-instantons in type IIB and F-theory}
\chaptermark{Fluxed instantons in type IIB and F-theory}
\label{sec:chap_instantons}
In this chapter we turn to a discussion of D- and M-brane instantons in type IIB orientifold compactifications and in the F-theory limit of M-theory compactifications on elliptically fibered fourfolds. We begin in section~\ref{sec:Dinstantons_IIB} with a brief review of some background on D-brane instantons and the reasons for their importance to the phenomenology of string compactifications, which will be largely based on~\cite{Blumenhagen:2009qh}. It will become apparent that there is an important interplay between D-brane instanton effects and the Abelian gauge symmetries discussed in chapter~\ref{sec:u(1)inIIBandFtheory}. To be more precise, the $U(1)$ symmetries affect the shape of the instanton contributions to the low energy action, while conversely the instantons can generate effective couplings in the low energy theory which break the $U(1)$ symmetries. In section~\ref{sec:flux_on_D3inst} we show that non-trivial gauge flux on the instanton can have important consequences for the $U(1)$ charges of the instanton, and discuss how these results affect the question of moduli stabilisation in type IIB in section~\ref{sec:modstab_fluxed_inst}. We then turn our attention to $M5$-instantons wrapped on vertical divisors in an elliptically fibered fourfold, which in the F-theory limit are related to $D3$-brane instantons in type IIB. The partition function of such $M5$-instantons, which encodes their contribution to the low energy effective action, is derived in section~\ref{sec:M5partition} and compared to the corresponding IIB results in section~\ref{sec:partition_fct_match}. In section~\ref{sec:zeroModesG4} we consider how switching on non-trivial $G_4$ flux affects the form of the $M5$ partition function. The selection rules that must be fulfilled by an $M5$-instanton in the presence of $G_4$ flux in order to generate low energy couplings of uncharged fields are discussed in a general setting in section~\ref{sec:selectionRulesGen}. Finally, these selection rules are evaluated in a specific example and compared to the known type IIB results.

\section{Euclidean $D3$-brane instantons with world-volume fluxes}
\sectionmark{Euclidean $D3$ instantons with world-volume fluxes}
\label{sec:fluxed_D3instantons}

\subsection{D-brane instantons in type IIB compactifications}
\label{sec:Dinstantons_IIB}

A well-known feature of the quantum effective action of a perturbative quantum field theory is that it receives two conceptually different types of contributions. One part can be obtained by summing up contributions that essentially arise from Feynman diagrams with more and more loops, which yields a power series in the coupling constant $g$ of the theory. In addition, non-perturbative effects can give rise to terms whose dependence on $g$ is schematically given by $e^{-1/g^2}$. At weak coupling, perturbative contributions will in general of course dominate over the non-perturbative terms. However, non-perturbative contributions can still be extremely important for some couplings if the structure of the theory is such that only a finite number of perturbative contributions to these couplings are non-zero. This situation in particular arises in the context of 4-dimensional theories with $\cN=1$ supersymmetry, in which the superpotential and gauge kinetic functions receive perturbative corrections only up to 1-loop level~\cite{Grisaru:1979wc, Seiberg:1993vc}. The low energy theories obtained by compactifying string theory to 4 dimensions on suitable compactification manifolds fall precisely into this class. It is thus natural to expect that non-perturbative effects can play an important role in string theory compactifications, and this subject has received widespread interest in the past two decades. Our discussion will be based on the reviews~\cite{Blumenhagen:2009qh, Akerblom:2007nh}, to which we refer the reader for references to much of the original literature.

One way to understand non-perturbative corrections to the effective action in field theory is via the path integral formula for the computation of correlation functions, which schematically takes the form\footnote{In order to obtain a convergent integral it is often helpful to carry out a Wick rotation such that the exponential factor becomes $\exp(-S)$.}
\be
\left< \cO(\phi) \right> = \frac{1}{Z} \int \cD\phi \, \cO(\phi) \exp\left(iS[\phi]\right).
\label{correlator_path_integral}
\ee
Here the various fields of the theory are collectively denoted by $\phi$, while $\cO(\phi)$ is the operator whose expectation value is computed and $Z$ is the path integral without insertion of any operators. The key point is that the path integral will be dominated by the region in field space which is close to field configurations that describe stationary points of the action $S$~\cite{belavin1975pseudoparticle, t1976computation}. The standard perturbative expansion corresponds to an expansion around one particular such stationary point, namely the vacuum of the theory. However, additional contributions can arise if there exist stationary points of $S$ which correspond to field configurations that are topologically non-trivial and cannot be continuously deformed into the vacuum. The original examples of such solutions to the equations of motion were constructed in a gauge theory framework and correspond to configurations of the field strength $F$ which are localised in space and time~\cite{belavin1975pseudoparticle, t1976computation}. Due to this localisation property, such configurations are known as instantons. The contribution to the correlation function due to such an instanton can be evaluated using a saddle point approximation, in which the fields are expanded around the instanton configuration and the path integral is approximated by a Gaussian integral.

As already mentioned in section~\ref{sec:action_IIB_bulk}, the calculation of string theory amplitudes on non-trivial background spacetimes actually involves two separate perturbative expansions. This of course also means that two different types of non-perturbative effects may appear. The first perturbative expansion arises in the quantum field theory on the string worldsheet, which is an interacting theory known as a non-linear sigma model for strings propagating in non-trivial backgrounds~\cite{Polchinski:1998, Green:87, Becker:2007zj, Blumenhagen:2006ci}. The interaction terms appear when the background fields appearing in the worldsheet action are expanded into Taylor series about the rest position of the string. The terms of this Taylor series are suppressed by powers of $\sqrt{\alpha'}/R$, with $R$ the typical scale over which the background fields vary. Higher order terms of this type on the worldsheet lead to corrections to the string theory effective action which are known as $\alpha'$ corrections. In analogy with the discussion around~\eqref{correlator_path_integral}, we expect corrections that are non-perturbative in $\alpha'$ to arise due to configurations of the string worldsheet that cannot be continuously shrunken to a point. Such configurations can arise in string theory compactifications if the worldsheet is wrapped around a non-trivial 2-cycle of the compactification manifold. As such configurations appear pointlike from the perspective of the non-compact external spacetime they are known as worldsheet instantons. The corrections to the low energy action due to worldsheet instantons were initially studied in heterotic string theory in~\cite{Wen:1985jz, Dine:1986zy, Dine:1987bq}, and applied to the type II setting e.g. in~\cite{Ooguri:1999bv, Kachru:2000ih, Aganagic:2000gs, Aganagic:2001nx, Lerche:2001cw}.

The second perturbative expansion in string theory is an expansion in terms of the string coupling $g_s$. It arises from the fact that the contribution to a string theory amplitude from a worldsheet of genus $g$ includes a factor\footnote{This formula is valid for the case of oriented closed strings, for open or unoriented strings also the number of boundaries and crosscaps of the worldsheet enter in addition to the genus.} of $g_s^{2g-2}$. Polchinski showed in~\cite{Polchinski:1995mt} that the non-perturbative corrections in $g_s$ can be described in terms of the theory of open strings ending on D-branes. As was extensively motivated in the previous chapters, spacetime-filling D-branes are an essential ingredient in the construction of phenomenologically interesting models from the compactification of type II string theory or F-theory. In addition, it is possible to include D-branes wrapped fully on non-trivial cycles of the compactification manifold~\cite{Witten:1995im, Witten:1995gx, Douglas:1995bn}. These so-called D-instantons are pointlike from the perspective of 4-dimensional spacetime and thus do not break Lorentz invariance. The picture that emerges in the context of string theory compactifications with D-branes is very similar to the field theory picture discussed around~\eqref{correlator_path_integral}. Namely, the chosen compactification manifold and configuration of spacetime-filling branes is viewed as the vacuum of the theory, and corrections arise from the different ways in which D-instantons on non-trivial cycles of the internal manifold can be added to this configuration~\cite{Blumenhagen:2009qh}.

Following their discovery, D-instanton effects have been extensively studied in the type II setting. In particular, it was soon realised that D-instantons can give a microscopic description for certain gauge instantons in the low energy theory, as discussed e.g. in~\cite{Billo:2002hm}. Beyond this, D-brane instantons can give rise to effects which do not scale with the coupling of a visible gauge group, as would be the case for a classical gauge instanton. For this reason, D-instantons have also been referred to as stringy or exotic instantons~\cite{Florea:2006si, Argurio:2007vqa}. For example, D-instanton effects can give rise to supersymmetry breaking effects whose scale is naturally small due to the non-perturbative nature of the instanton contribution~\cite{Blumenhagen:2009qh, Aharony:2007db} and is not directly related to the size of the gauge coupling.
The fact that D-brane instantons can lead to non-trivial effective F- or D-term potentials was used in~\cite{Kachru:2003aw} to show that they can play a crucial role for moduli stabilisation. This can be seen particularly easily in type IIB compactifications, where the specific form of the K\"ahler potential guarantees that some of the K\"ahler moduli do not appear in the perturbative scalar potential. This means that D-instantons, despite being non-perturbative in $g_s$, actually form the leading\footnote{Here we mean leading in the sense that there are no dominant perturbative contributions. Different non-perturbative contributions, e.g. from gaugino condensation on $D7$-branes, can in principle contribute at a similar scale~\cite{Kachru:2003aw}.} terms which depend on bespoke moduli and are crucial in ensuring that they are stabilised as required. 

A further crucial property of D-instantons is that their action contains fields which shift non-trivially under the $U(1)$ symmetries of the model. Such charged instantons appear in the low energy effective action together with fields of the opposite $U(1)$ charge, as required in order to guarantee gauge invariance of the total action. When some or all of the moduli fields in the instanton action gain a mass and are fixed around their vacuum expectation value, this leads to effective interactions in the low energy theory which are not invariant under the $U(1)$s~\cite{KashaniPoor:2005si, Ibanez:2006da, Florea:2006si, Blumenhagen:2006xt}. In other words, charged instantons can effectively break the low energy $U(1)$ symmetries. Their contribution is thus particularly important from a model builder's perspective if some phenomenologically required couplings are forbidden at the perturbative level by $U(1)$ symmetries, as was the case for example for the top Yukawa coupling in type IIB $SU(5)$ GUT models discussed in section~\ref{sec:FtheoryGUTs}. The generation of certain Yukawa couplings including $\mu$-terms and neutrino masses via D-brane instanton effects has been extensively studied and is reviewed e.g. in~\cite{Blumenhagen:2009qh, Akerblom:2007nh, Cvetic:2007sj}. 

To determine the contribution of a given instanton to the effective action, it is in principle necessary to directly evaluate correlators of fields in the instanton background by summing over the relevant string theory scattering amplitudes~\cite{Blumenhagen:2009qh}. The result can then be compared to the known field theory expressions of these correlators, which makes it possible to extract the effective superpotential and K\"ahler potential. In~\cite{Blumenhagen:2006xt} it was shown that the instanton contribution to the 4-dimensional effective action can be written in a form which is reminiscent of the field theory expression~\eqref{correlator_path_integral}, namely\footnote{Here and in the following, we use an index $E$ for quantities associated with a D-brane instanton, as in~\cite{Grimm:2011dj}. This notation stems from the fact that D-instantons are Euclidean D-branes.}
\be
S^{(4d)}_E \sim \int \cD \cM_0 \exp\left[-\hat{S}_E(\cM_0)\right].
\label{instanton_contrib}
\ee
Here $\cM_0$ stands for the massless fields on the instanton world-volume, which are known as zero modes. Whereas in the field theory expression~\eqref{correlator_path_integral} the path integral runs over all fields including the massive parts~\cite{t1976computation}, the contribution due to the massive string modes is already included in~\eqref{instanton_contrib} in the effective instanton action $\hat{S}_E$. In other words, $\hat{S}_E$ can be viewed as consisting of the usual classical action $S_E^{cl}$ of the instanton plus an additional term $S_E^{int}$, which describes interaction terms of the zero modes and encodes the full contribution of the massive modes to the string theory correlator~\cite{Blumenhagen:2009qh}. For D-brane instantons, the bosonic part of the classical action $S_E^{cl}$ is simply the sum of the Dirac-Born-Infeld and Chern-Simons actions of section~\ref{sec:Dbranes}. On the other hand, determining the interaction terms in $S_E^{int}$ is somewhat less straight forward, and requires explicitly evaluating the relevant string theory amplitudes in the chosen instanton background~\cite{Blumenhagen:2006xt}.

The zero modes $\cM_0$ have been extracted explicitly in~\eqref{instanton_contrib} because this makes it easy to see that in particular the fermionic zero modes play a crucial role for the instanton contribution. This fact is familiar already from the field theoretic setting~\cite{t1976computation}, and can be seen as a consequence of the integration over Grassmann-valued fermionic fields in~\eqref{instanton_contrib}. Recall that an integral over a Grassmannian variable vanishes unless the integrand is linear in the integration variable. This means that if there is a fermionic zero mode which does not appear in $S_E^{int}$, the entire instanton contribution~\eqref{instanton_contrib} vanishes~\cite{Blumenhagen:2009qh}. 
To illustrate schematically what happens when fermionic zero modes do appear in $S_E^{int}$ let us consider a representative pair\footnote{As $S_E^{int}$ is an ordinary number, the number of Grassmannian variables in each individual term in $S_E^{int}$ must be even.} $\lambda_i , \ i=1,2,$ of fermionic zero modes. These zero modes may appear in the interaction Lagrangian in terms of the schematic form
\be
S_E^{int} \supset \int \lambda_1 \Phi_{12} \lambda_2,
\label{inst_interaction_schematic}
\ee
with $\Phi_{12}$ a bosonic operator. Now the path integrals over $\lambda_i$ in~\eqref{instanton_contrib} lead to
\be
\int \cD \lambda_1\cD\lambda_2  \exp\left[-\hat{S}_E\right] = \Phi_{12} \exp\left[-S^{cl}_E - \tilde{S}_E^{int}\right],
\label{superpot_soaked_up_zeros}
\ee
with $\tilde{S}_E^{int}$ denoting the part of $S_E^{int}$ not involving the $\lambda_i$. In this situation, one says that the fermionic zero modes $\lambda_i$ have been soaked up or saturated by the interaction~\eqref{inst_interaction_schematic}.

From the discussion above, the calculation of the contribution of a D-instanton to the effective action can be roughly split into three distinct steps. First, one specifies the actual instanton configuration. This includes identifying the cycle it wraps in the compactification manifold, its position relative to other D-branes in the model and any world-volume gauge flux that may be present on the instanton. In a second step one can then identify the different types of zero modes that appear on the instanton. Finally, the string scattering amplitudes in the given background must be evaluated to explicitly determine the couplings~\eqref{inst_interaction_schematic}. For later use, let us simply note at this point that instanton corrections to the superpotential involve string theory amplitudes only up to one-loop order~\cite{Blumenhagen:2009qh}. Nevertheless, the final step is generally the most involved from a computational perspective. Once the types of zero modes have been identified it is however often possible to see directly which type of string worldsheets can contribute to which types of couplings between the zero modes~\cite{Blumenhagen:2009qh}. In particular, it can often be determined whether or not all fermionic zero modes can in principle be soaked up via couplings of the form~\eqref{inst_interaction_schematic} without having to explicitly evaluate the couplings. For many phenomenological purposes it is sufficient to know that a term of the form~\eqref{superpot_soaked_up_zeros} is induced, without requiring the exact form of the prefactor $\Phi_{12}$. Therefore it is often only checked whether an interaction as in~\eqref{inst_interaction_schematic} is generically allowed, and then the interaction is assumed to be present without working out the explicit scattering amplitudes. Throughout this thesis, we will follow this approach. However, it is important to remember the caveat that the interaction term~\eqref{inst_interaction_schematic}, and hence the entire instanton contribution, can in principle still vanish if an unfortunate cancellation between the contributing scattering amplitudes takes place.

Let us now turn to the types of zero modes that can appear, following~\cite{Blumenhagen:2009qh, Blumenhagen:2010ja}. For definiteness we focus on the case of a Euclidean $D3$, or $E3$, instanton wrapped on a non-trivial 4-cycle $D_E$ in a type IIB orientifold. Regardless of the geometry of $D_E$, one always obtains a set of zero modes that can be identified as the Goldstinos due to the breaking of super-Poincar\'e invariance by the introduction of the instanton. These zero modes are known as the universal zero modes. On the bosonic side, they are simply given by the coordinates $x^\mu$ of the instanton in 4-dimensional space. The number of fermionic zero modes depends on the number of supersymmetries that are broken by the instanton. If the instanton is placed on an arbitrary cycle, it will generically break supersymmetry completely, and one obtains 8 fermionic zero modes corresponding to the 8 broken SUSY generators~\cite{Blumenhagen:2009qh}. However, just like a spacetime-filling D-brane a D-instanton will leave an $\cN=1$ supersymmetry unbroken if it fulfills the calibration conditions~\eqref{calibration_vol} and~\eqref{calibration_F}. In this case one speaks of a 1/2-BPS instanton, and generically obtains 4 universal fermionic zero modes. 

Due to its localisation in spacetime an instanton actually breaks a different combination of supersymmetry generators than a spacetime-filling $D$-brane wrapped on the same cycle~\cite{Blumenhagen:2009qh}. To be precise, two of the 4 universal fermionic zero modes of a 1/2-BPS instanton, conventionally denoted by a spinor $\theta^\alpha$, can be related to the generators of the 4-dimensional $\cN=1$ supersymmetry algebra which remains unbroken after the introduction of the spacetime-filling D-branes. The path integral factor $\int d^2 \theta$ corresponding to these zero modes can therefore be identified with a part of the usual superspace measure of the 4-dimensional $\cN=1$ effective action~\cite{Blumenhagen:2009qh}. The other two universal fermionic zero modes of a  1/2-BPS instanton are usually denoted by $\bar{\tau}^{\dot{\alpha}}$. Depending on the geometry of the instanton cycle $D_E$, some of the universal fermionic zero modes can be removed by the orientifold projection. For a single instanton one distinguishes two distinct cases, known as $U(1)$ and $O(1)$ instantons\footnote{This can also be generalised to stacks of multiple instantons. In this case also the possibility of symplectic Chan-Paton groups arises and the different cases are referred to as $U(N)$, $O(N)$ and $Sp(N)$ instantons~\cite{Blumenhagen:2009qh}. However, in this thesis we will not consider multi-instanton effects and will stick to the case $N=1$.}. A $U(1)$ instanton is characterised by a cycle $D_E$ which is not invariant under the orientifold involution $\sigma$. Of course, one must strictly speaking consider instanton-image-instanton pairs in this case in order to obtain a well-defined orientifold model. None of the universal zero modes are projected out in the case of a $U(1)$ instanton, which in particular means that the $\bar{\tau}$ modes must be soaked up by a suitable interaction in order for the instanton to contribute to the low energy action. This difficulty is avoided for $O(1)$ instantons, for which the $\bar{\tau}$ modes are projected out by the orientifold projection. An instanton is of type $O(1)$ if the cycle $D_E$ is invariant under the orientifold action in the sense that $\sigma(D_E) = D_E$. In the following, we will mostly focus on $O(1)$ instantons as they are generically more likely to be relevant for the low energy phenomenology than $U(1)$ instantons and in addition admit a clearer F-theory uplift.

The fact that a 1/2-BPS instanton breaks only two of the 4-dimensional supersymmetry generators $\theta^\alpha, \, \bar{\theta}^{\dot\alpha}$ means that it can only contribute to terms in the effective action which are integrated over only half of superspace. In other words, these are the holomorphic quantities of an $\cN=1$ supersymmetric theory, namely the superpotential~\cite{Blumenhagen:2009qh}
\be
S_W = \int d^4 x d^2\theta W(\Phi)
\ee
and the gauge kinetic function
\be
S_{gauge} = \int d^4x d^2\theta f(\Phi) \mathrm{tr}(W^\alpha W_\alpha).
\ee
Corrections to non-holomorphic quantities like the K\"ahler potential or D-terms are given by terms which are integrated over the full $\cN=1$ superspace and thus involve the integral measure $\int d^2\theta d^2\bar\theta$. Such corrections can therefore be induced only by non-BPS instantons, which break supersymmetry completely. Such non-BPS instantons of course also generically include 4 additional fermionic zero modes $\tau^\alpha$, $\bar{\tau}^{\dot \alpha}$. These must again be soaked up by suitable interactions or projected out by the orientifold action in order to obtain a non-vanishing correction. Further comments regarding corrections from non-BPS instantons as well as additional references are given in~\cite{Blumenhagen:2009qh}. All instantons appearing in the remainder of this thesis will be taken to be 1/2-BPS instantons, although it will be obvious that many results can also be easily carried over to the non-BPS case.

To identify additional instanton zero modes, recall that zero modes are given by the massless fields which propagate on the world-volume of the instantonic D-brane. As already briefly discussed in section~\ref{sec:Dbranes}, a set of bosonic fields and their fermionic superpartners can be associated to deformations of the brane world-volume. Such fields remain massless if the associated deformations leave the calibration conditions~\eqref{calibration_vol} and~\eqref{calibration_F} intact~\cite{Blumenhagen:2006ci}. In this case they are referred to as deformation moduli resp. modulini. Note that the universal zero modes discussed above fall exactly into this category, corresponding to deformations of the instanton in the external 4-dimensional spacetime. Additional moduli and modulini may arise from deformations of the internal cycle $D_E$. The number of bosonic and fermionic zero modes that arise in this manner is related to the Hodge numbers $h^{2,0}(D_E)$ resp. $h^{2,0}_{\pm}(D_E)$ in the case of $E3$-instantons~\cite{Blumenhagen:2010ja}. Further massless fields are given by the Wilson line moduli of the instanton world-volume gauge field and their fermionic superpartners. These fields can be viewed as arising from a Kaluza-Klein expansion of the gauge field and are counted for $E3$-instantons by $h^{1,0}(D_E)$ resp. $h^{1,0}_{\pm}(D_E)$~\cite{Blumenhagen:2010ja}. This completes the specification of the zero modes which can appear on isolated $U(1)$ or $O(1)$ instantons, i.e. instantons which do not intersect any other spacetime-filling or instantonic D-branes.

If the instanton intersects another D-brane, further massless modes appear at the intersection. Microscopically, they can be understood as arising from strings stretching from the instanton to the D-brane, which shrink to zero size and lead to massless states at the intersection locus. Note that in particular zero modes can appear at the intersection locus between two instantons, for example at the intersection between a $U(1)$ instanton and the associated image instanton~\cite{Blumenhagen:2009qh}. However, the zero modes that arise at intersections between instantons and spacetime-filling D-branes will be of more interest to us in the following. This is because such zero modes are charged with respect to the gauge group of the spacetime-filling D-brane, which is visible in four dimensions. They are therefore also known as charged zero modes, in contrast to all the previously discussed zero modes which are uncharged with respect to the 4-dimensional gauge group. 

Let us schematically denote the charged fermionic zero modes between the chosen instanton and a stack of spacetime-filling D-branes by $\lambda_{Ai}$, where $A$ labels the different D-brane stacks and $i$ the different zero modes at each intersection. The interaction terms in the effective instanton action involving $\lambda_{Ai}$ then take the schematic form~\cite{Blumenhagen:2009qh}
\be
S_E^{int} \supset \int \lambda_{Ai} \Phi_{(Ai)(Bj)} \lambda_{Bj}.
\label{inst_interaction_charged}
\ee
It can be shown that such interactions are always induced in such a way that the overall term~\eqref{inst_interaction_charged} is gauge invariant~\cite{Blumenhagen:2009qh}. As before, we focus on the case where the visible gauge group due to the spacetime-filling D-branes is $U(N_A)$. Then if $q_{Ai}$ denotes the charge of $\lambda_{Ai}$ with respect to the diagonal $U(1)$ on stack $A$, the operator $\Phi_{(Ai)(Bj)}$ carries charges $(-q_{Ai}, -q_{Bj})$ with respect to the diagonal $U(1)$s on the two brane stacks. $\Phi_{(Ai)(Bj)}$ can be seen as originating from a charged matter field localised at the intersection of the brane stacks $A$ and $B$~\cite{Blumenhagen:2009qh}. Performing the path integral over $\lambda_{Ai}$ and $\lambda_{Bj}$ and assuming that all other zero modes are also saturated by appropriate interactions one obtains a contribution to the low energy action of the form
\be
S^{(4d)} \sim \int d^4 x \; d^2\theta \prod_{(Ai,Bj)} \Phi_{(Ai)(Bj)} \exp[-S^{cl}_E].
\label{charged_interaction_superpot}
\ee

Although each matter field $\Phi_{(Ai)(Bj)}$ is charged, it is possible that the product appearing in~\eqref{charged_interaction_superpot} is neutral with respect to $U(1)_A$. This can happen if the zero modes between the instanton and the D-brane can be grouped into pairs of opposite charge. In other words, this can happen if and only if the net chirality of the zero modes at the instanton-brane intersection is zero. In the following we will be mostly interested in the intersection between an $E3$-instanton on $D_E$ and a stack of spacetime-filling $D7$-branes on $D_A$, in which case the net chirality is measured by the integral $I_{AE}$ of section~\ref{sec:Stueckelberg}. Note that if the net chirality $I_{AE}$ is non-zero, the classical instanton action $S^{cl}_E$ also transforms under gauge transformations of $U(1)_A$, so that the overall term~\eqref{charged_interaction_superpot} is gauge invariant. However, as discussed above the $U(1)$ symmetries of the low energy action are spontaneously broken when the fields in the instanton action are fixed around a certain vacuum value, so that $\exp[-S^{cl}_E]$ in~\eqref{charged_interaction_superpot} is viewed as a constant. In particular, charged instantons can thus give rise to effective couplings that would perturbatively be forbidden by the $U(1)$ symmetries of the action, as already mentioned above.

\subsection{The importance of instanton gauge flux}
\label{sec:flux_on_D3inst}

The previous chapters have shown that a type IIB compactification with D-branes but no extra structure in general leads to a low energy effective action involving a number of massless scalar fields or moduli. As these scalars often appear in the couplings of the matter and gauge fields, we will only obtain experimentally testable predictions for the visible couplings if the moduli fields are fixed to certain values in some way. This problem is known as the problem of moduli stabilisation, and in principle affects all types of string compactifications. Nevertheless, specifically in the type IIB setting significant process has been made and a certain standard approach to moduli stabilisation has emerged. As we will briefly review in the following, this approach to moduli stabilisation relies heavily on the existence of certain non-perturbative contributions due to $E3$-instantons. Throughout this chapter, we will use the notation introduced in section~\ref{sec:U(1)inIIB} for the bases of the cohomology groups on the Calabi-Yau orientifold $X_3$, the expansions of the various fields and Poincar\'e duals of the brane divisors into these bases and for the relevant intersection numbers.

The moduli in a type IIB orientifold compactification with $D7$-branes can be grouped into 3 distinct groups~\cite{Grimm:2004uq, Jockers:2004yj, Grimm:2005fa}, comprising of the complex structure moduli $z^M$, the $D7$-brane moduli and the K\"ahler moduli\footnote{In compactifications with $h^{1,1}_-\neq0$ one additionally obtains the $G^a$ of~\eqref{def_Ga}. As will become apparent below, the $G^a$ behave very similarly to the $T_\alpha$ from the point of view of moduli stabilisation, which can be traced back to the fact that they both involve axionic scalars with perturbative shift symmetries. In the general discussion of moduli stabilisation we will therefore slightly inaccurately refer to both $T_\alpha$ and $G^a$ as 'K\"ahler moduli'.} $T_\alpha$ of~\eqref{def_Talpha}. As discussed in~\cite{Jockers:2004yj, Kerstan:2011dy}, the bulk and brane moduli spaces are actually somewhat entangled due to the fact that the brane calibration conditions~\eqref{calibration_vol} and~\eqref{calibration_F} depend on the bulk fields. To avoid these difficulties, we focus on the stabilisation of the bulk moduli and assume for simplicity that the $D7$-branes are wrapped on rigid divisors without Wilson lines, so that no brane moduli arise.

The standard approach to type IIB bulk moduli stabilisation relies on two crucial ingredients, namely bulk fluxes and $E3$-instantons. As noted in~\cite{Gukov:1999ya}, non-trivial bulk fluxes of $H_3=dB_2$ or $F_3 = dC_2$ induce a tree-level superpotential of the form
\be
W_{GKV} \propto \int_{X_3} \Omega_3 \wedge (F_3 - \tau H_3).
\label{GKV_superpot}
\ee
This superpotential is known as the Gukov-Vafa-Witten superpotential and clearly depends on the complex structure moduli $z^M$ through the holomorphic $(3,0)$-form $\Omega_3$. For suitably chosen fluxes, this superpotential can stabilise the complex structure moduli and the dilaton $\tau$ completely\footnote{It also leads to a non-trivial warp factor, which is important in showing that the compactification can lead to a 4-dimensional de Sitter vacuum~\cite{Kachru:2003aw}.}~\cite{Giddings:2001yu}. However, the characteristic no-scale structure of the tree-level IIB K\"ahler potential~\cite{Grimm:2004uq, Grimm:2005fa} shows that at leading order the perturbative superpotential leaves the K\"ahler moduli unfixed~\cite{Kachru:2003aw}. This can be remedied by including one (or several) $E3$-instantons\footnote{In order to contribute to the low energy action the instantons must of course have the correct zero mode structure. This can be achieved e.g. for $O(1)$ instantons wrapped on rigid divisors with no Wilson line moduli, as discussed in the previous section.}, whose classical action $S_E^{cl}$ explicitly depends on the K\"ahler moduli. Due to their perturbative shift symmetries, the K\"ahler moduli can in fact only enter the effective action through the exponential of the classical instanton action~\cite{Blumenhagen:2009qh}. Therefore, the overall effective superpotential can be written schematically as~\cite{Kachru:2003aw, Balasubramanian:2005zx}
\be
W = W_{GKV}(z, \tau) + \sum_E A_E(z,\tau) e^{-S_E^{cl}},
\label{superpot_GKV_instantons}
\ee
where the prefactors $A_E$ encode the string one-loop contribution due to $S_E^{int}$ in~\eqref{instanton_contrib}. Note that a superpotential similar to~\eqref{superpot_GKV_instantons} can also be induced by gaugino condensation on $D7$-branes~\cite{Kachru:2003aw}. However, as noted in~\cite{Balasubramanian:2005zx} this can lead to a dependence of the prefactors $A_E$ on the K\"ahler moduli, which would complicate the evaluation of the minima of the potential. In comparison with the generation of non-perturbative effects via gaugino condensation, $E3$-instanton effects also give more flexibility as the instantons can be placed on divisors which do not correspond to spacetime-filling $D7$-branes. Therefore, we focus in the following on the case where the non-perturbative contributions to the superpotential are generated by $E3$-instantons. 

To find the vacua in the presence of the superpotential~\eqref{superpot_GKV_instantons}, the authors of~\cite{Kachru:2003aw, Balasubramanian:2005zx} use a two-step approach and assume that the Gukov-Vafa-Witten superpotential can be used to first stabilise the complex structure moduli at a high mass scale. In the second step,~\eqref{superpot_GKV_instantons} can then be viewed as a function of the K\"ahler moduli only, with $W_{GKV}$ and the $A_E$ treated as constants. The minimisation of the resulting F-term potential with respect to the K\"ahler moduli was explicitly discussed in~\cite{Kachru:2003aw, Balasubramanian:2005zx}, where it was shown that consistent minima can indeed be found. In particular, the authors of~\cite{Balasubramanian:2005zx} showed that, after taking perturbative $\alpha'$ corrections to the K\"ahler potential into account and choosing suitable instanton configurations, the K\"ahler moduli can be stabilised in such a way that the Calabi-Yau volume $\cV$ is exponentially large. This is important because it shows that it is possible to remain within the domain of validity of the supergravity approximation, as required for the self-consistency of the minimisation procedure. For completeness, let us briefly mention that the moduli stabilisation proposal presented here leads to a supersymmetric vacuum with a negative cosmological constant, which therefore describes an anti-de-Sitter spacetime. In~\cite{Kachru:2003aw} it was proposed that this can be uplifted to a phenomenologically more interesting de Sitter vacuum by introducing a number of anti-$D3$-branes. These anti-branes break supersymmetry completely and add a positive term to the potential. The fact that the bulk fluxes introduce a non-trivial warp factor is crucial in order to show that this positive contribution to the potential energy can be made small enough that the original vacuum is not destabilised~\cite{Kachru:2003aw}.

The general analysis reviewed above shows that it is in principle possible to stabilise the moduli in a phenomenologically acceptable regime using the F-term constraints obtained from the superpotential\eqref{superpot_GKV_instantons}, provided that suitable instanton contributions to the superpotential are present. However, the authors of~\cite{Blumenhagen:2007sm} noted that this assumption is more problematic than it seems at first sight in models with a phenomenologically viable chiral matter sector. The problem arises from the fact that in light of the chirality formulae given in section~\ref{sec:Stueckelberg} a chiral matter sector requires non-trivial fluxes to be present on the spacetime-filling $D7$-branes hosting the visible sector. Such fluxes lead to a gauging of the K\"ahler moduli with respect to the diagonal $U(1)$s on the $D7$-branes, as described by~\eqref{gauging2}. It was shown in~\cite{Blumenhagen:2007sm} that the classical action of the simplest $E3$-instantons which would be necessary to stabilise the K\"ahler moduli describing the volume of the $D7$-brane divisors $D_A$ would involve a combination of the K\"ahler moduli with non-trivial gauging. As discussed at the end of section~\ref{sec:Dinstantons_IIB}, this means that the prefactor of $\exp[-S_E^{cl}]$ in the effective superpotential must involve charged matter fields $\Phi_{(Ai)(Bj)}$. These matter fields must in particular be charged under the visible sector gauge group in order to cancel the charge of the $\exp[-S_E^{cl}]$ factor. In order for the visible gauge group to remain unbroken the vacuum expectation value of these charged fields must therefore vanish. However, this means that the instanton does not contribute to the effective superpotential for the K\"ahler moduli which remains after all other fields in~\eqref{superpot_GKV_instantons} have been stabilised. As a consequence, the K\"ahler moduli of the $D7$-brane cycles will not be fixed by the resulting F-term constraints~\cite{Blumenhagen:2007sm}. On the other hand, the gauging of the $T_\alpha$ leads to non-trivial D-term constraints as in~\eqref{DtermIIB}. The authors of~\cite{Blumenhagen:2007sm} argued that it is still possible for all moduli to be stabilised by a suitable interplay between the F- and D-term constraints. However, this stabilisation mechanism is necessarily much more complicated than the purely F-term-based mechanism envisioned in~\cite{Kachru:2003aw, Balasubramanian:2005zx}. Furthermore, the D-terms typically drive the K\"ahler moduli to the boundary of the K\"ahler cone, which means that a certain cycle collapses to zero size and higher order $\alpha'$ corrections may become important~\cite{Blumenhagen:2007sm}.

As noted in~\cite{Grimm:2011dj}, the problems in reconciling moduli stabilisation with a chiral matter sector can be alleviated by taking into account the fact that gauge flux can be switched on not only on the $D7$-branes but also on the $E3$-instantons. As we will review below, the instanton fluxes contribute to the charge of the instanton with respect to the visible $U(1)$s. In particular, by switching on suitable instanton flux it is possible to obtain gauge-invariant configurations even for instantons involving the K\"ahler moduli associated with the $D7$-brane cycles. This means that such fluxed instantons can appear in the effective superpotential\eqref{superpot_GKV_instantons} without necessarily involving charged matter fields in the prefactors. This therefore once again opens up the possibility of fixing the moduli completely via the F-terms and avoiding the problems discussed in~\cite{Blumenhagen:2007sm}. To verify these claims we will explicitly analyse the superpotential contribution of a given instanton with non-vanishing world-volume fluxes, following the presentation of~\cite{Grimm:2011dj}.

The first step is to specify which type of instanton flux may be switched on. Note that as the gauge flux is a purely internal mode in the sense that it does not depend on the 4-dimensional spacetime, the orientifold projection acts on instanton flux in exactly the same manner as it would on gauge flux of a spacetime-filling $D7$-brane wrapping the same divisor~\cite{Grimm:2011dj}. In other words, flux on a $U(1)$ instanton-image-instanton pair wrapped on the divisors $D_E$, $D_E'$ must obey
\be
\tilde{\cF}^{E'} = -\sigma^* \tilde{\cF}^E.
\ee 
Here we are using the notation\footnote{As we focus on the case of single instantons and do not consider stacks, we may drop the subscript that was used to label the generators of the gauge group.} that was introduced in~\eqref{def_tilde_cF} for the gauge invariant combination of fluxes and the Kalb-Ramond field. For an $O(1)$ instanton on a single invariant divisor, the flux must be purely orientifold-odd such that $\tilde{\cF}^E = -\sigma^*\tilde{\cF}^E$.
Furthermore, the instanton fluxes are subject to the Freed-Witten quantisation condition
\be
 {\tilde {\cal F}}^E + \iota^* B_+ + \frac{1}{2}c_1(K_{D_E}) \in H^2(D_E, \mathbb Z),
 \label{FWanom_inst}
\ee
in full analogy with $D7$-brane fluxes. As emphasised in~\cite{Grimm:2011dj}, this constraint can be particularly restrictive for $O(1)$ instantons. The reason for this is that the fluxes are restricted to be orientifold-odd, so that a possible half-integer\footnote{Recall from section~\ref{sec:IIBreduction} that $B_+$ can essentially only take the values 0 or 1/2.} $B_+$ contribution can only be cancelled if the first Chern class of the instanton divisor has some odd periods. Conversely, if $B_+=0$ then the term $c_1(K_{D_E})$ must be even. In practice, the Freed-Witten condition can rule out a significant number of candidate instanton divisors which would otherwise be expected to contribute to the superpotential~\cite{Grimm:2011dj, Blumenhagen:2008zz, Collinucci:2008sq}.

Next let us consider how the introduction of non-trivial fluxes affects the fermionic zero mode structure. For simplicity, let us focus on instantons on rigid divisors with no deformation or Wilson line moduli. This means we must consider only the universal zero modes. As we want the instanton to yield a superpotential contribution, no $\bar{\theta}$ zero modes should appear and the instanton should remain 1/2-BPS. This means that the flux must satisfy both F- and D-term supersymmetry conditions. The F-term condition requires $\tilde{\cF}^E \in H^{1,1}(D_E)$, however this is trivially satisfied as we have assumed $D_E$ to be rigid\footnote{More generally, in the non-rigid case switching on flux can lift zero modes corresponding to deformations of the instanton divisor~\cite{Bianchi:2011qh}. Roughly speaking, only those deformations can remain unobstructed which do not change the Hodge type of the instanton flux.} and thus $H^{2,0}(D_E) = H^{0,2}(D_E)=0$~\cite{Grimm:2011dj}. The D-term condition~\eqref{calibration_vol} can be written as~\cite{Jockers:2004yj, Grimm:2011dj}
\be
\label{DSUSY1}
\int _{D_E} J \wedge \tilde {\cal F}^E  = \frac12  \int_{D_E^+} J \wedge \tilde {\cal F}^{E, +}   +   \frac12  \int_{D_E^-} J \wedge \tilde {\cal F}^{E, -} =  0.
\ee
As in section~\ref{sec:U(1)inIIB}, superscripts $\pm$ are used to indicate the orientifold-even and -odd components of the gauge flux and the instanton divisor. Equation~\eqref{DSUSY1} is the analogue of the D-term~\eqref{DtermIIB} induced by gauge flux on a spacetime-filling $D7$-brane. Note that for an $O(1)$ instanton this condition is trivially fulfilled, as in this case $D_E^- = \tilde{\cF}^{E, +} = 0$. In other words, an $O(1)$ instanton on a rigid divisor will remain 1/2-BPS for any orientifold-odd flux that is switched on, regardless of the value of the K\"ahler moduli~\cite{Grimm:2011dj}. Furthermore, as the configuration of $O(1)$ instanton with flux $\tilde{\cF}^{E, -}$ is orientifold-invariant as a whole, the orientifold action is unchanged from the fluxless case. In particular, the $\bar{\tau}$ modes will still be projected out~\cite{Grimm:2011dj}. Therefore, in the absence of deformation or Wilson line modulini an $O(1)$ instanton with orientifold-odd flux can contribute to the effective superpotential, provided that any charged fermionic zero modes that may be present can be soaked up by suitable interaction terms. The expectation that the only effect of the introduction of instanton flux on the fermionic zero mode spectrum is a possible lifting of deformation modulini was confirmed by a careful analysis of the fermionic part of the instanton action in~\cite{Bianchi:2011qh}.

Having established that fluxed instantons can give rise to terms in the effective superpotential, let us now analyse how the fluxes affect the explicit form of the instanton contribution. The fluxes appear in the classical instanton action $S_E^{cl}$, which is given (up to an overall sign originating from the different Wick rotation) by the sum of the Chern-Simons and Dirac-Born-Infeld actions for a spacetime-filling $D3$-brane given in~\eqref{S_CS} and~\eqref{S_DBI}. The dependence of the instanton action on the 4-dimensional fields can be derived in a straightforward manner using techniques similar to those used in the case of spacetime-filling $D7$-branes in~\cite{Jockers:2004yj}. More precisely, we use the calibration condition~\eqref{calibration_vol} valid for a 1/2-BPS instanton and insert the expansions~\eqref{JBC_expand} of the bulk fields and~\eqref{def_tilde_cF} of the instanton flux into the action. In the case of a $U(1)$ instanton, it is clear that as they form equivalent descriptions of the same physics we must add the actions of instanton and image instanton and divide by 2. As discussed in~\cite{Kerstan:2012cy}, it is actually necessary to include this factor of 1/2 even for $O(1)$ instantons on a single invariant divisor. This takes into account the fact that the physical instanton couplings involve intersections on the orientifold quotient, which are only half as large as the intersections computed on the threefold $X_3$. In the case where the instanton can be seen as arising during the recombination of an instanton-image-instanton pair it is also clear that we must divide the action of the $O(1)$ instanton by 2 to maintain consistency over the recombination process. As in section~\ref{sec:IIBreduction} it is helpful to introduce the even and odd combinations of the instanton cycle and its orientifold image for a $U(1)$ instanton 
\be
D_E^+ = D_E+D_E' = C^\alpha_E \, \omega_\alpha,\qquad D_E^- = D_E - D_E'= C^a_E \, \omega_a.
\ee
For an $O(1)$ instanton, we follow~\cite{Kerstan:2012cy} and choose $D_E^+ = D_E = C^\alpha_E \omega_\alpha$.
Furthermore, as in~\eqref{flux_expansion_twoforms} we distinguish between fluxes that can be written in terms of forms that arise by pullback from $X_3$ to $D_I$ and variable fluxes in the orthogonal complement
\be
\tilde{\cF}^E = \tilde{\cF}^{E, \alpha} \omega_\alpha + \tilde{\cF}^{E, a} \omega_a + \tilde{\cF}^E_{\tv}.
\ee
The integrals over the instanton cycle can then be evaluated using the relation~\eqref{dual_cohom_base} and the intersection numbers~\eqref{triple_intersections}. Putting all the components together, one obtains~\cite{Grimm:2011dj}
\be
\begin{aligned}
\label{instanton_class_action}
 S_E^{cl}  =&\  \pi \Big( C^\alpha_E ( T_{\alpha} + i  \Delta^E_{ \alpha}) 
                     +  i C^a_E  \Delta^E_{a} +    i  \Delta_{\tv}^{E}      \Big),  \\
\Delta^E_{ \alpha} =& \   \cK_{\alpha b c} \, G^b \, \tilde{ \cal F}^{E, c}  + \frac{\tau}{2} \Big( \cK_{\alpha b c}   \tilde{{ \cal F}}^{E, b} \,   \tilde{{ \cal F}}^{E, c}         
 + \cK_{\alpha \beta \gamma}  {\tilde{ \cal F}}^{E, \beta}  \,  {\tilde{ \cal F}}^{E, \gamma} \Big),   \\
\Delta^E_{ a} =&\ \cK_{a b \gamma} G^b \, {\tilde{ \cal F}}^{E, \gamma} +    \tau \, \cK_{a b \gamma} \,   { \cal F}^{E, b}  \, {\tilde{ \cal F}}^{E, \gamma}, \\
\Delta^{E}_{\tv} =&\  \tau \, \int_{D_E} \tilde{{\cal F}}^E_\tv \wedge  \tilde{{\cal F}}^E_\tv .
\end{aligned}
\ee

A crucial aspect of~\eqref{instanton_class_action} is that in the presence of instanton fluxes $\tilde{ \cal F}^{E}$ the action involves the moduli $G^a$. Under a gauge transformation of a $U(1)$ symmetry associated with a stack of $D7$-branes, both $G^a$ and $T_\alpha$ transform due to~\eqref{gauging1} and~\eqref{gauging2}. Using the charges~\eqref{Qsodd} and~\eqref{Qseven}, on finds that the overall variation of the instanton action under a gauge transformation $A^A \rightarrow A^A + d\Lambda^A$ of the diagonal $U(1)$ on brane stack $A$ takes the form
\be
e^{-S_E} \rightarrow \ e^{ -i q_A \Lambda^A} e^{-S_E}, 
\ee
where the charge is given by~\cite{Grimm:2011dj}
\be
\begin{aligned}
\label{U1-trafo}
q_A = \  - \frac{1}{2} N_A \Big( \, & \cK_{\alpha b c } \, C^{\alpha}_E  \, C^b_A \, ({\cal F}^{A, c} 
- {\cal F}^{E, c}) \\ & +  \, \cK_{\alpha \beta \gamma} C^\alpha_E \, C^\beta_A \, {\tilde{\cal F}}^{A, \gamma}    - \cK_{a b \gamma} C^a_E C^b_A {\tilde{\cal F}}^{E, \gamma} \Big).
\end{aligned}
\ee
This implies that for a suitable choice of flux the instanton can be gauge invariant even if the action of an unfluxed instanton on the same divisor $D_E$ would transform non-trivially. If the instanton flux is chosen in such a way that $q_A=0$, the superpotential contribution of the instanton therefore does not necessarily involve matter fields charged under the gauge group on the brane stack $D_A$. In particular, the contribution of the fluxed instanton to the effective superpotential for the K\"ahler moduli obtained after the other moduli have been stabilised has a chance of being non-zero, despite the fact that the vacuum expectation values of the charged matter fields must vanish\footnote{Note that even if the instanton is uncharged, vectorlike pairs of charged zero modes can still cause pairs of charged matter fields with opposite charges to appear in the superpotential contribution. Therefore the vanishing of the instanton charge in itself does not guarantee that it contributes to the effective superpotential for the K\"ahler moduli, and the absence of vectorlike pairs of charged zero modes must also be explicitly checked~\cite{Grimm:2011dj}.}. Fluxed instantons may therefore provide a way around the problems noted in~\cite{Blumenhagen:2007sm} and can make it possible to stabilise all K\"ahler moduli using F-term constraints even in models with a chiral matter sector. Let us emphasise that cancellation of the instanton charge $q_A$ depends crucially on the existence of orientifold-odd 2-forms on $X_3$, and in particular on the appearance of geometric gauging due to non-zero orientifold-odd wrapping numbers $C_A^a$ for the $D7$-brane divisors $D_A$. Therefore, the discussion of~\cite{Blumenhagen:2007sm} regarding the complications in K\"ahler moduli stabilisation in models with chiral charged matter remains valid for compactifications with $H^{1,1}_-(X_3)=0$.

From the general discussion of section~\ref{sec:Dinstantons_IIB}, we expect that instanton contributions which do not involve any matter fields charged under $U(1)_A$ can arise only if the net chirality of the instanton zero modes with respect to $U(1)_A$ vanishes. Note that zero modes with a non-zero $U(1)_A$ charge can arise at the intersection $D_E\cap D_A'$ between the instanton and the image brane stack, as well as at $D_E\cap D_A$. The net number of zero modes with $U(1)$ charges $(-1_E, 1_A)$ at the locus $D_E\cap D_A$ is given by the chiral index $I_{EA}$ defined in~\eqref{chiral_index_AB}~\cite{Blumenhagen:2008zz}. Similarly, $I_{EA'}$ counts the zero modes with charges $(-1_E, -1_A)$ localised at $D_E\cap D_A'$. Combining the two, and taking into account that the number of zero modes is multiplied by $N_A$ for a stack of $D7$-branes\footnote{This factor can be viewed as arising from a trace over the gauge indices, noting that the generator of $U(1)_A$ is simply the $N_A\times N_A$ unit matrix.}, one obtains the net $U(1)_A$ chirality of the charged zero modes~\cite{Grimm:2011dj}
\be
\begin{aligned}
\label{net_chirality_E3inst}
I_{EA} - I_{EA'} =&\ \frac{1}{2} N_A\Big( \, \cK_{\alpha b c } \, C^{\alpha}_E  \, C^b_A \, (\tilde{{\cal F}}^{A, c} - \tilde{{\cal F}}^{E, c})  \\
&  \;\; + \cK_{\alpha \beta \gamma} C^\alpha_E \, C^\beta_A \, {\tilde{\cal F}}^{A, \gamma}  - \cK_{a b \gamma} C^a_E C^b_A {\tilde{\cal F}}^{E, \gamma} \Big).
\end{aligned}
\ee
This equation is valid for $U(1)$ instantons, with $[D_E] = \frac12 C^\alpha_E \omega_\alpha + \frac12 C^a_E \omega_a$. For $O(1)$ instantons with $[D_E] = C^\alpha_E\omega_\alpha$ the zero modes at the the intersections $D_E \cap D_A$ and $D_E\cap D_A'$ should be identified, so the physical number of zero modes is actually $\frac12(I_{EA} - I_{EA'})$. The net result for an $O(1)$ instanton is still given by the expression on the right side of~\eqref{net_chirality_E3inst} with $C^a_E = \tilde{\cF}^{E,\alpha} = 0$. Note that for both $U(1)$ and $O(1)$ instantons the net chirality is precisely the negative of the $U(1)_A$ charge of the classical instanton action obtained in~\eqref{U1-trafo}. The minus sign is as expected, because it ensures that the charge of $\exp(-S_E)$ in the path integral
\be
\int \cD \{\lambda_{Ai}\} e^{-S_E}
\ee
exactly cancels the charge of the zero modes $\lambda_{Ai}$. As expected, a vanishing net charge of the classical instanton action is therefore equivalent to zero net chirality of the charged instanton zero modes. In other words, from a microscopic perspective flux which leads to an uncharged instanton actually lifts some of the charged fermionic zero modes at the instanton-brane intersection.

\subsection{Moduli stabilisation with fluxed $D3$-brane instantons}
\label{sec:modstab_fluxed_inst}

In the discussion of the preceding sections, we have largely focused on the contribution of a single arbitrary but fixed instanton configuration. From the field theory path integral picture discussed around~\eqref{correlator_path_integral}, it is clear that the overall non-perturbative D-instanton correction to the 4-dimensional effective action involves a sum over all possible instanton configurations\footnote{Of course, many of the instanton configurations that could in principle be considered are expected to involve fermionic zero modes that are not soaked up, so that their contribution to the effective action vanishes.}. The instanton sum in particular involves a sum over the number $n_I$ of instantons. However, contributions from configurations with multiple instantons are clearly suppressed ever more strongly with growing $n_I$. In the following we will therefore focus on the single-instanton sector. In the spirit of~\cite{Kachru:2003aw, Balasubramanian:2005zx}, we assume that the dilaton and complex structure moduli are stabilised by bulk fluxes at a high scale. The remaining effective superpotential for the K\"ahler moduli can then be written in the form~\cite{Grimm:2011dj}
\be
\begin{aligned} \label{Wmod}
   W =  W_0 + \sum_{E,\, \tilde{\cF}^E} A_E(\tilde{{\cal F}}^E) \, e^{- \pi C^\alpha_{E}  T_\alpha -\tilde{q}_{E a} G^a } , \\
   \tilde{q}_{E a} = i\pi\cK_{\alpha a b}  \left(C^\alpha_E \, \tilde{\cF}^{E, b} + C^b_E \, \tilde{\cF}^{E, \alpha}\right).
\end{aligned}
\ee
Here the sum over $E$ is effectively a sum over all the possible divisors $D_E\subset X_3$ on which the instanton may be wrapped. For every fixed instanton geometry we then sum over the lattice of possible instanton fluxes $\cF^E$ that are compatible with the Freed-Witten quantisation constraint~\eqref{FWanom_inst}. As in~\cite{Grimm:2011dj}, we have absorbed the part of the instanton action~\eqref{instanton_class_action} that is quadratic in the fluxes into the constants $A_E(\tilde{{\cal F}}^E)$.

The fact that charged matter fields are expected to have vanishing vacuum expectation values in order to leave the visible gauge group unbroken means that the sum over fluxes in~\eqref{Wmod} actually runs only over the sublattice of instanton fluxes for which the instanton charge of equation~\eqref{U1-trafo} vanishes. To be precise, the charge $q_A$ has to vanish simultaneously for all indices $A$ which correspond to a stack of $D7$-branes hosting a visible gauge group. Whether suitable instanton configurations which achieve this exist at all depends on the intersection numbers of $X_3$ as well as on the chosen configuration of $D7$-brane divisors and fluxes. Let us also emphasise once more that the vanishing of the instanton charge in~\eqref{U1-trafo} is only a necessary but not sufficient condition for the instanton contribution to the effective superpotential for the K\"ahler moduli to be non-vanishing. The contribution of an uncharged instanton can still vanish due to the appearance of vectorlike pairs of charged zero modes with zero net charge, or indeed due to other uncharged zero modes such as the universal $\bar{\tau}$ modes for $U(1)$ instantons or zero modes associated with deformations and Wilson lines. Nevertheless, for a generic orientifold $X_3$ with a suitable $D7$-brane configuration it is natural to expect that the sum over instanton fluxes and divisors in~\eqref{Wmod} will include some instantons with the right zero mode structure to contribute to the superpotential.

It was argued in~\cite{Grimm:2011dj} that a non-vanishing instanton contribution is only possible for divisors with $C_E^\alpha\neq 0$. In other words, the classical instanton action~\eqref{instanton_class_action} must always depend on the K\"ahler moduli $T_\alpha$. This is because an instanton independent of $T_\alpha$ could lead to an effective superpotential for the $G^a$ which survives in a decompactification limit $T_\alpha\rightarrow \infty$, in which all internal volumes are taken to infinity. However, this would be in contradiction with the fact that there is no superpotential in the decompactified 10-dimensional theory~\cite{Grimm:2011dj}. Therefore, instantons with $C_E^\alpha=0$ must automatically have some fermionic zero modes which cannot be soaked up by any interactions. As noted in~\cite{Blumenhagen:2007bn}, such behaviour is expected of $U(1)$ instanton-image-instanton pairs whose geometry does not allow them to recombine into a single invariant $O(1)$ instanton (see also~\cite{Blumenhagen:2010ja}). It is thus natural to expect that $C_E^\alpha=0$ describes a case in which instanton recombination is impossible, although we will not attempt to check this explicitly. 

Let us now consider the minima of the effective scalar potential that arise in the presence of suitable instanton contributions as in~\eqref{Wmod}. We will follow~\cite{Kachru:2003aw, Balasubramanian:2005zx, Grimm:2011dj} and consider vacua with unbroken supersymmetry. This amounts to requiring the D-terms $D_A$ of~\eqref{DtermIIB} and the F-terms $F_i \equiv (\partial_i + K_i) W,\ i = G^a, T_\alpha $ to vanish in the vacuum. Note that in the case where $h^{1,1}_-(X_3)=0$, no moduli $G^a$ appear and~\eqref{Wmod} reduces to the superpotential considered in~\cite{Kachru:2003aw, Balasubramanian:2005zx}. In this situation supersymmetric Anti-de-Sitter vacua with all moduli stabilised were shown to exist\footnote{To be precise, such vacua were shown to exist provided suitable instanton contributions are present, and as discussed above this proviso is generically not fulfilled in models with $h^{1,1}_-(X_3)=0$ and a chiral matter spectrum~\cite{Blumenhagen:2007sm}.} if $W_0<0$. In the following we will discuss how this picture changes upon allowing for $h^{1,1}_-(X_3)\neq 0$.

When considering compactifications with $h^{1,1}_-(X_3)\neq 0$,~\eqref{Wmod} gives rise to two types of F-terms corresponding to the fields $T_\alpha$ and $G^a$. Using the partial derivatives of the K\"ahler potential given in~\eqref{derivs_kaehler_pot}, one can directly evaluate the K\"ahler covariant derivatives of the superpotential to obtain~\cite{Grimm:2011dj}
\begin{align}
\label{ftermT}
F_{T_\alpha}  = &\ - \frac{v^\alpha}{2\cV} W -\pi\sum_{E,\, \tilde{\cF}^E} C^\alpha_E A_E  e^{- \pi C^\alpha_{E}  T_\alpha -\tilde{q}_{E a} G^a } , \\
F_{G^a} =  &\ - \frac{i}{2\cV}\cK_{\alpha a b} v^\alpha b^b W - \sum_{E,\, \tilde{\cF}^E} \tilde{q}_{E a}A_E  e^{- \pi C^\alpha_{E}  T_\alpha -\tilde{q}_{E a} G^a }.
\label{ftermG}
\end{align}
As in~\cite{Kachru:2003aw, Balasubramanian:2005zx} we focus on vacua in which the vacuum expectation value $\left<W\right>$ of the superpotential does not vanish. In order to discuss the moduli stabilisation problem in such vacua it is sufficient to focus on the conditions obtained from the vanishing of the F-terms. This is because the D-terms of\footnote{We consider vacua where the visible gauge group is taken to remain unbroken. This means that the expectation values of any charged matter fields that may be present are zero, such that also the additional contributions to the D-terms generated by the charged matter fields vanish in the vacuum.}~\eqref{DtermIIB} vanish automatically if the F-terms are zero~\cite{Grimm:2011dj}. To check this explicitly using the expressions~\eqref{ftermT} and~\eqref{ftermG}, it is crucial to note that the instanton sum runs only over flux configurations which cause the net instanton charge of~\eqref{U1-trafo} to vanish. This implies~\cite{Grimm:2011dj}
\be
\label{contract_charge}
C^a_A \tilde{q}_{E a} = i\pi C^\alpha_E \left(\cK_{\alpha a b} C^a_A \tilde{\cF}^{A, b} + \cK_{\alpha\beta\gamma} C^\beta_A \tilde{\cF}^{A, \gamma} \right),
\ee
which can be used to show that
\be
\left<W\right> \left<D_A\right> \propto -\left(\cK_{\alpha a b} C^a_A \tilde{\cF}^{A, b} + \cK_{\alpha\beta\gamma} C^\beta_A \tilde{\cF}^{A, \gamma} \right)\left<F_{T_\alpha}\right> - i C^a_A \left<F_{G^a}\right>.
\ee
It was noted in~\cite{Grimm:2011dj} that this result is no coincidence. In fact, it is a specific example of the general result that F- and D-term equations are always proportional in vacua with non-vanishing superpotential vev~\cite{wess1992supersymmetry, Villadoro:2005yq}. This general result can be most easily checked by performing a K\"ahler transformation $K \rightarrow K + \log |W|^2$, which is well-defined in the vicinity of the vacuum as long as $\left<W\right>\neq0$.

As each of the equations $\left<F_{T_\alpha}\right> = 0$ is linear in $v^\alpha$, it is clear that the real parts\footnote{Here we assume for simplicity that $\left<W\right>$ is real. The same discussion goes through in the general case with complex $\left<W\right>$, one must simply be slightly more careful when referring to real and imaginary parts of the equations.} of this set of equations fix the K\"ahler moduli $v^\alpha$ completely. This in turn fixes one part of the chiral fields $T_\alpha$, which according to~\eqref{def_Talpha} also involve the fields $c_\alpha$, $c^a$ and $b^a$. Let us assume for the moment that the fields $c^a$ and $b^a$ appearing in $G^a$ are fixed completely by the vanishing of $\left<F_{G^a} \right>$, which will be checked explicitly below for the simplest case involving only a single instanton. Then the vanishing of the imaginary part of $\left<F_{T_\alpha}\right>$ effectively yields a set of linear equations for the axion fields $c_\alpha$. Note that the number of axions $c_\alpha$ that is fixed in this manner is given by the rank of the matrix $C^\alpha_E$. In particular, some axions will in general remain unfixed at this level because generically we expect that not all independent divisor classes on $X_3$ will lead to the right zero mode structure to host non-vanishing instanton contributions. In the case where the $T_\alpha$ exhibit a non-trivial St\"uckelberg gauging as in~\eqref{gauging2}, some of the axions will be eaten up in the process by which the $U(1)$ symmetries become massive.
The remaining axions are expected to be fixed by subdominant contributions to the superpotential such as those associated with multi-instanton configurations~\cite{Bobkov:2010rf}. Some of the axions are thus expected to be significantly lighter than the other stabilised moduli fields, which can have interesting phenomenological consequences as discussed e.g. in~\cite{Arvanitaki:2009fg}. Here we will not discuss these axion-related effects any further and will simply assume that all axion fields are stabilised once higher order effects are taken into account, fixing the $T_\alpha$ completely.

In order for the supergravity approach used above to be self-consistent, we require all K\"ahler moduli to be fixed in the interior of the K\"ahler cone so that no non-trivial cycles of $X_3$ shrink to zero size. To analyse the relevant conditions for this, let us follow~\cite{Grimm:2011dj} and assume that the basis $\omega_\alpha$ is chosen in such a manner that the K\"ahler cone corresponds to $\left\{v^\alpha > 0\ \forall\, \alpha \right\}$. Equation~\eqref{ftermT} then clearly shows that a necessary condition for the $v^\alpha$ to lie inside the K\"ahler cone is that for every index $\alpha$ there must be an instanton contributing to the superpotential with $C^\alpha_E \neq 0$~\cite{Grimm:2011dj}. Note that this is true regardless of whether $h^{1,1}_-(X_3)= 0$ or not. The advantage of considering compactifications with $h^{1,1}_-(X_3)\neq 0$ and fluxed instanton configurations is that the number of divisors that can contribute to the superpotential~\eqref{Wmod} is enlarged, as the scope for finding uncharged instanton configurations is increased. In this sense, non-trivial instanton fluxes play an important role in making it generically possible for the $v^\alpha$ to be stabilised inside the K\"ahler cone in compactifications with orientifold-odd 2-forms~\cite{Grimm:2011dj}. 

The general ideas regarding moduli stabilisation with $E3$-instantons can be illustrated already in the simplest case in which only one instanton divisor contributes to the superpotential~\eqref{Wmod}. This case was considered in~\cite{Bobkov:2010rf, Grimm:2011dj}. Taking $\left<W\right> < 0$ as in~\cite{Kachru:2003aw, Balasubramanian:2005zx}, the discussion above and the F-term~\eqref{ftermT} shows that the $v^\alpha$ will lie inside the K\"ahler cone if and only if\footnote{Actually, $C^\alpha_E <0 \ \forall \, \alpha$ is also possible as long as it is accompanied by a suitable shift of the axions $c_\alpha$~\cite{Grimm:2011dj}.} the instanton divisor obeys $C^\alpha_E >0 \ \forall \, \alpha$. In other words, the instanton must be wrapped on an ample divisor. Of course, the assumption that an instanton configuration on an ample divisor exists which is uncharged with respect to all $D7$-brane gauge groups as required to contribute to~\eqref{Wmod} is highly non-trivial. This is especially true in the case with $h^{1,1}_-(X_3)= 0$ in which no charges can be absorbed by instanton fluxes. However, let us assume for the sake of the present argument that such configurations can indeed be found in the given model by suitable choices of instanton flux. After inserting~\eqref{ftermT} into~\eqref{ftermG}, and noting that the term involving $T_\alpha$ can be cancelled in the case where only a single instanton divisor contributes, one obtains
\be
\label{fixing_b_1inst}
i\pi \cK_{\alpha a b} C^\alpha_E b^b = \frac{\sum_{ \tilde{\cF}^E } \tilde q_{E a} A_E  e^{- \tilde q_{E c} G^c }}{\sum_{ \tilde{\cF}^E }  A_E  e^{- \tilde q_{E c} G^c }}.
\ee
This shows that the $G^a$ can be stabilised independently from the $T_\alpha$. As noted in~\cite{Grimm:2011dj}, the quantity $\cK_{\alpha a b} v^\alpha$ is proportional to the kinetic metric of the $G^a$ and must therefore be non-degenerate for all values of $v^\alpha$ in the K\"ahler cone. This guarantees that the imaginary part of~\eqref{fixing_b_1inst} is sufficient to fix all moduli $b^b$ provided the instanton divisor is ample and $C^\alpha_E >0 \ \forall \, \alpha$. The real part of~\eqref{fixing_b_1inst} yields a set of equations for the axions $c^a$, in analogy with the case of the axions $c_\alpha$ discussed above. The number of axions that are fixed in this manner is given by the number of linearly independent vectors $\tilde q_{E c}$ that appear as a consequence of the sum over instanton fluxes in~\eqref{fixing_b_1inst}. As above, it is assumed that the remaining axions are either eaten up by massive $U(1)$ symmetries due to the geometric St\"uckelberg mechanism or fixed at lower masses by subleading effects. In this case the discussion presented here shows that it is in principle possible to stabilise all moduli with $v^\alpha$ inside the K\"ahler cone with just a single instanton, generalising the results of~\cite{Bobkov:2010rf} to the case with $h^{1,1}_-(X_3)> 0$. The success of moduli stabilisation of course hinges upon the existence of a suitable instanton configuration which gives rise to the required terms in the superpotential, and as argued above we expect that it is easier to find such configurations if $h^{1,1}_-(X_3)> 0$ and one takes instanton fluxes into account.

Let us briefly mention one further aspect relating to the stabilisation of the axions $c^a$ and $c_\alpha$ that was pointed out in~\cite{Grimm:2011dj}. Namely, each axion which is unfixed will have a superpartner that is tachyonic~\cite{Conlon:2006tq}. Note that this is not incompatible with the stability of the superpartners as discussed above, because we are working in an Anti-de-Sitter setting where tachyonic fields can still be Breitenlohner-Friedmann stable. However, it is crucial that this stability is not destroyed by the uplift mechanism that is used to uplift the vacuum to a configuration with 4-dimensional Minkowski or de Sitter spacetime. This requirement can of course restrict the uplift mechanisms that may be considered.

In this section we have seen that Euclidean $D3$-brane instantons can play an important role for moduli stabilisation in type IIB orientifolds. However, only instantons with the right zero mode structure can contribute to the effective superpotential for the K\"ahler moduli. Especially in models with a chiral 4-dimensional charged matter spectrum it can be difficult to achieve complete moduli stabilisation using only unfluxed instanton configurations. Instanton fluxes can help to avoid these problems as they can lift both charged and uncharged fermionic zero modes. This makes it crucial to take into account the different flux configurations that can arise on the instanton. In other words, even after fixing the instanton divisor it is necessary to consider the full instanton partition function including the sum over all possible fluxes instead of focusing on just one fixed flux configuration. In the following, we will discuss how similar effects arise in the F-theory setting, in which the role of the type IIB $E3$-instantons is taken over by $M5$-brane instantons wrapped on vertical divisors.

\section{Fluxed $M5$-instantons in F-theory and the IIB limit}
\label{sec:fluxedM5sInFtheory}
The study of instanton effects in M/F-theory was initiated by Witten~\cite{Witten:1996bn}, who showed that $M5$-branes wrapped on 6-cycles of the compactification manifold contribute to the effective superpotential in compactifications of M-theory to three dimensions.
Just like D-instantons in the type IIB case discussed above, the $M5$-branes couple directly to axions, i.e. to scalar fields which enjoy a shift symmetry at the perturbative level. These axions were identified by Witten as being the scalars dual in three dimensions to the 3-dimensional vectors $A^\Lambda$ appearing in the dimensional reduction~\eqref{expansion_C3} of $C_3$. In order to describe the axionic coupling directly, it is convenient to use a democratic formulation of the 11-dimensional supergravity involving a 6-form field $C_6$ in addition to $C_3$. This democratic formulation is discussed in more detail in appendix~\ref{sec:app_democ_Mtheory}. The fact that $M5$-branes form magnetic sources for $C_3$ is described by an electric coupling to $C_6$, which appears in the classical world-volume action of an instantonic $M5$-brane~\cite{Kerstan:2012cy}
\be
\label{S_M5_class}
S_{M5} \supset 2\pi \int d^6 x \sqrt{\det g} + 2\pi i \int C_6.
\ee
The duality between $C_6$ and $C_3$ is described by the fact that their field strengths are related by $\ast G_7 =  G_4$, mirroring the situation in the democratic formulation of type IIB supergravity discussed in section~\ref{sec:action_IIB_bulk}. $M5$-brane instantons induce interactions in the 3-dimensional effective actions which involve the factor $e^{-S_{M5}}$ times a one-loop determinant of the world-volume fields. As in the type II setting, this one-loop determinant is independent of the K\"ahler moduli of the compactification fourfold and the axionic scalars~\cite{Witten:1996bn}.

In this thesis we are primarily interested in an M-theory compactification on an elliptically fibered fourfold $Y_4$, which leads to a 4-dimensional effective theory in the F-theory limit of vanishing fiber volume. The action of an $M5$-instanton stays finite in the F-theory limit if and only if the divisor $D_M$ wrapped by the instanton is a vertical divisor~\cite{Witten:1996bn}, and we focus on instantons on such vertical divisors in the following. In other words, $D_M$ must be related to a divisor $D_M^b$ in the base $B_3$ by $D_M = \pi^{-1}(D_M^b)$, where $\pi$ is the projection defining the elliptic fibration. If the model admits a smooth type IIB limit, the $M5$-instanton on $D_M$ is expected to be dual to a type IIB $E3$-instanton wrapped on $D_M^b$~\cite{Witten:1996bn}.

Despite the fact that it is possible to argue on general grounds that the low energy effects of $M5$-instantons are similar to those of D-brane instantons in type II string theory, the treatment of $M5$-instantons is somewhat more complicated at the technical level. The underlying reason for this is the fact that the fundamental quantum theory behind the 11-dimensional supergravity approximation to M-theory is not explicitly known. This is in contrast to the type II case, where aspects such as the instanton zero mode spectrum and interactions between the zero modes can be studied directly by quantising open strings and computing string scattering amplitudes. This means that the zero mode spectrum of $M5$-brane instantons has to be determined by more indirect methods, for example by studying the fermionic part of the world-volume action. By studying a possible world-volume anomaly, Witten was able to determine that different types of uncharged fermionic zero modes are counted by the holomorphic cohomology groups $H^{p,0}(D_M)$.
Despite the fact that the world-volume action is rather complicated due to the existence of a self-dual 2-form field which makes a purely Lagrangian description problematic~\cite{Pasti:1997gx, Bandos:1997ui}, significant progress towards understanding zero modes in the presence of background fluxes has been made by studying the fermionic action~\cite{Kallosh:2005yu, Saulina:2005ve, Kallosh:2005gs, Martucci:2005rb, Bergshoeff:2005yp, Tsimpis:2007sx}. The effects of instanton fluxes and charged zero modes have been studied using an approach based on the IIB-F-theory duality in~\cite{Blumenhagen:2010ja, Bianchi:2011qh}. By computing and matching the cohomology groups of $D_M$ and $D_M^b$ in specific examples, the authors of~\cite{Blumenhagen:2010ja} were able to show how exactly the neutral zero mode spectrum is uplifted from IIB to F-theory. Further work on instanton zero modes in M/F-theory using techniques based for example on F-theory-heterotic duality or anomaly inflow arguments can be found in~\cite{Marsano:2008py, Cvetic:2009mt, Cvetic:2009ah, Donagi:2010pd, Cvetic:2011gp, Marsano:2011nn}. 

In this section, which is largely based on reference~\cite{Kerstan:2012cy}, we aim to use the duality with $E3$-instantons on $D_M^b$ to derive information about $M5$-instantons on the vertical divisor $D_M$. Throughout, we will attempt to clarify how these results can be obtained or interpreted in a purely F-theoretic setting without making recourse to the type IIB picture. It is clear that, just as in type IIB, the full contribution to the low energy action due to $M5$-instantons will involve calculating the full instanton partition function. This partition function, which can be split into a one-loop quantum part and a classical part involving a sum over all instanton fluxes in the classical instanton action, is the object of study in subsection~\ref{sec:M5partition}. Due to the complicated form of the instanton action the evaluation of the partition function is highly non-trivial. We follow the prescription of Witten~\cite{Witten:1996hc}, who suggests that the partition function is given by one of the factors obtained from holomorphic factorisation of an auxiliary partition function derived from a simpler auxiliary instanton action. Identifying the correct factor directly using the prescription of~\cite{Witten:1996hc} is rather complicated, however we will see in subsection~\ref{sec:partition_fct_match} that a match with the type IIB partition function allows us to identify this factor in an elegant way. In fact, we will see that the prescription that emerges for picking the correct factor can be formulated directly in the M-theory language, so that it can be used to calculate the $M5$-instanton partition function without requiring explicit knowledge (or existence) of the type IIB limit. 

In the following subsections, we focus on charged chiral zero modes in the presence of non-trivial $G_4$ flux. In subsection~\ref{sec:selectionRulesGen}, we will attempt to construct certain integrals which may be used to explicitly check for the existence of chiral charged zero modes in concrete examples. These integrals can in some sense be viewed as the F-theory uplift of the chiral indices of~\eqref{net_chirality_E3inst}, which characterise the chiral charged zero mode spectrum in type IIB. Finally, we consider the resulting selection rules for the absence of chiral charged zero modes in a specific example with a known type IIB limit in subsection~\ref{sec:selectionRulesExample}. The comparison with the type IIB results will provide evidence that chiral charged zero modes can be linked to instanton charges under $U(1)$ symmetries. In particular, we will see that it is important to take into account not only massless $U(1)$s but also the geometrically massive diagonal $U(1)$ which was the focus of section~\ref{sec:U(1)inFtheory}.

\subsection{The classical $M5$-instanton partition function}
\label{sec:M5partition}
In this subsection we review the derivation of the classical partition function of an $M5$-instanton wrapped on a divisor $D_M$ of an elliptically fibered fourfold $Y_4$, using the technique of holomorphic factorisation as in~\cite{Witten:1996hc, Kerstan:2012cy}. Although we will later be interested mostly in the case where $D_M$ is a vertical divisor, such that the instanton can be related to an $E3$-instanton in type IIB, the results of this section are valid also for a more general divisor. To avoid any complications that may arise due to the appearance of charged zero modes, we assume that the instanton divisor is isolated in the sense that it does not intersect any of the $D7$-branes that may be present in the model. Although we will discuss generally how the instanton partition function splits into classical and quantum pieces, we focus on the calculation of the classical part only. We will therefore simply assume that the structure of neutral instanton zero modes is such that the quantum partition function\footnote{The quantum partition function is nothing other than the one-loop determinant of the world-volume fields which gives rise to the prefactor $A_E$ in~\eqref{superpot_GKV_instantons}.} yields a non-zero result, such that the instanton contributes to the effective superpotential.

The world-volume theory of an $M5$-brane involves a chiral 2-form $\cB$, which means that the associated 3-form field strength $\cH = d\cB$ is restricted to be imaginary self-dual~\cite{Witten:1996hc}
\be
\ast \cH = i \cH.
\label{selfduality_cH}
\ee 
Here we are focusing specifically on $M5$-brane instantons, whose 6-dimensional world-volume has a metric of Euclidean signature. In this case, our Hodge star conventions and in particular equation~\eqref{hodge_star_square} shows that the Hodge star applied to 3-forms obeys $\ast^2 = -1$, such that the eigenvalues are $\pm i$. For future use, let us note that any 3-form $Q$ on $D_M$ can be split into chiral and anti-chiral pieces by~\cite{Kerstan:2012cy}
\be
Q^{\pm} = \frac12 (1 \mp i \ast) Q \,,  \qquad Q = Q^+ + Q^-.
\ee

The self-duality constraint~\eqref{selfduality_cH} means that a purely Lagrangian description of the $M5$-brane world-volume theory involving only true physical fields is not possible. One way around this problem is to introduce additional auxiliary fields, whose equations of motion reproduce the self-duality constraint. The relevant Lagrangian was discussed in~\cite{Howe:1996yn, Howe:1997fb, Howe:1997vn, Pasti:1997gx, Bandos:1997ui, Ko:2013dka}, and takes a relatively complicated form. Therefore, we will follow~\cite{Witten:1996hc, Kerstan:2012cy} and use a different approach. Namely, we first consider a pseudo-action $S_\cB$ involving a non-chiral field, and then implement the self-duality constraint in an appropriate manner when deriving physical quantities such as the equations of motion and the partition function. In this sense, our approach is analogous to the treatment of type IIB supergravity presented in section~\ref{sec:action_IIB_bulk}, where we also used a pseudo-action supplemented by an additional constraint to account for the self-duality of $G_5$.

The part of the pseudo-action of an $M5$-brane which involves $\cB$ can be written as~\cite{Witten:1996hc, Kerstan:2012cy}
\be
\label{M5action1}
S_{\cB} = - 2 \pi \,  \int_{D_M} \big[ ({\cal H} + a \iota^* C_3) \wedge \ast  ({\cal H} + a \iota^* C_3)   + b \,  {\cal H} \wedge \iota^* C_3 \big],
\ee
where $a, b$ are suitable constants. In the following, we will often suppress the explicit pullback $\iota^*$ of $C_3$ to the $M5$-brane world-volume. The pseudo-action~\eqref{M5action1} shares a number of features with the type IIB D-brane actions discussed in section~\ref{sec:Dbranes}, with $\cB$ playing an analogous role to the D-brane world-volume gauge field. For example, it involves the combination ${\cal H} + a \iota^* C_3$, which is gauge-invariant under an 11-dimensional gauge transformation of $C_3$
\be
C_3 \rightarrow C_3 - d \chi_2 , \qquad \cB \rightarrow \cB + a \iota^* \chi_2.
\label{gauge_trf_C3}
\ee
The analogue for type IIB D-branes is the fact that the D-brane action involves the combination $F/2\pi - \iota^* B_2$ of the brane gauge field and the Kalb-Ramond 2-form. Furthermore, the second term in~\eqref{M5action1} shows that in the presence of non-trivial $\cH$-flux the $M5$-brane carries a non-zero $M2$-brane charge. The analogous effect for fluxed D-branes was discussed in the context of tadpole cancellation in section~\ref{sec:IIB_tadpoles}. 

Let us emphasise that despite these similarities, we cannot directly compare the $M5$-brane action~\eqref{M5action1} dimensionally reduced on the elliptic fiber of $Y_4$ to that of a $D3$-brane in type IIB. The underlying reason is of course that~\eqref{M5action1} does not yet take the chiral nature of $\cB$ into account. In particular, it is dangerous to directly attempt to identify specific D-instanton world-volume flux configurations $\tilde{\cF}_E$ with 3-form flux $\cH$ on an $M5$-instanton. A central result of~\cite{Kerstan:2012cy}, whose derivation we review in the following, is that an explicit correspondence between $E3$- and $M5$-instantons on vertical divisors can nevertheless be established at the level of the classical partition functions. 

The first step towards deriving the abovementioned correspondence is to actually define the partition function of the chiral 2-form $\cB$. Witten argued in~\cite{Witten:1996hc} that the desired chiral partition function can be extracted from the non-chiral partition function which can be obtained in the usual manner by evaluating the path integral over the pseudo-action~\eqref{M5action1}. More precisely, the non-chiral partition function can be split into a sum of terms which are each a product of a factor involving only the chiral part $\cH^+$ of $\cH$ and a second factor which involves only the anti-chiral piece $\cH^-$. This process is known as holomorphic factorisation~\cite{Witten:1996hc, Henningson:1999dm}. The partition function of the chiral 2-form is then identified with the chiral factor of one of these summands. A prescription for identifying the correct summand was given by Witten~\cite{Witten:1996hc}. However, this prescription relies on a rather complicated mathematical construction\footnote{Witten's approach uses the form of the 11-dimensional Chern-Simons interaction to construct a specific line bundle on the intermediate Jacobian $H^3(D_M,\mathbb C)/ H^3(D_M, \mathbb Z)$ on the $M5$-brane. This line bundle can then be used to identify the physical partition function.} and is difficult to carry out in practice. As we will see, the comparison with the known type IIB instanton partition function makes it possible to avoid this complicated construction and to identify the correct summand much more easily.

In order to facilitate the identification of the chiral partition function, it is helpful to choose the coefficients $a, b$ in~\eqref{M5action1} in such a way that only the chiral part $\cH^+$ couples to $C_3$. After holomorphic factorisation it is then clear that only the chiral factor involves $C_3$. Due to the fact that
\be
\int_{D_M} \cH^\pm \wedge C_3^\pm = \int_{D_M} \ast \cH^\pm \wedge \ast C_3^\pm = (\pm i)^2 \int_{D_M} \cH^\pm \wedge C_3^\pm ,
\ee
it is clear that requiring only $\cH^+$ to couple to $ C_3$ is equivalent to imposing that only $C^-_3$ should couple to $\cH$. Inserting the expansion $C_3 = C_3^+ + C_3^-$ into~\eqref{M5action1}, it is easy to see that this is the case if and only if $b= -2ai$~\cite{Kerstan:2012cy}. This allows us to eliminate one of the parameters in favour of the other. The value of the remaining parameter can be fixed by comparing the final result for the partition function with the known type IIB expressions. The result of this match is~\cite{Kerstan:2012cy}
\be \label{bparameter}
b=-i,    \qquad  a=1/2,
\ee
as will be explicitly confirmed in the next subsection.

Let us point out that after setting $b= -2ai$ the specific values~\eqref{bparameter} of $a$ and $b$ can be determined up to a sign also directly in the M-theory setting by demanding the gauge invariance of the overall $M5$-brane action under the gauge transformation~\eqref{gauge_trf_C3}. To demonstrate this gauge invariance, it is crucial to note that under a gauge transformation $\delta C_3 = - d\chi_2$, the dual field $C_6$ must shift according to~\cite{Kerstan:2012cy}
\be
\delta C_6 = \frac12 \chi_2 \wedge G_4.
\label{shift_C6}
\ee
This behaviour follows from the fact that the field strength $G_7$ must be gauge invariant and the specific form of $G_7$ derived in appendix~\ref{sec:app_democ_Mtheory}. It implies that the part of the instanton action involving $C_6$, which is given in~\eqref{S_M5_class}, shifts as
\be
\delta \left( 2\pi i\int_{D_M} C_6 \right) = \pi i \int_{D_M} \chi_2 \wedge G_4.
\ee
Using partial integration, it is straightforward to check that this shift is cancelled by
\be
\delta S_{\cB} =  \pi i b^2 \int_{D_M} \chi_2 \wedge G_4
\ee
provided that $b=\pm i$. In particular, the values given in~\eqref{bparameter} guarantee gauge invariance of the overall action, as required\footnote{Gauge invariance of the $M5$-instanton action is also discussed in~\cite{Grimm:2011sk, Marsano:2011nn}.}. The fact that the part $S_\cB$ itself is not gauge invariant shows that the partition function derived using only $S_\cB$ transforms as a section of a non-trivial line bundle over the configuration space of the background 3-form $C_3$. This famous result plays a crucial role in the construction used by Witten to obtain the correct chiral partition function~\cite{Witten:1996hc}. However, it will not be important in the context of this thesis because as mentioned above we will identify the correct chiral partition function by using the duality with the partition function of type IIB $E3$-instantons. Let us also mention at this point that although we have only demonstrated gauge invariance of the pseudo-action~\eqref{M5action1}, the full covariant action including auxiliary fields shares this property~\cite{Pasti:1997gx}.

While our focus in this thesis is on the partition function of the chiral 2-form $\cB$, it is important to keep in mind that the full partition function of the $M5$-brane involves also contributions from other degrees of freedom. These include the fermionic degrees of freedom that are associated with the superpartner of $\cB$, as well as the bosonic and fermionic degrees of freedom originating from the embedding of the $M5$-instanton into superspace~\cite{Kerstan:2012cy}. Much of the discussion in section~\ref{sec:Dinstantons_IIB}, which centered on D-instantons in type IIB, can be directly transferred to the case of $M5$-instantons. In particular, some of the zero modes associated with the superembedding of $M5$-instantons can be identified with the superspace measure of the low energy effective action~\cite{Witten:1996bn}. Just as  before, additional fermionic zero modes can cause the instanton contribution to the effective action to vanish if they are not soaked up by suitable interactions. A necessary condition for the instanton contribution to be non-vanishing is that the holomorphic Euler characteristic $\chi(D_M, \cO_{D_M})$ of the instanton divisor is equal to 1~\cite{Witten:1996bn}. In the type IIB language, this condition is necessary for the absence of uncharged fermionic zero modes. The type IIB analogy further suggests that non-trivial $G_4$ flux on the instanton world-volume can lead to the appearance of chiral charged zero modes. In fact, the M-theory version of the Freed-Witten anomaly cancellation condition~\cite{Freed:1999vc} implies that $[\iota^* G_4 ]\neq 0$ is only possible if there are suitable $M2$-branes ending on the $M5$-instanton which cancel the induced tadpole for $\cB$~\cite{Kerstan:2012cy}. These $M2$-branes are natural candidates to describe the expected charged zero modes. For the moment, we leave these zero mode considerations aside and focus on the evaluation of the partition function of $\cB$. In other words, we take $[\iota^* G_4]=0$ and furthermore implicitly assume that the structure of the neutral zero modes is such that a superpotential contribution is induced. The additional effects that appear upon the inclusion of non-trivial $G_4$ flux will be discussed further in subsection~\ref{sec:zeroModesG4}.

We now turn to the computation of the non-chiral gauge invariant partition function of $\cB$ as in~\cite{Kerstan:2012cy}. As discussed above, this part of the partition function can be written as
\begin{align}
W_{M5}^{tot.} = &\ \int \cD \cB \ e^{-S_{M5}}, \\
S_{M5} = &\ 2\pi \int d^6 x \sqrt{\det g} + 2\pi i \int C_6 + S_{\cB} .
\label{S_M5}
\end{align}
The superscript in $W_{M5}^{tot.}$ is included to indicate that this is the total, non-chiral partition function and contains contributions from both the chiral and anti-chiral parts of $\cH$.
Following the usual field theory logic of~\cite{t1976computation}, the path integral can be approximated by taking a sum over the saddle points of the integral and a Gaussian integral over the fluctuations around these saddle points. Without loss of generality we can choose a gauge for $C_3$ in which $C_3$ is co-closed, such that $d\ast \iota^* C_3 = 0$. For vanishing pullback of $G_4$, $C_3$ can thus be taken to be harmonic, which can be used to show that the solutions to the equations of motion of $\cB$ obtained from~\eqref{M5action1} are given by harmonic flux configurations $\cH_0$~\cite{Kerstan:2012cy}. Although a full translation of the analysis carried out by Freed and Witten in~\cite{Freed:1999vc} to M-theory is beyond the scope of this thesis, it is clear\footnote{As in~\cite{Freed:1999vc}, we expect the argumentation to be based on the fact that $M2$-branes ending on the $M5$-instanton couple to $\cB$. If we take the 2-cycle on which the $M2$-brane ends around a closed loop in the $M5$-brane world-volume, single-valuedness of the $M2$ partition function requires $\cH$ to be quantised.} that a quantisation condition should be present which restricts $\cH_0$ to discrete values. By analogy with type IIB, we expect that up to a possible half-integer shift depending on the geometry of the instanton divisor, $\cH_0$ will lie in $H^3(D_M,\bZ)$. For simplicity, we will assume in the following that $\cH_0$ is integer-quantised, and briefly comment on what is expected to change in the case where a half-integer shift is present in subsection~\ref{sec:partition_fct_match}. In analogy with the type IIB calibration condition~\eqref{calibration_F}, we also expect that F-term supersymmetry constraints may arise which limit the summation over $\cH_0$ to a sublattice of $H^3(D_M,\bZ)$. However, to simplify the notation we will not indicate this sublattice explicitly in the following, and will instead write a sum simply over the full lattice $H^3(D_M,\bZ)$. We will briefly return to this point at the end of the next subsection.

Decomposing the path integral into fluctuations around the consistently quantised, supersymmetric classical solutions such that $\cH = \cH_0 + d\delta\cB$ leads to
\be
W_{M5}^{tot.} = \sum_{\cH_0 \in H^3(D_M,\bZ)} e^{-S_{M5}[\cH_0]} \int \cD \delta \cB \ e^{-S'_{M5}[d \delta \cB, {\cal H}_0]}.
\label{partition_fct_cB}
\ee
As in~\cite{Kerstan:2012cy}, we use a prime in the second factor to signify that it involves $\cH_0$ only through the cross-term in the kinetic part of the action. Furthermore note that the path integral over $\cD \delta \cB$ should run over non-closed 2-forms only, because in the absence of non-trivial $G_4$ flux any closed components of $\delta\cB$ can be removed by a gauge transformation\footnote{At first sight it may seem that this leaves a residual gauge freedom of shifting $\cB$ by non-closed 2-forms $c_2$ obeying $d\ast d c_2 =0$. However, this condition would imply that $d c_2$ is harmonic, and thus vanishes by exactness. Therefore the only permissible gauge transformations of $\cB$ indeed consist of shifts by a \textit{closed} 2-form $c_2$.} of $\cB$~\cite{Kerstan:2012cy}.

In the following, we focus on the classical part of the total partition function including chiral and non-chiral contributions,
\be
W^{tot.,\, cl.}_{M5} = \sum_{\cH_0 \in H^3(D_M,\bZ)} e^{-S_{M5}[\cH_0]}.
\label{class_partition_fct}
\ee
For a rigid instanton in a fixed background geometry, the harmonicity of the fluxes $\cH_0$ ensures that the fluctuation-dependent piece in~\eqref{partition_fct_cB} actually decouples completely from the classical partition function~\cite{Kerstan:2012cy}. This follows from the fact that harmonicity implies
\be
d\ast_0 \cH_0 = 0,
\ee
where $\ast_0$ is the Hodge star built using the background metric. Clearly, the decoupling in general breaks down as soon as fluctuations around the background geometry are taken into account. In type IIB language this can be seen as a manifestation of the fact that the overall moduli space of a theory with D-branes cannot be easily split into parts corresponding to open and closed strings, see e.g.~\cite{Jockers:2004yj, Kerstan:2011dy, Grimm:2011dx}. Put differently, the presence of fluxes $\cH_0$ can induce an effective potential for some of the would-be moduli, in particular for possible deformation moduli of the instanton divisor and for K\"ahler and complex structure moduli of the bulk space $Y_4$. The moduli that remain unobstructed are those which leave the supersymmetry of the fluxed instanton solution unaffected, and which in particular maintain harmonicity of $\cH_0$. In section~\ref{sec:flux_on_D3inst} we saw that for certain types of fluxed $E3$-instantons, namely for $O(1)$ instantons on rigid divisors with orientifold-odd fluxes, no bulk moduli are obstructed and the sum over fluxes can be evaluated separately without having to take into account any background fluctuations. It is natural to expect that similar statements hold at least for the class of fluxed $M5$-instantons which describe the F-theory uplift of such rigid $O(1)$ $E3$-instantons. While in the general case the interplay between instanton fluxes and bulk moduli can in principle be obtained from the second factor in~\eqref{partition_fct_cB}, the explicit evaluation of this path integral is technically challenging\footnote{Note that even in the type IIB case the evaluation of the quantum partition function is highly non-trivial, and we simply absorbed the entire one-loop contribution into the possibly flux-dependent prefactors $A_E(\tilde{\cF})$ in~\eqref{Wmod}.}. Therefore we will follow~\cite{Kerstan:2012cy} and focus on the classical sum over integer quantised harmonic instanton fluxes $\cH_0$ in~\eqref{class_partition_fct}, assuming that we are in a situation where fluctuations around the background geometry can be decoupled at least in a first approximation. With this understanding, we drop the index $0$ in the following and denote the 3-form instanton flux simply by $\cH$.

The classical partition function for an action closely related to the form of~\eqref{M5action1} was computed in~\cite{Henningson:1999dm}. These results were adapted to the case of an $M5$-instanton in~\cite{Kerstan:2012cy}, whose presentation we now briefly review. The part of $S_{M5}[\cH]$ which actually depends on $\cH$ can be split off from the remainder by
\be
S_{M5} = -2\pi \int_{D_M} \cH \wedge \ast \cH - 4\pi i \int_{D_M} \cH \wedge Q + S_{M5}'.
\label{SM5_H_independent}
\ee
Here $S_{M5}'$ contains the volume of the instanton divisor as well as the couplings to $C_6$ and $C_3$ which do not depend on the instanton flux $\cH$. Clearly, $e^{-S_{M5}'}$ simply appears as an overall prefactor of the classical partition function~\eqref{class_partition_fct}. We therefore focus on the non-trivial part of the partition function first and reinstate the prefactor at a later stage. As explained above, we will later take $b=-2ai$ such that $\cH$ couples only to the anti-chiral part. This will lead to $Q=-ib C_3^-$, however for the moment let us consider a general $Q\in H^3(D_M, \bC)$.

To evaluate
\be \label{WQ1}
W(Q) = \sum_{\cH \in \ H^3(D_M, \, \mathbb{Z})}{\rm exp} \left( 2 \pi \int \cH \wedge \ast \cH +4 \pi i \int {\cal H} \wedge Q \right),
\ee
we follow~\cite{Henningson:1999dm, Kerstan:2012cy} and introduce a real symplectic basis $(E_M, F^N)$  of $H^3(D_M, \mathbb{Z})$. This basis is taken to satisfy
\begin{align}
\label{integer_basis}
 \int_{D_M} E_M \wedge F^N = &\ \delta_M^N, \\
  \int_{D_M} E_M \wedge E_N = &\ 0 =  \int_{D_M} F^M \wedge F^N,  \qquad  M, N = 1, \ldots , \frac12 b^3(D_M).
\end{align}
Any form in $H^3(D_M,\bZ)$ can be expanded into this basis with integer expansion coefficients. In particular, the sum over $\cH$ in~\eqref{WQ1} becomes a double sum over $\bZ^{\frac12 b^3(D_M)}$ after expanding $\cH$ into this basis\footnote{In the case where the sum over $\cH$ is restricted to a sublattice of $H^3(D_M,\bZ)$, we of course also need only consider a basis $E_M, F^M$ of this sublattice. Apart from the reduced range of the indices $M,N$, this has no effect on the results derived in the following.}.
As a linear operator, the Hodge star restricted to $H^3(D_M, \mathbb{Z})$ can be represented in the basis $(E_M, F^N)$ by a matrix. Following~\cite{Kerstan:2012cy}, we actually introduce two real matrices $X, Y$ of dimension $\frac12 b^3(D_M)\times \frac12 b^3(D_M)$ such that
\be
F^N = X^{MN} E_N + Y^{MN}(\ast E_N).
\ee
It is easy to check that $X$ and $Y$ are symmetric. The fact that the Hodge star operator evaluated on 3-forms in 6 Euclidean dimensions is negative definite according to equation~\eqref{hodge_star_inner_prod} furthermore implies that $Y$ is negative definite and in particular invertible. It is sometimes convenient to use a complex self-dual or anti-self-dual basis of $H^3(D_M)$, made up of forms $E^\pm_M$ which satisfy $\ast E^\pm_M = \pm i E^\pm_M$. The relation between the $E_M^\pm$ and the previous basis $(E_M, F^N)$ is given by~\cite{Kerstan:2012cy}
\be
\label{self_dual_basis}
E^+_M = - \frac{i}{2} {\I}\,Z_{MN}(F^N - \ov Z^{NP} \, E_P), \quad E^-_M = \frac{i}{2} {\I}\,Z_{MN}(F^N - Z^{NP} \, E_P).
\ee
Here we have combined $X$ and $Y$ into the complex matrix
\be
Z^{MN} = X^{MN} + i Y^{MN},
\ee
while ${\I}\,Z_{MN}$ with lowered indices is used to denote the inverse of the imaginary part of $Z$. Splitting $Q$ into self-dual and anti-self-dual components leads to
\be
Q = Q^+ + Q^-  = Q^{+, M}  E^+_M + Q^{-, M} E^-_M.
\ee

The sum in~\eqref{WQ1} was evaluated in~\cite{Henningson:1999dm} for the case where $Q$ is real and $Q^+ = (Q^-)^*$. Using a Poisson resummation, the result can be written as a sum of products of theta functions up to an overall anomalous prefactor, which is expected to cancel against the quantum part of the partition function. Leaving aside this prefactor, the result was adapted to the case of complex $Q$ in~\cite{Kerstan:2012cy}, with the result\footnote{In the case where $Q$ is real, the two factors in each summand of~\eqref{WQ} are related by complex conjugation.}
\be
\label{WQ}
W(Q) \simeq \sum_{\alpha, \beta}  \Theta [   {   \tiny  \begin{matrix} \alpha   \\  \beta    \end{matrix} } ] (- Z, - Q^+, - Q^-)  \, \,    \Theta [   {   \tiny  \begin{matrix} \alpha   \\  \beta    \end{matrix} } ] ( \ov{Z}, Q^-, Q^+) .
\ee
The sum runs over $ \frac12 h^3(D_M)$-dimensional vectors $\alpha_M, \beta^N$ with entries $0, \frac{1}{2}$. These vectors label a choice of a line bundle on the intermediate Jacobian \\ $H^3(D_M,\mathbb C)/ H^3(D_M, \mathbb Z)$. The theta functions can be viewed as sections of these line bundles~\cite{Kerstan:2012cy}. Explicitly, they are given by
\be
\begin{aligned}
&  \Theta [   {   \tiny  \begin{matrix} \alpha   \\  \beta    \end{matrix} } ] (Z, Q^+, Q^-)   =  \exp \left[ \frac{\pi}{2} Q^{+,M}  \I{Z}_{MN}  ( Q^{+,N} - Q^{-,N}) \right]  \\
& \ \ \times \sum_{k_M \in \, \mathbb Z} \exp \Big[  i \pi ( (k+\alpha)_M Z^{MN} (k+ \alpha)_N   + 2 (k + \alpha)_M (Q^{+,M} + \beta^M) )  \Big].   
\end{aligned}
\ee

The overall partition function~\eqref{WQ} clearly contains contributions both from the chiral and anti-chiral parts of $\cH$. As argued by Witten in~\cite{Witten:1996hc}, the physical partition function of the chiral field $\cB$ can be identified with one factor of one of the summands. The correct factor can be identified using the natural covariant derivatives
\be
\begin{aligned}
\frac{D}{D Q^{+,M}} &= \frac{\delta}{\delta Q^{+,M}} - \frac{\pi}{2} Q^{-,N} \I\, Z_{NM} \ , \\ \frac{D}{D Q^{-,M}} &= \frac{\delta}{\delta Q^{-,M}} - \frac{\pi}{2} Q^{+,N} \I\, Z_{NM}, 
\end{aligned}
\ee
which are defined on the intermediate Jacobian. The two factors in~\eqref{WQ} are respectively holomorphic and anti-holomorphic with respect to these covariant derivatives, in the sense that~\cite{Kerstan:2012cy}
\be
\frac{D}{D Q^{+, M}} \Theta [   {   \tiny  \begin{matrix} \alpha   \\  \beta    \end{matrix} } ] ( \ov{Z}, Q^-, Q^+) = 0 = \frac{D}{D Q^{-,M}} \Theta [   {   \tiny  \begin{matrix} \alpha   \\  \beta    \end{matrix} } ] (- Z, - Q^+, - Q^-).
\ee
As argued before, the chiral part $\cH^+$ of the field strength couples only to $Q^-$. Therefore, the desired partition function of the chiral 2-form is given by one of the anti-holomorphic factors in~\eqref{WQ}, which is annihilated by $D/D Q^{+, M}$.

The problem of finding the chiral partition function is therefore reduced to the problem of picking the correct summand in~\eqref{WQ}. In other words, it is necessary to choose the correct line bundle labelled by the vectors $\alpha_M, \beta^N$. The prescription to construct this line bundle is given in~\cite{Witten:1996hc}. However, as already mentioned above carrying out this construction in practice is highly non-trivial. Therefore for the moment we will sidestep this issue and simply pick one of the terms in~\eqref{WQ}, leaving $\alpha_M$ and $\beta^N$ undetermined. As we will see in the next subsection, the comparison with the type IIB partition function of an $E3$-instanton will allow us to identify the correct partition function without having to use Witten's construction.

It is a simple matter to apply these general results to the case of the pseudo-action of the $M5$-instanton given in~\eqref{S_M5}. Choosing $b=-2ai$ in~\eqref{M5action1} leads to $Q=-ib C_3^-$. After reinstating the overall prefactor of the partition function coming from the $\cH$-independent part of the action, we obtain the non-chiral partition function~\cite{Kerstan:2012cy}
\be
W_{M5}^{tot.,\, cl.} =  e^{-2\pi \left( \mathrm{Vol}(D_{M5}) + i \int C_6 + \frac{b^2}{4}\int C_3\wedge \ast C_3 \right)} W(-i b C_3^-).
\ee
The physical classical partition function of the chiral 2-form is obtained by choosing one of the anti-holomorphic factors in $W(-i b C_3^-)$, which finally leads to~\cite{Kerstan:2012cy}
\begin{align}
\label{W_M5}
&\hspace{-0.9cm} W^{cl.}_{M5} =  e^{-2\pi \left( \mathrm{Vol}(D_{M5}) + i \int C_6 \right)} {\cal Z} [   {   \tiny  \begin{matrix} \alpha   \\  \beta    \end{matrix} } ] \, , \\
& \hspace{-0.9cm}{\cal Z} [   {   \tiny  \begin{matrix} \alpha   \\  \beta    \end{matrix} } ] =  \exp \left[ \frac{\pi}{2}b^2 C^{-,M} \I Z_{MN} (C^{-,N} - C^{+,N}) \right]  \\
& \hspace{-0.7cm}\  \times \sum_{k_M \in\  \mathbb{Z}} \exp \left[ i\pi \left( (k+\alpha)_M \ov{Z}^{MN} (k+\alpha)_N + 2(k+\alpha)_M (\beta^M - ib C^{-,M}) \right) \right]. \nonumber
\end{align}
In the next subsection, we will compare this result to the partition function of a suitable $E3$-instanton. As we will see, the match not only provides an important consistency check for the result~\eqref{W_M5}, but also allows us to identify the correct values of $\alpha_M, \beta^N$.

\subsection{Matching the $M5$ and $D3$ partition functions}
\label{sec:partition_fct_match}

Let us now focus on the case where the $M5$-instanton is wrapped on a vertical divisor $D_M = \pi^{-1}(D_M^b)$ of an elliptically fibered, resolved fourfold $\hat{Y}_4$. Following the general logic of the F-theory/type IIB duality, such an instanton is expected to be dual to an $E3$-instanton in the double cover $X_3$ of $B_3$. More precisely, the divisor $D_E\subset X_3$ wrapped by the $E_3$-instanton is the preimage of $D_M^b\subset B_3$ under the orientifold projection. 
As part of the duality, the classical partition functions of the $E3$- and $M5$-instantons should match. The aim of this subsection is to explicitly demonstrate this match, following the discussion in~\cite{Kerstan:2012cy}. We will focus on the case where the type IIB instanton is of $O(1)$ type, such that $D_E$ is a single divisor invariant under the orientifold involution.

The classical partition functions discussed in sections~\ref{sec:modstab_fluxed_inst} and~\ref{sec:M5partition} are essentially determined by sums over instanton fluxes. As these fluxes are classified by cohomology groups on the instanton divisors, it is clear that it is necessary to understand how the cohomology groups of $D_E$ and $D_M$ are related. The uplift of the cohomology groups of $D_E$ was discussed in general terms in~\cite{Kerstan:2012cy}. Further evidence for the expected relationships is provided in reference~\cite{Blumenhagen:2010ja}, in which the relevant Hodge numbers of $D_E$ and $D_M$ are computed and compared for the case where $X_3$ is a hypersurface in a weighted projective space. We focus on the cohomology group $H^3(D_M)$, which is the relevant group to describe non-trivial 3-form flux $\cH$. 

References~\cite{Blumenhagen:2010ja, Kerstan:2012cy} argue that $H^3(D_M)$ can be split into several distinct subspaces comprising forms with different origin in terms of the geometry of $D_E$. To understand this split, it is important to note that forms on $D_E$ with different orientifold parities have a qualitatively different uplift behaviour. To be precise, the orientifold-even cohomology groups $H^{p,q}_+(D_E)$ should directly contribute to $H^{p,q}(D_M)$, while the orientifold-odd forms in $H^{p,q}_-(D_E)$ are expected to contribute to $H^{p+1,q}(D_M)\oplus H^{p,q+1}(D_M)$~\cite{Donagi:2010pd, Blumenhagen:2010ja, Kerstan:2012cy}. To understand this, note that orientifold-even forms have a well-defined orientifold projection to a 3-form on $D_M^b$, which can in turn be pulled back to yield a 3-form on $D_M$. Orientifold-odd forms on the other hand cannot easily be defined directly on $D_M^b$. Nevertheless, orientifold-odd forms can be combined with the 1-forms $dx, dy$ of the elliptic fiber\footnote{As $D_M$ is assumed to be a vertical divisor, it contains the entire elliptic fiber.} over $D_M^b$. As already discussed around equation~\eqref{uplift_omega}, this is expected to yield well-defined forms on $D_M$ as the monodromy of the elliptic fiber cancels the orientifold parity of the base form. Schematically, the uplift can be depicted as~\cite{Donagi:2010pd, Blumenhagen:2010ja, Kerstan:2012cy}
\be
\label{uplift_odd_forms_instanton}
\omega\in H^{p,q}_-(D_E) \rightarrow \omega\wedge dx, \ \omega\wedge dy \in H^{p+1,q}(D_M)\oplus H^{p,q+1}(D_M).
\ee

Continuing the analogy with the discussion of section~\ref{sec:KKreduxFtheory}, we expect that the intuitive uplift~\eqref{uplift_odd_forms_instanton} fails if the form $\omega$ has non-trivial pullback to the discriminant locus over which the fibration degenerates. In the type IIB language, the discriminant describes the locations of the orientifold planes and $D7$-branes of the model. Note that an $O(1)$ instanton necessarily has an intersection with the $I_1$-part of the discriminant locus which describes the $O7$-plane~\cite{Kerstan:2012cy}. However, this intersection does not impact on the uplift of the cohomology groups because the orientifold-odd forms have vanishing pullback to the orientifold plane. Therefore, problems may arise only on the intersection between $D_E$ and a $D7$-brane divisor $D_A$. As $D_E\cap D_A$ is a non-trivial holomorphic 2-cycle in $X_3$, the only forms whose uplift can be affected are forms in $H^{1,1}_-(D_E)$ which are furthermore non-trivial on $X_3$ in the sense that they can be viewed as arising by pullback from a bulk form\footnote{This explains why no mismatch in the dimensions of the cohomology groups is observed in the model considered in~\cite{Blumenhagen:2010ja}, which is built on an orientifold $(X_3, \sigma)$ with $H^{1,1}_-(X_3)$.}~\cite{Kerstan:2012cy}. In the following we continue to use the notation of sections~\ref{sec:IIBreduction} and~\ref{sec:flux_on_D3inst}, which allows us to identify the forms with potentially problematic uplift as the pullbacks $\iota^* \omega_a,\ \omega_a \in H^{1,1}_-(X_3)$. Using $D_E = C^\alpha_E \omega_\alpha$ and $D_A = \frac12 (D_A^+ + D_A^-) = \frac12(C^\alpha_A\omega\alpha + C^a_A\omega_a)$ immediately allows us to deduce that the uplift~\eqref{uplift_odd_forms_instanton} of $\iota^*\omega_a$ can fail if~\cite{Kerstan:2012cy}
\be
\label{pullback_brane_instanton}
0 \neq \int_{D_E\cap D_A} \iota^* \omega_a = \frac12 \cK_{a b \alpha} C^b_A C^\alpha_E.
\ee
In other words, we see that we can ensure that the uplift of the cohomology groups goes through without problems at the harmonic level if we take $C^a_A = 0$ for all $D7$-brane stacks in the model. This nicely matches the discussion of section~\ref{sec:massiveU(1)sFtheory}, in which we saw that it is necessary to include non-harmonic forms on the F-theory side to explicitly account for all known type IIB effect precisely if $C^a_A \neq 0$. To avoid any potential problems with the instanton cohomology uplift it would be sufficient to require that the intersections~\eqref{pullback_brane_instanton} vanish. However, we follow~\cite{Kerstan:2012cy} and impose the stronger condition $C^a_A = 0$, which simplifies the notation because we don't have to be so careful about the distinction between the bulk forms and their pullbacks to the instanton divisor. We will briefly comment on the general case with some non-vanishing intersections~\eqref{pullback_brane_instanton} at the end of subsection~\ref{sec:selectionRulesExample}.

The discussion above can be used to relate $H^3(D_M)$ to the cohomology groups of $D_E$. An obvious contribution to $H^3(D_M)$ arises from forms in $H^3_+ (D_E)$. As the Hodge type should not change in the uplift of orientifold-even forms and $H^3_+ (D_E)=H^{2,1}_+(D_E)\oplus H^{1,2}_+(D_E)$, the associated 3-forms on $D_M$ lie in $H^{2,1}(D_M)\oplus H^{1,2}(D_M)$. Let us denote the subspace of $H^{2,1}(D_M)$ arising by pullback from $H^{2,1}_+(D_E)$ by $H^{2,1}_{pb}(D_M)$.
Additional contributions to $H^3(D_M)$ come from the uplift of the orientifold-odd forms in $H^2_-(D_E)$. More precisely, we expect that $H^{2,0}_-(D_E)$ contributes to $H^{3,0}(D_M)$ and $H^{2,1}(D_M)$ while $H^{1,1}_-(D_E)$ contributes to $H^{2,1}(D_M)$ and $H^{1,2}(D_M)$~\cite{Blumenhagen:2010ja, Kerstan:2012cy}. Note that forms in $H^{(2,0)}_-(D_E)$ are in one-to-one correspondence with holomorphic, orientifold-even sections of the normal bundle of $D_E$, where the concrete link is obtained by contracting such a section of the normal bundle with the orientifold-odd holomorphic 3-form $\Omega_3$ on $X_3$~\cite{Jockers:2004yj}. In other words, $H^{(2,0)}_-(D_E)$ corresponds to deformations of $D_E$ which maintain the structure of $D_E$ as a holomorphic, orientifold-invariant object. We therefore denote the subspace of $H^{2,1}(D_M)$ related to $H^{2,0}_-(D_E)$ by $H^{2,1}_{def.}(D_M)$ and the remaining part of $H^{2,1}(D_M)$, which is related to $H^{1,1}_-(D_E)$, by\footnote{In reference~\cite{Kerstan:2012cy} the spaces $H^{2,1}_{def.}(D_M)$ and $H^{2,1}_{flux}(D_M)$ are called $H^{2,1}_{hor.}(D_M)$ and $H^{2,1}_{ver.}(D_M)$, respectively. The reason for our different choice of nomenclature will become apparent at the end of this subsection.} $H^{2,1}_{flux}(D_M)$. Finally, the analysis of~\cite{Blumenhagen:2010ja} suggests that additional 3-forms on $D_M$ can arise if an intersection between the instanton and a $D7$-brane has $h^{1,0}_-(D_E\cap D_A) \neq 0$. In the following, we will always assume as in~\cite{Kerstan:2012cy} that $h^{1,0}_-(D_E\cap D_A) = 0$, so that this possible contribution is absent.
In summary, we have
\be
\begin{aligned}
H^{2,1}(D_M) &= H^{2,1}_{pb}(D_M) \oplus H^{2,1}_{def.}(D_M) \oplus H^{2,1}_{flux}(D_M) \\
& \simeq H^{2,1}_+(D_E)\oplus H^{(2,0)}_-(D_E) \oplus H^{1,1}_-(D_E).
\label{uplift_H21_Dm}
\end{aligned}
\ee

As discussed in section~\ref{sec:Dinstantons_IIB}, $h^{1,0}(D_E)\neq 0$ or $h^{2,0}(D_E)\neq 0$ leads to the appearance of fermionic zero modes which often cause the entire instanton contribution to vanish. Therefore the main phenomenological interest lies on rigid instantons with no deformations or Wilson lines. In the following, we will assume that the instanton divisor $D_E$ is rigid in this sense, so that 
\be
\label{cohom_uplift_rigid_inst}
H^3(D_M) = H^{2,1}_{flux}(D_M) \oplus H^{1,2}_{flux}(D_M) \simeq H^{1,1}_-(D_E) \oplus H^{1,1}_-(D_E).
\ee
Recall from section~\ref{sec:flux_on_D3inst} that the space of  fluxes on $O(1)$ $E3$-instantons is precisely given by $H^{1,1}_-(D_E)$. Therefore the general IIB/F-theory duality anyway leads us to expect that even in the non-rigid case all the viable 3-form fluxes on $D_M$ should lie in the part of $H^{3}(D_M)$ given in~\eqref{cohom_uplift_rigid_inst}. We will briefly consider relaxing the rigidity condition at the end of this subsection. 

The relationship between the cohomology groups discussed above strongly suggests that the sum over fluxes in the classical type IIB instanton partition function
\be
W_{E3}^{cl.} = \sum_{\tilde{\cF}^E} e^{-S_E^{cl.}}
\ee
can be identified with the sum over 3-form fluxes which contributes to the classical $M5$-instanton partition function~\eqref{W_M5}. In computing~\eqref{W_M5} we had chosen a basis of $H^3(D_M)$ in which the expansion coefficients of $\cH$ were integer. To make the correspondence with the $E3$ case more precise, it is helpful to take a similar approach on the IIB side. In other words, we choose a basis $\{\omega_M\}$ of $H^{1,1}_-(D_E)$ such that $\tilde{\cF}^E \equiv \cF^M \omega_M$ with integer $\cF^M$. It was pointed out in~\cite{Kerstan:2012cy} that the Freed-Witten quantisation condition~\eqref{FWanom_inst} actually requires the integral of $\tilde{\cF}^E$ over any 2-cycle on the physical instanton divisor $D_M^b$ in the orientifold quotient to be integer. As the integrals evaluated on the double cover $X_3$ are twice as large as on $B_3$, this means that the integral of $\tilde{\cF}^E$ over a 2-cycle in $D_E$ must actually be an even integer. In order for this to be compatible with $\cF^M $ being integer, we normalise the basis $\{\omega_M\}$ by~\cite{Kerstan:2012cy}
\be
\label{def_omega_M}
\int_{D_E} \omega_M \wedge \omega_N = 2 \delta_{MN}.
\ee
As discussed in section~\ref{sec:flux_on_D3inst}, the instanton flux can be split into flux which can be described by a form pulled back from $X_3$ and the orthogonal so-called variable flux $\cF^{\tv}$. To take this into account, it is helpful to split the set $\{\omega_M\}$ into a subset $\{\omega_m\}$ which spans the subspace $\iota^* H^{1,1}_- (X_3) \cap H^{1,1}_- (D_E)$ and a basis $\{\omega_{\hat{m}} \}$ of the orthogonal complement. For integer quantised fluxes we therefore have the expansion~\cite{Kerstan:2012cy}
\be
\tilde{\cF}^E = \cF^m \omega_m + (\cF^{\tv})^{ \hat m} \omega_{\hat m} \equiv \cF^M \omega_M \ , \qquad \cF^M \in \ \mathbb{Z} \ .
\ee

When evaluating the classical $E3$-instanton action in~\eqref{instanton_class_action} we had used a different expansion of $\tilde{\cF}^E$ in terms of the pullbacks $\iota^* \omega_a$. As discussed around equation~\eqref{triple_intersections}, the $\omega_a$ are normalised to have even integer intersection numbers on $X_3$. Nevertheless, it is important to note that their pullbacks to $D_E$ do not necessarily share the same property~\cite{Kerstan:2012cy}. We may of course expand the $\iota^*\omega_a$ into the basis $\omega_m$, such that $\iota^*\omega_a = M_a^m \omega_m$. Using $D_E =C^\alpha_E \omega_\alpha$ and the intersection numbers~\eqref{triple_intersections} it is easy to derive~\cite{Kerstan:2012cy}
\be
\label{M_and_C}
C^\alpha_E \,  \cK_{\alpha a b} = \int_{D_E} \omega_a \wedge \omega_b = 2 M_a^m M_b^n \delta_{mn}.
\ee
$M_a^m$ can be used to relate the expansion coefficients of $\tilde{\cF}^E$, $\iota^*C_2$ and $\iota^*B_2$ into the $\iota^*\omega_a$, which were used in~\eqref{instanton_class_action}, to expansion coefficients with respect to the basis $\{\omega_m\}$. The result of this is~\cite{Kerstan:2012cy}
\be
G^a M^m_a  \equiv G^m  , \qquad \tilde{\cF}^{E,a} M_a^m = \cF^m.
\ee
Using the new basis, we may write the classical partition function of the $E3$-instanton as~\cite{Kerstan:2012cy}
\be
\begin{aligned}
\label{ZE3_rewrite}
W^{cl.}_{E3} =&\,  \exp \left[-\pi \left( \frac12 C^\alpha_E \cK_{\alpha \beta\gamma}v^\beta v^\gamma + iC^\alpha_E(c_\alpha - \frac12 \cK_{\alpha ab}c^a b^b)\right)\right]  \\ 
& \times  \exp \left[ -\frac{i\pi}{\tau - \ov{\tau}}\delta_{mn}G^m(G^n-\ov{G}^n) \right] \\
&\times \sum_{\cF^M \in\ \mathbb{Z}} e^{-i\pi \left(  2 \delta_{mn} G^m \cF^n + \tau \delta_{MN}\cF^M\cF^N  \right) }.
\end{aligned}
\ee

Comparing this expression with the $M5$-instanton result of~\eqref{W_M5}, let us first focus on the overall prefactor 
\be
W^{cl.}_{M5} \propto \exp\left[-2\pi \left( \mathrm{Vol}(D_{M5}) + i \int_{D_M} C_6 \right)\right].
\ee
To extract the moduli dependence of this factor, we insert the expansion~\eqref{kaehler_exp_fourfold} of the K\"ahler form and the expansion of $C_6$ given in appendix~\ref{sec:app_democ_Mtheory}. The Poincar\'e dual of $D_M$ in $\hat Y_4$ is formally given by the same expression $[D_M]=C_E^\alpha \omega_\alpha$ as the Poincar\'e dual of $D_E$ in $X_3$, although as discussed in section~\ref{sec:massiveU(1)sFtheory} the intersection forms of the $\omega_\alpha$ on $\hat Y_4$ and $X_3$ differ by a factor of 2. Using the intersection numbers~\eqref{intersection_numbers_fourfold} and~\eqref{dual_bases} immediately leads to
\be
W^{cl.}_{M5} \propto \exp\left[-2\pi \left( \frac{1}{12} C^\alpha_E \cK_{\alpha \Lambda\Sigma\Theta}v^\Lambda v^\Sigma v^\Theta + i \frac12 \tilde{c}_\alpha \right)\right].
\ee
The match of the two factors now follows from the relationship $t_\alpha \rightarrow \frac12 T_\alpha$ between the type IIB and F-theory K\"ahler moduli discussed in section~\ref{sec:FtheoryLimit} and appendix~\ref{sec:app_democ_Mtheory}. Explicitly, the fact that the intersection of 4 vertical divisors in $\hat Y_4$ vanishes and the identification~\eqref{kaehler_mod_IIB_ftheory} leads to 
\be
\frac{1}{12} C^\alpha_E \cK_{\alpha \Lambda\Sigma\Theta}v^\Lambda v^\Sigma v^\Theta = \frac14 C^\alpha_E \cK_{\alpha \beta \gamma} v_{IIB}^\beta v_{IIB}^\gamma + \cO(\epsilon).
\ee
The crucial shift in the relation
\be
\tilde{c}_\alpha = c_\alpha - \frac12 \cK_{\alpha a b} c^a b^b
\ee
between the M-theory axion $\tilde{c}_\alpha$ and the type IIB axion $c_\alpha$ was derived directly in appendix~\ref{sec:app_democ_Mtheory} by considering the F-theory limit of the low energy action obtained by dimensional reduction of the 11-dimensional supergravity in the democratic formulation.

We now turn to the match of the remaining parts of the $M5$-instanton partition function~\eqref{W_M5} and $W^{cl.}_{E3}$.
It is natural to expect that $E_M$ and $F^M$ used in the calculation of $W^{cl.}_{M5}$ can be viewed as arising from the forms $\omega_M$ via the construction in~\eqref{uplift_odd_forms_instanton}. Taking into account the usual factor of 2 difference between intersections on $X_3$ and $\hat{Y}_4$, we see that the normalisation~\eqref{def_omega_M} is in agreement with~\eqref{integer_basis}. In particular, the symplectic structure of the 1-forms on the elliptic fiber yields a natural explanation for the symplectic structure of $E_M$ and $F^M$. 
Just like in the IIB case it is helpful to distinguish between forms in $H^3(D_M)$ that arise by pullback from $H^3(\hat Y_4)$ and forms in the orthogonal complement. In other words, we split the set $E_M$ used in the previous subsection into $E_{m}, \ E_{\hat{m}}$, and similarly for the $F^M$. Of course, this means that in all parts of~\eqref{W_M5} which involve $C_3$, the index should run only over the subset $M = m$ corresponding to the pullback forms. Alternatively, we can simply set $C^{\pm,\hat{m}}=0$. Then comparing~\eqref{W_M5} to~\eqref{ZE3_rewrite} reveals a precise match, provided that we identify~\cite{Kerstan:2012cy}
\be
\label{identifications_for_partition_function}
\cF^M \leftrightarrow \delta^{MN} k_N \ ; \quad G^m \leftrightarrow  C^{-,m} \ ; \quad \ov{G}^m \leftrightarrow  C^{+,m} \ ; \quad \ov{Z}^{MN} = - \tau \delta^{MN}.
\ee
Furthermore, we must set the parameter $b$ appearing in the $M5$-brane pseudo-action to $b=-i$. A key result of the match is that we are able to unambiguously identify~\cite{Kerstan:2012cy}
\be
\alpha_M = 0 =\beta^M.
\label{identify_line_bundle}
\ee
In other words, we have been able to identify the line bundle on the intermediate Jacobian of $D_M$ which is associated to the physical partition function of the chiral 2-form. As discussed in the previous subsection, this identification is highly non-trivial directly in M-theory.

Let us now briefly discuss this result in the context of other investigations of the $M5$-brane partition function, following~\cite{Kerstan:2012cy}. Recall that the entries of the vector $\alpha_M$ are 0 or 1/2. As it enters the $M5$-brane partition function exclusively through the combination $k_M + \alpha_M$, it seems to account for a possible shift in the quantisation of the fluxes. For $O(1)$ instantons the fact that the fluxes are orientifold-odd while the divisor is orientifold-even implies that the fluxes are always integer-quantised~\cite{Kerstan:2012cy}, so that $\alpha_M=0$ as stated above. Intriguingly, this hints that type IIB $U(1)$ instantons with half-integer quantised fluxes could be related to $M5$-instanton configurations with $\alpha_M \neq 0$. However, we are not able to confirm this explicitly, because it is unclear how the orientifold-even fluxes on a $U(1)$ instanton, which uplift to 2-forms on $D_M$, should be described in terms of the 3-form $\cH$~\cite{Kerstan:2012cy}. For this reason, the uplift of $U(1)$ instantons will not be considered any further in this thesis. The $M5$ partition function has previously been evaluated in the case where the brane divisor is a flat torus by direct computation in~\cite{Dolan:1998qk} and by using modular invariance or comparison with $D4$-branes in~\cite{Gustavsson:2000kr, Gustavsson:2011ur}. In all cases, these references also obtain the result $\alpha_M = 0 =\beta^M$ in full agreement with~\eqref{identify_line_bundle}. It is also compatible with analogous results for $NS5$-instantons, which were studied in~\cite{Dijkgraaf:2002ac}. It was pointed out in~\cite{Gustavsson:2000kr} that arguments based on modular invariance of the partition function can be used to find~\eqref{identify_line_bundle} whenever the $M5$ world-volume takes the form $T^2\times M_4$, with $M_4$ simply connected. Let us emphasise that the derivation based on~\cite{Kerstan:2012cy} which we have presented here is applicable in a much more general setting. In particular, as we will discuss momentarily, we expect that it is possible to relax the assumption of rigidity of the instanton divisor without qualitatively changing the results. Therefore we expect the result~\eqref{identify_line_bundle} to hold as long as $D_M$ is a vertical divisor, provided that no non-trivial intersections of the form~\eqref{pullback_brane_instanton} with orientifold-odd components of $D7$-brane divisors appear\footnote{From the discussion in section~\ref{sec:U(1)inFtheory} it is clear that in the case where $D7$-branes with $C^a_A\neq 0$ are present we expect that some non-harmonic forms should be taken into account in the F-theory description. This suggests that a non-vanishing intersection~\eqref{pullback_brane_instanton} might signal that we have to take into account non-harmonic $\cH$ flux on $D_M$~\cite{Kerstan:2012cy}. We will briefly return to this speculative suggestion at the end of section~\ref{sec:selectionRulesExample}.}. Note that this does not mean that the instanton divisor is not allowed to intersect the discriminant. In fact, such an intersection and hence a degeneration of the elliptic fibration in the instanton world-volume is necessary in order for the instanton to be of $O(1)$ type~\cite{Kerstan:2012cy}.  

It is important to verify that the identifications made in~\eqref{identifications_for_partition_function} are consistent with the various field identifications that were established in section~\ref{sec:FtheoryLimit}. In particular, we had found that the fields $c^a$ and $b_a$ appearing in $ C_3 \supset c^a \alpha_a + b_a \beta^a$ can be directly identified with the corresponding type IIB fields which appear in $G^a = c^a - \tau b^a$. To see that this is consistent with $G^m \leftrightarrow  C^{-,m}$, we note that $E_m$, $F^m\in H^3(D_M)$ are obtained from the $\omega_m \in H^{1,1}_-(D_E)$ by the same schematic uplift~\eqref{uplift_odd_forms_instanton} which we had used to relate the forms $\alpha_a$, $\beta^b \in H^3(\hat Y_4)$ to the $\omega_a \in H^{1,1}_-(X_3)$. This allows us to translate the relation $\omega_a = M^m_a \omega_m$ into~\cite{Kerstan:2012cy}
\be
\iota^* \alpha_a = M_a^m E_m, \qquad\quad \iota^* \delta_{ab}\beta^b = M_a^m \delta_{mn} F^n.
\ee
Using $\ov{Z}^{MN} = - \tau \delta^{MN}$ in the definition~\eqref{self_dual_basis} of the (anti-)self-dual basis $E_m^\pm$ leads to~\cite{Kerstan:2012cy}
\be
\begin{aligned}
\iota^* C_3 \supset c^a \iota^* \alpha_a + b_b \iota^* \beta^b & =  M_a^m (c^a - \tau \delta^{ab}b_b) E_m^- + M_a^m (c^a - \bar{\tau} \delta^{ab}b_b)E_m^+  \\
\label{pullback_C3_and_Cm}
& =  M_a^m G^a E_m^- + M_a^m \bar{G}^a E_m^+,
\end{aligned}
\ee
matching the expectation from~\eqref{identifications_for_partition_function}.

To conclude this subsection, let us consider how the match of the partition functions presented above is expected to change if some of the assumptions that were made about the cohomology of $D_E$ are relaxed. In particular, we might allow for $h^{2,1}_+(D_E) \simeq h^{1,0}_+(D_E) \neq 0$ as well as for the presence of holomorphic deformations counted by $h^{2,0}_-(D_E)$ and additional 1-forms on intersections with $D7$-branes counted by $h^1_-(D_E\cap D_A)$. As discussed around equation~\eqref{uplift_H21_Dm}, $H^3(D_M)$ then contains additional forms beyond the subspace $H^{2,1}_{flux}(D_M)$ that corresponds to $H^{1,1}_-(D_E)$. In order to obtain a match with the type IIB instanton partition function, it is clear that the sum over $\cH$ in the $M5$ partition function~\eqref{class_partition_fct} must nevertheless be restricted to run over $H^{2,1}_{flux}(D_M)$ only. It is not immediately clear how putative 3-form fluxes related to $h^{2,1}_+(D_E)$ and $h^1_-(D_E\cap D_A)$ would be related to IIB fluxes, which are 2-forms on $D_E$. It is therefore difficult to gain intuition about the relevant restriction mechanism in these cases from the type IIB side, and we will not attempt to answer this puzzle in the following. 

The problem is more tractable for $\cH$-fluxes in $H^{3,0}(D_M) \oplus H^{2,1}_{def.}(D_M)\subset H^3(D_M)$, which are associated with $H^{2,0}_-(D_E)$ and can thus be related to type IIB instanton fluxes. On the type IIB side we know that the instanton fluxes are restricted to lie in the subspace $H^{1,1}_-(D_E) \subset H^2_-(D_E)$ by the F-term supersymmetry condition~\eqref{calibration_F}. As already alluded to in subsection~\ref{sec:M5partition}, we expect that there must be a corresponding supersymmetry condition relating to the $M5$-brane which rules out fluxes lying in the subspace $H^{3,0}(D_M) \oplus H^{2,1}_{def.}(D_M)\subset H^3(D_M)$~\cite{Kerstan:2012cy}.

To gain intuition for the form of this supersymmetry condition we consider first the type IIB case in slightly more detail. Here the relevant supersymmetry condition~\eqref{calibration_F} can be obtained from the superpotential term\footnote{A similar form for the superpotential was suggested in~\cite{Kerstan:2012cy}; however, there the deformation moduli $\zeta^A$ were omitted.}
\be 
\label{FtermE3}
W_{\tilde{\cF}^E} = \int_{D_E}  \zeta^A i_{v^A} \Omega_3 \wedge \tilde{\cF}^{E}.
\ee
In this expression the $\zeta^A$ are the deformation moduli of the brane, counted by $h^{2,0}_-(D_E)$. The $v^A$ are holomorphic normal vector fields to $D_E$, while $i_{v^A}$ denotes the insertion of a vector field into a differential form. Note that by definition the deformation moduli are zero at the rest position of the brane, so that the vacuum expectation value of this superpotential term vanishes
\be
\left<\zeta^A \right> = 0 \Rightarrow \left< W_{\tilde{\cF}^E} \right> = 0.
\ee
Nevertheless, the superpotential~\eqref{FtermE3} yields a non-trivial restriction on the fluxes because the F-term supersymmetry condition involves the K\"ahler covariant derivatives $D_{\zeta^A} W = \partial_{\zeta^A} W + W \partial_{\zeta^A} K$. More precisely, requiring the F-term to vanish in the vacuum yields\footnote{Note that $W_{\tilde{\cF}^E}$ also depends on the complex structure moduli $z^M$ of $X_3$ through $\Omega_3$. However, the K\"ahler covariant derivatives of $W_{\tilde{\cF}^E}$ with respect to the complex structure moduli are proportional to the $\zeta^A$ and therefore vanish automatically in the vacuum.}
\begin{align}
\left< D_{\zeta^A} W_{\tilde{\cF}^E} \right> = & \int_{D_E} i_{v^A} \Omega_3 \wedge \tilde{\cF}^{E} + \left< \int_{D_E}  \zeta^A i_{v^A} \left[ \partial_{\zeta^A} \Omega_3 + \Omega_3\partial_{\zeta^A} K \right] \wedge \tilde{\cF}^{E} \right> \nonumber \\
= & \int_{D_E} i_{v^A} \Omega_3 \wedge \tilde{\cF}^{E} = 0,
\end{align}
where we again used $\left<\zeta^A \right> = 0$. Because the set $i_{v^A} \Omega_3$ spans $H^{2,0}_-(D_E)$~\cite{Jockers:2004yj}, this implies that only fluxes in $H^{1,1}_-(D_E)$ can be switched on in a supersymmetric vacuum.

Let us now use this type IIB intuition to make a proposition for the form of the corresponding superpotential in F-theory. It is natural to assume that the flux in~\eqref{FtermE3} should be replaced by $\cH_3$ in the case of an $M5$-instanton, while the $z^A$ should be taken to be the deformation moduli of $D_M$. Thus only the fate of $\Omega_3$ in~\eqref{FtermE3} remains to be determined. As $\Omega_3$ is a form of negative orientifold parity on $X_3$, the arguments outlined at the beginning of this subsection lead us to expect that in the F-theory uplift one obtains a (3,1)-form $\Omega_{3,1}$ in addition to the holomorphic (4,0)-form $\Omega_4$ of $Y_4$. We propose that the $M5$-brane superpotential can then be written as\footnote{Note that the suggestion for the superpotential presented  in~\cite{Kerstan:2012cy} was missing the term involving $\Omega_{3,1}$.}
\be 
\label{FtermM5}
W_{\cH} = \int_{D_M} \zeta^A i_{v^A} \left( \Omega_4 + \Omega_{3,1}\right) \wedge \cal H.
\ee
Using $\left<\zeta^A \right> = 0$ we obtain the corresponding F-term supersymmetry condition 
\be
\label{FtermM5_eval}
\int_{D_M} i_{v^A} \left( \Omega_4 + \Omega_{3,1}\right) \wedge \cH = 0. 
\ee
Because $D_M$ wraps the elliptic fiber, the normal vector fields $v^A$ can be considered to lie in the tangent space of the base $B_3$ of $Y_4$. Furthermore the 3 holomorphic indices of $\Omega_{3,1}$ can be intuitively viewed as base indices because of its relation to the holomorphic 3-form of $X_3$, while the anti-holomorphic index is an index in the elliptic fiber. This means that $i_{v^A} \Omega_{3,1}$ is a (2,1)-form on $D_M$ with one index in the elliptic fiber, and $i_{v^A} \Omega_{3,1} \wedge \cH$ can be non-vanishing for $\cH \in H^{2,1}_{def.}(D_M)$. Similarly, $i_{v^A} \Omega_4$ is a (3,0)-form with one index in the elliptic fiber, which gives it the right index structure to have a non-zero wedge product with elements of $H^{3,0}(D_M)$. Due to the different index structures there can be no cancellations between the two terms in~\eqref{FtermM5_eval}. Therefore the proposed superpotential~\eqref{FtermM5} rules out exactly those $\cH$-fluxes related to $H^{2,0}_-(D_E)$, just as expected from the type IIB picture.

This discussion, together with the expectation that the $M5$ and $E3$ partition functions should match, forms compelling evidence for the assumption that a superpotential term of the form~\eqref{FtermM5} arises for fluxed $M5$-instantons. Let us emphasise that in a general setting there may be additional forms related to $h^{2,1}_+(D_E)$ and $h^1_-(D_E\cap D_A)$, which are not ruled out by the superpotential proposed above. It would be interesting to study whether these forms are ruled out by additional superpotential terms or whether a completely different mechanism must be invoked.

\subsection{$M5$-instanton zero modes in backgrounds with $G_4$ flux}
\label{sec:zeroModesG4}

In the previous two subsections we have focused on $M5$-instantons in backgrounds with vanishing $G_4$ flux. As extensively discussed in the earlier parts of this thesis, non-vanishing $G_4$ flux is an essential part of many approaches to F-theory model building and plays an important role in particular for moduli stabilisation and in generating chiral matter spectra. It is therefore very important to consider how the presence of $G_4$ flux affects the contribution of $M5$-instantons to the low energy effective action. We turn to this point in the present subsection, which is again based on reference~\cite{Kerstan:2012cy}.

To obtain some intuition, let us briefly recall from section~\ref{sec:fluxed_D3instantons} how fluxes besides the instanton world-volume flux affect $E3$-instantons in the type IIB setting. We distinguish between $D7$-brane world-volume fluxes $\tilde{\cF}^A$ and bulk fluxes corresponding to topologically non-trivial configurations of $G_3 \sim dC_2 -\tau dB_3$, which lead to qualitatively different effects. Bulk fluxes $G_3$ can induce interaction terms in the effective instanton world-volume action, which can play a crucial role in lifting neutral fermionic zero modes~\cite{Blumenhagen:2009qh, Gorlich:2004qm, Tripathy:2005hv, Bergshoeff:2005yp, Lust:2005cu, Lust:2006zg, Blumenhagen:2007bn, Billo:2008sp, Park:2010hj}. On the other hand, $D7$-brane fluxes influence the charged zero mode spectrum originating from open strings localised at the intersection between brane and instanton~\cite{Blumenhagen:2009qh, Blumenhagen:2006xt, Florea:2006si, Ibanez:2006da}. As discussed in detail in section~\ref{sec:flux_on_D3inst}, the net chirality of the charged zero mode spectrum in fact depends on an interplay between the pullback of the $D7$-brane flux and the instanton world-volume flux to the intersection curve~\cite{Grimm:2011dj}.

Moving back to the F/M-theory setting, both types of type IIB fluxes are described by $G_4$ fluxes. Whereas type IIB $D7$-brane fluxes correspond to $G_4$ fluxes along forms localised near the discriminant locus, bulk fluxes correspond to $G_4$ fluxes that are spread out over $B_3$. The effects of the analogues of bulk fluxes on $M5$-instantons have been considered in references~\cite{Kallosh:2005yu, Saulina:2005ve, Kallosh:2005gs}, who find that the flux lifts neutral fermionic $M5$ zero modes just as in the IIB picture. Although we will not always mention it explicitly, we follow~\cite{Kerstan:2012cy} in this thesis and focus on analysing the effects of fluxes that are associated with $D7$-brane gauge fluxes. Nevertheless it will be obvious that many results do not rely on the distinction between the different types of $G_4$ fluxes and are in particular valid also for bulk fluxes.

The first step towards finding the effect of $G_4$ flux on the partition function of $M5$-instanton is to determine how it appears in the world-volume pseudo-action. This can be derived unambiguously from the form used in~\eqref{M5action1} by using partial integration to rewrite the coupling involving $C_3$ as~\cite{Kerstan:2012cy}
\be
S_B \supset 2\pi i\int \cH\wedge \iota^* C_3 = - 2\pi i \int \cB \wedge \iota^* dC_3, 
\ee
and then promoting $dC_3$ to the full non-trivial field strength
\be
\iota^* dC_3 \rightarrow \iota^*G_4 + \iota^*d C_3.
\label{split_g4_dc3}
\ee
We will always make the explicit distinction as in~\eqref{split_g4_dc3}, so that when writing $\iota^* G_4$ we mean only the part which is topologically non-trivial and harmonic on $D_M$, while the other parts appearing in $S_\cB$ will by written as $\iota^*dC_3$. With this understanding, we therefore have the pseudo-action~\cite{Kerstan:2012cy}
\be
S_{M_5, G_4} = 2\pi \int d^6 x \sqrt{\det g} + 2\pi i \int C_6 + S_{\cal B} -  2 \pi i   \int {\cal B} \wedge \iota^*G_4.
\label{action_with_G4}
\ee
Note that of course it is possible for $\iota^* G_4$ to be zero even if $[G_4]\neq0$ in the cohomology of the bulk $\hat Y_4$.

One obvious consequence of the additional term in the action is that the group of large gauge transformations of $\cB$ is broken to a discrete subgroup\footnote{Strictly speaking, only the part of the group generated by 2-forms in $\iota^* H^2(\hat Y_4)$ is broken. Shifts along elements of $H^2(D_M,\bR)$ which are orthogonal to $\iota^* H^2(\hat Y_4)$ are still possible if such forms are present, however we ignore this unimportant distinction in the following.}~\cite{Kerstan:2012cy}. If $G_4$ is integer quantised, it is clear that the unbroken discrete subgroup is simply $H^2(D_M,\bZ)$, such that $\delta S_{{M_5, G_4}} \in 2\pi i \bZ$. If the quantisation of $G_4$ is shifted to include some half-integer periods, the unbroken group of large gauge transformations of $\cB$ will also be shifted accordingly. However, to avoid complicating the notation any further we will simply denote the unbroken group by $H^2(D_M,\bZ)$, as in~\cite{Kerstan:2012cy}. The partial breaking of the gauge symmetry implies that in the path integral over $\cB$ one must now integrate not only over non-closed 2-forms, but also over $H^2(D_M, \mathbb R)/ H^2(D_M, \mathbb Z)$. In other words, we have
\be
\int \cD \cB e^{-S_{M_5, G_4}} \propto \int_{H^2(D_M, \mathbb R)/ H^2(D_M, \mathbb Z)} \cD \cB' \ \exp\left[ 2 \pi i   \int {\cal B}' \wedge \iota^*G_4 \right].
\label{S_M5-G}
\ee
In the absence of any other effects cancelling the tadpole for $\cB'$, the integral is proportional to the delta function $\delta(\iota^* G_4)$~\cite{Donagi:2010pd}, so that the contribution of the $M5$-instanton vanishes for non-trivial pullback $\iota^* G_4$~\cite{Kerstan:2012cy}.

As pointed out in~\cite{Kerstan:2012cy}, the fact that the partition function vanishes in the presence of non-trivial flux if no other effects are taken into account is actually just as one would expect from the type IIB perspective. This is because in the IIB language non-trivial gauge flux would lead to a chiral charged zero mode spectrum, which causes the instanton contribution to vanish if no suitable interaction terms with open string fields are included to soak up the zero modes. In fact, the correspondence between~\eqref{S_M5-G} and the type IIB instanton contribution can be made even more precise~\cite{Kerstan:2012cy}. Schematically denoting the charged zero modes by $\lambda_{Ai}$ as in section~\ref{sec:Dinstantons_IIB}, the type IIB instanton contribution includes an integral over these zero modes of the form
\be
\int \cD\{ \lambda_{Ai}\}  \exp\left[-S_E\right].
\ee
This suggests that the integral over $\cD\{ \lambda_{Ai}\}$ can be identified with the integral of $\int \cD\cB' \exp\left[ 2 \pi i   \int {\cal B}' \wedge \iota^*G_4 \right]$ over $H^2(D_M, \mathbb R)/ H^2(D_M, \mathbb Z)$ in~\eqref{S_M5-G}~\cite{Kerstan:2012cy}. As argued in section~\ref{sec:flux_on_D3inst}, in the presence of a chiral charged zero mode spectrum the instanton action $S_E$ is charged with respect to the $U(1)$ symmetries of the model. This $U(1)$ charge is cancelled by the path integral measure $\cD\{ \lambda_{Ai}\}$. At the end of this subsection and in the next subsection we will analyse the charges of the action of the $M5$-instanton with respect to $U(1)$ symmetries. Although the actual path integral measure $\cD \cB'$ is gauge invariant, it is clear from~\eqref{gauge_trf_C3} that $\int \cD \cB' \exp\left[ 2 \pi i   \int {\cal B}' \wedge \iota^*G_4 \right]$ does carry a $U(1)$ charge. Due to~\eqref{shift_C6} this charge precisely cancels the charge of $S_{M5}$ originating from the shift of $C_6$~\cite{Kerstan:2012cy}. This observation again supports the intuitive identification of the new contribution in the $M5$ path integral with the IIB charged zero mode path integral measure $\cD\{ \lambda_{Ai}\}$.

Of course, at least in the type IIB setting it is well known that non-trivial superpotential contributions can be induced even in the presence of a chiral charged zero mode spectrum. As discussed in more detail in section~\ref{sec:Dinstantons_IIB}, these contributions involve charged open string fields $\Phi_{(Ai)(Bj)}$ and can in fact be crucial for the low energy phenomenology as they can induce couplings between charged fields which are forbidden at the perturbative level. At the technical level the charged open string fields help to saturate the Grassmann integral over $\cD\{ \lambda_{Ai}\}$ via interaction terms of the form
\be
S_E^{int} \supset \int \lambda_{Ai} \Phi_{(Ai)(Bj)} \lambda_{Bj}
\label{inst_interaction_charged_2}
\ee
in the effective instanton action.

The corresponding mechanism for $M5$-instantons, which allows a non-vanishing contribution even in the presence of topologically non-trivial $G_4$ flux, was investigated in~\cite{Kerstan:2012cy}. The crucial ingredient which effectively takes over the role of the open string fields is the appearance of $M2$-branes ending on the $M5$-instanton. There are several ways to see the appearance of these $M2$-branes. From a pure supergravity perspective, one can consider the equation of motion of $\cB$ that follows from the action~\eqref{action_with_G4}
\be
d \ast \cH \propto \iota^* G_4.
\label{eom_H_with_G4}
\ee
The full background solution can thus not simply involve a harmonic instanton field strength $\cH$. Roughly speaking, we can decompose $\cH = \cH_0 + \cH'$, where $\cH_0$ is one of the usual harmonic solutions that would be present in the absence of $G_4$. $\cH'$ on the other hand is an inhomogeneous special solution of~\eqref{eom_H_with_G4}. Crucially, in a supersymmetric configuration such a solitonic solution $\cH'$ must be accompanied by a non-trivial profile of the deformation modes of the $M5$-brane~\cite{Howe:1997ue, Gauntlett:1998wb}. This profile describes a sharp spike in the directions normal to the bulk of the $M5$-brane which ends on the 2-cycle $\mathrm{PD}_{D_M}[\iota^* G_4]$ that is Poincar\'e-dual to $G_4$ in $D_M$. From the supergravity perspective, this configuration can be viewed as an $M2$-brane ending on the 2-cycle $\mathrm{PD}_{D_M}[\iota^* G_4]\subset D_M$. It was pointed out in~\cite{Kerstan:2012cy} that this picture provides an alternative way to understand the vanishing of the naked $M5$ contribution to the superpotential in the presence of $G_4$ flux. Namely, the vanishing can be seen as a consequence of the appearance of additional zero modes that describe deformations of the solitonic solution $\cH'$, or more precisely of the associated fermionic superpartner to $\cH$~\cite{Kerstan:2012cy}. In this sense the supergravity explanation of the vanishing of the $M5$ contribution is somewhat similar to the IIB picture.

The presence of the $M2$-branes can also be deduced without analysing the supergravity solution by considering anomaly cancellation in the spirit of Freed and Witten~\cite{Freed:1999vc, Marsano:2011nn}. Anomaly cancellation in the path integral of a probe $M2$-brane requires that the net tadpole for the field $\cB$, which couples to the boundary of the $M2$-brane, must vanish. In other words, there must be physical $M2$-branes ending on $D_M$ which cancel the tadpole~\eqref{eom_H_with_G4}. The world-volumes of these $M2$-branes are therefore given by 3-chains $\Gamma_i$ of the form~\cite{Marsano:2011nn, Kerstan:2012cy} $\Gamma_i = I \times \gamma_i$, with $I$ a semi-infinite time interval such that $\partial \Gamma_i = \gamma_i$ and
\be
\label{M2_cycles}
\sum_i \gamma_i = \mathrm{PD}_{D_M}[\iota^* G_4 ].
\ee
Clearly, this exactly matches the supergravity picture discussed above.

The key to obtaining a non-vanishing $M5$-instanton contribution in the presence of $\iota^* G_4\neq 0$ is to take the $M2$-branes seriously as an integral part of the background configuration. In other words, the full path integral must involve the exponential of the $M2$-brane action on an equal footing with the action of the $M5$-instanton~\cite{Marsano:2011nn}. These additional factors, which may be written as~\cite{Bergshoeff:1987cm}
\be
\label{M2_vertex_factors}
V_{M2,i} \propto \exp \left[ -2\pi i \left( \frac12 \int_{\Gamma_i} C_3  + \int_{\gamma_i} \cB \right) \right],
\ee
can be viewed in string theory language as the vertex factors of the $M2$-brane states~\cite{Kerstan:2012cy}. The full and potentially non-vanishing path integral over $\cB$ then takes the form
\be
\label{superpot_with_M2}
W_{M5} \propto \int \cD \cB \ \prod_i V_{M2,i} \ e^{-S_{M_5, G_4}}.
\ee

When comparing this expression to the type IIB picture, the $M2$-brane vertex factors can be seen as the analogue of the interaction term~\cite{Kerstan:2012cy}
\be
e^{-\int \lambda_{Ai} \Phi_{(Ai)(Bj)} \lambda_{Bj}}
\ee
in the type IIB path integral. Note that in both cases the total insertion is uncharged with respect to any $U(1)$ symmetries that may be present in the model. To see this explicitly for the $V_{M2,i}$, recall from the expansion of $C_3$ in section~\ref{sec:U(1)inFtheory} that a gauge transformation $A^\Lambda \rightarrow A^\Lambda + d\chi^\Lambda$ of a massless $U(1)$ leads to a shift $C_3 \rightarrow C_3 + d\chi^\Lambda \wedge \omega_\Lambda$. According to equation~\eqref{gauge_trf_C3}, this is accompanied by a shift $\cB\rightarrow \cB -\frac12 \iota^* \chi^\Lambda \omega_\Lambda$. In fact, for a massless $U(1)$ this is simply a large gauge transformation of $\cB$, because $\chi^\Lambda$ depends only on the coordinates of the external 3-dimensional space and so $\iota^* \chi^\Lambda$ is constant while $\omega_\Lambda$ is harmonic. Plugging these shifts into the $M2$-brane action~\eqref{M2_vertex_factors}, and using $\partial\Gamma_i = \gamma_i$ immediately shows that the vertex factor is gauge invariant. Although we will focus mostly on massless $U(1)$s in the following, let us note that the gauge invariance also extends to gauge transformations along massive diagonal $U(1)$s. The crucial point is that because the $M2$-brane world-volumes have one dimension along the external spacetime, the forms $\tw_{0A}$ are actually closed when pulled back to $\Gamma_i$ despite the fact that $d\tw_{0A}\neq 0$ in $\hat Y_4$. In particular, after pulling back to $\Gamma_i$ the shift $\delta C_3 = d\chi^{0A}\wedge \tw_{0A}$ is a shift by an exact form and we can again use partial integration to show the gauge invariance of the vertex factor~\eqref{M2_vertex_factors}.

\subsection{Selection rules for a non-chiral charged zero mode spectrum}
\label{sec:selectionRulesGen}

As we have seen, the form of the $M5$-instanton contribution to the low energy effective action depends on whether the flux pullback $\iota^* G_4$ vanishes or not, because non-vanishing flux implies that $M2$-brane vertex operators must be included. 
From a model building perspective it is of course important to be able to determine the form of the instanton contribution at least schematically, which makes it necessary to check in particular whether $\iota^* G_4 = 0$. 

Although this criterion looks deceptively simple, it is actually non-trivial to evaluate in practice. Of course, vanishing of the pullback is equivalent to requiring that the integral
\be
\int_{C_{(4)}} \iota^* G_4
\label{int_G4_4cycles}
\ee
vanishes for all 4-cycles $C_{(4)}\in H_4(D_M)$. The problem of determining $\iota^* G_4$ is thus equivalent to the problem of constructing the 4-cycles in $H_4(D_M)$. 

Of course, constructing all possible 4-cycles on the instanton divisor $D_M$ is in general a challenging mathematical problem. However, it was pointed out in~\cite{Kerstan:2012cy} that due to the properties of the $G_4$ flux only a very restricted class of 4-cycles has any chance of yielding a non-zero integral~\eqref{int_G4_4cycles}. For example, as $\iota^* G_4$ arises by pullback from $\hat Y_4$ it is clear that the cycle must be topologically non-trivial when viewed as a cycle in $\hat Y_4$. A further simplification arises if we consider M-theory compactifications that can be uplifted to 4-dimensional F-theory vacua. In this case 4-dimensional Lorentz invariance requires that, figuratively speaking, $G_4$ has exactly one leg in the elliptic fiber~\cite{Grimm:2011sk, Kerstan:2012cy}. In other words, F-theory fluxes have vanishing integrals over 4-cycles which either lie completely in the base or wrap the elliptic fiber. We will stick to fluxes of this type in the following. Our discussion of the 4-cycles which may yield a non-vanishing integral~\eqref{int_G4_4cycles} and can thus be used to detect a possible flux pullback is based on reference~\cite{Kerstan:2012cy}.

It is once again possible to gain some useful intuition from the type IIB picture. The discussion in the previous subsection leads us to expect that in the type IIB language $\iota^* G_4 \neq 0$ signals the existence of a chiral charged zero mode spectrum on the instanton. It was explicitly demonstrated in section~\ref{sec:flux_on_D3inst} that in the type IIB setting the net chirality of the charged zero modes is equal to the $U(1)$ charge of the instanton action. The IIB superpotential of an instanton with charged zero modes takes the form
\be
W \propto \cO e^{-S_E},
\label{charged_inst_IIB}
\ee 
with $\cO$ an operator built from charged matter fields whose $U(1)$ charge exactly cancels that of the instanton action. Let us point out that actually the $M5$ contribution~\eqref{superpot_with_M2} can be interpreted in a similar way. Defining $\hat{V}_{M2,i} = \exp \left[ -\pi i \int_{\Gamma_i} C_3 \right]$, we can cancel the $G_4$-dependent contribution in~\eqref{superpot_with_M2} using~\eqref{M2_cycles} to rewrite
\be
\label{superpot_with_M2_2}
W_{M5} \propto \int \cD \cB \ \prod_i \hat{V}_{M2,i} \ e^{-S_{M5}}.
\ee
Note that although the $V_{M2,i}$ are gauge invariant as mentioned at the end of the previous subsection, this is not true of the $\hat{V}_{M2,i}$.
In complete analogy with~\eqref{charged_inst_IIB} we thus have the usual contribution $e^{-S_{M5}}$ that would also be present in the absence of $G_4$ flux, times a charged operator $\prod_i \hat{V}_{M2,i}$ whose $U(1)$ charge cancels that of the instanton action. 

Let us take a brief detour at this point to mention an important observation that was made in~\cite{Grimm:2011dj} regarding the implications of chiral charged zero modes in IIB and F-theory GUT models. Recall from section~\ref{sec:flux_on_D3inst} that instantons whose superpotential contributions take the form~\eqref{charged_inst_IIB} cannot participate in K\"ahler moduli stabilisation in type IIB GUT models. The problem is that the operator $\cO$ with non-vanishing $U(1)$ charge will automatically be charged also under the non-Abelian part of the GUT group, so that the vacuum expectation value of $\cO$ must vanish in order to leave the visible gauge group unbroken. In F-theory models on the other hand it is possible to construct operators which have a non-zero $U(1)$ charge while being uncharged with respect to the GUT group. To illustrate the principle, let us consider a model with an $SU(5)$ stack and two additional $U(1)$ $D7$-branes as in~\cite{Grimm:2011dj}. At the intersection of the three branes the $SU(5)$ singularity experiences a rank two enhancement to $SO(12)$, $SU(7)$ or $E_6$. Following the general logic of section~\ref{sec:FtheoryGUTs}, these rank 2 enhancements encode the possible matter field interactions or in other words the ways in which operators $\cO$ as in~\eqref{charged_inst_IIB} can be constructed as products of matter fields. Focusing on the $E_6$ case, we recall the decomposition of the adjoint from~\eqref{Yukawa_enhancement}
\be
\begin{aligned}
 \mathbf{78} \rightarrow  \mathbf{24}^{(0,0)} \oplus \left[ \big(  \right. &  \mathbf{1}^{(0,0)} \oplus \mathbf{1}^{(-5,-3)}   \left.\left. \oplus  \mathbf{5}^{(-3,3)} \oplus \mathbf{10}^{(-1,-3)} \oplus \mathbf{10}^{(4,0)} \right) \oplus \text{ c.c.} \right].
\end{aligned}
\ee
The states denoted $\mathbf{1}^{(-5,-3)}$ are the sought-after couplings which are GUT singlets while nevertheless carrying non-vanishing $U(1)$ charges. These operators are not prevented from acquiring a non-zero vacuum expectation value. An instanton contribution of the form~\eqref{charged_inst_IIB} involving these GUT singlets can thus play a role in K\"ahler moduli stabilisation despite the non-vanishing $U(1)$ charge of the instanton. Note that $SU(5)$ GUT models with an $E_6$ point do not admit a smooth type IIB limit~\cite{Krause:2012yh}, so this is not in conflict with the IIB discussion of section~\ref{sec:flux_on_D3inst}. 

Returning to the $M5$-instanton contribution in~\eqref{superpot_with_M2_2}, the IIB intuition discussed above leads us to expect that vertex factors appear if the $M5$-instanton action experiences a non-vanishing shift under $U(1)$ gauge transformations. As shown in appendix~\ref{sec:app_democ_Mtheory}, the shift of the fields $\tilde{c}_\alpha$ appearing in the expansion of $C_6$ under such a gauge transformation is\footnote{Here and in the following we use the notation of section~\ref{sec:U(1)inFtheory} and appendix~\ref{sec:app_democ_Mtheory}.}~\cite{Kerstan:2012cy}
\be
\label{gauge_trf_A}
A^\Lambda \rightarrow A^\Lambda + d\Lambda^\Lambda \ \Rightarrow \ \delta_{\Lambda^\Lambda} \tilde{c}_\alpha =  4 \Theta_{\alpha  \Lambda} \Lambda^\Lambda.
\ee 
The coefficient matrix appearing in $\delta_{\Lambda^\Lambda} \tilde{c}_\alpha$ is given as before by
\be
\Theta_{\alpha\Lambda} = -\frac12 \int \omega_\alpha \wedge \omega_\Lambda \wedge G_4.
\ee
Applying this to the $M5$-instanton action $S_{M5} = \pi i C^\alpha_E \tilde{c}_\alpha +...$ we have
\be
\delta_{\Lambda^\Lambda} S_{M5} = 4 \pi i C^\alpha_E \Theta_{\alpha \Lambda} \Lambda^\Lambda = - 2 \pi i\int_{D_M} \iota^* \omega_\Lambda \wedge \iota^* G_4.
\label{shift_M5_action}
\ee
As $G_4$ is orthogonal to surfaces wrapping the elliptic fiber or lying completely in the base, the classification of the various 2-forms $\omega_\Lambda$ given in section~\ref{sec:KKreduxFtheory} shows that non-vanishing shifts are possible if~\eqref{gauge_trf_A} corresponds to a gauge transformation of a massless low energy $U(1)$ symmetry. In other words, we have identified one possible set of 4-cycles $C_{(4)}$ which give a non-vanishing integral~\eqref{int_G4_4cycles}. These 4-cycles can be constructed as the Poincar\'e dual of $\iota^* \omega$ in $D_M$ or of $[D_M]\wedge \omega$ in $\hat Y_4$, with $\omega$ a 2-form corresponding to either a Cartan $U(1)$ or a massless additional non-Cartan $U(1)$. In phenomenologically interesting models Cartan fluxes are usually switched off to avoid breaking the non-Abelian part of the low energy gauge group. However even in this case we obtain a non-trivial selection rule from~\eqref{shift_M5_action} if additional non-Cartan massless $U(1)$s are present in the model.

Let us briefly mention at this point that the supergravity argumentation can be extended directly to gauge transformations of massive $U(1)$ by including the non-harmonic forms $\tw_{0A}$ of section~\ref{sec:massiveU(1)sFtheory}. This strongly suggests that the well-known type IIB selection rules due to geometrically massive $U(1)$ continue to hold in F-theory~\cite{Kerstan:2012cy}. In other words, we expect an instanton charged under a geometrically massive $U(1)$ to lead to effective contributions involving $M2$-brane vertex operators as in~\eqref{superpot_with_M2_2}. We will present additional evidence for this expectation in the next subsection.

Of course, it must be possible to recover the selection rules due to geometrically massive $U(1)$s even in the massless F-theory reduction by considering suitable 4-cycles $C_{(4)}$ without explicitly using the non-harmonic form $\tw_{0A}$. 
At a technical level it is anyway desirable to understand the relevant 4-cycles directly, because it is not clear how to construct the form $\tw_{0A}$ explicitly. Furthermore, there is no reason to expect that all selection rules in F-theory have a connection with a type IIB gauge symmetry~\cite{Kerstan:2012cy}, so in general we expect that $D_M$ can contain additional 4-cycles. It was argued in~\cite{Kerstan:2012cy} that these 4-cycles are not described by algebraic surfaces for general values of the complex structure moduli and can thus in particular not be viewed as an intersection of two divisors like the surfaces $D_M\cap [\omega_\Lambda]$ considered above.
Roughly speaking, two types of cycles may be considered. One type is related to the matter surfaces located at the intersection of two stacks of 7-branes~\cite{Kerstan:2012cy}. The second type is related to the Higgsing of a massless $U(1)$ symmetry, which is previously described by a harmonic form $\omega$ and has an associated $G_4$ flux $G_4 \sim \cF \wedge \omega$. As explicitly discussed in~\cite{Braun:2011zm}, the $G_4$ flux is transformed into a non-algebraic 4-form which can no longer be written as a wedge product of two 2-forms when one moves into a locus in complex structure moduli space in which the $U(1)$ is Higgsed and $\omega$ ceases to exist. The non-algebraic cycles mentioned here are difficult to describe in full generality due to the fact that their Poincar\'e dual 4-form does not factorise. For this reason, rather than attempting a general description we follow~\cite{Kerstan:2012cy} and consider a specific model which suffices to illustrate the essential features. In particular, it will be clear how to generalise the results from the $SU(5)\times U(1)_X$ model which we will consider to models with $SU(N), N\neq 5$ or to models with multiple $U(1)$s.

\subsection{Selection rules in the $SU(5)\times U(1)_X$ model and comparison to the type IIB results}
\label{sec:selectionRulesExample}

In this subsection we consider an $SU(5)\times U(1)$ model which can globally be described as the vanishing locus of a Tate polynomial
\be
P_T = x^3 - y^2 + a_1 x y z + a_2 x^2 z^2 + a_3 y z^3 + a_4 x z^4 + a_6 z^6 = 0.
\label{P_Tate_sel_rule}
\ee
As in~\eqref{a_i_SU(5)} we ensure that the model exhibits an $SU(5)$ singularity along a divisor $\cW^b= \{w=0\}$ in the base $B_3$ by assuming $a_i = a_{i,i-1} w^{i-1}$, with $a_{i,i-1}$ not containing an overall factor of $w$. We furthermore set $a_6 = 0$, in which case the model exhibits a massless $U(1)$ symmetry in addition to the $SU(5)$~\cite{Grimm:2010ez, Krause:2011xj}. Of course, as discussed in section~\ref{sec:U(1)geometry} this is not the only way in which an additional $U(1)$ symmetry can be engineered, and to distinguish it from other possibilities the $U(1)$ obtained by setting $a_6=0$ is often referred to as $U(1)_X$~\cite{Grimm:2010ez, Krause:2011xj}.

The singularities of the $SU(5)\times U(1)_X$ model can be resolved by introducing a set additional coordinates $s, e_i, 1=1,...,4$ together with appropriate scaling relations, and considering the transformed Tate equation~\cite{Krause:2011xj}
\be
\label{tate_proper_transform}
 \begin{aligned}
P_T: \{  y^2\,s\,e_3\,e_4 &- a_1\,x\,y\,z\,s - a_{3,2}\,y\,z^3\,e_0^2\,e_1\,e_4 \\
                   &= x^3\,s^2\,e_1\,e_2^2\,e_3 + a_{2,1}\,x^2\,z^2\,s\,e_0\,e_1\,e_2 + a_{4,3}\,x\,z^4\,e_0^3\,e_1^2\,e_2\,e_4\}.
 \end{aligned}
\end{equation}
The loci $E_i \equiv \{e_i=0\}$ and $S \equiv \{s=0\}$ are the resolution divisors corresponding to the Cartan generators of the $SU(5)$ and to the additional $U(1)$ singularity, respectively. $E_0 = \{e_0 = 0\}$ describes the proper transform of the original singular divisor $\cW = \{w=0\}$, such that $\cW = \sum_{I=0}^4 E_I $. The 2-form $\tw_{X}$, which describes the additional $U(1)_X$ in the expansion of $C_3$, is related to but not identical to the resolution form $S$. The precise expression for $\tw_{X}$ was worked out in~\cite{Krause:2011xj} using the requirement that $\tw_X$ must be orthogonal to the intersection of three vertical divisors, to the section $Z$ of the elliptic fibration intersected with two vertical divisors, and to the $SU(5)$ Cartan generators. The result is~\cite{Krause:2011xj}
\be \label{twX}
 \tw_X = 5(S - Z - \bar{K}_B) + (2,4,6,3)_i E_i,
\ee
with $\bar{K_B}$ denotes the first Chern class of the anticanonical bundle of the base.

We now turn to the types of $G_4$ flux that can be constructed in the present model. Quite generally it is possible to distinguish between 'vertical' fluxes which can be written as a sum of products of 2-forms\footnote{In mathematical terms such fluxes are said to lie in the primary vertical subspace $H^{2,2}_{ver.}(\hat Y_4)$~\cite{Greene:1993vm, Krause:2012yh}.} and 'horizontal' non-algebraic fluxes that cannot be factorised in this manner. Fluxes which admit a factorisation into a sum of products of 2-forms are not only much easier to describe explicitly than the non-algebraic fluxes, but also automatically fulfill the F-term constraints regardless of how one moves around the complex structure moduli space~\cite{Krause:2012yh}. We will therefore focus on vertical fluxes in the following. The possible vertical $G_4$ fluxes in the $SU(5)\times U(1)_X$ model were classified in~\cite{Krause:2012yh}, where it was shown that in addition to the Cartan fluxes $G_4 \sim \cF \wedge E_i$ exactly two types of consistent flux can be constructed. One is the standard flux associated with the generator of the additional massless $U(1)_X$, given by\footnote{Our normalisation of the fluxes differs from that used in~\cite{Krause:2012yh} by an overall factor of 5, to match the normalisation of the $\tw_X$ in~\cite{Krause:2011xj}.}
\be
\label{G_4_X}
G^X_4  =  - \cF^{b} \wedge \tw_X
\ee
with $\cF^b$ a 2-form pulled back from $B_3$. The second type of flux, 
\be
G^\lambda_4 = \lambda ( 5E_2\wedge E_4 + (2,-1,1,-2)_i E_i \wedge \bar{K}_B ),
\label{G_4_lambda}
\ee
is not directly associated with a massless $U(1)$ symmetry. As argued in~\cite{Krause:2012yh} it describes a type of flux that is often considered in the spectral cover approach to F-theory model building, which is why it will sometimes be referred to as spectral cover flux. The parameters $\cF^b$ and $\lambda$ must of course be chosen in an appropriate manner in order to fulfill the Freed-Witten quantisation condition~\cite{Collinucci:2010gz, Collinucci:2012as}
\be
G_4 + \frac12 c_2(\hat Y_4) \in H^4 (\hat Y_4, \mathbb{Z}).
\ee
We will implicitly assume in the following that this is the case.

The 2-form $\tw_X$ associated with the massless $U(1)_X$ immediately allows us to construct a surface which can be used to test for a non-vanishing pullback of $G_4$ along the lines of~\eqref{int_G4_4cycles}, by choosing~\cite{Kerstan:2012cy}
\be
[C_{(4)}^X] = -  D_M \wedge \tw_X.
\label{cycle_U1_inst}
\ee
The integrals of the fluxes over this 4-cycle can be evaluated using the various intersection numbers given in~\cite{Krause:2012yh}, with the result~\cite{Kerstan:2012cy}
\begin{align}
\label{sel_rule_G^X_over_C^X}
\int_{C_{(4)}^X} \iota^* G_4^X  &= - 10 \int_{B_3} (5\bar{K}_B - 3 {\cal W}^{b} )\wedge D^{b}_M \wedge \cF^{b} , \\
\int_{C_{(4)}^X} \iota^* G_4^\lambda &= -  5 {\lambda} \int_{B_3} \bar{K}_B \wedge {\cal W}^{b} \wedge D_M^{b} \label{sel_rule_G^lambda_over_C^X}.
\end{align}

A second type of 4-cycles that can be considered is related to the matter surfaces, which consist of additional $\bP^1$'s fibered over codimension 2 loci in the base over which the singularity type enhances. Before explicitly considering surfaces $C_{(4)} \subset D_M$, let us briefly review the construction of matter surfaces in the chosen $SU(5)\times U(1)_X$ model. Recent investigations into the matter surface structure in the context of $SU(5)$ GUT models include~\cite{Marsano:2011hv, Esole:2011sm, Esole:2011cn, Lawrie:2012gg}, but we stick here to the approach and notation of~\cite{Krause:2011xj}. One matter surface in the $SU(5)\times U(1)_X$ model is fibered over the curve of $SU(2)$ enhancement, which after resolution gives rise to the additional divisor $S$ and is denoted by $\cC_{1_{-5}}$. Further matter surfaces are fibered over matter curves lying inside the GUT divisor $\cW^b$, over which the singularity enhances beyond $SU(5)$. More precisely, one finds an enhancement to $SO(10)$ over the curve $\cC_{10_1} = {\cal W}^{b} \cap \{ a_1 = 0 \}$ and enhancement to $SU(6)$ over the two curves $\cC_{5_3} = \cW^b\cap \{a_{3,2} = 0\}$ and $\cC_{5_{-2}} = \cW^b\cap\{a_{3,2}a_{2,1} - a_1 a_{4,3}=0\}$. Note that there are in general multiple matter surfaces associated to a given matter curve, corresponding to different choices of the linear combination of $\bP^1$'s which is fibered over the matter curve. However, all matter surfaces associated with a fixed matter curve lead to states in the same $SU(5)$ representation and carrying the same $U(1)_X$ charge. As a consequence, the integral of $G_4$ flux over a matter surface depends only on the matter curve, and not on the explicit choice of the matter surface over the matter curve~\cite{Krause:2011xj, Krause:2012yh}. The naming of the matter curves follows the transformation behaviour of the associated matter states, so e.g. matter surfaces over $\cC_{10_1}$ lead to states in the $\mathbf{10}$ representation of $SU(5)$ with unit $U(1)_X$ charge.

Following~\cite{Kerstan:2012cy}, we focus on the matter surfaces $C_{10_1}^k, k=1,...,10$ associated with the matter curve $\cC_{10_1}$. As mentioned above, we can pick an arbitrary representative matter surface to evaluate the $G_4$ integrals, because this integral does not depend on the choice of $k$. A convenient choice is the matter surface denoted by $C_{24}$ in~\cite{Krause:2011xj}, which can be written as a total intersection in the ambient 5-fold
\be
\label{P124}
{C}_{24}: \{e_2=0\} \cap \{e_4=0\} \cap \{ a_1=0  \}.
\ee
It is straightforward to check that this surface lies inside the fourfold $\hat Y_4$ described by~\eqref{tate_proper_transform}.
Note that it is not possible to rewrite~\eqref{P124} as an intersection of two further equations with the Tate equation. In other words, the Poincar\'e dual to $C_{24}$ in $\hat Y_4$ cannot be written as a product of two 2-forms. Nevertheless, the integrals of $G_4^X$ and $G_4^\lambda$ over $C_{24}$ can be evaluated without too much difficulty because it is given by an intersection of 5 divisors in the ambient 5-fold $X_5$. In this way, the authors of references~\cite{Krause:2011xj, Krause:2012yh} obtain
\be
\label{chiralities_10_matter}
\begin{aligned}
\int_{C_{24}} G_4^X &= - \int_{\cC_{10_1}} \cF^b, \qquad \qquad
\int_{C_{24}} G_4^\lambda &= \lambda \int_{\cC_{10_1}} 5\cW^b - 6\bar{K}_B. 
\end{aligned}
\ee

Let us now return to the instanton wrapped on $D_M$. Note that as $D_M$ is a vertical divisor, it is described inside both $X_5$ and $\hat Y_4$ by the vanishing locus $\{ D_M = 0\}$ of a polynomial depending only on the base coordinates. This suggests that we can build a surface analogous to $C_{24}$ but contained inside $D_M$ by replacing $a_1$ in~\eqref{P124} with the polynomial $D_M$. In other words, we consider the intersection
\be
\label{c_24_dm}
{C}_{24}|_{D_M}: \{e_2=0\} \cap \{e_4=0\} \cap \{ D_M=0  \}
\ee
in the ambient 5-fold $X_5$.
A look at the integrals~\eqref{chiralities_10_matter} suggests that upon replacing $a_1 \rightarrow D_M$ one obtains~\cite{Kerstan:2012cy}
\begin{align}
\label{sel_matter_surface_Glambda}
 \int_{{C}_{24}|_{D_M}} \iota^* G^\lambda_4 &=  \lambda \int_{B_3}  D_M^{\rm b} \wedge {\cal W}^{\rm b} \wedge (6 \bar{\cK} - 5 {\cal W}^{\rm b}), \\    \int_{{C}_{24}|_{D_M}} \iota^* G^X_4 & =  \int_{B_3}  D_M^{\rm b} \wedge {\cal W}^{\rm b} \wedge {\cal F}^{\rm b}.
 \label{sel_matter_surface_Gx}
\end{align}
It thus looks very much as if the surface ${C}_{24}|_{D_M}$ can indeed be used to derive a non-trivial selection rule for the absence of chiral charged zero modes on the $M5$-instanton. However, a potential subtlety was pointed out in~\cite{Kerstan:2012cy}. The issue is that while ${C}_{24}|_{D_M}$ is clearly contained in $D_M \subset X_5$, it will in general not lie within $D_M \cap \hat Y_4$. It was argued in~\cite{Kerstan:2012cy} that despite this, a closely related non-algebraic surface is expected to exist within $D_M \cap \hat Y_4$ and that the integral of $G_4$ over this non-algebraic surface equals the integral over ${C}_{24}|_{D_M} \subset X_5$. In fact, as we will review in the following this can be shown explicitly at least if certain mathematical assumptions are fulfilled.

The basic idea is that ${C}_{24}|_{D_M}$ clearly lies inside $\hat Y_4$ if $E_i\cap D_M$ is contained in $E_i \cap \{a_1\}$ for $i=2$ or $i=4$. As the divisors $E_i$ are all fibered over $\cW^b$, this is equivalent to requiring~\cite{Kerstan:2012cy}
\be
D_M^b \cap \cW^b \subset a_1.
\label{4cycle_inside_instanton}
\ee 
Of course, for general choices of $a_1$ or in other words for general choices of the complex structure moduli of $\hat Y_4$ equation~\eqref{4cycle_inside_instanton} will not be fulfilled. However, as argued in~\cite{Kerstan:2012cy} it is sufficient if there is a surface $\{\tilde{a}_1=0\}$ in the same class $\bar{K}_B$ as $\{a_1=0\}$ inside which $D_M^b \cap \cW^b$ is contained. At the level of the defining functions, this is the case if there exists an integer $n$ and suitable sections $\alpha$ and $\beta$ on $B_3$ such that~\cite{Kerstan:2012cy}
\be
(\tilde{a}_1)^n = \alpha D_M^b  + \beta \cW^b.
\label{specialisation1}
\ee
Note that we can allow for meromorphic $\alpha$ and $\beta$ as long as all poles lie outside of the locus $D_M^b \cap \cW^b$. If equation~\eqref{specialisation1} is fulfilled, ${C}_{24}|_{D_M}$ is contained in the auxiliary fourfold $\{ \tilde{P}_T =0\}$, where $\tilde{P}_T$ is obtained from $P_T$ in~\eqref{tate_proper_transform} by replacing $a_1$ with $\tilde{a}_1$. Because this auxiliary fourfold is obtained from $\hat Y_4$ by a smooth deformation from $a_1$ to $\tilde{a}_1$, a 4-cycle $C_{(4)}$ related to ${C}_{24}|_{D_M}$ must have existed already in $\hat Y_4$~\cite{Kerstan:2012cy}. In the course of the complex structure deformation taking $\tilde{a}_1\rightarrow a_1$ the cycle $C_{(4)}$ can become non-holomorphic. At the level of the Poincar\'e dual 4-form this means that it acquires pieces of Hodge type (1,3) and (3,1) in addition to keeping the original piece of type (2,2). The crucial point is that the deformation nevertheless does not change the values of the integrals~\eqref{sel_matter_surface_Glambda} and~\eqref{sel_matter_surface_Gx}, due to the fact that $G_4$ has Hodge type (2,2)~\cite{Kerstan:2012cy}.

The question remains how restrictive the assumption~\eqref{specialisation1} is. We follow~\cite{Kerstan:2012cy} and focus on the case where $B_3$ is a hypersurface (or total intersection) in a toric space, which enables us to intuitively picture the sections appearing in~\eqref{specialisation1} as homogeneous polynomials in the toric coordinates. This is anyway the case which lends itself most directly to model building purposes due to the calculational power of the toric approach. Most model building applications furthermore focus on rigid instanton divisors $D_M$ to avoid vanishing contributions due to neutral deformation zero modes. In the toric language it is clear that rigidity requires $D_M$ to be described by a polynomial of quite a low degree. In many cases this will also be true of the GUT divisor $\cW^b$, whose degree is restricted by the fact that $a_{2,1}$, which lies in the class $2\bar{K}_B-\cW^b$, must be a holomorphic polynomial. If the degrees of both $\cW^b$ and $D_M$ are lower than the degree of $a_1$ with respect to every one of the toric scaling relations, then it is clear that~\eqref{specialisation1} can be fulfilled with $n=1$. Slightly more generally, it was pointed out in~\cite{Kerstan:2012cy} that in the toric setting $\bar{K}_B$ is often a so-called big divisor. Essentially, this means that the polynomial $a_1$ has a positive charge with respect to every toric scaling relation. For any arbitrary but fixed divisor $D$ one can then always find an integer $n$ such that $a_1^n/D$ has positive toric charges, such that $a_1^n/D$ corresponds to an effective divisor class and can be represented by a holomorphic polynomial. This means that it is possible to write~\cite{Kerstan:2012cy}
\be
(a_1)^n = \alpha D^{ b}_M + \beta {\cal W}^{b} + \delta.
\ee
The question is therefore whether by a suitable choice of $\alpha$, $\beta$ and $n$ it is possible to achieve
\be
(a_1)^n - \delta = (\tilde{a}_1)^n,
\ee
with $\tilde{a}_1$ in the same class $\bar{K}_B$ as $a_1$. Due to the large amount of freedom in choosing $\alpha$, $\beta$ and $n$ we generically expect this to be possible~\cite{Kerstan:2012cy}. It would of course be desirable to prove this rigorously and also to extend the analysis to the case where $[a_1]$ is not a big divisor, but this is beyond the scope of this thesis.

To summarise, we expect that beyond the surfaces~\eqref{cycle_U1_inst} associated to massless $U(1)$ symmetries there do exist additional 4-cycles in $D_M\subset \hat{Y}_4$ which can support a non-vanishing integral of $G_4$. For general choices of the complex structure moduli these 4-cycles are not holomorphic, so that the Poincar\'e dual 4-forms are not purely of Hodge type (2,2). However, by a suitable smooth deformation of the geometry we expect that it is possible to relate them to holomorphic 4-cycles. The deformations do not change the (2,2) part of the cycles and thus the value of the integral of $G_4$, so that it is possible to evaluate the selection rules at the point in complex structure moduli space at which the cycles are holomorphic. Furthermore, we have argued that the relevant 4-cycles are related to fibrations of $\bP^1$'s over suitable curves lying inside the loci of singularity enhancement. We explicitly demonstrated the relevant construction for a 4-cycle related to the matter surface $C_{24}$ in the $SU(5)\times U(1)_X$ Tate model of~\cite{Krause:2011xj}. Similar constructions are of course possible starting with one of the other matter surfaces over $\cC_{10_1}$ or indeed over one of the other matter curves~\cite{Kerstan:2012cy}. It is natural to expect that the cycle that is obtained when starting from the matter surface over $\cC_{1_{-5}}$, which is the locus corresponding to the extra $U(1)_X$, is related to the 4-cycle $C_{(4)}^X$ given in~\eqref{cycle_U1_inst}, although we will not attempt to show this explicitly. This is in particular relevant if the geometry is deformed in such a way that the $U(1)_X$ is Higgsed, such that the form $\tw_X$ ceases to exist and both $G_4^X$ and $[C_{(4)}^X]$ can no longer be taken to lie in $H^{2,2}_{ver.}(\hat Y_4)$~\cite{Kerstan:2012cy}. It is clear that at least in principle the discussion of the present subsection can be adapted to other models with gauge group $SU(N), N\neq 5$ or to models with multiple massless $U(1)$ symmetries.

Many of the 4-cycles that can be described using the construction outlined above will of course lead to equivalent selection rules. This is in particular clear for the various cycles related to different matter surfaces over the same matter curve, because the integral of $G_4$ over all these 4-cycles will be equal~\cite{Krause:2011xj, Krause:2012yh}. Furthermore, even cycles constructed from matter surfaces associated with different matter curves are not necessarily linear independent~\cite{Kerstan:2012cy}. This means that to check whether $\iota^* G_4=0$ it is generically not necessary to evaluate the integral of $G_4$ over every 4-cycle $C_{(4)}$ that can possibly be constructed. However, unfortunately there does not seem to be an easily available method of determining when one has obtained a complete set of selection rules for the absence of charged chiral zero modes. In models with a dual type IIB description, one can gain some information about the number and the underlying physical reason for the selection rules by comparing the F-theory results to the known type IIB selection rules. This will now be demonstrated for the $SU(5)\times U(1)_X$ model considered above, whose type IIB limit has been discussed very explicitly in~\cite{Krause:2012yh}. As before, our exposition will be based on referene~\cite{Kerstan:2012cy}.

The type IIB selection rules for the absence of chiral charged zero modes are obtained by requiring the chiral indices~\eqref{net_chirality_E3inst} to vanish for every $D7$-brane stack in the model. To evaluate these chiral indices we require knowledge of the type IIB $D7$-brane divisors as well as the gauge flux configuration on the $D7$-branes. For the orientifold dual of the $SU(5)\times U(1)_X$ model discussed above, with fluxes $G_4^X,\, G_4^\lambda$ as in~\eqref{G_4_X} and~\eqref{G_4_lambda}, the relevant data was derived in~\cite{Krause:2012yh}. In the double cover $\pi_O: X_3 \rightarrow B_3$ of the base the $SU(5)$ symmetry arises from a stack of 5 $D7$-branes wrapped on a divisor $D_A$ and the image stack on the divisor $D_A'$. The divisors $D_A, D_A'$ are related to the F-theory GUT divisor by~\cite{Kerstan:2012cy} $[\pi_O(D_A+D_A')] = \cW^b$. To satisfy the $D7$-brane tadpole cancellation condition the IIB model in addition contains a single $D7$-brane\footnote{Of course we must also include an image brane on the divisor $D_B'$.} on the divisor~\cite{Krause:2012yh}
\be
D_B = 4 D_{O7} - 2 {D_A} - 3 {D'_A}.
\ee
Note that the class of the $O7$-plane corresponds to the anticanonical class of the base
\be
D_{O7} = \pi_O^*\bar{K}_B.
\ee
As before, the type IIB $O(1)$ $E3$-instanton is wrapped on the divisor $D_E = \pi_O^* D_M^b$.

The type IIB low energy supergravity contains the two independent $U(1)$ symmetries $U(1)_{A}$, $U(1)_B$ associated with the brane-image-brane pairs\footnote{In the case of the non-Abelian stack on $D_A$, $U(1)_A$ refers to the diagonal $U(1) \subset U(5)$.} on $D_A$ and $D_B$. Note that as neither of the divisors is orientifold-invariant, the two $U(1)$ symmetries are both geometrically massive according to the general discussion of section~\ref{sec:Stueckelberg}. However, a massless linear combination of the two $U(1)$ gauge bosons was identified in\footnote{To simplify the notation and in particular the $U(1)$ charges of the various fields we suppress the explicit factor of $2\pi$ between the M-theory and IIB gauge fields, see equation~\eqref{relation_AA}.}~\cite{Krause:2012yh}
\be
\label{massless_u1}
U(1)_X = \frac12 (U(1)_A - 5\ U(1)_B).
\ee
As the notation suggests, this massless $U(1)$ is the IIB equivalent of the massless $U(1)_X$ symmetry that was present in the F-theory model. Note that this correspondence also fixes the sign in~\eqref{massless_u1}, which is of course not fixed by masslessness. To see this we consider the $U(1)_{A,B}$ charges of the type IIB matter fields at the various intersections. For example, we find fields in the representation $\mathbf{5}_{1,-1}$ at the locus $D_A\cap D_B$ and fields in $\mathbf{5}_{1,1}$ at the locus $D_A \cap D_B'$~\cite{Blumenhagen:2006ci, Krause:2012yh}. Here the first and second indices denotes the charges with respect to $U(1)_A$ and $U(1)_B$, respectively. In order for the $U(1)$ charges to match the $U(1)_X$ charges of the representations $\mathbf{5}_3$ and $\mathbf{5}_{-2}$ that we had on the F-theory side we must use the identification~\eqref{massless_u1}.

The general non-trivial flux configurations along $U(1)_{A/B}$ that fulfill the $D5$-brane tadpole cancellation conditions were studied in~\cite{Krause:2012yh}, with the result that exactly two independent types of consistent fluxes exist. Furthermore, reference~\cite{Krause:2012yh} showed that the flux configurations have a direct correspondence with the F-theory $G_4$ fluxes $G_4^X$ and $G_4^\lambda$. The results may be summarised as~\cite{Krause:2012yh, Kerstan:2012cy}
\begin{align}
\label{fluxIIB_G_4^X}
G_4^X : \qquad & \tilde{\cF}^+_A = - \frac{1}{2} \pi_O^*(\cF^b), \quad & \tilde{\cF}^+_B = + \frac{5}{2} \pi_O^*(\cF^b),\quad \cF^-_{A,B}=0, \\
G_4^\lambda : \qquad & \tilde{\cF}^+_A = - 2  \lambda D_{O7}, \quad & \tilde{\cF}^+_B = 0,\quad\quad\quad \cF^-_{A,B}=0. \label{fluxIIB_G_4^lambda}
\end{align}
These results can be checked by comparing the chiral indices of the various matter fields to the F-theory results. For example, focus on the configuration $G_4^X$ and consider the chiral index of the fields at the intersection between $D_A$ and $D_B$
\be
\begin{aligned}
\chi(\mathbf{5}_{1,-1})  = I_{BA}& = - 5 \int_{X_3} D_A \wedge D_B \wedge (\tilde{\cF}^B - \tilde{\cF}^A) \\& = - \int_{B_3}(9 \bar{K}_B - 6 \cW^b)\wedge \cW^b\wedge \cF^b.
\label{chiral_5_1_-1}
\end{aligned}
\ee
Here we inserted~\eqref{fluxIIB_G_4^X} and used the cohomological relation~\cite{Krause:2012yh} $D_{O7}\wedge D_A = D_A\wedge D_A'$. Furthermore we included a factor of 5 to account for the trace over the diagonal $U(1)_A$ generator. On the F-theory side, the corresponding chiral index induced by the flux $G_4^X$ of~\eqref{G_4_X} is\footnote{The minus sign in~\eqref{chiral_5_3} is due to the fact that our $U(1)_X$ is normalised differently to the one used in~\cite{Krause:2011xj}, where the states on the curve $\{a_{3,2} =0\}$ carried charge $-3$. This minus sign is important to obtain the correct signs in~\eqref{fluxIIB_G_4^X}.}~\cite{Krause:2011xj}
\be
\label{chiral_5_3}
\chi(\mathbf{5}_3) = \int_{C^k_{5_3}} G_4^X = - q(\mathbf{5}_3) \int_{\cC_{5_3}} \cF^b.
\ee
Using $[\cC_{5_3}]=[\{a_{3,2} =0\}] = 3\bar{K}_B - 2\cW^b$ and $q(\mathbf{5}_3)=3$ we see that this matches~\eqref{chiral_5_1_-1}, which in particular confirms~\eqref{fluxIIB_G_4^X}.

Having obtained the fluxes~\eqref{fluxIIB_G_4^X} and~\eqref{fluxIIB_G_4^lambda} it is straightforward to evaluate the net chiralities of the charged zero modes at the intersections of the $D7$-branes with the $E3$-instanton on $D_E$. Let us focus first on the flux configuration corresponding to $G_4^X$. Possible fluxes on the instanton as considered in section~\ref{sec:flux_on_D3inst} are set to zero throughout this section. As we are considering an $O(1)$ instanton with $D_E=C^\alpha_E \omega_\alpha$ and the orientifold-odd parts of the fluxes are zero we find
\be
\begin{aligned}
\frac12(I_{EA}-I_{EA'}) = I_{EA} &= -\frac54\int_{X_3} (D_A+D_A')\wedge D_E \wedge\pi_O^*(\cF^b) \\&= -\frac52 \int_{B_3} \cW^b\wedge D_M^b \wedge \cF^b.
\end{aligned}
\ee
In a similar manner we find
\be
\begin{aligned}
\frac12(I_{EB}-I_{EB'})  = I_{EB}& = \frac54\int_{X_3} (8D_{O7}-5(D_A+D_A'))\wedge D_E \wedge\pi_O^*(\cF^b) \\& = \frac52 \int_{B_3} (8\bar{K}_B - 5\cW^b)\wedge D_M^b\wedge \cF^b.
\end{aligned}
\ee
Combining these results using~\eqref{massless_u1} yields the net charge of the instanton zero modes with respect to the massless $U(1)_X$
\be
\frac12(I_{EA}-5I_{EB}) = -10\int_{B_3} (5\bar{K}_B-3\cW^b)\wedge D_M^b\wedge \cF^b.
\label{net_chirality_U(1)_X}
\ee
This exactly matches~\eqref{sel_rule_G^X_over_C^X}, which strongly suggests that the integral $\int_{C_{(4)}^X} \iota^* G_4^X$ measures the net $U(1)_X$ charge of the $M5$-instanton zero modes also in F-theory. This is of course in agreement with our previous observation that according to~\eqref{shift_M5_action} the same selection rule is obtained by demanding invariance of the instanton action $S_{M5}$ with respect to $U(1)_X$ gauge transformations.

In the type IIB picture it is clear that requiring~\eqref{net_chirality_U(1)_X} to vanish is not a sufficient condition to ensure total absence of chiral charged zero modes, because this requires both $I_{EA}$ and $I_{EB}$ to vanish independently. To obtain a sufficient set of conditions in IIB we may simply pick any combination of $U(1)_A$ and $U(1)_B$ that is linear independent from $U(1)_X$ and require that the net chirality with respect to this combination must also vanish. However, recall that~\eqref{massless_u1} describes the only geometrically massless combination of $U(1)_A$ and $U(1)_B$. It is thus clear that the second selection rule in F-theory cannot be derived from the gauge invariance of the instanton action with respect to a massless $U(1)$~\cite{Kerstan:2012cy}. Therefore it must be described by a selection rule obtained from one of the non-algebraic 4-cycles $C_{(4)}\subset D_M$ discussed above. Indeed, a closer look at~\eqref{sel_matter_surface_Gx} shows that the vanishing of $\int_{{C}_{24}|_{D_M}} \iota^* G^X_4$ exactly corresponds to the vanishing of $I_{EA}$ in type IIB~\cite{Kerstan:2012cy}.

The discussion above confirms our expectation that the instanton selection rules associated with geometrically massive $U(1)$s in type IIB continue to play a role in F-theory despite the fact that the massive $U(1)$s are not directly visible in the harmonic dimensional reduction~\cite{Kerstan:2012cy}. Instead, the corresponding selection rules can be derived in F-theory from the Freed-Witten anomaly cancellation condition $\iota^* G_4 = 0$. Alternatively it is possible to introduce the non-harmonic forms $\tw_{0A}$ of section~\ref{sec:massiveU(1)sFtheory}, which as we have argued directly describe the F-theory analogues of the geometrically massive $U(1)_A$, $U(1)_B$ and not only of the massless combination~\eqref{massless_u1}. As shown in appendix~\ref{sec:app_democ_Mtheory}, the shift of the $M5$ action under a gauge transformation along such a massive $U(1)$ is still described by equation~\eqref{shift_M5_action}. Explicitly evaluating this shift using the intersection numbers and fluxes of section~\ref{sec:massiveU(1)sFtheory} shows that it exactly matches the shift of the IIB instanton action under the corresponding massive $U(1)$~\cite{Kerstan:2012cy}. In other words, we expect that at least in F-theory models with a well-defined orientifold dual all selection rules can be derived by demanding gauge invariance of the instanton action, provided that geometrically massive $U(1)$s and the non-harmonic forms necessary to describe them are taken into account.

We now turn to the flux configuration~\eqref{fluxIIB_G_4^lambda} corresponding to $G_4^\lambda$. The chiral indices of the instanton are found to be~\cite{Kerstan:2012cy}
\be
\begin{aligned}
\chi_{A,E}(G_4^\lambda) &= - 5 \lambda \int_{X_3} D_{O7} \wedge D_E \wedge D^+_A = - 10 \lambda\int_{B_3} \bar{\cal K} \wedge D_M^{\rm b} \wedge {\cal W}, \\
\chi_{B,E}(G_4^\lambda) & =0.
\end{aligned}
\ee
The net chirality of the charged zero modes with respect to $U(1)_X$ is therefore
\be
\frac12(I_{EA}-5I_{EB}) = - 5 \lambda\int_{B_3} \bar{\cal K} \wedge D_M^{\rm b} \wedge {\cal W},
\label{chiral_inst_U1X_Glambda}
\ee
which is again in perfect agreement with $\int_{C_{(4)}^X} \iota^* G_4^\lambda$ given in~\eqref{sel_rule_G^lambda_over_C^X}. Note that as $I_{EB}=0$ the vanishing of~\eqref{chiral_inst_U1X_Glambda} is the only selection rule that applies in the type IIB setting. At first sight, this is somewhat puzzling because in the F-theory setting we obtained a seemingly independent selection rule by requiring the integral $\int_{{C}_{24}|_{D_M}} \iota^* G^\lambda_4$ of~\eqref{sel_matter_surface_Glambda} to vanish. The resolution of this puzzle was given in~\cite{Kerstan:2012cy}. Namely, the IIB orientifold limit of the $SU(5)\times U(1)_X$ model with smooth GUT divisor $\cW^b$ is well-defined only if certain cohomological relations hold, in which case the two seemingly different F-theory selection rules are actually equivalent. The first relation, which is necessary for the smoothness of the orientifold limit, is the absence of the so-called conifold point~\cite{Kerstan:2012cy, Krause:2012yh}
\be
\label{conifold_point}
\{w=0\} \cap \{a_1 = 0 \} \cap \{ a_{2,1} = 0\} = 0.
\ee
This can be directly translated to the cohomological relation
\be
2 \cW^b \wedge \bar{K}_B^2 = (\cW^b)^2 \wedge \bar{K}_B
\ee
on $B_3$, or equivalently~\cite{Krause:2012yh}
\be
(D_A^+)^2 - (D_A^-)^2 = 2 D^+_A D_{O7}
\ee
on $X_3$. Furthermore, in order for the GUT divisor $\cW^b$ to be smooth $D_A$ and $D_A'$ must have no intersections away from the orientifold plane, as these would lead to self-intersections of the discriminant locus and thus to singularities in $\cW^b$~\cite{Kerstan:2012cy}. Starting with 
\be
D_A \wedge D_A' = \frac14 D_A^+ \wedge D_A^+ + \frac14 D_A^- \wedge D_A^- - \frac12 D_A^+ \wedge D_A^-,
\ee
we note that the part of this intersection that lies inside $D_{O7}$ can only be contained in the first term because by definition $D_A^-$ does not intersect $D_{O7}$. In other words, absence of an intersection point away from the orientifold plane leads to
\be
 D_A^- \wedge D_A^- = 2 D_A^+ \wedge D_A^- - D_A^+ \wedge D_A^+|_{\rm off \,  O7}.
\ee
We can now take the wedge product of this equation with $D_E$ and integrate it over $X_3$. The first term on the right hand side vanishes due to the orientifold parity of the integrand. The second term on the right side must also vanish, because as argued above we can smoothly deform the geometry in such a way that $D_A^+\cap D_E$ lies inside the orientifold plane~\cite{Kerstan:2012cy}. Plugging these cohomological relations into the integral~\eqref{sel_matter_surface_Glambda} leads to~\cite{Kerstan:2012cy}
\be
\begin{aligned}
\label{integralmatch}
\hspace{-0.2cm} \int_{B_3} D_M^{\rm b} \wedge {\cal W}^{\rm b} \wedge (5 {\cal W}^{ b} - 6 \bar{\cal K}) & =  2 \int_{X_3} D_E \wedge D_A^+ \wedge D_{O7} + \frac{5 \lambda}{2} D_E \wedge (D_A^-)^2 \\ &=  2 \int_{X_3} D_E \wedge D_A^+ \wedge D_{O7}.
\end{aligned}
\ee
This is proportional to $\int_{C_{(4)}^X} \iota^* G_4^\lambda$, so that we indeed only obtain one selection rule both in F-theory and in type IIB.

In a general F-theory model it is of course not necessary to assume that a smooth type IIB limit exists. The discussion above shows that in this case the number of independent F-theory selection rules may be larger than we would naively assume using our IIB intuition. An intuitive microscopic explanation for this observation was offered in~\cite{Kerstan:2012cy}. It is based on the fact that the conifold point located at the triple intersection~\eqref{conifold_point}, if present, describes a point where the singularity of the elliptic fiber enhances to type $E_6$. It is natural to expect that $M2$-branes wrapped on the additional fiber $\bP^1$'s that appear over the $E_6$ point can lead to the appearance of additional charged zero modes. The additional selection rule should then correspond to the absence of these additional zero modes. However, to check this explicitly requires a better understanding of the microscopic nature of the charged $M5$ zero modes, and we will not pursue this point any further.

To close this section, let us emphasise that despite the nice picture obtained above the understanding of $M5$-instanton selection rules in F-theory is far from complete. Indeed, even cases in which we have a very clear understanding in the type IIB setting can be somewhat confusing in F-theory. The most obvious example of this is the case of a $U(1)$ $E3$-instanton, where it is not clear how the orientifold-odd part of $D_E$ is accounted for in the F-theory uplift. The situation is even worse when allowing for instanton flux, because $U(1)$ instantons admit fluxes along orientifold even 2-forms, which uplift to 2-forms in F-theory and can thus not be described by 3-form flux $\cH$ on the $M5$-instanton. Even for $O(1)$ instantons, the uplift of the selection rules is not clear in the presence of instanton flux~\cite{Kerstan:2012cy}. As discussed in section~\ref{sec:flux_on_D3inst}, in the type IIB setting it is clear that orientifold-odd flux on an $O(1)$ instanton can lift charged zero modes and in particular contributes to the chiral index. However, the discussion above does not immediately show how instanton flux can affect the selection rules for the absence of chiral zero modes in F-theory. A possible solution for this puzzle was suggested in~\cite{Kerstan:2012cy}. The key point is that the IIB instanton fluxes which contribute to the chiral index are those for which the pullback to the intersection between instanton and $D7$-brane is non-vanishing. As discussed around~\eqref{pullback_brane_instanton}, such fluxes correspond to 2-forms on $D_E$ which are not expected to uplift to harmonic 3-forms on $D_M$. This suggests that it may be possible to describe chirality-inducing fluxes on the $M5$-instanton by means of non-harmonic $\cH$~\cite{Kerstan:2012cy}. Formally including such non-closed fluxes would lead to an additional contribution to the $U(1)$ gauge transformation of the instanton action, due to the term $\cH \wedge \iota^* C_3$ in $S_\cB$. This suggests that non-harmonic $\cH$ would contribute to the selection rules. Despite this intriguing observation, much more work is of course required to clarify the role of possible non-harmonic 3-form flux on $M5$-instantons and to investigate whether alternative explanations of the selection rules can also be found.

\chapter{Conclusion}
From a phenomenological perspective, F-theory vacua form one of the most interesting corners of the string theory landscape. Conceptually, the F-theory construction can be viewed as an improved method for constructing type IIB vacua, which naturally incorporates the back-reaction of $D7$-branes and the associated strong coupling effects. A famous manifestation of these strong coupling effects is the appearance of exceptional gauge symmetry, which has crucial implications for the phenomenology of GUT models. Despite these differences, it is often helpful to keep the familiar type IIB orientifold picture involving intersecting branes in mind when working with F-theory vacua. The reason is that some features of the effective theory can have a very intuitive explanation in the type IIB picture which is far less clear in the more abstract and geometric F-theory language. Two such aspects, which are well-studied in the type IIB setting but are as yet relatively poorly understood in F-theory, concern the form of instanton corrections to the effective action and the impact of additional Abelian gauge symmetries which are not part of the Cartan subalgebra of a non-Abelian gauge group. The main aim of this thesis is to investigate how the known type IIB results relating to these topics are reproduced in F-theory and to elucidate the similarities and differences between the type IIB and F-theory descriptions.

The first part of this thesis focuses on the F-theory uplift of $U(1)$ symmetries which are associated with the diagonal generator of a $U(N)$ gauge group that appears on a stack of $D7$-branes in type IIB. Such $U(1)$s are ubiquitous in type IIB compactifications, and often gain a mass via the so-called St\"uckelberg mechanism. More precisely, it is possible to distinguish between two qualitatively different cases which are referred to as the geometric and the flux-induced St\"uckelberg mechanism. In the first case, the $U(1)$ becomes massive even in the absence of fluxes, while in the second case the $U(1)$ mass matrix is quadratic in the world-volume fluxes on the $D7$-branes. The description of $U(1)$ symmetries in type IIB, and in particular the St\"uckelberg mechanism and the derivation of the mass matrix are reviewed in section~\ref{sec:U(1)inIIB}. Although the exact mass of course depends on the geometric details of the compactification manifold and the $D7$-brane divisors, the St\"uckelberg mass scale is generically comparable to the Kaluza-Klein scale.
This means that a $U(1)$ which becomes massive via the St\"uckelberg mechanism is no longer visible as a gauge symmetry in the low energy effective action below the compactification scale. Nevertheless, the underlying $U(1)$ symmetry has important consequences for the low energy phenomenology in the shape of selection rules on the couplings of the various fields that are charged under the $U(1)$. 

From the general logic of the F-theory/type IIB duality it is natural to expect that the F-theory analogue of the flux-induced St\"uckelberg mechanism involves $G_4$ flux, which describes the F-theory uplift of $D7$-brane world-volume fluxes. On the other hand, geometrically massive $U(1)$s should not appear at all in the F-theory Kaluza-Klein reduction at the massless, harmonic level. The main guiding principle when looking to get a handle on the F-theoretic description of geometrically massive $U(1)$s is thus that the low energy selection rules that are known from the type IIB setting should be reproduced. In particular, the fact that the K\"ahler moduli and the moduli fields associated with the fields $C_2$, $B_2$ in the type IIB setting can be charged with respect to the massive $U(1)$s should be reproduced in F-theory. Based on these considerations, it was suggested in~\cite{Grimm:2011tb} that the effects of geometrically massive $U(1)$s and the associated gauge fluxes can be derived in the F-theory setting if certain non-harmonic forms are taken into account in the dimensional reduction. The details of the relevant construction are discussed in section~\ref{sec:U(1)inFtheory} of this thesis. A careful analysis of the dimensional reduction involving the abovementioned non-harmonic forms including the F-theory limit explicitly shows that indeed all the known type IIB effects can be reproduced exactly. In other words, including the non-harmonic forms gives a possibility to make the underlying physical reason for the low energy selection rules directly visible in the shape of a massive $U(1)$ gauge symmetry. 

In order to make use of this construction to derive the selection rules in concrete F-theory models it is of course necessary to be able to construct the relevant non-harmonic forms explicitly. As a first step in this direction, a discussion of how the appearance of geometrically massive $U(1)$s manifests itself in the geometry of the F-theory compactification manifold is given in section~\ref{sec:U(1)geometry}. To identify the massive $U(1)$s it is necessary to focus on the local geometry in the vicinity of the singular locus in the compactification manifold which forms the F-theoretic description of the relevant stack of $D7$-branes. This is in contrast with massless $U(1)$s, which are associated with global features of the compactification manifold in the shape of additional sections of the elliptic fibration. The picture that emerges is thus fully consistent with the type IIB intuition, where also the $U(1)$s of individual brane stacks are generically massive, while massless $U(1)$s are global in nature in the sense that they correspond to combinations of $U(1)$s from different stacks of branes. Nevertheless, it would be desirable to further investigate the construction of the non-harmonic forms required for the description of geometrically massive $U(1)$s in explicit models. In particular, it would be interesting to understand how the relevant forms are distinguished from the other non-harmonic forms in the massive Kaluza-Klein tower, and whether $G_4$ fluxes associated with geometrically massive $U(1)$s always admit an alternative description in terms of harmonic forms as suggested by the analysis of~\cite{Krause:2012yh}.

The second part of this thesis is concerned with $D3$-brane instantons in type IIB and their F-theory uplift in the guise of $M5$-instantons wrapped on vertical divisors of the elliptic fibration. In particular, we focus on the interplay between instantons and the massless and massive $U(1)$ symmetries discussed in the first part of the thesis. 
The crucial point is that instantons can generate couplings that break the $U(1)$ symmetries if there is a chiral spectrum of charged zero modes on the instanton world-volume. In the type IIB setting it is known that the existence of a chiral charged zero mode spectrum is equivalent to a non-vanishing charge of the instanton action under the $U(1)$ symmetries. The question of whether an instanton has a non-vanishing $U(1)$ charge is very important for model building purposes because it crucially affects the form of the effective instanton contribution. More precisely, charged instantons lead to couplings that involve matter fields charged under the $U(1)$ symmetries. This can be crucial in order to generate couplings that are forbidden by the $U(1)$ symmetries at the perturbative level. On the other hand, because charged matter fields must typically have a vanishing vacuum expectation value for phenomenological reasons, charged instantons in type IIB usually do not help with regards to moduli stabilisation.

The description of Euclidean $D3$-instantons in type IIB is reviewed in section~\ref{sec:fluxed_D3instantons}. Particular emphasis is placed on the derivation of the $U(1)$ charge of the instanton action and the selection rules for the absence of chiral charged zero modes. As first pointed out in~\cite{Grimm:2011dj}, world-volume instanton flux can affect the instanton charge with respect to geometrically massive $U(1)$s. In particular, it is possible that a certain configuration of instanton flux leaves the instanton uncharged even in cases where the unfluxed instanton wrapped on the same divisor would have a non-zero $U(1)$ charge. This means that instanton fluxes can play a crucial role with regards to moduli stabilisation because as mentioned above charged instantons do not participate in moduli stabilisation. This observation also reinforces the fact that it is crucial to take into account the full instanton partition function, and in particular the sum over instanton fluxes, when computing the instanton contribution to the low energy effective action. 

The F-theory uplift of type IIB $D3$-instantons is given by $M5$-instantons on vertical divisors. The importance of considering the full partition function rather than a single instanton contribution is particularly obvious in F-theory. This is because, as reviewed in section~\ref{sec:fluxedM5sInFtheory}, the instanton world-volume action has quite a complicated form whereas the instanton partition function can be computed using a rather simpler auxiliary action. The computation of the $M5$-instanton partition is performed in section~\ref{sec:fluxedM5sInFtheory}, based on the analysis of~\cite{Kerstan:2012cy}. The standard method for this computation, due to Witten, actually leads to a number of candidate partition functions from which the correct physical partition function must be chosen. To fully determine the moduli dependence of the $M5$-instanton action we perform the dimensional reduction of the F-theory effective action in the democratic formulation in appendix~\ref{sec:app_democ_Mtheory}. In particular we extend the discussion presented in~\cite{Kerstan:2012cy} in order to explicitly identify the imaginary part of the F-theory K\"ahler moduli which appear in the $M5$ action. This allows us to explicitly match the $M5$ partition function with the type IIB result for the case of $O(1)$ instantons. In fact, the type IIB match allows us to identify the physical partition function of the $M5$-instanton from amongst the set of candidates, which is non-trivial to achieve directly in F-theory. The correspondence between the partition functions furthermore suggests the existence of a type of F-term condition which restricts the 3-form flux that can be switched on on the $M5$-instanton. A first suggestion for the form of this F-term condition was given in~\cite{Kerstan:2012cy}, and we present additional evidence for this suggestion in section~\ref{sec:fluxedM5sInFtheory}. 

Finally, we consider the selection rules for the absence of chiral charged zero modes in the F-theory picture. The relevant selection rules can be deduced from the Freed-Witten anomaly cancellation condition, which requires $\iota^* G_4 = 0$ in the absence of charged chiral zero modes. The actual selection rules are obtained by checking for non-zero integrals of $G_4$ over various 4-cycles in the instanton world-volume, and the form of the 4-cycles which can yield non-trivial selection rules is discussed at the end of section~\ref{sec:fluxedM5sInFtheory}. In the presence of massless $U(1)$s, one class of selection rules can be associated with the vanishing of the instanton charge with respect to these $U(1)$s, matching the type IIB picture. However, the number of selection rules in F-theory is generically larger than the number of massless $U(1)$ symmetries. Based on a comparison of the F-theory and type IIB selection rules in an explicit example, we argue that the additional selection rules can actually be understood as arising due to geometrically massive $U(1)$ symmetries, just as in the type IIB picture. This gives another strong indication that geometrically massive $U(1)$s can continue to play an important role in F-theory despite not being directly visible in the low energy effective action.

Although the discussion of chapter~\ref{sec:chap_instantons} sheds light on many aspects of the relationship between $M5$-instantons in F-theory and Euclidean $D3$-instantons in F-theory, several important questions remain which require further study. One important example concerns IIB configurations which involve a distinct instanton-image-instanton pair, known as a $U(1)$ instanton. Such instantons can support world-volume flux with even orientifold parity. By considering the relationship between the cohomology groups on the world-volume of the $U(1)$ instanton and on the divisor associated with the corresponding $M5$-instanton, it is easy to deduce that orientifold-even flux is uplifted to a 2-form on the $M5$-instanton. It is as yet unclear how to interpret this in the F-theory picture, where instanton world-volume fluxes are described by 3-forms on the $M5$ divisor rather than 2-forms. Furthermore, additional work is required to fully understand the F-theoretic analogue of the IIB observation that instanton fluxes can influence the instanton $U(1)$ charge or equivalently the charged zero mode spectrum. Some hints are discussed at the end of section~\ref{sec:fluxedM5sInFtheory}, which suggest that chirality-inducing instanton flux may be described in F-theory with the help of non-harmonic 3-forms on the instanton divisor. Nevertheless, a more precise understanding of the F-theoretic origin of these non-harmonic forms and the reason for their appearance in the $M5$ partition function would of course be very desirable.


\appendix

\chapter{Conventions and review of mathematical techniques}
\chaptermark{Conventions and mathematical definitions}
\label{sec:app_maths}

In this appendix we present a brief overview of the most important mathematical definitions and theorems used in this thesis. The material is standard and can be found in most textbooks on differential geometry or algebraic topology, such as~\cite{Hartshorne:1977, Griffiths:1978, spivak1979comprehensive, tu1982differential, hatcher2002algebraic}. Furthermore, many excellent reviews such as~\cite{Denef:2008wq, Greene:1996cy, Bouchard:2007ik} exist which focus the applications of these subjects to physics and string theory in particular, and these reviews form the basis of the short presentation given in this appendix.

\section{Conventions for differential form calculus}
\label{sec:conventions}

In this section we recall some basic facts regarding the calculus of differential forms, as well as fixing our conventions. A differential form $\omega_p$ of degree $p$ on a manifold $M$ can be viewed as a totally antisymmetric multilinear functional\footnote{The base field will be $\bR$ or $\bC$ throughout this thesis.} on the $p$-fold tensor product of the tangent space of $M$. In local coordinates, there exists a canonical basis of the tangent bundle of $M$ denoted by $\left\{ \partial / \partial x^i \right\}$. The dual basis, whose elements are denoted by $d x^i$, is a canonical basis of the cotangent bundle or the space of differential 1-forms. Any higher rank differential form can be written as a superposition of tensor products of these basis elements. This expansion takes the form
\be
\omega_p = \sum_{i_1,\ldots,i_p} \omega_{i_1 \ldots i_p} dx^{i_1} \otimes \ldots \otimes dx^{i_p},
\label{comp_expansion_p_form}
\ee
where the coefficient functions $ \omega_{i_1 \ldots i_p}$ are totally antisymmetric. To simplify the formulae for the various supergravity actions involving differential forms, we choose conventions in this thesis in which the coordinates spacetime $x^i$ are dimensionless. Furthermore, the coefficient functions $\omega_{i_1 \ldots i_p}$ of all $p$-forms $\omega$, including the $p$-forms used in the Kaluza-Klein reductions, will be taken to be dimensionless. This in particular means that integrals of $p$-forms over $p$-cycles are dimensionless. 

Note that the antisymmetry of~\eqref{comp_expansion_p_form} implies that forms of degree $p$ with $p$ larger than the dimension of $M$ must vanish. The bundle $\Omega^*(M)$ of differential forms of various degrees is given the structure of an algebra by the exterior or wedge product. The product of a $p$-form $\omega$ and a $q$-form $\eta$ is a $(p+q)$-form defined in local coordinates by
\be
\begin{aligned}
(\omega\wedge\eta)_{i_1\ldots i_{p+q}} & = \frac{(p+q)!}{p!q!}\omega_{[i_1 \ldots i_p} \eta_{i_{p+1} \ldots i_{p+q}]} \\
&  \equiv \frac{1}{p!q!} \sum_{\sigma} \mathrm{sgn}(\sigma) \omega_{i_{\sigma(1)} \ldots i_{\sigma(p)}} \eta_{i_{\sigma(p+1)} \ldots i_{\sigma(p+q)}}.
\end{aligned}
\ee
The exterior derivative of a $p$-form $\omega$ is defined by
\be
(d\omega)_{i_1\ldots i_{p+1}} = (p+1) \partial_{[i_1} \omega_{i_2\ldots i_{p+1}]}.
\label{def_ext_deriv}
\ee
These definitions immediately lead to the basic identities
\begin{align}
d^2 \omega_p & = 0 , \qquad \omega_p \wedge \eta_q = (-1)^{pq} \eta_q \wedge \omega_p , \\ 
 d (\omega_p \wedge \eta_q) & = (d \omega_p) \wedge \eta_q + (-1)^p \omega_p \wedge (d\eta_q).\label{ext_deriv_and_wedge}
\end{align}
A form annihilated by the exterior derivative is called a closed form, while a form which can itself be written as an exterior derivative of another form is called exact.


A (smooth) map $f: M\rightarrow N$ between two manifolds defines a corresponding pullback map $f^*: \Omega^p(N) \rightarrow \Omega^p (M)$. The pullback of a form $\omega$ can be defined in local coordinates by
\be
(f^* \omega)_{i_1 \ldots i_p} = \frac{\partial f^{j_1}}{\partial x^{i_1}} \ldots \frac{\partial f^{j_p}}{\partial x^{i_p}} \omega_{j_1 \ldots j_p}.
\ee
The pullback commutes with both the wedge product and the exterior derivative in the sense that
\be
f^* (\omega \wedge \eta) = (f^* \omega) \wedge (f^* \eta) , \qquad f^*(d\omega) = d (f^* \omega).
\ee

On a complex manifold $M$, the cotangent bundle can be split into a direct sum of holomorphic and anti-holomorphic pieces. This decomposition immediately extends to forms of higher degree, and forms with $p$ holomorphic and $q$ anti-holomorphic indices are commonly referred to as $(p,q)$-forms. The space of $(p,q)$-forms will be denoted by $\Omega^{p,q}(M)$. The exterior differential can be decomposed as $d = \partial + \bar{\partial}$, where $\partial$ ($\bar{\partial}$) takes a $(p,q)$-form to a form of degree $(p+1,q)$ ($(p,q+1)$). The wedge product is compatible with this decomposition in the sense that the product of a $(p,q)$-form and a $(p',q')$-form is a form of degree $(p+p',q+q')$. In index notation, barred indices are often used for anti-holomorphic indices in order to emphasise the distinction between holomorphic and anti-holomorphic parts, so that a $(p,q)$-form may be written as
\be
\omega_{i_1\ldots i_p \overline{j}_1\ldots\overline{j}_q}.
\ee

The nilpotence of the exterior derivative implies that the bundles of differential $p$-forms form a cochain complex ($\Omega^*(M),d$). The associated cohomology is known as the de Rham cohomology (or Dolbeault cohomology in the complex case). The corresponding cohomology groups are defined as usual by
\be
H^p (M) = \frac{\text{Ker } d|_{\Omega^p}}{\text{Im } d|_{\Omega^{p-1}}}, 
\ee
or the complex analog
\be
 H^{p,q}(M) = \frac{\text{Ker } \bar{\partial}|_{\Omega^{p,q}}}{\text{Im } \bar{\partial}|_{\Omega^{p,q-1}}}.
\ee
Given a differential form $\omega$, we denote its cohomology class by $[\omega]$, although the brackets are often omitted if it is clear from the context that we are working in cohomology. The dimensions of the de Rham cohomology groups are known as the Betti numbers and conventionally denoted $b^p(M) = \text{dim}(H^p(M))$, while the complex analogue are the so-called Hodge numbers $h^{p,q} = \text{dim}(H^{p,q})$. The property~\eqref{ext_deriv_and_wedge} implies that the wedge product of differential forms lifts to a product on cohomology. We also denote this product by the wedge symbol, although the explicit wedge symbol is sometimes left out if it is clear from the context that we are considering products of cohomology classes.

There is a natural pairing between differential $p$-forms $\omega_p$ and $p$-dimensional oriented submanifolds $\Gamma^p$ of $M$ defined by integration
\be
(\omega_p, \Gamma^p) \rightarrow \int_{\Gamma^p} \omega_p.
\label{integration_pairing}
\ee
This integration is well-defined\footnote{We assume for simplicity that $M$ and $\Gamma^p$ are compact, or that $\omega_p$ has compact support.} due to the antisymmetry of the differential form, which ensures that under a coordinate change the transformation of the form cancels that of the integration measure. In the context of integration of differential forms, the change of variables formula takes the form
\be
\int_{f(\Gamma)} \omega = s \int_{\Gamma} f^* \omega,
\ee
where $s=1$ if $f$ is orientation-preserving and $s=-1$ otherwise.
The spaces of $p$-dimensional oriented submanifolds together with the boundary operator form a chain complex whose homology groups $H_p(M)$ form the (singular) homology of $M$. As a consequence of Stokes' theorem
\be
\int_{\Gamma} d \omega = \int_{\partial \Gamma} \omega,
\ee
the pairing~\eqref{integration_pairing} lifts to a pairing on homology and cohomology. Poincar\'e duality implies that this is actually a duality pairing which induces an isomorphism $H^p(M) \simeq H_{d-p} (M), \ d = \text{dim}(M)$.

The previous definitions and results were purely topological in nature in the sense that they made no reference to a metric. However, given a metric $g$ on $M$ an additional isomorphism from the space of $p$-forms to the space of $(d-p)$-forms can be defined in the shape of the so-called Hodge star operator $\ast$. In order to define the Hodge star we will use the epsilon \emph{tensor} $\hat{\epsilon}$, which transforms non-trivially under coordinate changes and whose indices are raised and lowered using the metric $g$. We use the hat to distinguish it from the ordinary $\epsilon$-\emph{symbol}, which is defined by its antisymmetry and $\epsilon_{0\ldots (d-1)}=1$ and does not transform under coordinate redefinitions. We use the conventions of~\cite{Polchinski:1998, Blumenhagen:2006ci}, in which there holds
\begin{align}
\hat{\epsilon}^{\mu_1\ldots\mu_{d}} = \  & |\det g|^{-\frac12} \epsilon_{\mu_1\ldots\mu_d}, \\ 
\hat{\epsilon}_{\mu_1\ldots\mu_{d}} = \  & \mathrm{sgn}(g) |\det g|^{\frac12} \epsilon_{\mu_1\ldots\mu_d},
\end{align}
with $\mathrm{sgn}(g)$ denoting the signature of the metric.
The Hodge star operator $\ast$ now acts on a $p$-form $\eta$ by
\be
(\ast \eta)_{\mu_1\ldots\mu_{d-p}} = \frac{1}{p!}\left.\hat{\epsilon}_{\mu_1\ldots\mu_{d-p}}\right.^{\nu_1\ldots\nu_p} \eta_{\nu_1\ldots\nu_p}
\label{Hodge_star}
\ee
Defined in this way, it obeys
\begin{align}
\label{hodge_star_square}
**\omega_p & = \mathrm{sgn}(g) \ (-1)^{p(d-p)} \omega_p, \\
\eta_p \wedge \ast \omega_p &= \mathrm{sgn}(g) (-1)^{p(d-p)} \frac{1}{p!} |\det g|^{\frac12} \eta_{\mu_1\ldots\mu_p} \omega^{\mu_1\ldots\mu_p} d^d x,
\label{hodge_star_inner_prod}
\end{align}
where $d$ is the dimension of $M$ and $\mathrm{sgn}(g)$ is the signature of the metric.

The Hodge star defines an inner product on the space $\Omega^{p,q}(M)$ given by
\be
\left< \omega , \eta \right> = \int_M \omega \wedge \ast \overline{\eta}.
\ee
The codifferential $d^\dagger$ is defined as the formal adjoint of $d$ with respect to this inner product, so that it obeys
\be
 \left< \omega , d \eta \right> = \left<d^\dagger \omega , \eta \right>.
\ee
When acting on a $p$-form, $d^\dagger$ can be written in terms of the exterior differential $d$ and the Hodge star as
\be
d^\dagger = (-1)^{d p + d + 1} \text{sgn}(g) \ \ast d \ast.
\ee
From this it immediately follows that $(d^\dagger)^2 = 0$. In this sense the codifferential behaves just like the exterior differential, and one can also define the concepts of co-closed and co-exact forms in a completely analogous manner. The Laplace operator can be written in terms of the two differential operators as\footnote{In the complex case one can define similar operators out of the holomorphic and antiholomorphic partial derivatives $\partial$ and $\bar{\partial}$ and their adjoints. However, in the case of K\"ahler manifolds, which will be the primary case of interest for us, the resulting Laplacians differ from that built using the total exterior derivative only by an overall factor of 2.}  
\be
\Delta = d d^\dagger + d^\dagger d.
\ee
As in the case of flat space, forms annihilated by the Laplace operator are known as harmonic forms. It is immediately clear from the definition that a form which is both closed and co-closed is harmonic, and in fact the converse is also true. The theory of harmonic forms is closely tied to the theory of cohomology by the so-called Hodge decomposition, which states that every form $\omega$ can be written uniquely as a sum of a harmonic, an exact and a co-exact form
\be
\omega = \gamma + d \alpha + d^\dagger \beta, \qquad \Delta \gamma = 0.
\ee
If $\omega$ is closed, then $d^\dagger \beta$ is also closed and hence harmonic. This means that the co-exact contribution $d^\dagger \beta$ vanishes if $\omega$ is closed, as all harmonic contributions are already included in $\gamma$. It follows that $\gamma$ and $\omega$ lie in the same cohomology class. In other words, each cohomology class has a unique associated harmonic representative, giving an isomorphism from the space of harmonic $p$-forms to the cohomology group $H^p(M)$.

In general, the interplay between the Hodge star and the pullback or wedge product operations is non-trivial. However, let us conclude this appendix by mentioning two simpler special cases that are relevant at some points in this thesis. For one, if $\sigma: M\rightarrow M$ is an isometry, then the pullback along $\sigma$ commutes with the Hodge star up to a sign
\be
\ast (\sigma^* \omega) = s \sigma^* (\ast \omega).
\ee
Here $s=1$ if $\sigma$ is orientation-preserving and $s=-1$ if the orientation is reversed. Now assume that $M=M_1 \times M_2$ is a direct product of two manifolds $M_1$, $M_2$, and that the metric on $M$ is block-diagonal with the blocks corresponding to metrics on $M_1$ and $M_2$. In this case, letting $d_1 = \text{dim}(M_1)$ and given forms $\omega_i \in \Omega^{p_i}(M_i), \ i=1,2$, there holds
\be
\label{hodge_star_block_metric}
\ast_M (\omega_1 \wedge \omega_2) = (-1)^{p_2 (d_1 - p_1)} (\ast_{M_1} \omega_1) \wedge (\ast_{M_2} \omega_2).
\ee

\section{Calabi-Yau manifolds}
\label{sec:Calabi-Yau}

In this section we review some of the essential features of Calabi-Yau manifolds, which play a distinguished role in the theory of string theory compactifications. Calabi-Yau manifolds can be characterised in a number of equivalent ways. The most famous of these equivalent requirements is the existence of a certain form of metric, which was conjectured by Calabi and later proven by Yau~\cite{yau1977calabi, yau1978ricci}. 

A common theme of all the equivalent characterisations is that a Calabi-Yau manifold $M$ is a K\"ahler manifold. This means that $M$ is a complex manifold with a metric $g$ fulfilling certain additional conditions. The first condition is that the metric must be Hermitian, which means that the complex structure acting on the tangent space of $M$ leaves the inner product invariant. In complex coordinates, it can be shown that this is equivalent to all entries of $g$ with purely holomorphic or purely anti-holomorphic indices vanishing
\be
g_{i j } = g_{\ov{i} \ov{j}} = 0, \qquad g_{i \ov{j}} = g_{\ov{j} i } \neq 0.
\ee
Any complex manifold admits a Hermitian metric, so this requirement restricts the metric $g$ but not the underlying manifold $M$.
To a Hermitian metric one can associate a canonical $(1,1)$-form defined in local coordinates by
\be
J = i g_{i \ov{j}} dz^i \otimes d\bar{z}^{\ov{j}} - i g_{\ov{j} i } d\bar{z}^{\ov{j}} \otimes dz^i = i g_{i\ov{j}} dz^i \wedge d\bar{z}^{\ov{j}}.
\ee
If this form is closed, i.e. $d J =0$, then $J$ is known as a K\"ahler form and the pair $(M,g)$ is a K\"ahler manifold. 

On a K\"ahler manifold of $n$ complex dimensions the $n$-fold exterior product of $J$ is proportional to the volume form
\be
\text{vol}(M) = \frac{1}{n!} \int_M J^n.
\ee
This proves that the K\"ahler form represents a non-trivial class in cohomology. Furthermore, on a K\"ahler manifold the Laplace operators built from the different exterior differential operators $d, \partial, \bar{\partial}$ are all proportional, $\Delta_d = 2\Delta_\partial = 2 \Delta_{\bar{\partial}}$. This implies that the de Rham cohomology groups can be decomposed into the Dolbeault cohomology groups, i.e.
\be
H^p (M) = \bigoplus_{r+s=p} H^{r,s}(M).
\ee
The proportionality of the Laplace operators also shows that the complex conjugate of a harmonic form is again harmonic, so that the Hodge numbers obey $h^{p,q}=h^{q,p}$. The absence of mixed metric components on a K\"ahler manifold implies that Hodge duality~\eqref{Hodge_star} is compatible with the decomposition into holomorphic and anti-holomorphic indices, in the sense that a $(p,q)$-form is transformed by Hodge duality into a form with definite $(p',q')$ type. More precisely, if $\omega$ is a $(p,q)$-form then $\ast \omega$ is an $(n-q,n-p)$-form\footnote{Some authors absorb an extra complex conjugation into the definition of the Hodge duality in the complex case, which would lead to $\ast \omega \in \Omega^{n-p,n-q}(M)$.}.

The K\"ahler condition $dJ=0$ can be shown to imply that the only non-vanishing Christoffel symbols are those with all indices holomorphic or all indices anti-holomorphic. This in turn means that the holomorphic and anti-holomorphic parts of the tangent space of $M$ are not mixed under parallel transport along a closed curve using the Levi-Civita connection. In other words, the holonomy group of a K\"ahler manifold is restricted. The holonomy group of $M$ is the group of transformations of tangent vectors under parallel transport along closed curves. For a generic manifold of real dimension $d=2n$ the holonomy group is $SO(d)$. The fact that holomorphic and anti-holomorphic components are not mixed under parallel transport implies that for a K\"ahler manifold the holonomy group is restricted to a $U(n)$ subgroup of $SO(2n)$.

A Calabi-Yau manifold can be defined as a compact K\"ahler manifold of complex dimension $n$ whose holonomy group is restricted even further to $SU(n)\subset U(n)$. From a physical perspective, this definition of Calabi-Yau manifolds appears quite naturally. This is because unbroken $\cN=1$ supersymmetry after compactification requires the existence of a covariantly constant spinor, and $SU(n)$ is the largest\footnote{In principle a holonomy group that is contained within $SU(n)$ rather than filling out $SU(n)$ completely suffices for this condition, and often the definition of Calabi-Yau manifolds is slightly adjusted to allow holonomy groups contained in $SU(n)$. This definition would allow for several special cases which are not simply connected (such as products of tori) to be classed as Calabi-Yau, however we will generally take Calabi-Yau to mean that $M$ is simply connected and the holonomy is $SU(n)$.} holonomy group that admits the existence of such a spinor~\cite{Becker:2007zj}. There are a number of different equivalent characterisations of Calabi-Yau manifolds, which can be useful in different situations. The most famous is the existence of a Ricci-flat K\"ahler metric. A criterion that is much easier to check on a practical level is that the first Chern class of the (holomorphic) tangent bundle vanishes, or equivalently that the canonical bundle is trivial. Finally, a complex K\"ahler $n$-fold is Calabi-Yau if and only if it admits a nowhere-vanishing holomorphic $(n,0)$-form, which is conventionally denoted by $\Omega_n$. It is easy to check that the compactness of $M$ implies that $\Omega_n$ is unique up to multiplication by a constant.

The conditions of compactness and of having holonomy $SU(n)$ can be shown to imply that for a Calabi-Yau $n$-fold $M$ there holds~\cite{gross2003calabi}
\be
h^{p,0}(M) = h^{0,p}(M) = \genfrac{\{}{.}{0pt}{}{1, \qquad\quad p = 1, n,}{0, \qquad \ \ \, \text{else}. \qquad}
\ee
This fact, together with the symmetries between the various cohomology groups of K\"ahler manifolds induced by complex conjugation and Hodge duality, implies that only a few of the Hodge numbers of a Calabi-Yau manifold are independent. To illustrate the various symmetries, the Hodge numbers are usually arranged into a diamond. For the cases of Calabi-Yau 3- and 4-folds, which are the most important cases in the context of this thesis, the Hodge diamonds take the form~\cite{Becker:2007zj}
\be
\text{Calabi-Yau 3-fold:} \qquad \qquad \begin{matrix} & & & 1\ & & & \\ & & 0\ & & 0\ & & \\
 & 0\ & & h^{1,1} & & 0\ & \\ 1 \ & & h^{2,1} & & h^{2,1} & & 1\ \\ & 0\ & & h^{1,1} & & 0\ & \\ & & 0\ & & 0\ & & \\ & & & 1\ & & &  
\end{matrix}
\ee
\be
\text{Calabi-Yau 4-fold:} \qquad \qquad  \begin{matrix}
& & & & 1\ & & & & \\ & & & 0\ & & 0\ & & & \\ & & 0\ & & h^{1,1} & & 0\ & & \\
 & 0\ & & h^{2,1} & & h^{2,1} & & 0\ & \\ 1\ & & h^{3,1} & & h^{2,2} & & h^{3,1} & & 1\ \\ & 0\ & & h^{2,1} & & h^{2,1} & & 0\ &  \\ & & 0\ & & h^{1,1} & & 0\ & &  \\ & & & 0\ & & 0\ & & & \\ & & & & 1\ & & & & 
\end{matrix}
\ee
In the case of Calabi-Yau 4-folds there is an additional linear relation between the Hodge numbers which implies that~\cite{Becker:2007zj} $h^{2,2} = 2(22 + 2 h^{1,1} + 2 h^{3,1} - h^{2,1})$.

The independent Hodge numbers are important because they count the number of ways in which the manifold $M$ can be deformed without destroying the Calabi-Yau property. Each such deformation gives rise to a massless\footnote{Here we mean massless in the context of a compactification on a Calabi-Yau manifold without further structure. In a realistic compactification, further ingredients such as fluxes must be introduced to give masses to these would-be moduli fields.} field in the low energy action. This means that from a phenomenological perspective determining the Hodge numbers is one of the most important tasks when constructing a compactification manifold. To accomplish this, it is often helpful to consider the Euler characteristic defined by $\chi (M) = \sum_{p=0}^{2n} (-1)^p \  b_p(M)$. For Calabi-Yau 3- and 4-folds, the specific form of the Hodge diamond implies that the Euler characteristic can be written as~\cite{Becker:2007zj}
\be
\begin{aligned}
\chi  =  & \  2 (h^{1,1} - h^{2,1}) , & \qquad n = 3, \\ 
\chi = & \ 6 (8+h^{1,1} + h^{3,1} - h^{2,1}) , & \qquad n=4. 
\end{aligned}
\ee
The advantage of these formulae is that the Euler characteristic can sometimes be computed independently from the Hodge numbers, which reduces the number of Hodge numbers that must be determined directly. For example, for a closed and smooth manifold the Euler characteristic is equal to the Euler number, which is given by the integral of the top Chern class (also called the Euler class)~\cite{Bouchard:2007ik}
\be
\chi = \int_M c_n (M).
\ee

\section{Vector bundles and characteristic classes}
\label{sec:char_classes}
In this section we review some of the essential definitions and theorems relating to vector bundles (or more generally fiber bundles), which play an important role in the modern mathematical description of many physical concepts within string theory and beyond. We largely follow the presentations of references~\cite{hatcher2003vector, Bouchard:2007ik, Denef:2008wq}, to which we refer for proofs and further details. 

Roughly speaking, a fiber bundle is a space E which admits a continuous projection $\pi: E\rightarrow B$ to a base manifold $B$ and locally looks like a Cartesian product of $B$ and a fiber space $F$. More precisely, for each $p\in B$ there must exist a neighborhood $U$ and a homeomorphism $\phi_U:\pi^{-1}(U) \rightarrow U\times F$ which is compatible with the projection $\pi$. A vector bundle of rank $n$ is a special case of this definition where $F$ is an $n$-dimensional real or complex vector space (for $n=1$ one usually speaks of a line bundle). In addition, when restricted to $\pi^{-1}(p')$ with $p'$ an arbitrary point in $U$, the map $\phi_U$ must define a vector space isomorphism to $\{p'\}\times F$. In the complex case one may further require $\pi$ and $\phi_U$ to be holomorphic and biholomorphic, respectively, in which case one speaks of a holomorphic vector bundle. A covering of $B$ by a family $\{U_\alpha\}$ of open sets together with the associated maps $\phi_\alpha$ is called a local trivialisation of $E$. At the overlap of two patches $U_\alpha \cap U_\beta$, the composition $\phi_\alpha \circ \phi_\beta^{-1}$ defines a map $t_{\alpha\beta}$ from $U_\alpha \cap U_\beta$ to the group $GL(n)$ of isomorphisms of the fiber. These maps are known as the transition functions of the trivialisation, which obey the cocycle condition $t_{\alpha\beta} t_{\beta\gamma} = t_{\alpha\gamma}$.

A (global) section of a fiber bundle is a map $s: B\rightarrow E$ which inverts the projection in the sense that $\pi(s(p)) = p$ for all $p\in B$. The space of sections of $E$ is conventionally denoted by $\Gamma(E)$. Locally, any section $s$ can be viewed as a map from $U\subset B$ to the fiber $F$, but the value of $s$ is defined only up to multiplication by the transition functions. A vector bundle always admits a global section in the shape of the zero section or trivial section. In contrast, nowhere-vanishing global sections need not necessarily exist. In fact, if a vector bundle of rank $n$ has $n$ nowhere-vanishing sections $s_i$ such that for every $p\in B$ the $s_i(p)$ are linear independent in $F$, then the bundle is trivial and can be globally written as a Cartesian product $E=B\times F$. 

The operations of taking direct sums or tensor products of vector spaces can be extended fiberwise to vector bundles\footnote{Given a subbundle $E'$ of $E$, i.e. a bundle $E'$ and a map $f:E'\rightarrow E$ that allows the fibers of $E'$ to be identified with subspaces of the fibers of $E$, one can also construct the quotient bundle by applying the vector space quotient fiberwise. The rank of the quotient bundle $E/E'$ is given by rank($E)- \text{rank}(E')$.}. The direct sum of two vector bundles of ranks $n$ and $m$ yields a bundle of rank $n+m$, while the tensor product yields a bundle of rank $n\times m$. Clearly, the transition functions of a bundle obtained as a direct sum or tensor product of two vector bundles are simply the direct sums or tensor products of the transition functions of the original bundles. Given a line bundle $L$, we may construct the inverse line bundle $L^{-1}$ as the bundle whose transition functions are the inverses of the transition functions of $L$. Clearly, $L\otimes L^{-1}$ is the trivial line bundle, justifying the intuitive notation. Other powers $L^m$ of line bundles may be defined in a similar manner. Note that this construction does not generally work for vector bundles of higher rank, as the would-be transition functions do not necessarily commute, which can lead to a violation of the cocycle condition.

An interesting aspect of holomorphic line bundles on a manifold complex $M$ is that they can be identified with a certain class of codimension 1 surfaces in $M$ known as divisors. A divisor $D$ is a formal sum of the form $D = \sum_i n_i S_i$, where the $S_i$ are holomorphic irreducible hypersurfaces in $M$ and $n_i \in \bZ$. In 4-dimensional string theory compactifications, understanding the nature of divisors of the compactification manifold is important because supersymmetric $D7$-brane configurations are described by divisors. Each irreducible holomorphic surface $S_i$ can be represented locally in a patch $U_\alpha$ as the zero locus of a holomorphic function $f^i_\alpha$. The divisor is then represented  by the meromorphic function $f^D_\alpha = \prod_i (f^i_\alpha)^{n_i}$. The set of divisors (and by extension the set of homology classes of divisors) can be given the structure of a group, with $D+D'$ being the divisor whose defining functions are the products of the defining functions of $D$ and $D'$. In order for the surfaces $S_i$ to be well-defined, the ratios $f^i_\alpha/f^i_\beta$ defined on the overlap of two patches must not have any zeroes or poles. This means that the ratios $f^D_\alpha/f^D_\beta$ can be identified as the transition functions of a line bundle, which is conventionally denoted $\cO(D)$. Similarly, the locus of zeros and poles of a meromorphic section of $\cO(D)$ will define a divisor that is homologically equivalent to $D$.
In the case where $f^D$ can be written as a globally defined meromorphic function all transition functions are constant, so that $\cO(D)$ is the trivial line bundle. Crucially, it can also be shown that a divisor given by a globally defined function is trivial in homology. Hence the relation between divisors and line bundles described above is actually an isomorphism between the group of isomorphism classes of line bundles (endowed with the tensor product operation) and the group of homology classes of divisors.

In a similar manner to the direct sums and tensor products of vector bundles discussed above, one can define exterior powers of vector bundles. The $p$-th exterior power of a vector bundle $E$ is denoted by $\Lambda^p (E)$. In the case where $p = \text{rank}(E)$, $\Lambda^p (E)$ is a line bundle which is often called the determinant line bundle, because its transition functions are the determinants of the transition matrices of the original bundle $E$. Of particular significance is the case where $E = T^* (M)$ is the cotangent bundle of the underlying manifold $M$. In this case, $\Lambda^p (T^* (M)) \equiv \Lambda^p (M)$ is the bundle whose sections are differential $p$-forms on $M$. In the complex case, one may similarly construct the complex vector bundle $\Lambda^{p,q} (M)$ from exterior powers of the holomorphic and anti-holomorphic cotangent bundles. Note that this is a holomorphic vector bundle only if $q=0$. In the case where $p$ is the (complex) dimension of $M$, $\Lambda^{p,0}(M)$ defines a holomorphic line bundle which is known as the canonical bundle. This makes it clear that, as mentioned in appendix~\ref{sec:Calabi-Yau}, the existence of a nowhere-vanishing holomorphic $(p,0)$-form is equivalent to the triviality of the canonical bundle. 

Given a general vector bundle $E$ over $M$, one may define $E$-valued differential forms on $M$ as sections of the product bundle $E\otimes \Lambda^r(M)$. In complete analogy to the usual definition of de Rham cohomology one may then define $E$-valued cohomology groups $H^p(M,E)$.
Any (smooth) vector bundle $E$ over $M$ may be endowed with a connection, which can formally be viewed as a map $\nabla: \Gamma(E) \rightarrow \Gamma(E\otimes T^* (M))$ satisfying the product rule. Locally, a connection can be written as $\nabla = d + A\wedge$. Here $A$ is a 1-form with values in the endomorphism ring $End(E)$, i.e. locally the space of $r\times r$ matrices with $r = \text{rank}(E)$. Often $A$ itself is referred to as the connection of the vector bundle. As usual, such a connection can be used to construct a curvature form $F = dA + A\wedge A$, which is a 2-form with values in $End(E)$. One use of such connections is to define a notion of parallel transport in the vector bundle. However, of more interest to us is the fact that connections may be used to define certain characteristic classes which give measures of how severely the vector bundle is twisted, i.e. how much it deviates from the trivial bundle.

The most important characteristic class for the purpose of this thesis is the Chern class of a complex vector bundle\footnote{The real analogues are given by the so-called Stiefel-Whitney and Pontryagin classes, which, however, play a much smaller role in this thesis.}. Given a complex vector bundle $E$ of rank $r$ endowed with a connection $A$ with curvature $F$, the (total) Chern class is defined by
\be
c(E) = \det \left( 1+\frac{1}{2\pi} F \right).
\label{def_total_Chern_class}
\ee
The determinant is of course evaluated with respect to the matrix indices of the matrix-valued form $F$. Clearly, $c(E) = c_1(E) + c_2(E) + \ldots + c_r(E)$ can be written as a sum of forms of varying degrees. $c_i(E)$ is a form of degree $2i$, whose cohomology class is called the $i$-th Chern class. It is obvious that the representative forms $c_i(E)$ depend on the choice of connection through the curvature $F$. However, the actual Chern cohomology classes are independent of $F$ and are thus well-defined for a complex vector bundle without having to specify a connection. In the specific case where $E$ is the tangent bundle of the base manifold $M$ one often speaks simply of the Chern classes of $M$, denoted by $c_i(M)$. The top Chern class of a holomorphic vector bundle is also called the Euler class of the bundle.

The total Chern class is compatible with the operations of taking direct sums or quotients of vector bundles in the sense that
\be
c(E \oplus E') = c(E) \wedge c(E') , \qquad c(E) = c(E/E') \wedge c(E').
\ee
These formulae are known as the Whitney sum formulae. 
The Chern class can also be expressed in terms of the so-called Chern roots, which are the eigenvalues $\lambda_i , \ i=1,...,r$ of the $r\times r$ curvature form $\frac{1}{2\pi} F$. In terms of these eigenvalues, the determinant in~\eqref{def_total_Chern_class} becomes
\be
c(E) = \prod_{i=1}^r \left( 1+\lambda_i \right),
\ee
while the Euler class is simply the product of the Chern roots, $c_r(E) = \prod_{i=1}^r \lambda_i$. The Chern roots are particularly useful to define other characteristic classes and to illustrate their relationship with the Chern class. In the following we give the definitions of the most important characteristic classes used in this thesis.

The Chern character $ch(E)$ is given by the trace of the exponentiated curvature, and can be related to the Chern classes by
\be
\text{ch}(E) = \text{Tr}(e^{\frac{1}{2\pi} F}) = \sum_i e^{\lambda_i} = r + c_1 + \frac12 (c_1^2 - 2 c_2)+\frac16 (c_1^3 - 3 c_1 c_2 - 3 c_3)+ \ldots.
\ee
When considering direct sums or products of vector bundles, the Chern characters obey
\be
\text{ch}(E\oplus E') = \text{ch}(E) + \text{ch}(E') , \qquad \text{ch}(E\otimes E') = \text{ch}(E) \text{ch}(E').
\ee
The Todd class of a vector bundle can be defined by 
\be
\text{Td}(E) = \prod_i \frac{\lambda_i}{1-e^{-\lambda_i}} = 1+\frac12 c_1 + \frac{1}{12}(c_1^2+c_2) +\frac{1}{24}c_1c_2 + \ldots.
\ee
Like the Chern character, the Todd class obeys $\text{Td}(E\otimes E') = \text{Td}(E) \text{Td}(E')$.
The Todd class and Chern characters of a holomorphic vector bundle $E$ over $M$ can be used to calculate the holomorphic Euler characteristic, which is defined by
\be
\chi(M,E) = \sum_i (-1)^i h^i(M,E).
\ee
The famous Hirzebruch-Riemann-Roch theorem states that
\be
\chi(M,E) = \int_M \text{ch}(E) \text{Td}(T(M)).
\ee
Finally, two further characteristic classes that appear in the world-volume actions of D-branes are the A-roof genus and the Hirzebruch L-genus, defined by
\begin{align}
\hat{A}(E) = &  \prod_i \frac{\lambda_i}{2 \sinh(\lambda_i/2)} = 1 - \frac{1}{24}(c_1^2 - 2 c_2) + \ldots, \\
L(E) = & \prod_i \frac{\lambda_i}{\tanh{\lambda_i}} = 1 + \frac{1}{3}(c_1^2 - 2 c_2) + \ldots.
\end{align}
Note that the A-roof genus and the Hirzebruch L-genus can be defined in exactly the same manner in terms of the eigenvalues of the curvature forms for real bundles, although in the real case of course the expansion in terms of Chern classes should be ignored.

\section{Toric construction of Calabi-Yau spaces}
\label{sec:projective_varieties}
Many interesting qualitative predictions of string theory compactifications can be deduced from the general properties of Calabi-Yau manifolds. However, deriving quantitative predictions e.g. on the existence and magnitudes of couplings in the low energy theory often makes it necessary to explicitly construct the compactification manifold in order to be able to compute suitable integrals or intersection numbers. While no general recipe for the construction of Calabi-Yau manifolds is known, a class of Calabi-Yau spaces can be realised as algebraic subvarieties in so-called toric varieties. The toric framework allows for the calculation of quantities like intersection numbers in a straightforward manner and even lends itself to computerised calculations, allowing for computer-powered scans over large numbers of compactification manifolds. For these reasons, the overwhelming number of explicit Calabi-Yau manifolds in the string theory literature and all concrete examples used in this thesis are of a toric form. In this section, we provide a short introduction to the basic concepts and calculational tools of toric geometry, following~\cite{Denef:2008wq, Bouchard:2007ik}.

Part of the power of toric geometry comes from the fact that there are several different equivalent ways to describe a toric variety, which all have individual strengths depending on which type of question one is trying to answer. We will largely follow the approach of~\cite{Bouchard:2007ik, Cox:1993fz}, in which toric varieties are viewed as a generalisation of the concept of weighted projective spaces. In this approach, a toric variety is obtained from $\bC^n$ by quotienting out an action of an algebraic torus $(\bC^*)^p$. In order to obtain a well-defined quotient, it is necessary to subtract the set $Z\subset \bC^m$ which is left invariant by a continuous subgroup of $(\bC^*)^p$. A toric variety\footnote{Sometimes we will be sloppy and refer to toric spaces as manifolds, although strictly speaking they may include singularities and should thus formally be categorised as varieties.} $M$ of (complex) dimension $n=m-p$ can thus be written as
\be
M = (\bC^m - Z)/(\bC^*)^p.
\label{def_toric_variety}
\ee
This definition clearly includes all weighted projective spaces as well as similar objects generated by quotienting out by more than one rescaling action simultaneously. As suggested by the name, a toric variety is not necessarily smooth, but a singular toric variety always has a resolution which is itself a toric variety. Toric varieties inherit a complex structure and a K\"ahler form from the underlying space $\bC^m$, and a homogeneous holomorphic function of the homogeneous coordinates will (locally) define a holomorphic function on $M$.

Divisors on toric varieties can be defined as the zeroes and poles of homogeneous rational functions of the homogeneous coordinates. The group of divisors\footnote{See appendix~\ref{sec:char_classes} for the definition of the group of divisors and in particular the addition operation between divisors.} of a toric variety is spanned by the so-called toric divisors $D_i = \{ x_i = 0 \}$, where the $x_i$ are coordinates of $\bC^m$. Divisors defined by a globally defined function, i.e. a function which is invariant under the $(\bC^*)^p$ action, are homologically trivial. Two divisors differing by such a trivial divisor are said to be linearly equivalent. Thus e.g. on the projective space $\bP^n$ there is only one independent divisor class. Using the fact that intersection numbers are invariant under linear equivalence, the intersection between $n=\text{dim}(M)$ general divisors can often be reduced to a combination of simple intersections of $n$ different toric divisors $D_i$. These simple intersections can often be directly computed using the very simple definitions of the $D_i$ and the $(\bC^*)^p$ symmetry action, as demonstrated e.g. in~\cite{Denef:2008wq}. Intersections of $n'\leq n$ divisors may also be considered, which result in holomorphic submanifolds of $M$ of complex dimension $n - n'$. Choosing $n'=n-1$, we obtain curves or 2-cycles, and in fact intersections of divisors generate the entire cone of curves with holomorphic representatives, which is known as the Mori cone. The Mori cone in turn can be used to determine the K\"ahler cone of $M$. The K\"ahler cone is defined as the space in which the parameters of $M$ may be varied while keeping the volumes of all holomorphic curves, as evaluated by integrating the K\"ahler form over the curve, positive. As one varies the parameters defining $M$ and one reaches the boundary of the K\"ahler cone, one or more of the holomorphic curves collapses to zero size. However, under variation of the parameters the intersection properties of the divisors $D_i$ and hence the Mori cone may also change. It may then be possible that one still has a well-defined geometry with positive curve volumes after passing through the boundary of the original K\"ahler cone, taking into account that the Mori cone on both sides of the boundary need not be the same. Such a transition is known as a flop transition of the geometry.

The Chern class of a toric variety has a very simple expression in terms of the Poincar\'e dual cohomology classes to the toric divisor classes $D_i$, which we denote by the same symbol
\be
c(M) = \prod_i (1+D_i).
\ee
This shows that the first Chern class is given by the sum over all toric divisor classes $c_1(M) = \sum_i D_i$. In other words, a section of the canonical bundle is a homogeneous polynomial whose weights with respect to the toric rescalings matches those of the monomial $\prod_i x_i$. As discussed in~\cite{Bouchard:2007ik}, this sum is non-zero if $M$ is compact. In order to obtain Calabi-Yau spaces as required for string theory model building, it is therefore necessary to consider submanifolds of toric varieties. In this thesis we will focus on submanifolds $X$ which can be defined as the simultaneous vanishing locus of a set of homogeneous polynomials $P_i$ (or in other words, an intersection of a set of divisors) in a toric variety $M$
\be
X = P_1 \cap \ldots \cap P_s , \qquad P_j \subset M ,\ j=1,\ldots,s. 
\ee

The Chern class of such a submanifold can be computed using the Whitney sum formula given in appendix~\ref{sec:char_classes}, as the tangent bundle of $X$ can be written as the quotient of the tangent bundle of the ambient space $M$ by the normal bundle of $X$ in $M$. Using the fact that the Chern class of the normal bundle of a divisor is simply the divisor class, $c(N(P)) = 1 + P$, one obtains
\be
c(X) = \frac{c(M)}{\prod_{j=1}^s (1+P_j)} = 1 + \sum_{i=1}^m D_i - \sum_{j=1}^s P_s + \ldots.
\ee
This formula translates the Calabi-Yau condition for $X$ into a simple condition on the homogeneous degrees of the defining polynomials $P_j$. In many cases, it is also possible to derive similarly relatively simple expressions in terms of the toric divisors for other characteristic classes. These expressions may then be used with the Hirzebruch-Riemann-Roch index theorem to compute holomorphic Euler characteristics and obtain relations between different Hodge numbers of $X$. As the dimension of the complex structure moduli space of $X$ can often be computed by counting parameters appearing in the polynomials $P_j$, these index theorems can then be used to determine the cohomology of $X$. For further details, we refer the reader to~\cite{Denef:2008wq}.

While the construction of toric varieties presented above is very concrete, a different combinatorial approach is more suited for many calculational purposes. In this language, an $n$-dimensional toric variety $M$ can be described using a set of vectors in an $n$-dimensional lattice $\bZ^n$. The set of vectors defines a polytope in the lattice, while sets of different numbers of these vectors span cones of various dimensions. The definition a toric variety is completed by specifying a fan, which is a set of cones satisfying several conditions as described in~\cite{Bouchard:2007ik} (for example, the intersection of two cones in the fan must again be an element of the fan). When relating this description to the previous construction of toric varieties, each of the vectors corresponds to a homogeneous coordinate of the underlying space $\bC^m$. The linear relations between the vectors encode the action of the $(C^*)^p$ group by which $\bC^m$ is quotiented, while the fan specifies the excluded set $Z$ in~\eqref{def_toric_variety}.

An advantage of this formulation is that hypersurfaces in toric varieties can also be described by means of polyhedra in the lattice which is dual to the lattice used in the definition of the ambient variety $M$~\cite{Batyrev:1994hm}. More precisely, the polyhedron defining $M$ is required to be reflexive, which means that all vertices of the dual polyhedron must lie on lattice points of the dual lattice. The vertices of the dual polyhedron can then be interpreted as defining certain monomials built from the homogeneous coordinates of $M$. As discussed in more detail in~\cite{Bouchard:2007ik}, the hypersurface defined by the vanishing of the generic polynomial built from these monomials is automatically Calabi-Yau. The main advantage of this construction is that many properties of the resulting Calabi-Yau space $X$ can be read off directly from the properties of the defining polytopes. For example, it is possible to determine in this way whether the space is smooth, whether it describes an elliptic fibration, and what the Hodge numbers of $X$ are. Furthermore, the algorithmic combinatorial nature of the construction of lattice polyhedra is ideally suited to use of computers, and scans and classifications of large numbers of toric Calabi-Yau manifolds have been performed in this manner~\cite{Kreuzer:1998vb, Kreuzer:2000xy}.





\chapter{Dimensional reduction of the democratic M-theory action}
\chaptermark{Democratic M-theory reduction}
\label{sec:app_democ_Mtheory}

The dimensional reduction of M-theory on an elliptically fibered fourfold including the F-theory limit was discussed in detail in section~\ref{sec:U(1)inFtheory}. In that analysis we used the standard 11-dimensional supergravity action~\eqref{S11normal}, whose form we recall for the convenience of the reader
\be
\label{S11normal_app}
S_{11} = \frac{1}{2\kappa_{11}^2} \int d^{11} x \sqrt{-g} R - \frac{1}{2\kappa_{11}^2} \int \left( \frac12 G_4 \wedge \ast G_4 + \frac{1}{6} C_3 \wedge G_4\wedge G_4 \right) .
\ee
Just like in the type IIB case, it is advantageous to use a slightly different formulation including a dual 6-form potential $C_6$ in addition to $C_3$ as soon as $M5$-branes are included in the theory. The magnetic coupling of $C_3$ to $M5$-branes can then be described by an electric coupling involving $C_6$, which is described by a simple Chern-Simons type action as in~\eqref{S_M5_class}.

Recall from section~\ref{sec:U(1)inFtheory} that the Kaluza-Klein reduction of $C_3$ leads to a set of 3-dimensional vector fields $A^\alpha$. These vectors and their real scalar superpartners $\xi^\alpha$ originating from the dimensional reduction of the K\"ahler form were dualised into the complexified K\"ahler moduli $t_\alpha$ in subsection~\ref{sec:FtheoryLimit}. However, the Legendre transformation~\eqref{def_re_t_alpha} used in this dualisation process directly specifies only the real part of $t_\alpha$. General arguments based on supersymmetry can be used to show that the classical action of an $M5$-brane instanton should be a holomorphic linear function of the $t_\alpha$~\cite{Witten:1996bn}. The fact that the $M5$-brane action directly involves a scalar from the Kaluza-Klein reduction of $C_6$ makes it clear that the imaginary part of $t_\alpha$ should be related to this scalar. As we will see, performing the dimensional reduction of M-theory directly in the democratic formulation involving $C_6$ and $C_3$ allows us to determine the precise relationship and show how exactly the $t_\alpha$ enter the $M5$-brane action. In particular, we are able to confirm the relationship~\eqref{rel_t_T} between the $t_\alpha$ and the type IIB K\"ahler moduli, which is used in subsection~\ref{sec:partition_fct_match} to precisely match the moduli dependence of an $M5$-instanton on a vertical divisor to that of an $E3$-instanton in type IIB.

The authors of~\cite{Bandos:1997gd} showed that in order to be able to write down a covariant action involving both $C_6$ and $C_3$ it is necessary to introduce auxiliary fields. As in the other parts of this thesis, we will follow a different path and consider instead a pseudo-action without auxiliary fields, which is supplemented by suitable duality constraints at the level of the equations of motion. The basic idea behind this construction, initially used in the type II setting in~\cite{Fukuma:1999jt, Bergshoeff:2001pv}, is to replace the equations of motion of $C_3$ obtained from the original action by the Bianchi identity of a dual field strength. The equation of motion of $C_3$ from~\eqref{S11normal_app} is
\be
\label{eom_c3}
d \ast G_4 + \frac12 G_4\wedge G_4 = 0. 
\ee
This equation is solved by introducing a dual field strength $G_7$ related to $G_4$ by\footnote{In 11-dimensional space with Lorentzian signature this is equivalent to $G_7 = - \ast G_4$, see~\eqref{hodge_star_square}.} $\ast G_7 = G_4$. Inserting this into~\eqref{eom_c3} leads to
\be
G_7 = d C_6 + \frac12 C_3 \wedge G_4.
\ee
With these definitions, it is straightforward to check that the original Bianchi identities and equations of motion are reproduced by the action
\be
\label{S11democ}
S_D = \frac{1}{2\kappa_{11}^2} \int d^{11} x \sqrt{-g} R - \frac{1}{6\kappa_{11}^2} \int \left( G_4 \wedge \ast G_4 + \frac12 G_7 \wedge\ast G_7 \right).
\ee
The relative factor between the Einstein-Hilbert term and the term involving $G_4$ and $G_7$ can be fixed by requiring that the original action~\eqref{S11normal_app} is reproduced if we introduce a Lagrangian multiplier $-\int dC_6 \wedge dC_3$ to the part of the action involving $G_4$ and $G_7$ and integrate out $C_6$~\cite{Kerstan:2012cy}. The dimensional reduction of this democratic 11-dimensional supergravity on an elliptically fibered Calabi-Yau fourfold was first considered in~\cite{Kerstan:2012cy} and will be reviewed in the following.

In this appendix we focus on the dimensional reduction at the massless level, and will only comment briefly at the end of the section on the modifications that arise when non-closed forms such as those used in section~\ref{sec:massiveU(1)sFtheory} are included. Our notation follows the notation used in chapter~\ref{sec:u(1)inIIBandFtheory} and references~\cite{Kerstan:2012cy, Grimm:2011tb}.
In particular, we will use the bases $\{\omega_\Lambda\}$ of $H^{1,1}(\hat{Y}_4)$ and $\{\alpha_a, \beta^b\}$ of $H^{3}(\hat{Y}_4)$ which were introduced in section~\ref{sec:KKreduxFtheory}. Throughout this appendix we assume that any singularities that may have been present on $Y_4$ have been resolved and work with the resolved space $\hat{Y}_4$. In particular, the set $\{\omega_\Lambda\}$ may include forms $\omega_{IA}$ dual to resolution divisors corresponding either to non-Abelian gauge groups or additional massless Abelian gauge bosons. Beyond this, the set of (1,1)-forms includes the forms $\omega_\alpha$, which arise by pullback of $H^{1,1}(B_3)$, and the form $\omega_0$ dual to the zero section of the fibration. In the following, it will sometimes be convenient to group $\omega_0$ and $\omega_{IA}$ together into a set labelled by a common index for which we use the letters $r, s, ...$. We again assume that $h^{2,1}(B_3)=0$ to simplify the intersection numbers of the forms in $H^{3}(\hat{Y}_4)$. For the convenience of the reader, let us recall the relevant intersection numbers from subsection~\ref{sec:massiveU(1)sFtheory}
\begin{align}
\label{intersection_numbers_fourfold}
\int \omega_\Lambda\wedge\alpha_a\wedge\beta^b & = \ \genfrac{\{}{.}{0pt}{0}{\frac12 \cK_{\alpha a c}\delta^{cb}, \qquad \quad\qquad \Lambda = \alpha,}{0, \qquad\qquad\qquad\quad \quad \Lambda = r,} \\
\int \omega_\Lambda \wedge\alpha_a\wedge\alpha_b & = \int \omega_\Lambda \wedge \beta^a \wedge\beta^b  =  0, \\
\int \omega_\Lambda \wedge\omega_\Sigma \wedge \omega_\Pi \wedge \omega_\Theta & =  \frac12 \cK_{\Lambda\Sigma\Pi\Theta}, \quad\qquad \text{with } \cK_{0\alpha\beta\gamma} \equiv \cK_{\alpha\beta\gamma}.
\end{align}
For the Kaluza-Klein reduction of $C_6$, we will also require the dual bases $\{\tilde{\omega}^\Lambda \}$ of $H^{3,3}(\hat{Y}_4)$ and $\{ \tilde{\alpha}^a , \tilde{\beta}_b \}$ of $H^5(\hat{Y}_4)$, which are normalised by
\begin{align}
\label{dual_bases}
\int \omega_\Lambda \wedge \tilde{\omega}^\Sigma & = \frac12 \delta_\Lambda^\Sigma, \\
\int \alpha_a \wedge \tilde{\alpha}^b & = \int \beta^b \wedge \tilde{\beta}_a = \frac12 \delta_a^b , \\
 \int \alpha_a \wedge \tilde{\beta}_b & = \int \beta^a \wedge \tilde{\alpha}^b = 0 .
\end{align}
Finally, we follow~\cite{Kerstan:2012cy} in introducing a basis $\{ \eta^m \}$ of $H^4 (\hat{Y}_4)$ which satisfies\footnote{To summarise, the ranges of the different types of indices used in this appendix are
\be
\begin{aligned}
\alpha, \beta & = 1,...,h^{1,1}(B_3), & \quad r,s&=1,...,h^{1,1}(\hat{Y}_4)- h^{1,1}(B_3), &\quad \Lambda, \Sigma &= 1,...,h^{1,1}(\hat{Y}_4) , \\
a, b & = 1,..., h^{2,1}(\hat Y_4), & \quad i,j & = 1,..., b^{3}(\hat Y_4), &\quad 
n,m & = 1,...,b^{4}(\hat Y_4).
\end{aligned}
\ee}
\be
\int \eta^m \wedge \eta^n = \delta^{mn}.
\ee

Due to the term $C_3\wedge G_4$ appearing in $G_7$ we will also need to consider wedge products of these basis forms. As the product is of course well-defined in cohomology, we can expand the product of, say, a $p$- and a $q$-form into the basis of $H^{p+q}$ at the level of cohomology classes. For example, the intersection numbers above yield
\be
[\omega_\Lambda] \wedge[\omega_\Sigma ]\wedge [\omega_\Pi ]  =   \cK_{\Lambda\Sigma\Pi\Theta} [\tilde{\omega}^\Theta]
\ee
In the following we would like to use relations such as this at the level of differential forms, and not just at the level of cohomology classes. However, due to the fact that products of harmonic forms are not necessarily harmonic, this means that we may have to absorb an exact piece into the $\tilde{\omega}^\Theta$, and similarly into the bases of $H^4(\hat Y_4)$ and $H^5(\hat Y_4)$. With this caveat in mind, we have~\cite{Kerstan:2012cy}
\be
\begin{aligned}
\omega_\Lambda \wedge\omega_\Sigma \wedge \omega_\Pi  = &\  \cK_{\Lambda\Sigma\Pi\Theta} \tilde{\omega}^\Theta, \qquad & \alpha_a\wedge\beta^b & = \cK_{\alpha a c} \delta^{cb} \tilde{\omega}^\alpha, \\
\alpha_a\wedge \omega_\alpha  = &\ - \cK_{\alpha a c } \delta^{cb}\tilde{\beta}_b, \qquad & \beta^b \wedge \omega_\alpha & =  \cK_{\alpha a c} \delta^{cb} \tilde{\alpha}^a, \\
\alpha_a \wedge\alpha_b  = 0 = &\ \beta^a\wedge\beta^b, \qquad & \alpha_a\wedge\omega_r & = 0 =  \beta^a\wedge\omega_r,  \\
\omega_\Lambda\wedge\omega_\Sigma  \equiv &\ R_{m \Lambda\Sigma } \eta^m & \Rightarrow \omega_\Lambda \wedge \eta^m & =  2 \delta^{mn} R_{n \Lambda \Sigma} \tilde{\omega}^{\Sigma}.
\label{productsofforms}
\end{aligned}
\ee
Note that the first equation in the last line can be viewed as the definition of the constants $R_{m\Lambda\Sigma}$.

In the course of the Kaluza-Klein reduction, we expand $C_3$ and $C_6$ into the basis of forms introduced above. Following the notation of~\cite{Kerstan:2012cy}, we have
\begin{align}
\label{expansions_C3_C6}
C_3 &= A^\Lambda \wedge \omega_\Lambda + c^a \alpha_a + b_b \beta^b, \\
C_6 &= \tilde{c}_\Lambda \tilde{\omega}^\Lambda + \tilde{U}_a \wedge \tilde{\alpha}^a + \tilde{V}^b \wedge\tilde{\beta}_b + r_m \wedge \eta^m.
\end{align}
Note that the fields $r_m$ are spacetime 2-forms and are therefore non-dynamical in 3 dimensions. However, as we will see below it is necessary to include them in the dimensional reduction in order to consistently account for non-trivial $G_4$ fluxes, which appear in the Kaluza-Klein reduction via~\cite{Kerstan:2012cy}
\be
G_4 = d C_3 + \cF_m \eta^m. 
\ee

After inserting these expansions into the action~\eqref{S11democ} and integrating over $\hat{Y}_4$ one obtains the 3-dimensional low energy effective action. The kinetic terms of the 3-dimensional fields involve the matrices
\be
N_{\Lambda\Sigma} := \int \omega_\Lambda \wedge\ast \omega_\Sigma, \qquad
M_{ij} := \int \rho_i \wedge \ast \rho_j, \qquad  
P^{nm} := \int \eta^n \wedge\ast \eta^m.
\label{kinetic_matrices1}
\ee
Here we have grouped the 3-forms $\alpha_a, \beta^a$ into the common set $\{ \rho_i \}$ as in~\cite{Kerstan:2012cy}. Similarly, we write $\{ \tilde{\rho}^i \}$ for the overall set of 5-forms $\{ \tilde{\alpha}^a, \tilde{\beta}_b \}$.
Denoting the inverses of $N$ and $M$ by matrices with raised indices, the relations~\eqref{dual_bases} can be used to show~\cite{Kerstan:2012cy}
\be
\int \tilde{\omega}^\Lambda \wedge\ast \tilde{\omega}^\Sigma = \frac{1}{4} N^{\Lambda\Sigma}, \qquad \int \tilde{\rho}^i \wedge \ast \tilde{\rho}^j = \frac{1}{4} M^{ij}.
\ee
Using these abbreviations, it was shown in~\cite{Kerstan:2012cy} that the part of the action~\eqref{S11democ} depending on $C_3$ and $C_6$ can be rewritten as\footnote{As before, we work in conventions where all fields are dimensionless, so that $\kappa_3^2 = \kappa_{11}^2 = \frac{1}{4\pi}$.}
\be
\begin{aligned}
\label{S_redux_1}
\hspace{-0.4cm} S_C = -\frac{1}{6\kappa_{3}^2}& \int \Bigl[ \ N_{\Lambda\Sigma} F^\Lambda\wedge\ast F^\Sigma + M_{ij} dc^i\wedge\ast dc^j + \frac{1}{8} N^{\Lambda\Sigma} \cD \hat{c}_\Lambda \wedge \ast \cD \hat{c}_\Sigma \\
& + \frac{1}{8}M^{ij} \nabla U_i \wedge\ast \nabla U_j + \frac12 P^{mn} \cD r_m \wedge \ast \cD r_n + P^{mn} \cF_m \cF_n \ast 1 \Bigr].
\end{aligned}
\ee
Here the scalars $c^a$, $b_a$ in~\eqref{expansions_C3_C6} are collectively denoted by $c^i$, while $\hat{c}_\Lambda$ and $U_i$ in~\eqref{S_redux_1} are shifted in comparison with the corresponding fields in~\eqref{expansions_C3_C6} by
\be
\begin{aligned}
\label{def_tilde_c}
\hat{c}_\Lambda =&\ \genfrac{\{}{.}{0pt}{0} {\tilde{c}_{\Lambda} ,}{ \tilde{c}_\alpha - \frac12 \cK_{\alpha a c}\delta^{cb} c^a b_b,} & \qquad& \genfrac{.}{.}{0pt}{0} {\Lambda = r,}{\Lambda = \alpha,} \\ 
U_i  = &\ \genfrac{\{}{.}{0pt}{0} {\tilde{U}_a - \frac12 \cK_{\alpha a c}\delta^{cb}b_b A^\alpha ,}{ \tilde{V}^b + \frac12 \cK_{\alpha a c}\delta^{cb} c^a A^\alpha ,} & \qquad & \genfrac{.}{.}{0pt}{0} {U_i \cong \tilde{U}_a, }{U_i \cong \tilde{V}^b.}
\end{aligned}
\ee
Finally, we have introduced the covariant derivatives~\cite{Kerstan:2012cy}
\be
\begin{aligned}
\cD \hat{c}_{\Lambda}  = &\ d\hat{c}_\Lambda - 2 \Theta_{\Lambda\Sigma}A^\Sigma + \genfrac{\{}{.}{0pt}{0} {0  ,}{ \cK_{\alpha a c}\delta^{cb} b_b dc^a  ,} &\qquad& \genfrac{.}{.}{0pt}{0} {\Lambda = r,}{\Lambda = \alpha,}  \\
\nabla U_i  = &\ d U_i +  \genfrac{\{}{.}{0pt}{0} {\cK_{\alpha a c}\delta^{cb}b_b F^\alpha ,}{ - \cK_{\alpha a c}\delta^{cb} c^a F^\alpha ,} &\qquad& \genfrac{.}{.}{0pt}{0} {U_i \cong \tilde{U}_a, }{U_i \cong \tilde{V}^b,} \\
\cD r_m = &\ d r_m + \frac12 R_{m \Lambda\Sigma} A^\Lambda \wedge F^\Sigma. &&
\end{aligned}
\ee
Just as in section~\ref{sec:FtheoryLimit}, one thus finds a flux-induced gauging of the scalars controlled by the 'embedding tensor'
\be
\label{def_Theta}
\Theta_{\Lambda\Sigma} = -\frac12 \cF_m \delta^{mn} R_{n \Lambda\Sigma} = -\frac12 \int \omega_\Lambda \wedge \omega_\Sigma \wedge G_4.
\ee
Note that in the democratic formulation this flux-induced gauging appears directly in the dimensional reduction, whereas in section~\ref{sec:FtheoryLimit} the gauging appeared only after performing the Legendre transformation~\eqref{legendreTrf}.

Of course, the action~\eqref{S_redux_1} still contains redundancies as it includes degrees of freedom from both $C_3$ and $C_6$. These redundancies can be eliminated using the duality relation $\ast G_7 = G_4$. More precisely, we follow~\cite{Kerstan:2012cy} and translate the 11-dimensional duality relation into corresponding 3-dimensional relations using equation~\eqref{hodge_star_block_metric}. This leads to
\begin{align}
\label{duality_1}
2 M_{ij} \ast dc^j & = - \nabla U_i \quad& \Leftrightarrow &&\quad  \ast \nabla U_i &=\  2 M_{ij}d c^j, \\
\label{duality_2}
2N_{\Lambda\Sigma} \ast F^\Sigma &= - \cD \hat{c}_\Lambda \quad& \Leftrightarrow &&\quad \ast \cD \hat{c}_\Lambda &=\ 2N_{\Lambda\Sigma}  F^\Sigma, \\
\label{duality_3}
- \delta_{mn} P^{np} \cF_p \ast 1 & = \cD r_m \quad& \Leftrightarrow && \quad\cF_m &=\ \delta_{mn} P^{np} \ast \cD r_p.
\end{align}
At this point it is necessary to slightly shift the definition of the covariant derivative of $\hat{c}_\Lambda$ to
\be
\cD \hat{c}_{\Lambda}  :=  d\hat{c}_\Lambda - 4 \Theta_{\Lambda\Sigma}A^\Sigma + \genfrac{\{}{.}{0pt}{0} {0  ,}{ \cK_{\alpha a c}\delta^{cb} b_b dc^a ,} \qquad \genfrac{.}{.}{0pt}{0} {\Lambda = r,}{\Lambda = \alpha,}
\label{shifted_cov_deriv}
\ee
in order to achieve consistency of the duality relations with the equations of motion that one obtains from the action~\eqref{S_redux_1}~\cite{Kerstan:2012cy}. The fact that such a manual shift is required for consistency is typical of the pseudo-action formalism for democratic supergravity theories, and has been discussed in the type II setting in~\cite{Dall'Agata:2001zh, Jockers:2004yj, Kerstan:2011dy, Grimm:2011dx}.

After shifting $\cD \hat{c}_\Lambda$ as above, the duality relations~\eqref{duality_1} to~\eqref{duality_3} can be implemented by inserting suitable Lagrangian multiplier terms into the action and integrating out the redundant fields. Let us first consider the process of eliminating the non-dynamical 2-forms $r_m$ in favour of the flux quanta $\cF_m$.
As in~\cite{Kerstan:2012cy}, we add a Lagrangian multiplier $(6\kappa_3^2)^{-1} \int \delta^{mn} \cF_n d r_m $ to the action. Now varying with respect to $d r_m$ treated as an independent field leads to an equation of motion which is identical to the duality relation~\eqref{duality_3}. Therefore we can consistently eliminate $dr_m$ using this duality relation, leading to the action~\cite{Kerstan:2012cy}
\be
\begin{aligned}
\label{S_redux_2}
\hspace{-0.3cm} S_C = -\frac{1}{6\kappa_3^2}& \int_{\mathbb{R}^{1,2}} \Bigl[ \ N_{\Lambda\Sigma} F^\Lambda\wedge\ast F^\Sigma + M_{ij} dc^i\wedge\ast dc^j + \frac{1}{8} N^{\Lambda\Sigma} \cD \hat{c}_\Lambda \wedge \ast \cD \hat{c}_\Sigma \\
&\ + \frac{1}{8}M^{ij} \nabla U_i \wedge\ast \nabla U_j - \Theta_{\Lambda\Sigma}A^\Lambda \wedge F^\Sigma + \frac{3}{2}P^{mn} \cF_m \cF_n \ast 1 \Bigr]. 
\end{aligned}
\ee
The vectors $U_i$ can be eliminated in favour of the dual scalars $c^i$ in a completely analogous manner after introducing the Lagrange multiplier $(12\kappa_3^2)^{-1} \int d U_i \wedge dc^i$~\cite{Kerstan:2012cy}. 

In principle, we could use the same technique to eliminate either the vectors $A^\Lambda$ or the scalars $\hat{c}_\Lambda$. However, as discussed in subsection~\ref{sec:FtheoryLimit} it is advantageous for the uplift to 4 dimensions and the comparison with the type IIB results to keep the scalars $\hat{c}_\alpha$ and the vectors $A^r$, with $r=0, (IA)$. In other words, we would like to eliminate the scalars $\hat{c}_r$ and the vectors $A^\alpha$. A necessary condition for this to be possible is that the $A^\alpha$ do not appear in the gauging of the dual scalars $\hat{c}_\alpha$, or in other words~\cite{Kerstan:2012cy}
\be
\Theta_{\alpha\beta} = 0.
\label{vanishing_theta_alphabeta}
\ee
Recall from section~\ref{sec:MtheoryReduction} that this condition must anyway be fulfilled by the $G_4$ fluxes in order to allow for a consistent F-theory uplift. Therefore we assume in the following that~\eqref{vanishing_theta_alphabeta} is fulfilled. To eliminate the scalars $\hat{c}_r$ we follow~\cite{Kerstan:2012cy} and introduce the Lagrangian multiplier $(12\kappa_3^2)^{-1} \int  d \hat{c}_r \wedge F^r$. Varying with respect to $d\hat{c}_r$ leads to the duality relation
\be
F^r = \frac12 N^{r \Lambda} \ast \cD \hat{c}_\Lambda.
\label{duality_rel_cr}
\ee
Crucially, this relation suffices to determine $\cD \hat{c}_s$ completely in terms of $\cD \hat{c}_\alpha$ and $F^r$, without involving the vectors $A^\alpha$ which we also want to eliminate. To see this, note that the subblocks $N_{(1)} := (N_{\alpha\beta})$ and $N_{(2)} := (N_{rs})$ of $(N_{\Lambda\Sigma})$ are expected to be separately invertible and positive definite due to the differences in the geometric origin and index structure between the forms $\omega_\alpha$ and $\omega_r$. Denoting the respective inverses by $(N_{(1)}^{\alpha\beta})$ resp. $(N_{(2)}^{rs})$, we can construct the respective Schur complements
\be
S_{(1)rs} = N_{rs} - N_{r\alpha}N_{(1)}^{\alpha\beta} N_{\beta s}, \qquad S_{(2)\alpha\beta} = N_{\alpha\beta} - N_{\alpha r}N_{(2)}^{rs} N_{s \beta}.
\ee
By the matrix inversion lemma, these Schur complements are again positive definite and invertible, and form the inverses of the corresponding subblocks of the inverse matrix. In other words, we have
\be
S_{(1)rs} N^{sp} = \delta_r^p ,\qquad  S_{(2)\alpha\beta} N^{\beta\gamma} = \delta_\alpha^\gamma.
\ee
Multiplying~\eqref{duality_rel_cr} by $S_{(1)rs}$ leads to
\be
\cD \hat{c}_r = S_{(1)rs} N^{s\alpha} \cD \hat{c}_\alpha - 2 S_{(1)rs} \ast F^s,
\label{dual_rel_cr_explicit}
\ee
which as claimed does not involve $A^\alpha$ due to~\eqref{vanishing_theta_alphabeta}. As above, the duality relation involving the $A^\alpha$ can be obtained as the equation of motion of $F^\alpha$ after introducing the Lagrangian multiplier $-(6\kappa_3^2)^{-1} \int F^\alpha \wedge d \hat{c}_\alpha$~\cite{Kerstan:2012cy}. Using~\eqref{dual_rel_cr_explicit} and the matrix inversion lemma $N^{\alpha\beta} - N^{\alpha r}S_{(1)rs} N^{s\beta} = N_{(1)}^{\alpha\beta}$, this duality relation can be written as
\be
F^\alpha = \frac12 N_{(1)}^{\alpha\beta} \ast \cD \hat{c}_\beta + N^{\alpha r} S_{(1)rs}  F^s,
\ee
which does not involve $\hat{c}_r$. This guarantees that we may consistently eliminate both $\hat{c}_r$ and $A^\alpha$ from the action using the duality relations. 
The final result of the dualisation procedure is~\cite{Kerstan:2012cy}
\be
\begin{aligned}
\label{S_redux_3}
 S_C = -\frac{1}{4\kappa_3^2} \int_{\mathbb{R}^{1,2}}& \Bigl[ \ S_{(1)rs} F^r\wedge\ast F^s - N_{(1)}^{\alpha\beta} N_{\beta r} \cD \hat{c}_\alpha \wedge F^r  - 2 \Theta_{rs}A^r \wedge F^s  \\ 
 &\ + \frac{1}{4} N_{(1)}^{\alpha\beta} \cD \hat{c}_\alpha \wedge \ast \cD \hat{c}_\beta
 + M_{ij} dc^i\wedge\ast dc^j  + P^{mn} \cF_m \cF_n \ast 1 \Bigr]. 
\end{aligned}
\ee

So far we have neglected the Einstein-Hilbert term of the 11-dimensional action~\eqref{S11democ}. Clearly, the dimensional reduction of this term is not changed when passing from the usual formulation of the 11-dimensional supergravity to the democratic formulation. We may therefore use the results of reference~\cite{Haack:1999zv} for the reduction of the curvature scalar. As already mentioned in subsection~\ref{sec:KKreduxFtheory}, the fluctuations of the 11-dimensional metric give rise to the complex structure moduli $z^M$ and the $v^\Lambda$ which appear in the expansion of the K\"ahler form. While the $z^M$ are complex fields, the $v^\Lambda$ are real and combine with the $\hat{c}_\alpha$ and the $A^r$ into 3-dimensional chiral and vector multiplets, respectively~\cite{Kerstan:2012cy}. Combining the dimensionally reduced Einstein-Hilbert term from~\cite{Haack:1999zv} with~\eqref{S_redux_3} leads to~\cite{Kerstan:2012cy}
\be
\begin{aligned}
\label{S_redux_full}
 S_D = &\ \frac{1}{2\kappa_3^2} \int_{\mathbb{R}^{1,2}}d^3x \sqrt{-g} R  -\frac{1}{4\kappa_3^2} \int_{\mathbb{R}^{1,2}} \Bigl[ \cV S_{(1)rs} (d\xi^r\wedge\ast d\xi^s + F^r\wedge\ast F^s)  \\ &\ - N_{(1)}^{\alpha\beta} N_{\beta r} \cD \hat{c}_\alpha \wedge F^r 
 + \frac{1}{4\cV} N_{(1)}^{\alpha\beta} \cD \hat{c}_\alpha \wedge \ast \cD \hat{c}_\beta + \frac{1}{\cV} N_{(1)}^{\alpha\beta}d \cV_\alpha\wedge \ast d\cV_\beta  \\ &\
  + 4 G_{M\bar{N}}d z^M \wedge\ast d\bar{z}^{\bar{N}}  + \frac{1}{\cV} M_{ij} dc^i\wedge\ast dc^j 
 - 2 \Theta_{rs}A^r \wedge F^s \\ &\ + \frac{1}{\cV^3} P^{mn} \cF_m \cF_n \ast 1 \Bigr]. 
\end{aligned}
\ee
Here we have rescaled the 3-dimensional metric by $g_{\mu\nu}\rightarrow \cV^{-2} g_{\mu\nu}$, with $\cV$ the volume of $\hat Y_4$, in order to obtain a canonically normalised 3-dimensional curvature term. As in subsection~\ref{sec:MtheoryReduction} we have defined $\xi^r = v^r/\cV$, while
\be
\cV_\Lambda := \partial_{v^\Lambda} \cV = \frac{1}{3!} \int_{\hat Y_4} \omega_\Lambda \wedge J^3.
\ee
We refrain from giving the explicit expression for the kinetic metric $G_{M\bar{N}}$, as this would require introducing additional notation and the result can be found in~\cite{Haack:1999zv}.

The action~\eqref{S_redux_full} can be cast into the form of a general 3-dimensional $\cN=2$ with non-trivial gauging~\cite{Kerstan:2012cy}
\be
\begin{aligned}
\label{kinetic_lin_gen_app}
  S^{(3)}_{\cN=2} = \frac{1}{\kappa_3^2} \int \Big[ &\ \frac{1}{2}R_3 *1  
 - K_{I \bar J }\, \nabla M^I \wedge * \nabla \bar M^{J}
  + \frac{1}{4}  K_{A B}\, 
  d\xi^{A}\wedge * d\xi^{B} \\ 
  \phantom{\int \Big[} & \ + \frac{1}{4}  K_{A B}\, F^{A} \wedge * F^{B}
     + \,  F^{A} \wedge \I ( K_{A I} \, \nabla M^I) 
      \\ 
     &\ + \frac12 \Theta_{A B} A^{A} \wedge F^{B} - (V_\cT + V_{\mathrm{F}}) * 1 \Big]. 
\end{aligned}
\ee
Here, the indices $I,J$ label complex scalar fields while the pairs $(\xi^A, F^A)$ denote 3-dimensional vector multiplets. The covariant derivatives account for possible gaugings $\nabla M^I = dM^I + X^I_A A^A$. Comparison with~\eqref{S_redux_full} makes it clear that the vector multiplets in the dimensional reduction are given by $(\xi^r, A^r)$ up to a possible normalisation factor. The part of the action involving the complex structure moduli and the real scalars $\{c^i\}\equiv c^a, b_a$ is identical to the corresponding part in the action obtained from the dimensional reduction of the standard 11-dimensional supergravity~\eqref{S11normal_app}. The discussion of this reduction in~\cite{Haack:1999zv, Haack:2001jz, Grimm:2010ks} shows that the $z^M$ can directly be regarded as chiral fields. As in section~\ref{sec:MtheoryReduction}, the $c^a$ and $b_b$ combine into chiral fields $N^a = c^a -i f^{ab}b_b$, where $f^{ab}$ is a suitable function of the complex structure moduli. 
 
Equation~\eqref{S_redux_full} also strongly suggests that the remaining chiral fields involve $\hat{c}_\alpha$ and $\cV_\alpha$ through the combination
\be
t_\alpha = \cV_\alpha + \frac{i}{2} \hat{c}_\alpha + ...,
\label{kaehler_mod_democratic}
\ee
as was already noted in~\cite{Kerstan:2012cy}. In the following we extend the analysis of~\cite{Kerstan:2012cy} to explicitly confirm~\eqref{kaehler_mod_democratic} and also identify the remaining contributions. In order to do this, we require the 3-dimensional K\"ahler potential $K(z^M, N^a, \tau_\alpha, \xi^r)$. 
From the discussion in subsection~\ref{sec:FtheoryLimit} it is clear that the K\"ahler potential can be obtained from the kinetic potential~\eqref{kinPot} by the Legendre transformation~\eqref{legendreTrf}. 
Furthermore, the Legendre transform also allowed us to derive
\be
\R\ t_\alpha = \cV_\alpha + \frac{i}{4} d_\alpha
\label{re_t_alpha}
\ee
up to terms that vanish in the F-theory limit, with $d_\alpha$ given in~\eqref{def_d_alpha}. The advantage of the democratic formulation is that it is possible to compute not only the real part of $t_\alpha$, but also the imaginary part as we will illustrate in the following.

In order to compare the action~\eqref{S_redux_full} obtained from the dimensional reduction to the general form~\eqref{kinetic_lin_gen_app}, we require the derivatives of the K\"ahler potential $K$. The defining equations of the Legendre transformation relating $K$ to the kinetic potential $\tilde{K}$ can be used to express the derivatives of $K$ in terms of derivatives of $\tilde{K}$. Explicitly, we require the relations~\cite{Grimm:2011tb}
\be
\begin{aligned}\label{kaehler_pot_derivs}
K_{t_\alpha\bar{t}_\beta}& = -\frac14 \tilde K^{\xi^{\alpha}\xi^{\beta}}, \qquad K_{t_\alpha \bar{M}^{\bar{I}}} = \frac12 \tilde K^{\xi^{\alpha}\xi^{\beta}} \tilde{K}_{\xi^{\beta}\bar{M}^{\bar{I}}},\\
K_{M^I\bar{M}^{\bar{J}}} &= \tilde K_{M^I\bar{M}^{\bar{J}}} - \tilde{K}_{M^I\xi^{\alpha}} \tilde{K}^{\xi^{\alpha}\xi^{\beta}} \tilde{K}_{\xi^{\beta}\bar{M}^{\bar{J}}}.
\end{aligned}
\ee
Here $\tilde{K}^{\xi^{\Lambda}\xi^{\Sigma}}$ is the inverse of the submatrix $\tilde{K}_{\xi^{\Lambda}\xi^{\Sigma}}$, while $M^I$ can be any of the fields $\xi^r, N^a, z^M$ that are not dualised. Using the kinetic potential
\be
\label{kinPot_app}
\tilde{K}(z^M,N^a,\xi^\Lambda) = -3\log \cV + \frac{i}{4 }\xi^{\Lambda} (N^a - \bar{N}^a)(N^b- \bar{N}^{b})\int_{\hat Y_4}\omega_{\Lambda}\wedge \Psi_a \wedge\bar{\Psi}_{\bar{b}} + {K}_{CS} 
\ee
of section~\ref{sec:MtheoryReduction} and the expression
\be
\cV^{-3} =  \frac{1}{4!\cV^4}\int J^4 = \frac12 \frac{1}{4!} \cK_{\Lambda\Sigma\Pi\Theta} \xi^\Lambda\xi^\Sigma\xi^\Pi\xi^\Theta
\ee
for the volume it is straightforward to derive
\be
\tilde{K}_{\xi^{\Lambda}\xi^{\Sigma}} = \frac{\cV}{4} \cK_{\Lambda\Sigma\Pi\Theta} \xi^\Pi\xi^\Theta - \cV_{\Lambda}\cV_{\Sigma}.
\label{kaehler_derivs_xi}
\ee
We will also require the derivatives of $\tilde{K}$ with respect to $N^a$ and the mixed derivatives involving $\xi^r$ and $z^M$. Using the expression~\eqref{Psi_exp} to express $\Psi_a$ in terms of $\alpha_a$ and $\beta^b$ together with the intersection numbers~\eqref{intersection_numbers_fourfold}, it is simple to check that
\be
\begin{aligned}
\tilde{K}_{\xi^\Lambda N^a} & = \overline{\tilde{K}_{\xi^\Lambda \bar{N}^a}} = -\frac{i}{4} \delta_{\Lambda}^\alpha \cK_{\alpha a b}\delta^{bc}b_c, &\qquad \tilde{K}_{N^a\bar{N}^b} &= -\frac18 \xi^\alpha \cK_{\alpha a d} \delta^{dc}\R f_{cb}.
\label{kaehler_derivs_na}
\end{aligned}
\ee
When evaluating the partial derivatives of $\tilde{K}$ with respect to $z^M$, it is crucial that the other fields are held fixed. In particular, $\partial_{z^M}$ does not act on the function $f^{ab}$ appearing in the definition of $N^a$, however it does act on $f^{ab}$ when $f^{ab}$ appears as part of the forms $\Psi_a$. Keeping this in mind, one obtains 
\be
\tilde{K}_{\xi^r z^M} =0 \qquad\quad \tilde{K}_{\xi^\alpha z^M} = \frac{i}{4} \cK_{\alpha a b} b_c b_d \delta^{bc'} \R\, f^{ac}\R \, f_{c' d'}\partial_{z^M}(\I\, f^{d'd}).
\label{kaehler_derivs_z}
\ee
In the second term we used $\partial_{z^M}\R\, f_{ab} = - \R\, f_{ac} \partial_{z^M}(\R\, f^{cd}) \R\, f_{db}$, as well as the fact that $f^{ab}$ can be chosen to be holomorphic in the complex structure moduli~\cite{Grimm:2010ks}. Holomorphicity of $f^{ab}$ will be assumed here and in the following, and implies that we may replace $\partial_{z^M} \R \, f^{ab} =  i \partial_{z^M} \I \, f^{ab}$.

As shown in~\cite{Haack:1999zv}, the Hodge dual of a (1,1)-form on a Calabi-Yau fourfold can be expressed in terms of the K\"ahler form as
\be
\ast \omega_\Lambda =  \frac16 \frac{\cV_\Lambda}{\cV}J^3 - \frac12 \omega_\Lambda \wedge J^2  .
\ee
Plugging this into~\eqref{kinetic_matrices1} immediately leads to
\be
N_{\Lambda\Sigma} = -\frac{1}{\cV} \tilde{K}_{\xi^\Lambda\xi^\Sigma},  \text{ and in particular } N^{\alpha\beta}_{(1)} = -\cV \tilde{K}^{\xi^\alpha\xi^\beta}.
\label{rel_N_Ktilde}
\ee
Using~\eqref{kaehler_pot_derivs} we find
\be
K_{\xi^r\xi^s} = -\cV N_{rs} + \cV N_{r\alpha}N^{\alpha\beta}_{(1)} N_{\beta_s} = -\cV S_{(1)rs},
\ee
which after comparing~\eqref{S_redux_full} and~\eqref{kinetic_lin_gen_app} confirms that $(\xi^r, A^r)$ are the correctly normalised 3-dimensional vector multiplets. This means that we can identify $\I\ t_\alpha$ by comparing the term proportional to 
\be
\frac14 N_{(1)}^{\alpha\beta} N_{\beta r} \cD \hat{c}_\alpha \wedge F^r
\label{FwedgeImDc}
\ee 
in the action obtained by dimensional reduction to the term 
\be
F^{r} \wedge \I (\tilde K_{r I} \, \nabla M^I)
\label{FwedgeImNabla}
\ee 
in the general form of the action. Our strategy will be to compute the contributions of $N^a$ and $z^M$ to~\eqref{FwedgeImNabla}, to see which terms in~\eqref{FwedgeImDc} are already accounted for. The remaining terms can then be identified with contributions to $\I \, t_\alpha$.

Explicitly evaluating the term~\eqref{FwedgeImNabla} leads to
\be
\begin{aligned}
F^{r} &\wedge \I (\tilde K_{r I} \, \nabla M^I)  = \frac12 \tilde{K}^{\xi^\alpha\xi^\beta}\tilde{K}_{\xi^\beta \xi^r} F^r\wedge \nabla(\I \, t_\alpha) \\
&  +\frac{1}{4}\tilde{K}^{\xi^\alpha\xi^\beta}\tilde{K}_{\xi^\beta \xi^r} \cK_{\alpha a b}\delta^{bc}b_c F^r \wedge \left( d c^a + (\I\, f^{ad})b_d + b_d d(\I\, f^{ad})\right)\\
& -\frac{1}{8}\tilde{K}^{\xi^\alpha\xi^\beta}\tilde{K}_{\xi^\beta \xi^r} \cK_{\alpha a b} b_c b_d \delta^{bc'} \R\, f^{ac}\R \, f_{c' d'} F^r \wedge d(\I\, f^{d'd}).
\label{fwedgeNablaM}
\end{aligned}
\ee
Here we made use of equations~\eqref{kaehler_pot_derivs},~\eqref{kaehler_derivs_na} and~\eqref{kaehler_derivs_z}, and in particular the fact that $K_{\xi^r t_\alpha}$ is real while $K_{\xi^r N^a}$ is purely imaginary. Furthermore we used $\partial_{z^M} \I\, f^{ad}dz^M+\partial_{\bar{z}^M} \I\, f^{ad} d\bar{z}^M = d (\I\, f^{ad})$ in order to rewrite the total contribution of the complex structure moduli in terms of $d (\I\, f^{ad})$.
The result~\eqref{fwedgeNablaM} may be compared to
\be
\begin{aligned}
\frac14 N_{(1)}^{\alpha\beta} N_{\beta r} \cD \hat{c}_\alpha \wedge F^r & = \frac14 \tilde{K}^{\xi^\alpha\xi^\beta} \tilde{K}_{\xi^\beta \xi^r} \cD \hat{c}_\alpha \wedge F^r \\
& = \frac14 \tilde{K}^{\xi^\alpha\xi^\beta} \tilde{K}_{\xi^\beta \xi^r} F^r \wedge \left(\nabla \hat{c}_\alpha + \cK_{\alpha a b}\delta^{bc}b_c d\hat{c}_\alpha \right),
\end{aligned}
\ee
with $\nabla \hat{c}_\alpha = d\hat{c}_\alpha - 4\Theta_{\alpha r}A^r$. In this way we deduce
\be
\begin{aligned}
d\, \I & \, t_\alpha  = \frac12 d\hat{c}_\alpha - \frac14 \cK_{\alpha a b} \delta^{ac}d \left(\I\, f^{bd} b_c b_d\right)\\
&-\frac{1}{8}\tilde{K}^{\xi^\alpha\xi^\beta}\tilde{K}_{\xi^\beta \xi^r} \cK_{\alpha a b}b_c b_d \left(\delta^{bc}d(\I\, f^{ad})  - \delta^{bc'} \R\, f^{ac}\R \, f_{c' d'} d(\I\, f^{d'd}) \right)
\label{d_im_t_alpha}
\end{aligned}
\ee
In order for this to be consistent, the right hand side must of course be a total derivative. This is in general not clear for the term in the second line of~\eqref{d_im_t_alpha}. However, in the special case where $f^{ab}$ is proportional to $\delta^{ab}$ it is easy to see that the second line vanishes and so $d\, \I \, t_\alpha$ is a total derivative as required. In other words, we require 
\be
f^{ab} = f \delta^{ab}.
\label{relation_f_delta_app}
\ee
It is beyond the scope of this thesis to explicitly prove that this is consistent with the required holomorphy of the forms $\Psi_a$, however we will assume that this is the case and use~\eqref{relation_f_delta_app} in the following.
Recall that we had independently argued at the end of section~\ref{sec:FtheoryLimit} that we expect
\be
f^{ab} = -i\tau \delta^{ab}
\label{relation_f_tau_app}
\ee
to hold in the type IIB limit. In other words we expect $f = -i\tau$ up to terms that vanish in the F-theory limit of vanishing fiber volume. Note however that equation~\eqref{d_im_t_alpha} is valid without any reference to a type IIB limit, suggesting that~\eqref{relation_f_delta_app} is valid already in the full M-theory picture.

Combining~\eqref{d_im_t_alpha} with the real part of $t_\alpha$ in~\eqref{re_t_alpha} and explicitly evaluating $d_\alpha$ using the various intersection numbers leads to\footnote{This expression is valid up to corrections which vanish in the F-theory limit. Here and in the following we also use the obvious notation $b^a = \delta^{ab}b_b$.}
\be
t_\alpha = \cV_\alpha - \frac{1}{4} \cK_{\alpha a b}  f b^a b^b + i \frac12\hat{c}_\alpha.
\label{t_alpha_full}
\ee
Note that in particular the flux-induced gauging of the $\hat{c}_\alpha$ encoded in the covariant derivatives~\eqref{shifted_cov_deriv} precisely leads to the expected gauging of the $t_\alpha$ given in~\eqref{flux_induced_gauging_Mtheory}. 
Comparing the expression~\eqref{t_alpha_full} to the type IIB K\"ahler moduli $T_\alpha$ defined in~\eqref{def_Talpha} shows that it is exactly compatible with $t_\alpha \rightarrow \frac12 T_\alpha$ if $f\rightarrow -i\tau$ in the type IIB limit. In other words this yields an additional independent piece of evidence for~\eqref{relation_f_tau_app}.
Furthermore, we are able to read off the relationship between the scalars $\hat{c}_\alpha$ (or equivalently the $\tilde{c}_\alpha$ appearing directly in the expansion~\eqref{expansions_C3_C6} of $C_6$) and the type IIB axion $c_\alpha$
\be
\tilde{c}_\alpha = c_\alpha - \frac12 \cK_{\alpha a b} c^a b^b = \hat{c} + \frac12 \cK_{\alpha a b}c^a b^b.
\label{rel_c_tilde_c}
\ee
Therefore our explicit calculation of $\I \, t_\alpha$ has allowed us to confirm the relationship between $\tilde{c}_\alpha$ and $c_\alpha$ which was already guessed in~\cite{Kerstan:2012cy}. Let us emphasise that the shift in~\eqref{rel_c_tilde_c} is crucial in order to obtain a match between the moduli dependence of the partition functions of $M5$- and $E3$-instantons, as discussed in section~\ref{sec:partition_fct_match}.

Having obtained explicit expressions for all the chiral fields as well as for the K\"ahler potential, it is reasonably straightforward to verify that also the remaining terms in~\eqref{S_redux_full} can be cast into the expected form of~\eqref{kinetic_lin_gen_app}. In addition to the expressions~\eqref{kaehler_derivs_na} and~\eqref{kaehler_derivs_z} we will use
\be
\tilde{K}_{z^M \bar{z}^N} = (K_{CS})_{z^M\bar{z}^N}+\frac12 \cK_{\alpha a b}\xi^\alpha b^a b^b \frac{1}{\R\, f} \partial_{z^M}(\R\, f)\partial_{\bar{z}^N}(\R\, f).
\ee
Here and in the following we assume that $f^{ab} = f \delta^{ab}$ as in~\eqref{relation_f_delta_app} and furthermore take $f$ to be holomorphic, such that
\be
\partial_{z^M}\partial_{\bar{z}^N}(\R\, f) = 0 , \qquad \partial_{z^M}(\R\, f) = i \partial_{z^M}(\I\, f).
\ee
Using the relations~\eqref{kaehler_pot_derivs} we can now evaluate all the derivatives of the K\"ahler potential $K$. Inserting this into the the general expression for the kinetic term in~\eqref{kinetic_lin_gen_app} and making use of the cancellations between the different terms leads to
\be
\begin{aligned}
\hspace{-0.5cm} - K_{I \bar J }\, &\nabla M^I \wedge * \nabla \bar M^{J} = \frac14 \tilde{K}^{\xi^\alpha\xi^\beta}\left[ d\cV_\alpha \wedge\ast d\cV_\beta + \frac14 \nabla \hat{c}_\alpha \wedge\ast \nabla \bar{\hat{c}}_\beta \right] \\
& +\frac18 \tilde{K}^{\xi^\alpha\xi^\beta} \cK_{\alpha a b}b^b dc^a \wedge\ast \left[ \nabla \hat{c}_\beta +\frac12 \cK_{\alpha cd}b^c dc^d \right] \\ & - (K_{CS})_{z^M\bar{z}^N} dz^M\wedge\ast d\bar{z}^N \\
& +\frac18 \frac{\xi^\alpha}{\R\, f} \cK_{\alpha a b} \left[|f|^2 db^a\wedge\ast db^b+2\I\, f\, dc^a\wedge\ast db^b + dc^a\wedge\ast dc^b\right].
\end{aligned}
\ee
To see that this exactly matches the remaining terms in~\eqref{S_redux_full} we use~\eqref{rel_N_Ktilde} and the definition of $\cD \hat{c}_\alpha$ given in~\eqref{shifted_cov_deriv} as well as the fact that $G_{M\bar{N}} = (K_{CS})_{z^M\bar{z}^N}$~\cite{Grimm:2010ks, Haack:1999zv}. To account for the kinetic terms of $c^a$ and $b^a$, we explicitly evaluate the entries of the matrix $M_{ij}$ in~\eqref{kinetic_matrices1} using $\ast \Psi_a = i J\wedge \Psi_a$~\cite{Grimm:2010ks, Haack:1999zv}, which leads to
\be
\begin{aligned}
\int \alpha_a\wedge\ast\alpha_b = &\ -\frac12 \cV \xi^\alpha \cK_{\alpha a b} \frac{1}{\R\, f}, \\
\int \alpha_a\wedge\ast\beta^b = &\ -\frac12 \cV\xi^\alpha \cK_{\alpha a c}\delta^{bc} \frac{\I\, f}{\R\, f},\\
\int \beta^a\wedge\ast\beta^b = &\ -\frac12 \cV\xi^\alpha \cK_{\alpha  cd}\delta^{ad}\delta^{bc} \frac{|f|^2}{\R\, f}.
\end{aligned}
\ee
This completes our discussion of the dimensional reduction of the democratic 11-dimensional supergravity at the massless level. Having cast the 3-dimensional action into the general form~\eqref{kinetic_lin_gen_app} we could of course consider the F-theory uplift to 4 dimensions as in section~\ref{sec:FtheoryLimit}. However, we refrain from explicitly performing this uplift as the essential features were already discussed in~\cite{Grimm:2010ks} and in section~\ref{sec:FtheoryLimit}. Let us only note that while the M-theory reduction considered here allows for a 3-dimensional Chern-Simons term with $\Theta_{rs} \neq 0$, an F-theory uplift is only possible for $G_4$ fluxes which obey $\Theta_{rs} = 0$.

To close this appendix, let us follow~\cite{Kerstan:2012cy} and briefly consider what changes arise when the non-harmonic forms introduced in section~\ref{sec:massiveU(1)sFtheory} are included in the dimensional reduction. In other words, we extend the set $\omega_r$ to include the non-closed forms $\tw_{0A}$ with $A$ labelling the different stacks of $D7$-branes with unitary gauge group. Of course, to carry out the democratic reduction consistently we must then also include the 6-forms $\tilde{\tw}^{0A}$ which are dual to the $\tw_{0A}$ in the sense of~\eqref{dual_bases}. Recalling\footnote{Although we still use indices $a, b$ for the 3-forms $\alpha_a$, $\beta^b$, it is understood that the index range is enlarged compared to the earlier parts of this appendix, as some or all of the 3-forms $\alpha_a,\ \beta^a$ are necessarily non-harmonic due to~\eqref{dtw_0_app}.}
\be
\label{dtw_0_app}
d\tw_{0A} =  N_A C_A^a \alpha_a
\ee
from equation~\eqref{dtw_0}, it is obvious that the field strengths $dc^i$ in all relevant expressions above must be replaced by the covariant derivatives~\cite{Kerstan:2012cy}
\be
\label{geometric_gauging_app}
\nabla c^i = \genfrac{\{}{.}{0pt}{}{d c^a - N_{A}C^a_A A^{0A}, }{d b_b,}\qquad\quad  \genfrac{.}{.}{0pt}{}{c^i = c^a,}{c^i = b_b.}
\ee
Using the intersection numbers presented in section~\ref{sec:massiveU(1)sFtheory} we also find
\be
\omega_\Lambda \wedge d\beta^b = \genfrac{\{}{.}{0pt}{}{-\delta^{bc}\cK_{\alpha a c}N_A C^a_A \tilde{\omega}^\alpha , }{0,} \qquad \genfrac{.}{.}{0pt}{}{\Lambda = 0A,\ }{\text{otherwise.}}
\label{omega_wedge_dbeta}
\ee
The non-closed forms lead to additional terms in $G_7$, which using~\eqref{productsofforms} and~\eqref{omega_wedge_dbeta} can be written as
\be
G_7 \supset \frac12 b_a A^{0A}\wedge\tw_{0A}\wedge d\beta^a - \frac12 b_a \beta^a\wedge A^{0A} \wedge d\tw_{0A} = - \cK_{\alpha ab}N_A C^a_A b^b A^{0A}\tilde{\omega}^\alpha.
\label{shifted_g7}
\ee
This implies that the covariant derivatives of $\hat{c}_\alpha$ are also shifted and contain an extra term, namely\footnote{The second term in~\eqref{shifted_g7} was missed in~\cite{Kerstan:2012cy}, leading to a mismatch by a factor of 2 in~\cite{Kerstan:2012cy}.}~\cite{Kerstan:2012cy}
\be
\cD \hat{c}_{\alpha}  =  d\hat{c}_\alpha - 4\Theta_{\alpha r}A^r  + \cK_{\alpha a b} b^b d c^a  -  \cK_{\alpha a c}\delta^{cb} b_b N_A C^a_A A^{0A}.
\label{shifted_cov_deriv_hatcalpha}
\ee
As extensively discussed in section~\ref{sec:U(1)inFtheory}, the additional terms in the covariant derivatives imply that the shift symmetry of the axions $c^a$ is gauged even in the absence of fluxes. More precisely, we read off from~\eqref{geometric_gauging_app} and~\eqref{shifted_cov_deriv_hatcalpha} that the shift of $c^a$ under a gauge transformation $A^{0A}\rightarrow A^{0A} + d\Lambda^A$ is given by~\cite{Kerstan:2012cy}
\be
\delta_{\Lambda^A} c^a = N_A C_A^a \Lambda^A.
\ee
On the other hand, as the gauge transformations of the last two terms in~\eqref{shifted_cov_deriv_hatcalpha} cancel, the axions $\hat{c}_\alpha$ and hence the chiral fields $t_\alpha$ transform only in the presence of fluxes
\be
\delta_{\Lambda^A} \hat{c}_\alpha = 4\Theta_{\alpha(0A)} \Lambda^A.
\ee
Note that in light of~\eqref{rel_c_tilde_c} both the type IIB axions $c_\alpha$ and the fields $\tilde{c}_\alpha$ appearing directly in the reduction of $C_6$ do experience a non-trivial shift under a gauge transformation of $A^{0A}$. The fact that the overall chiral field $T_\alpha$ is gauge invariant despite the fact that $c_\alpha$ and $c^a$ are charged is well known in the type IIB setting, see equation~\eqref{gauging2}. The discussion above shows that also this detail is accounted for correctly in the F-theory reduction involving non-harmonic forms to describe massive diagonal $U(1)$s.



\bibliography{biblio}
\bibliographystyle{utphys}  

\end{document}